\documentclass[longauth]{aa}
\usepackage[varg]{txfonts}
\usepackage{bbding}
\usepackage{amssymb}
\usepackage{xcolor}
\usepackage{graphicx}

\usepackage{natbib,twoopt}
\usepackage[breaklinks=true]{hyperref} 
\bibpunct{(}{)}{;}{a}{}{,} 
\makeatletter
\newcommandtwoopt{\citeads}[3][][]{\href{http://adsabs.harvard.edu/abs/#3}%
{\def\hyper@linkstart##1##2{}%
\mod\hyper@linkend\@empty\citealp[#1][#2]{#3}}}
\newcommandtwoopt{\citepads}[3][][]{\href{http://adsabs.harvard.edu/abs/#3}%
{\def\hyper@linkstart##1##2{}%
\mod\hyper@linkend\@empty\citep[#1][#2]{#3}}}
\newcommandtwoopt{\citetads}[3][][]{\href{http://adsabs.harvard.edu/abs/#3}%
{\def\hyper@linkstart##1##2{}%
\mod\hyper@linkend\@empty\citet[#1][#2]{#3}}}
\newcommandtwoopt{\citeyearads}[3][][]{\href{http://adsabs.harvard.edu/abs/#3}%
{\def\hyper@linkstart##1##2{}%
\mod\hyper@linkend\@empty\citeyear[#1][#2]{#3}}}
\makeatother

\defcitealias{Falstad2021}{Paper~I}

\begin{document}

\title{CON-quest}

\subtitle{
  II. Spatially and spectrally resolved HCN/HCO$^+$ line ratios 
  in local luminous and ultraluminous infrared galaxies
}

\author{
  Y.~Nishimura \inst{\ref{DoA_UT},\ref{IoA_UT},\ref{NAOJ}}
  \and S.~Aalto \inst{\ref{Chalmers}} 
  \and M.~D.~Gorski \inst{\ref{Chalmers}}
  \and S.~K\"{o}nig \inst{\ref{Chalmers}}
  \and K.~Onishi \inst{\ref{Chalmers}}
  \and C.~Wethers \inst{\ref{Chalmers}}
  \and C.~Yang \inst{\ref{Chalmers}}
  \and L.~Barcos-Mu\~{n}oz \inst{\ref{NRAO}}
  \and F.~Combes \inst{\ref{LERMA}}
  \and T.~D\'{i}az-Santos \inst{\ref{FORTH},\ref{Cyprus}}
  \and J.~S.~Gallagher \inst{\ref{WM},\ref{Macalester}}
  \and S.~Garc\'{i}a-Burillo \inst{\ref{OAN}}
  \and E.~Gonz\'{a}lez-Alfonso \inst{\ref{Alcala}}
  \and T.~R.~Greve \inst{\ref{DTU},\ref{DAWN}}
  \and N.~Harada \inst{\ref{NAOJ},\ref{SOKENDAI}}
  \and C.~Henkel \inst{\ref{MPIfR},\ref{Abdulaziz}}
  \and M.~Imanishi \inst{\ref{NAOJ}}
  \and K.~Kohno \inst{\ref{IoA_UT}}
  \and S.~T.~Linden \inst{\ref{Steward}}
  \and J.~G.~Mangum \inst{\ref{NRAO}}
  \and S.~Mart\'{i}n \inst{\ref{ESO},\ref{JAO}}
  \and S.~Muller \inst{\ref{Chalmers}}
  \and G.~C.~Privon \inst{\ref{NRAO},\ref{Florida},\ref{Virginia}}
  \and C.~Ricci \inst{\ref{DiegoPortales},\ref{KIAA}}
  \and F.~Stanley\inst{\ref{IRAM}}
  \and P.~P.~van~der~Werf \inst{\ref{Leiden}}
  \and S.~Viti \inst{\ref{Leiden},\ref{UCL}}
}

\institute{
Department of Astronomy, The University of Tokyo, 
  7-3-1, Hongo, Bunkyo, Tokyo 113-0033, Japan\\
  \email{nishimura@astron.s.u-tokyo.ac.jp} \label{DoA_UT}
\and Institute of Astronomy, The University of Tokyo, 
  2-21-1, Osawa, Mitaka, Tokyo 181-0015, Japan \label{IoA_UT}
\and National Astronomical Observatory of Japan, 
  2-21-1, Osawa, Mitaka, Tokyo 181-8588, Japan \label{NAOJ}
\and Department of Space, Earth and Environment, 
  Onsala Space Observatory, Chalmers University of Technology, 
  SE-439 92 Onsala, Sweden \label{Chalmers}
\and National Radio Astronomy Observatory, 520 Edgemont Road, 
  Charlottesville, VA 22903, USA \label{NRAO}
\and Observatoire de Paris, LERMA, Coll\`{e}ge de France, 
  CNRS, PSL University, Sorbonne Universit\'{e}, Paris, France \label{LERMA}
\and Institute of Astrophysics, Foundation for Research 
  and Technology-Hellas (FORTH), Heraklion, 70013, Greece \label{FORTH}
\and School of Sciences, European University Cyprus, 
  Diogenes street, Engomi, 1516 Nicosia, Cyprus \label{Cyprus}
\and Department of Astronomy, University of Wisconsin-Madison, 
  5534 Sterling, 475 North Charter Street, 
  Madison WI 53706, USA \label{WM}
\and Department of Physics and Astronomy, Macalester College, 
  1600 Grand Abe, St. Paul, MN 55105 USA \label{Macalester}
\and Observatorio Astron\'{o}mico Nacional 
  (OAN-IGN)-Observatorio de Madrid, 
  Alfonso XII, 3, 28014-Madrid, Spain \label{OAN}
\and Universidad de Alcal\'{a}, Departamento 
  de F\'{i}sica y Matem\'{a}ticas, Campus Universitario, 
  E-28871 Alcal\'{a} de Henares, Madrid, Spain \label{Alcala}
\and DTU-Space, National Space Institute, Technical University of Denmark, 
  Elektrovej 327, DK-2800 Kgs. Lyngby, Denmark \label{DTU}
\and Cosmic Dawn Center (DAWN), DTU-Space, 
  Technical University of Denmark, 
  Elektrovej 327, DK-2800 Kgs.~Lyngby; Niels Bohr Institute, 
  University of Copenhagen, Juliane Maries Vej 30, 
  DK-2100 Copenhagen \label{DAWN}
\and Astronomical Science Program, Graduate Institute for Advanced Studies, 
  SOKENDAI, 2-21-1 Osawa, Mitaka, Tokyo 181-1855, Japan \label{SOKENDAI}
\and Max-Planck-Institut f\"{u}r Radioastronomie, 
  Auf dem H\"{u}gel 69, 53121, Bonn, Germany \label{MPIfR}
\and Astron.~Dept., King Abdulaziz University, P.O.~Box 80203, 
  21589 Jeddah, Saudi Arabia \label{Abdulaziz}
\and Steward Observatory, University of Arizona, 
  933 N. Cherry Ave., Tucson, AZ 85721, USA \label{Steward}
\and European Southern Observatory, Alonso de C\;{o}rdova 3107, 
  Vitacura 763 0355, Santiago, Chile \label{ESO}
\and Joint ALMA Observatory, Alonso de C\'{o}rdova 3107, 
  Vitacura 763 0355, Santiago, Chile \label{JAO}
\and Department of Astronomy, University of Florida, 
  P.O. Box 112055, Gainesville, FL 32611, USA \label{Florida}
\and Department of Astronomy, University of Virginia, 
  530 McCormick Road, Charlottesville, VA 22904, USA \label{Virginia}
\and N\'{u}cleo de Astronom\'{i}a de la Facultad de Ingenier\'{i}a, 
  Universidad Diego Portales, Av.~Ej\'{e}rcito Libertador 441, 
  Santiago, Chile \label{DiegoPortales}
\and Kavli Institute for Astronomy and Astrophysics, 
  Peking University, Beijing 100871, China \label{KIAA}
\and George Mason University, Department of Physics \& Astronomy, 
  MS 3F3, 4400 University Drive, Fairfax, 
  VA 22030, USA \label{GeorgeMason}
\and Institut de Radioastronomie Millim\'{e}trique (IRAM), 
  300 Rue de la Piscine, F-38400 Saint-Martin-d'H\`{e}res, France \label{IRAM}
\and Leiden Observatory, Leiden University, PO Box 9513, 
  2300 RA Leiden, The Netherlands \label{Leiden}
\and Department of Physics and Astronomy, 
  University College London, 
  Gower Street, London WC1E 6BT, UK \label{UCL}
}

\date{Received October 20, 2023 / Accepted February 23, 2023}

\abstract
{Nuclear regions of ultraluminous and luminous infrared galaxies 
(U/LIRGs) are powered by starbursts and/or active galactic nuclei (AGNs). 
These regions are often obscured by extremely high columns of gas and dust. 
Molecular lines in the submillimeter windows have the potential 
to determine the physical conditions of these compact obscured nuclei (CONs).}
{We aim to reveal the distributions of HCN and HCO$^+$ emission 
in local U/LIRGs and investigate whether and how they are related 
to galaxy properties.}
{Using the Atacama Large Millimeter/submillimeter Array (ALMA), 
we have conducted sensitive observations of 
the HCN $J$=3--2 and HCO$^+$ $J$=3--2 lines 
toward 23 U/LIRGs in the local Universe ($z<0.07$) 
with a spatial resolution of $\sim$0.3\arcsec ($\sim$50--400\,pc).}
{We detected both HCN and HCO$^+$ in 21 galaxies, 
only HCN in one galaxy, and neither in one galaxy. 
The global HCN/HCO$^+$ line ratios, 
averaged over scales of $\sim$0.5--4\,kpc, 
range from 0.4 to 2.3, with an unweighted mean of 1.1. 
These line ratios appear to have no systematic trend 
with bolometric AGN luminosity or star formation rate. 
The line ratio varies with position and velocity 
within each galaxy, with an average interquartile range of 0.38 
on a spaxel-by-spaxel basis. 
In eight out of ten galaxies known to have outflows and/or inflows, 
we found spatially and kinematically symmetric structures 
of high line ratios. These structures appear as a collimated bicone 
in two galaxies and as a thin spherical shell in six galaxies.}
{Non-LTE analysis suggests that the high HCN/HCO$^+$ line ratio 
in outflows is predominantly influenced by the abundance ratio. 
Chemical model calculations indicate that the enhancement of 
HCN abundance in outflows is likely due to high-temperature chemistry 
triggered by shock heating. These results imply that 
the HCN/HCO$^+$ line ratio can aid in identifying 
the outflow geometry when the shock velocity of the outflows 
is sufficiently high to heat the gas.}


\maketitle

\section{Introduction}
\label{sect:introduction}

Luminous infrared galaxies (LIRGs; 
$L_\mathrm{IR\,(8-1000\,\mu\mathrm{m})}$\,=\,$10^{11}$--$10^{12}L_\sun$) 
and ultraluminous infrared galaxies 
\citep[ULIRGs: $L_\mathrm{IR}$\,$\geq$\,$10^{12}L_\sun$; see e.g.,][for reviews]{Sanders1996,Perez-Torres2021} emit most of their energy 
at infrared wavelengths and are powered by nuclear starbursts and/or 
active galactic nuclei (AGNs) in their central regions. 
Observational and theoretical studies have both proposed that 
gas-rich galaxy mergers are one of the most important mechanisms 
to trigger starbursts and fuel supermassive black holes (SMBHs), namely 
by funneling large amounts of gas and dust into the nuclei 
\citep[e.g.,][]{Sanders1988,Hopkins2006}. 
These studies have also pointed out that the central SMBHs are 
deeply embedded by high columns of obscuring material 
during the process of mass accretion and 
that U/LIRGs eventually evolve into optically visible quasars 
when nuclear feedback (i.e., outflow) 
disperses the surrounding material. 
Hence, the evolution of U/LIRGs is of key importance 
to account for the large number of luminous quasars 
at high redshift, as merger events are considered 
to have been more frequent in the early Universe \citep[e.g.,][]{Romano2021}. 

There is currently mounting evidence that U/LIRGs often host 
compact ($\lesssim$100\,pc) and highly enshrouded 
($N_\mathrm{H_2}$\,$\gtrsim$\,$10^{24}$\,cm$^{-2}$) nuclei 
\citep[e.g.,][]{Sakamoto2013,Martin2016,Aalto2019,Ricci2021}. 
These compact obscured nuclei (CONs) exhibit 
bright emission of rotational transition of HCN, 
which is vibrationally excited by 
the mid-infrared continuum emitted from dust 
\citep[henceforth referred to as HCN-vib; e.g.,][]{%
Sakamoto2010,Costagliola2013,Imanishi2013,Aalto2015b,Aalto2019}. 
Using HCN-vib emission, \citet[hereafter \citetalias{Falstad2021}]%
{Falstad2021} conducted a systematic survey of CONs, 
and revealed that $\sim$50\% of the ULIRGs 
and $\sim$20\% of the LIRGs host CONs. 

Because of the high obscuration by dust, it is often difficult for 
observations at many wavelengths to probe the embedded nuclear activities 
in the center of U/LIRGs \citep[e.g.,][]{Lutz1996}. 
In particular for CON-host galaxies, we need 
probes free from severe extinction to know the physical properties 
of the nuclear regions, such as, if and how much the buried AGNs 
contribute to the total energy of the source. 
Molecular lines at (sub)millimeter wavelengths are 
less affected by dust extinction and thus have been explored 
for useful diagnostic methods. 
As a best practice, enhanced HCN/HCO$^+$ line ratios%
\footnote{Throughout this paper, ``line ratio'' refers to 
the line intensity ratio or, equivalently, the line luminosity ratio. 
The line luminosity $L'$ is expressed in units of K\,km\,s$^{-1}$\,pc$^2$.} 
($\gtrsim$1) have been proposed as being characteristic of 
the AGN-dominated galaxies \citep[e.g.,][]{Kohno2001,Krips2008,Imanishi2009}. 
The high HCN/HCO$^+$ line ratio 
was at first interpreted as a high abundance of HCN due to 
X-ray ionization in the close vicinity of an AGN \citep{Lepp1996}. 
However, there are some composite and starburst-dominated galaxies that 
show line ratios comparable to or higher than those of AGN-dominated galaxies, 
suggesting that not only X-ray irradiation but also other processes, 
such as optical depths, can elevate the line ratios 
\citep{Costagliola2011,Privon2015}. 
High-resolution observations toward nearby AGN-host galaxies 
have revealed that the HCN/HCO$^+$ line ratios vary 
within a few hundred parsec of circumnuclear disks around the AGNs 
and peak at off-centered locations in the disks (NGC\,1068: 
\citealt{Garcia-Burillo2014,Viti2014,Izumi2016}; NGC\,1097: 
\citealt{Martin2015}). There are also some cases where 
the elevated line ratio is totally unrelated to the AGNs 
(NGC\,3256: \citealt{Harada2018}; NGC\,4194: \citealt{Koenig2018}), 
implying that there are several mechanisms that can elevate 
the line ratios in the disks. Given these observational facts, 
further investigation is needed regarding how to use this line ratio 
as a diagnostic tool for hidden nuclear activities. 

There has been extensive discussion on the possible mechanisms 
for enhancing HCN abundances. As mentioned earlier, 
the effect of X-ray ionization was first served as a reason for 
HCN enhancements in AGNs \citep{Lepp1996}. 
Subsequent studies, however, have painted a more complex picture 
that would likely take place in the centers of active galaxies. 
Chemical models developed by \citet{Meijerink2005} 
have explored a wide range of physical conditions 
in X-ray dominated regions \citep[XDRs,][]{Maloney1996} 
and photon dominated regions \citep[PDRs,][]{Tielens1985}. 
Based on these models, \citet{Meijerink2007} noted that 
the high HCN/HCO$^+$ line ratio is not exclusively seen in XDRs, and thus 
AGN contribution may be hard to recognize only by the HCN/HCO$^+$ line ratio. 
Furthermore, \citet{Viti2014} pointed out that the observed 
molecular line ratios in the center of NGC\,1068, 
including species other than HCN and HCO$^+$, 
could not be reproduced by a single model per region.

Chemical models have also suggested that another important 
chemical process would be high-temperature gas-phase chemistry 
that can form HCN via the reaction 
$\mathrm{CN} + \mathrm{H_2} \rightarrow \mathrm{HCN} + \mathrm{H}$ 
with an activation barrier of $820$\,K \citep{Harada2010,Harada2013}. 
A high temperature can be generated by Coulomb heating in regions 
affected by XDRs, PDRs, and cosmic-ray dominated regions (CRDRs) 
with electrons produced by X-ray, UV, and cosmic-ray ionization, respectively, 
as well as mechanical heating by shocks in the vicinity of a nucleus. 
Indeed, shock heating can reasonably explain the bright HCN emission 
in the line wings (i.e., the outflowing components) of Mrk\,231 
\citep{Aalto2015a,Lindberg2016}. Such bright HCN emission in the outflow 
has also been reported for the western nucleus of Arp\,220 
\citep{Barcos-Munos2018}.

Not only chemical processes regarding molecules but 
the elemental abundance ratio is also 
a matter of concern for the HCN/HCO$^+$ line ratio. 
\citet{Bayet2008} modeled molecular abundances 
in hot cores with different initial elemental abundances 
and showed that the abundance of HCN, 
along with other nitrogen-bearing species such as HNC, 
tends to be roughly scaled with total nitrogen abundance. 
Observations toward low-metallicity molecular clouds also showed that 
the elemental N/O ratio crucially affects the HCN/HCO$^+$ abundance ratio 
in nitrogen-poor subsolar-metallicity galaxies 
\citep[e.g.,][]{Nishimura2016a,Nishimura2016b,Braine2017}. 
This scaling effect, however, has not been fully examined for solar- 
and supersolar-metallicity galaxies where the elemental N/O ratio 
can be elevated by differential galactic winds \citep[e.g.,][]{Vincenzo2016}. 

To enhance the HCN/HCO$^+$ line ratio, the excitation condition 
also plays an important role. Because HCN and HCO$^+$ have 
different excitation properties, the HCN/HCO$^+$ line ratio 
can be a function of local gas density and temperature. 
Several studies have highlighted that the higher critical density 
required for HCN excitation compared to that of HCO$^+$ 
could contribute to the high line ratios in denser regions 
\citep[e.g.,][]{Krips2008,Costagliola2011,Privon2015,Imanishi2019}. 
However, relying only on the traditional definition of critical density 
may lead to inaccuracies. Factors that have been suggested to affect 
line emission include radiative trapping in dense regions 
\citep{Shirley2015} and weak extended emission in diffuse regions 
\citep[e.g.,][]{Kauffmann2017,Nishimura2017,Pety2017}. 
It is also crucial to account for other excitation mechanisms, 
such as collision with free electrons \citep{Goldsmith2017} 
and mid-infrared pumping \citep{Aalto2007}.

This is the second paper of a series named ``CON-quest,'' 
whose aim is to understand the evolution of infrared luminous galaxies, 
in particular those that host CONs. In this paper, we present the results of 
the HCN $J$=3--2 and HCO$^+$ $J$=3--2 line observations toward 23 U/LIRGs 
using the Atacama Large Millimeter/submillimeter Array (ALMA), 
focusing on how the HCN and HCO$^+$ line fluxes are related to 
galaxy properties. Specifically, we tackle the following questions: 
Can the HCN/HCO$^+$ line ratio by itself be used as a diagnostic tool 
for the AGN strength in moderate-resolution (several tens to hundreds of parsec) 
observations? What are the determining factors of the HCN/HCO$^+$ line ratio? 
Which of these factors are most important for the line ratio, 
and in what condition?

The paper is organized as follows. We introduce the sample in 
Sect.~\ref{sect:sample} and describe the ALMA observations 
and data reduction in Sect.~\ref{sect:observations}. 
The results are presented in Sect.~\ref{sect:line} 
and \ref{sect:ratio}, where we focus on the line emission 
and the line ratios, respectively. 
In Sect.~\ref{sect:discussion}, we discuss  
the line ratios in terms of excitation conditions 
and abundances of molecules. The main conclusions of 
this work are summarized in Sect.~\ref{sect:conclusions}. 

\section{Sample}
\label{sect:sample}

The sample of this study consists of eight ULIRGs and 15 LIRGs 
in the local Universe ($z<0.07$), as listed in Table \ref{table:sample}. 
This is a subset of galaxies from the parent sample presented 
in \citetalias{Falstad2021} with six additional galaxies. 
The scope of this paper is limited to the newly obtained data 
from dedicated ALMA observations 
(project ID: 2017.1.00759.S and 2018.1.01344.S). 
We do not address the sub-LIRGs 
($L_\mathrm{IR}$\,$=$\,$10^{10}$--$10^{11}L_\sun$) 
included in the parent sample, and the results of the HCN and HCO$^+$ 
observations for these galaxies will be published in a separate paper 
(Onishi et al.,~in~prep.). 
This sample selection was done to ensure that the source properties 
could be compared properly by only studying sources that have 
HCN and HCO$^+$ data with similar sensitivities and spatial resolutions.
In addition to the targeted galaxies described above, our sample contains 
another six galaxies: IRAS\,F05189$-$2524, IRAS\,F10565$+$2448, 
IRAS\,19542$+$1110, ESO\,148$-$IG002, NGC\,6240, and UGC\,11763. 
These galaxies were also observed in the same ALMA projects 
but were not included in the sample of \citetalias{Falstad2021} 
because they did not meet the selection criteria 
for \citetalias{Falstad2021} due to the slightly larger distances. 
Inclusion of these galaxies does not change the range of 
the infrared luminosity of the sample as a whole 
but breaks the completeness of the sample held in \citetalias{Falstad2021}. 

In Table \ref{table:sample}, we list the main properties of the sample 
galaxies: redshift, luminosity distance, and infrared luminosity. 
All of these quantities are calculated consistently with 
\citetalias{Falstad2021}, as detailed in the footnotes of 
Table \ref{table:sample}. Additionally, we compiled several 
galaxy properties from the literature that are of particular interest 
in the context of the evolution of U/LIRGs: bolometric AGN fraction, 
merger stage, and presence of CONs and molecular in- and outflow.
For the bolometric AGN fraction, we employed the relative AGN contribution 
to the bolometric total luminosity calculated by \citet{Diaz-Santos2017} 
based on \emph{Spitzer}/IRS spectroscopic data. 
\citet{Diaz-Santos2017} combined up to five mid-infrared diagnostics, 
depending on the availability of data: 
the [\ion{Ne}{v}]$_{14.3}$/[\ion{Ne}{ii}]$_{12.8}$ 
and [\ion{O}{iv}]$_{25.9}$/[\ion{Ne}{ii}]$_{12.8}$ line ratio, 
the equivalent width (EQW) of the 6.2\,$\mu$m polycyclic aromatic hydrocarbons 
(PAH), the $S_{30}/S_{15}$ dust continuum ratio, and the Laurent diagram 
\citep{Laurent2000}. They then applied corrections to derive the bolometric 
AGN fraction \citep{Veilleux2009}. We used these values as best estimates 
while keeping in mind the limitations of mid-infrared 
diagnostics when dealing with obscured nuclei. 
For instance, the [\ion{Ne}{v}] and [\ion{O}{iv}] line 
fluxes tend to be smaller for deeply buried AGNs \citep{Yamada2019}. 
In addition, the 6.2\,$\mu$m PAH EQW diagnostics 
may overestimate the AGN fraction for starburst galaxies 
due to the brighter continuum used as a baseline value \citep{Privon2020}. 
For the merger stage, we adopted the visual classification by 
\citet{Stierwalt2013} based on \textit{Spitzer}/IRAC 3.6\,$\mu$m images. 
The existence of a CON is based on the HCN-vib surface density 
measured by \citetalias{Falstad2021}. 
For the galaxies not covered in \citetalias{Falstad2021}, 
the assessment of CON was not conducted in the same method. 
For IRAS\,F10565$+$2448, IRAS\,19542$+$1110, ESO\,148$-$IG002, and UGC\,11763, 
we found no feature of HCN-vib emission, suggesting that 
they are most likely classified as non-CONs. 
While IRAS\,F05189$-$2524 shows faint HCN-vib emission, 
it does not meet the criteria for CON classification 
in the earlier work \citep{Falstad2019}. 
For NGC\,6240, the broad line width of the neighboring HCO$^+$ emission 
complicates its evaluation. Alternative methods to evaluate 
the existence of a CON would be necessary for NGC\,6240. 
The presence of outflows and inflows was inferred from 
far-infrared spectroscopy of the OH lines and/or 
(sub)millimeter interferometry of the CO lines. 
We note that these tables are just a compilation of published information. 
The unlisted galaxies may have hitherto unknown outflows and inflows. 

\section{Observations and data reduction}
\label{sect:observations}

The HCN 3--2 and HCO$^+$ 3--2 line observations were carried out 
with ALMA using the band 6 receivers during Cycle 6 and 7 as two projects: 
2017.1.00759.S and 2018.1.01344.S (PI: S.~Aalto). 
Here, we outline the fundamental properties of each program.

\emph{2017.1.00759.S:} 
The nine U/LIRGs IRAS\,17208$-$0014, IRAS\,09022$-$3615, IRAS\,13120$-$5453, 
IRAS\,F14378$-$3651, IRAS\,F05189$-$2524, IRAS\,19542$+$1110, ESO\,148$-$IG002, 
NGC\,6240, and UGC\,11763 were observed in September 2018 
with the array configuration C43-5. 
The baseline lengths span between 15~m and 2~km, resulting in 
an angular resolution of $\sim$0.3\arcsec\ and 
a maximum recoverable scale (MRS) of $\sim$3--4\arcsec. 

\emph{2018.1.01344.S:} 
As part of this project, 
the 14 U/LIRGs IRAS\,F10565$+$2448, IRAS\,F17138$-$1017, IRAS\,17578$-$0400, 
ESO\,173$-$G015, NGC\,3110, IC\,4734, NGC\,5135, ESO\,221$-$IG10, IC\,5179, 
UGC\,2982, NGC\,2369, ESO\,286$-$G035, ESO\,320$-$G030, and NGC\,5734 
were observed between October 2018 and October 2019 
with the array configurations C43-4 and C43-5. 
The baseline lengths span between 15\,m and 2.5\,km, resulting in 
an angular resolution of $\sim$0.3\arcsec\ and an MRS of $\sim$3--4\arcsec. 

In both projects, the total integration time per field 
was $\lesssim$2 hours. 
Bright quasars were used as bandpass and flux calibrators, 
ensuring flux accuracy better than 10\% at Band 6. 
Fainter quasars close to each target were used for phase calibration. 
We used a single pointing with a field of view 
of $\sim$20\arcsec\ for each target. 
While the nominal MRS for the employed array configuration 
is $\sim$9\arcsec, as mentioned above, the effective MRS can be 
as small as $\sim$3--4\arcsec in our observations, 
depending on antenna availability. 
Given that the LIRG sample is located at smaller distances compared to 
the ULIRG sample, the potential impact of interferometric missing flux 
may be systematically more significant for the LIRG samples. 
This issue could affect galaxy-to-galaxy comparisons, 
as discussed in Sect.~\ref{subsect:Gao-Solomon}. 
At this moment, single-dish measurements are not homogeneously available 
for all sample galaxies to calibrate this missing flux issue. 
Nevertheless, we acknowledge the importance of addressing 
this problem in future studies. 

The correlator setup was the same for all targets: 
Two spectral windows of 1.875\,GHz width were placed 
in the upper sideband with native channel widths of 3.9\,MHz, 
centered at each of HCN $J$=3--2 (rest frequency 265.886\,GHz) and 
HCO$^+$ $J$=3--2 (rest frequency 267.558\,GHz). 
Two more spectral windows were placed in the lower sideband 
for better continuum identification. 

Calibration of the interferometric data was done with 
CASA\footnote{\url{http://casa.nrao.edu/}} 
\citep[][version 5.4.0]{CASAteam2022}. 
The continuum subtraction was performed using the CASA ``uvcontsub'' task, 
except for IRAS\,17578$-$0400. For IRAS\,17578$-$0400, it is hard to 
find line-free channels due to the large number of emission lines 
and the complex line profiles; hence, the continuum was not subtracted. 
The imaging of the calibrated visibility sets was also performed 
in CASA using the ``tclean'' task with a natural weighting. 
The synthesized beams are $\sim$0.3\arcsec, corresponding to 
spatial scales of $\sim$50--400\,pc for the range of the target distances, 
as detailed in Table~\ref{table:observations}. 
The data cubes were smoothed to a velocity resolution of 20\,km\,s$^{-1}$. 
The resulting rms noise levels are given in Table~\ref{table:observations}. 
For further analysis, the data cubes were exported in FITS format. 

We analyzed the exported cubes with our own codes, making use of the following Python packages: 
astropy\footnote{\url{http://www.astropy.org}} \citep{Astropy2013,Astropy2018}, 
spectral-cube,\footnote{\url{https://spectral-cube.readthedocs.io}} 
and pvextractor.\footnote{\url{https://pvextractor.readthedocs.io}} 
In the analysis, we adopted galaxy kinematics roughly estimated 
from the velocity field seen outside of the most nuclear region 
derived from either HCN or HCO$^+$ lines. 
The parameters are summarized in Table~\ref{table:observations}, 
although these may not be robust enough for a more detailed kinematic analysis. 
More accurate parameters could be found in the literature and/or 
derived from alternative molecular tracers such as CO emission. 
However, at this moment, such datasets are not homogeneously 
available for all target galaxies. 
Kinematic modeling that takes into account vertical components and 
non-circular motions is planned for future studies. 

\section{Line emission}
\label{sect:line}

\subsection{Spectra}
\label{subsect:spectra}

To obtain an overview of the HCN and HCO$^+$ line emission 
in each galaxy, we extracted spectra from two kinds of apertures: 
an aperture the size of the synthesized beam (hereafter ``resolved'') 
and an elliptical aperture $10\times10$ times the size of 
the synthesized beam (hereafter ``global''). 
Both are centered at the position listed in Table \ref{table:observations}. 
The extracted spectra are presented in Appendix~\ref{appendix:figures}. 
The resolved aperture corresponds to the physical scale of 
$\sim$200--400\,pc for ULIRGs and $\sim$50--100\,pc for LIRGs. 
Consequently, the global aperture corresponds to 
$\sim$2--4\,kpc for ULIRGs and $\sim$0.5--1\,kpc for LIRGs. 
The global aperture covers all, or nearly all, of the line-emitting 
regions of each galaxy in most cases (Sect.~\ref{subsect:moments}), 
but it should be noted that a considerable fraction of emission can 
also be seen outside of the aperture in NGC\,6240 (Fig.~\ref{figure:6240}); 
IRAS\,17578$-$0400 (Fig.~\ref{figure:17578-0400}); 
ESO\,173$-$G015 (Fig.~\ref{figure:173-G015}); 
IC\,4734 (Fig.~\ref{figure:4734}); 
NGC\,5135 (Fig.~\ref{figure:5135}); 
and NGC\,2369 (Fig.~\ref{figure:2369}). 

In the spectra, we confirmed a large number of line detections. 
Toward all galaxies but UGC\,11763, UGC\,2982 and NGC\,5734, 
we detected both HCN and HCO$^+$ lines at more than $5\sigma$ 
significance regardless of the aperture size for spectral extraction. 
For UGC\,11763, the HCN and HCO$^+$ lines were marginally detected 
at $\sim$3$\sigma$ and $\sim$5$\sigma$ significance, respectively, 
only on the velocity-integrated intensity 
extracted from the resolved aperture (Fig.~\ref{figure:11763}). 
For NGC\,5734, HCN was clearly detected, but HCO$^+$ is totally absent 
(Fig.~\ref{figure:5734}). Neither HCN nor HCO$^+$ was detected 
in UGC\,2982 (Fig.~\ref{figure:2982}). 
We note that these non-detections are in the sources 
with the lowest $L_\mathrm{IR}$ in our sample. 
In addition to HCN and HCO$^+$, 
other molecular species were detected in several sources: 
HC$_3$N-vib ($v_7$=2, $J$=29--28, $l$=$2e$; 265.462\,GHz); 
CH$_2$NH ($4_{1,3}$--$3_{1,2}$; 266.270\,GHz); 
HCN-vib ($v_2$=1, $J$=3--2; 267.199\,GHz); 
CH$_3$OH ($5_2$--$4_1$ E1; 266.838\,GHz); 
and HOC$^+$ ($J$=3--2; 268.451\,GHz). 
Detailed analysis of these molecular species is beyond the scope 
of this paper and would need to be presented in dedicated works 
\citep[as in][for CH$_2$NH]{Gorski2023}. 

In most target galaxies, the HCN and HCO$^+$ lines exhibit 
complex line profiles that cannot be properly fitted by a single Gaussian, 
similar to what has been reported in other studies 
\citep[e.g.,][]{Martin2016,Imanishi2019}. 
Among them, there are several galaxies with 
asymmetric double-peaked profiles. 
Such features are more clearly seen in the resolved spectra 
than the global ones.  
Although we cannot exclude the possibility that 
these features are caused by clumpy and non-axisymmetric circumnuclear disks, 
the double-peaked profiles are more likely to be absorption features 
resulting from self-absorbing gas near the nuclei (i.e., 
outflows or inflows), as explained below. 
The absorption peak was found to be slightly blueshifted relative 
to the systemic velocity in IRAS\,17208$-$0014 and IRAS\,17578$-$0400 
(Fig.~\ref{figure:17208-0014} and Fig.~\ref{figure:17578-0400}, 
respectively), while it is redshifted in ESO\,173$-$G015 and ESO\,320$-$G030 
(Fig.~\ref{figure:173-G015} and Fig.~\ref{figure:320-G030}, 
respectively). 

The absorption features more blueshifted than about $-50$\,km\,s$^{-1}$ 
can be interpreted as signposts of outflows \citep[e.g.,][]{Veilleux2013}. 
For IRAS\,17208$-$0014 (Fig.~\ref{figure:17208-0014}), 
it is consistent with the fact that molecular outflows are found 
by interferometric observations of the CO line \citep{Garcia-Burillo2015}. 
For IRAS\,17578$-$0400 (Fig.~\ref{figure:17578-0400}), 
the low-velocity elongation of the HCN and HCO$^+$ emission 
along its kinematic minor axis would be indicative of an outflow, 
as pointed out in \citetalias{Falstad2021} 
(see Fig.~8 of \citetalias{Falstad2021}). 
In higher-resolution observations of the same HCN and HCO$^+$ transitions, 
an elongation in the minor-axis direction was also found
(\citealt{Yang2023}, Yang et al.,~in prep). 

On the other hand, the redshifted absorption features could be 
indicative of inflowing motions. 
In ESO\,173$-$G015 (Fig.~\ref{figure:173-G015}), 
the absorption peaks are redshifted to 
$\sim$50\,km\,s$^{-1}$ in the resolved spectra (extracted from the 
$51\times47$\,pc elliptical aperture) and are at near the systemic 
velocity in the global spectra ($510\times470$\,pc). Regarding
ESO\,173$-$G015, it does not have any published evidence of 
inflow (nor outflow), but detection of HCN-vib has been reported in its center, 
although not bright enough to be categorized 
as a CON \citepalias{Falstad2021}. The nuclear properties and the gas kinematics of this galaxy would be worth investigating in more detail. 
The absorption features in ESO\,320$-$G030 (Fig.~\ref{figure:320-G030}) 
are clearly seen in the resolved spectra ($62\times53$\,pc) 
but are only marginally seen in the global spectra ($620\times530$\,pc). 
This is reasonably understood considering that the inflows take place 
on smaller scales \citep[130\,pc and 230--460\,pc;][]{Gonzalez-Alfonso2021}, 
while outflows are present on a much larger scale 
\citep[$\sim$2500\,pc;][]{Pereira-Santaella2016} in ESO\,320$-$G030. 

\subsection{Moment maps}
\label{subsect:moments}

The integrated intensity (moment~0), velocity field (moment~1), 
and velocity dispersion (moment~2) maps of the HCN and HCO$^+$ lines 
for all targets are presented in Appendix~\ref{appendix:figures}. 
The velocity ranges we considered in order to derive the moment maps are listed 
in Table \ref{table:luminosities}. These ranges were determined 
through visual inspection of the global spectra, 
and we examined the velocity at which the intensity drops 
below the noise level. 
We did not correct for potential line overlapping. 
As the lines of CH$_2$NH, HC$_3$N-vib, and HCN-vib sit 
close to the HCN and HCO$^+$ lines (separations are $<$500\,km\,s$^{-1}$), 
the moment maps of a line-rich source may be 
affected by the neighboring lines 
when the velocity range exceeds about $\pm$300\,km\,s$^{-1}$. 
No clipping was applied for the integrated intensity maps in order 
to salvage the faint emission, while we used a $3\sigma$ threshold 
clipping to properly derive the velocity field and the dispersion. 
The velocity dispersion was plotted as the square root of 
the intensity-weighted second moment of the spectrum. 
To help identify the kinematic center and compare panels, 
the kinematic major and minor axes are drawn in each panel. 

\subsubsection{Integrated intensity maps}
\label{subsubsect:mom0}

The integrated intensity maps show that the HCN and HCO$^+$ line emission 
mostly emerges from a central region in each galaxy. 
The apparent size of the line-emitting region is typically 1--3\,kpc 
in the ULIRGs and 0.5--1\,kpc in the LIRGs. Thus, the global aperture covers 
the major line-emitting region in each galaxy. 
The most remarkable exceptions are NGC\,5135 and NGC\,2369, as NGC\,5135 shows 
a spiral arm-like structure extended over $\sim$2\,kpc from the center 
(Fig.~\ref{figure:5135}), and NGC\,2369 shows an edge-on disk 
with a $\sim$2\,kpc radius (Fig.\,\ref{figure:2369}). 
Attention should also be paid to four other galaxies: 
NGC\,6240 (Fig.~\ref{figure:6240}); 
IRAS\,17578$-$0400 (Fig.~\ref{figure:17578-0400}); 
ESO\,173$-$G015 (Fig.~\ref{figure:173-G015}); 
and IC\,4734 (Fig.~\ref{figure:4734}), 
which show significant emission outside of the global aperture. 
We may have failed to recover the emission extended beyond the MRS 
because of the lack of short baseline in interferometric observations. 
The MRS is $\sim$3--4\arcsec, which corresponds to $\sim$2--3\,kpc 
and $\sim$0.6--1.5\,kpc on average for ULIRGs and LIRGs, respectively. 
Hence the LIRGs at relatively smaller distances could be affected 
by the missing fluxes. We further infer this 
from line luminosities in Sect.~\ref{subsect:Gao-Solomon}. 

\subsubsection{Velocity field and dispersion maps}
\label{subsubsect:mom1and2}

The velocity field and dispersion maps clearly indicate 
that the rotating motions are dominant in the sample galaxies, 
although some galaxies exhibit indications of irregular motions 
(e.g., IRAS\,09022$-$3615 (Fig.~\ref{figure:09022-3615}); 
ESO\,148$-$IG002 (Fig.~\ref{figure:148-IG002}); 
IRAS\,F17138$-$1017 (Fig.~\ref{figure:F17138-1017}); 
NGC\,5135 (Fig.~\ref{figure:5135}); 
and NGC\,2369 (Fig.~\ref{figure:2369})). 
Considerably disordered motion was found in NGC\,6240 
(Fig.~\ref{figure:6240}), 
which is a merger hosting two AGNs separated by $\sim$1.7\arcsec 
\citep[projected; e.g.,][]{Komossa2003}. 
For NGC\,6240, the moment maps of HCN and HCO$^+$ are overall 
similar to those of CO $J$=2--1 \citep{Treister2020,Saito2018} and 
[\ion{C}{i}] $^{3}$P$_{1}$--$^{3}$P$_{0}$ \citep{Cicone2018}, 
indicating the ubiquity of HCN and HCO$^+$ in the molecular gas 
and the complex nature of the gas kinematics in the system.  

While the velocity fields derived from HCN and HCO$^+$ are generally 
similar to each other, subtle differences are found in certain galaxies. Specifically, in the central regions of 
IRAS\,13120$-$5453 (Fig.~\ref{figure:13120-5453}); 
IRAS\,F14378$-$3651 (Fig.~\ref{figure:F14378-3651}); 
IRAS\,F05189$-$2524 (Fig.~\ref{figure:F05189-2524}); 
IRAS\,F10565$+$2448 (Fig.~\ref{figure:F10565+2448}); 
IRAS\,19542$+$1110 (Fig.~\ref{figure:19542+1110}); and 
ESO\,320$-$G030 (Fig.~\ref{figure:320-G030}), 
the velocity dispersion of HCN is slightly larger 
than that of HCO$^+$. 
Remarkably, all of these galaxies are known to have molecular outflows, 
as listed in Table~\ref{table:sample}.
In these galaxies, the superposition of the outflow velocities 
may be responsible for the increased dispersion of HCN 
because the abundance of HCN can be enhanced in the outflows, 
as discussed in Sect.~\ref{subsect:non-circular} and \ref{subsect:chemistry}. 
Conversely, for IRAS\,17208$-$0014, NGC\,6240, and IRAS\,17578$-$0400, where outflows are known, the line-broadening effect for HCN is not as evident when compared with HCO$^+$ (Fig.\ref{figure:17208-0014}, Fig.\ref{figure:6240}, and Fig.\ref{figure:17578-0400}, respectively). 
This is likely due to the distorted velocity field of HCO$^+$ caused by the neighboring HCN-vib line in IRAS\,17208$-$0014 and IRAS\,17578$-$0400, 
resulting in the inaccurate measurement of the dispersion of HCO$^+$. 
However, a different explanation may be considered for the outflow in NGC\,6240, where the shock velocity is relatively small \citep[$\sim$10\,km\,s$^{-1}$;][]{Meijerink2013} and the HCN abundance is not enhanced in the outflow. 
This point is discussed in Sect.~\ref{subsect:chemistry}. 

\subsection{Line luminosities}
\label{subsect:luminosities}

Based on the extracted spectra (Sect.~\ref{subsect:spectra}), 
we derived the resolved and global luminosities of HCN and HCO$^+$ used in this study, 
and we list them in Table \ref{table:luminosities}. 
These luminosities were calculated in units of K\,km\,s$^{-1}$\,pc$^2$ 
following the equation from \citet{Solomon2005}: 
$L'$\,=\,$3.25\times10^7\ S\Delta v\ \nu_\mathrm{obs}^{-2}\ %
D_\mathrm{L}^2\ (1+z)^{-3}$, 
where $S\Delta v$ is the velocity-integrated line flux density 
in Jy\,km\,s$^{-1}$, $\nu_\mathrm{obs}$ is the observed frequency 
in GHz, and $D_\mathrm{L}$ is the luminosity distance in Mpc.
The ranges for velocity integration are the same as those used to 
derive the moment maps, as listed in Table \ref{table:luminosities}. 
Due to blending with the neighboring 
CH$_2$NH ($-432$\,km\,s$^{-1}$ wrt.~HCN), 
HC$_3$N-vib ($+480$\,km\,s$^{-1}$ wrt.~HCN), and 
HCN-vib ($+402$\,km\,s$^{-1}$ wrt.~HCO$^+$) lines, 
the HCN and HCO$^+$ line flux density may be overestimated. 
For simplicity, we took into account all channels within 
the integration range and did not apply any correction for line blending. 
Hence, the values should be taken with caution, particularly for 
the galaxies with broad line profiles ($\gtrsim$300\,km\,s$^{-1}$), 
namely, IRAS\,17208$-$0014 (Fig.~\ref{figure:17208-0014}); 
IRAS\,09022$-$3615 (Fig.~\ref{figure:09022-3615}); 
IRAS\,13120$-$5453 (Fig.~\ref{figure:13120-5453}); 
and NGC\,6240 (Fig.~\ref{figure:6240}). 
Particularly, the resolved spectra of IRAS\,17208$-$0014 
and the global spectra of NGC\,6240 exhibit broad and complex line shapes 
that prevented us from obtaining precise measurement of the line luminosities. 
For a more accurate estimation excluding line blending effects, 
we suggest consulting references such as the CO line profiles 
detailed in \citet{Garcia-Burillo2015} for IRAS\,17208$-$0014 
and in \citet{Saito2018} for NGC\,6240. 
We note that the values for IRAS\,17578$-$0400 have large uncertainties 
because we could not identify line-free channels, and 
continuum subtraction was done by assuming continuum levels of 
0.02\,Jy and 0.07\,Jy for the resolved and global spectra, 
respectively, based on the lowest intensity channels 
(Fig.~\ref{figure:17578-0400}). 

\subsection{Correlation with infrared luminosity}
\label{subsect:Gao-Solomon}

A correlation between infrared luminosity ($L_\mathrm{IR}$) 
and molecular line luminosity ($L'_\mathrm{mol}$) 
was found by early studies \citep[e.g.,][]{Kennicutt1998,GaoSolomon2004}, 
and interpreted as a relation between star formation rate (SFR) 
and molecular gas content, as proposed earlier by \citet{Schmidt1959}. 
The observations were extensively conducted toward nearby galaxies 
\citep[e.g.,][]{Gracia-Carpio2008,Privon2015,Zhang2014,%
Jimenez-Donaire2019,Israel2023} 
as well as toward high-redshift systems \citep[e.g.,][]{Oteo2017}. 
The observed relation was found to be the power-law form of 
$L_\mathrm{IR}$\,$\propto$\,$L'^N_\mathrm{mol}$, 
where the index $N$ depends on the molecular transition 
\citep[e.g.,][]{Bussmann2008,Juneau2009}. 
The indices for commonly observed transitions of CO, HCN, and HCO$^+$ 
have been quantitatively predicted by simulations 
\citep[e.g.,][]{Krumholz2007,Narayanan2008}. 

\begin{figure*}
\includegraphics[width=0.5\hsize]{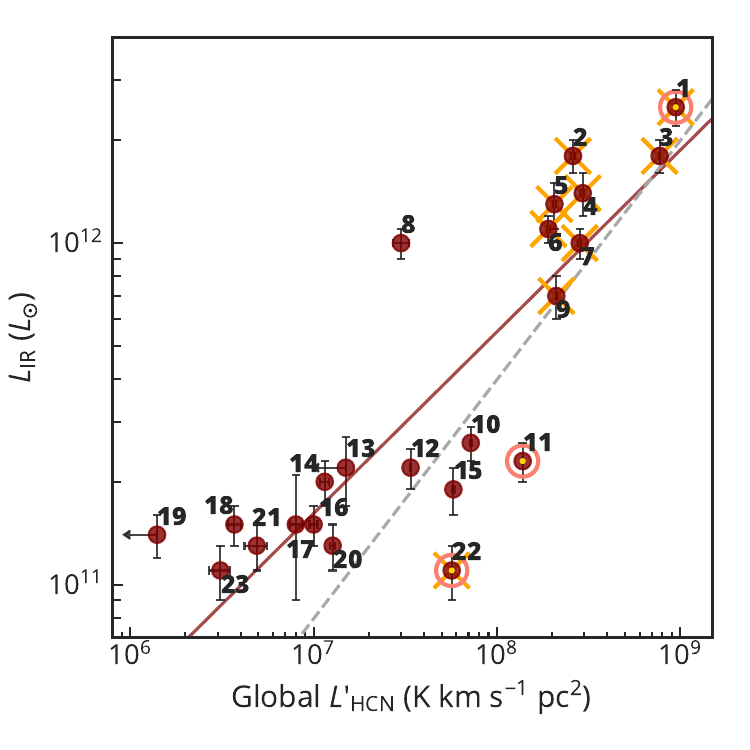}
\includegraphics[width=0.5\hsize]{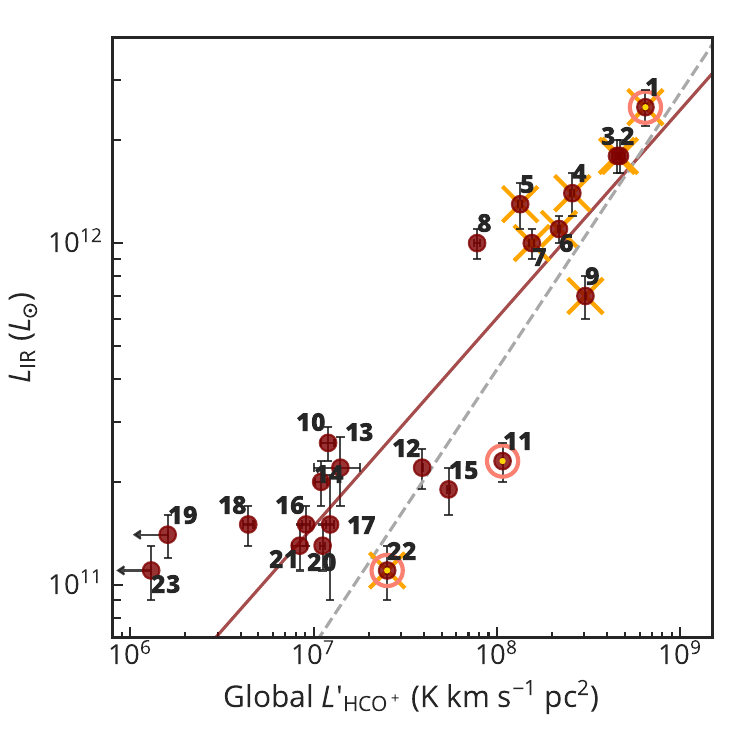}
\caption{$L_\mathrm{IR}$ as a function of 
(\emph{left}) global $L'_\mathrm{HCN}$ 
and (\emph{right}) global $L'_\mathrm{HCO^+}$. 
In both panels, the numbers indicate the galaxy ID as defined 
in Table \ref{table:sample}. 
Small yellow dots, crosses, and open circles indicate 
the presence of CONs, outflows, and inflows, respectively. 
The solid lines are the fitting results: 
$\log L_\mathrm{IR}=0.53\log L'_\mathrm{HCN}+7.5$ and 
$\log L_\mathrm{IR}=0.61\log L'_\mathrm{HCO^+}+6.9$. 
As reference, the dashed lines are the best-fit relations 
taken from \citet{Juneau2009}. }
\label{figure:Gao-Solomon}
\end{figure*}

In Fig.~\ref{figure:Gao-Solomon}, we present the relation between 
the global HCN and HCO$^+$ luminosities ($L'_\mathrm{HCN}$ and 
$L'_\mathrm{HCO^+}$ as calculated in Sect.~\ref{subsect:luminosities}) 
and $L_\mathrm{IR}$ (calculated from IRAS fluxes taken from 
\citet{Sanders2003}; see Sect.~\ref{sect:sample} and 
Table \ref{table:sample} for details) for our sample. 
Both $L'_\mathrm{HCN}$ and $L'_\mathrm{HCO^+}$ 
appear to correlate well with $L_\mathrm{IR}$ 
(Pearson correlation coefficients $r$\,=\,0.85 and 0.90, respectively). 
If the $L_\mathrm{IR}$ is corrected for the AGN contribution by 
$(1-\alpha_\mathrm{AGN})L_\mathrm{IR}$, the correlations are not 
largely changed ($r$\,=\,0.86 for both $L'_\mathrm{HCN}$ and 
$L'_\mathrm{HCO^+}$), implying that 
the contribution from AGN to the total $L_\mathrm{IR}$ is rather limited. 
Thus, the correlations would basically reflect 
the relation between the SFR and the line luminosities. 
We also note that if we use a much larger aperture 
for spectral extraction, for instance 15\arcsec 
(corresponds to 2.5--20\,kpc for our sample), 
to include the emission outside the global aperture 
(Sect.~\ref{subsubsect:mom0}), the observed trend is 
not significantly changed. 

The fitting results of the $L'_\mathrm{mol}$--$L_\mathrm{IR}$ relations 
are $\log L_\mathrm{IR}=(0.53\pm0.03)\log L'_\mathrm{HCN}+(7.5\pm0.3)$ 
and $\log L_\mathrm{IR}=(0.61\pm0.03)\log L'_\mathrm{HCO^+}+(6.9\pm0.2)$, 
as plotted by solid lines in Fig.~\ref{figure:Gao-Solomon}. 
As reference, we also plot the fitting results 
for single-dish measurements of the $J$$=$3--2 transitions 
in nearby galaxies with a range of 
$10^{10} L_\odot\,<\,L_\mathrm{IR}\,<\,10^{12.5} L_\odot$ 
reported by \citet{Juneau2009}. 
Our indices (i.e., the best-fit slopes) are slightly smaller 
than the fitting results of \citet{Juneau2009} 
(0.70$\pm$0.09 for HCN 3--2 and 0.81$\pm$0.21 for HCO$^+$ 3--2) 
and the model prediction \citep[$\sim$0.7 for HCN 3--2;][]{Narayanan2008}. 

As an explanation for the smaller slopes in our fitting results, 
we note that we may systematically underestimate the line flux in the LIRGs 
at smaller distances. Given that the LIRG sample is chosen to have 
smaller distances than the ULIRG sample, the limited aperture size 
for spectral extraction ($\sim$3\arcsec) and the interferometric missing flux 
from the extended gas components ($\gtrsim$3--4\arcsec) could lead to 
the underestimation of the line flux in the LIRG sample. 
The MRS of the observation of $\sim$3--4\arcsec corresponds to 
a physical scale of $\sim$0.6--1.5\,kpc and $\sim$2--3\,kpc on average 
for LIRGs and ULIRGs, respectively. For more accurate measurements 
of the galaxy-integrated $L'_\mathrm{mol}$, observations 
of a larger sample with a single-dish telescope would be essential. 
Alternatively, the lower gas temperature and/or 
the lower mean density in our LIRG sample 
could be responsible for the inefficient excitation of HCN and HCO$^+$ 
resulting in the lower $L'_\mathrm{HCN}$ and $L'_\mathrm{HCO^+}$. 

\section{HCN/HCO$^+$ line ratio}
\label{sect:ratio}

\subsection{Global and resolved line ratio}
\label{subsect:velocity-integrated}

We calculated the HCN/HCO$^+$ line ratio 
($L'_\mathrm{HCN}/L'_\mathrm{HCO^+}$), 
based on the line luminosities extracted from the global and resolved 
apertures (Sect.~\ref{subsect:luminosities}). The results are 
listed in Table \ref{table:integratedratio}. 
The global $L'_\mathrm{HCN}/L'_\mathrm{HCO^+}$ ranges from 0.4 to 2.3 
among the detected sample, with an unweighted mean of 1.1. 
This range is roughly comparable to that observed in the local U/LIRGs 
\citep[e.g.,][for the 1--0 transitions]{Privon2015}. 
The resolved $L'_\mathrm{HCN}/L'_\mathrm{HCO^+}$ is slightly higher 
than the global one in most cases (16 of 20 detected cases), 
ranging from 0.4 to 2.7, with an unweighted mean of 1.2. 
This may indicate that HCN would be increased and/or more efficiently 
excited in the central regions compared to the galaxy average. 
However, we note that a high $L'_\mathrm{HCN}/L'_\mathrm{HCO^+}$ 
is also found in other parts of galaxies and is not uniquely associated with 
the central regions (e.g., the case of ESO\,320-G030, 
Fig.~\ref{figure:320-G030}). 
Depending on the galaxy inclination, 
even the resolved $L'_\mathrm{HCN}/L'_\mathrm{HCO^+}$ can be 
a reflection of multiple velocity components along the line of sight. 
As detailed in Sect.~\ref{subsect:resolved}, spectrally resolved 
analysis would be helpful to disentangle the nuclear region from 
the surrounding disk regions.  

\subsection{Comparison with galaxy properties}
\label{subsect:galproperties}

To consider whether $L'_\mathrm{HCN}/L'_\mathrm{HCO^+}$ would be influenced 
by galaxy properties such as AGN dominance and star formation activity, 
we compared the global and resolved $L'_\mathrm{HCN}/L'_\mathrm{HCO^+}$ 
with mid- and far-infrared diagnostics available in the literature. 
As indicators of AGN dominance and strength, we used the bolometric AGN 
fraction ($\alpha_\mathrm{AGN}$; as listed in Table~\ref{table:sample}) 
and the bolometric AGN luminosity ($L_\mathrm{AGN}^\mathrm{bol}$; calculated as 
$L_\mathrm{AGN}^\mathrm{bol}$\,=\,$\alpha_\mathrm{AGN}L_\mathrm{bol}$, 
where $L_\mathrm{bol}$\,=\,$1.15L_\mathrm{IR}$ is assumed; 
\citealt{Veilleux2009}). 
We emphasize that $\alpha_\mathrm{AGN}$ was 
inferred from mid-infrared diagnostics, and this estimation should be 
treated with caution, particularly in the case of heavily obscured nuclei. 
The star formation activity was basically inferred from $L_\mathrm{IR}$ 
and quantified as SFR in units of $M_\sun\ \mathrm{yr}^{-1}$ via the relation 
$\mathrm{SFR}$\,=\,$(1-\alpha_\mathrm{AGN})\times10^{-10}L_\mathrm{IR}$ 
on the same assumption as \citet{Sturm2011}, 
that is, the SFR--$L_\mathrm{IR}$ relation calibrated by 
\citet{Kennicutt1998} but using a Chabrier initial mass function. 

In addition, we considered the possible influence by the dust temperature 
and the elemental N/O ratio. As a proxy of the temperature of 
the warm dust component, we used the IRAS 60\,$\mu$m/100\,$\mu$m 
flux density ratio ($S_{60}/S_{100}$; taken from \citealt{Sanders2003}). 
The $S_{60}/S_{100}$ color ratio of our sample ranges from 0.4 to 1.2, 
roughly corresponding to a dust temperature from 30 to 50\,K 
\citep[assuming dust emissivity proportional to 
$(\mathrm{wavelength})^{-1}$; e.g.,][]{Helou1988}. 
For the elemental N/O ratio, we used 
the far-infrared fine-structure line ratio of 
[\ion{N}{ii}]$_{122}$/[\ion{O}{iii}]$_{88}$ 
(calculated from galaxy-integrated line fluxes; 
taken from \citealt{Diaz-Santos2017}) as a rough estimator. 
We note that the [\ion{N}{ii}]$_{122}$/[\ion{O}{iii}]$_{88}$ line ratio 
is roughly scaled by the elemental N/O abundance ratio, 
but it would highly depend on the ionization parameter and 
the effective temperature of the ionizing source 
\citep[e.g.,][]{Pereira-Santaella2017,Herrera-Camus2018}. 
We only considered 12 galaxies in which both the [\ion{N}{ii}]$_{122}$ 
and [\ion{O}{iii}]$_{88}$ lines are detected 
with good significance, and hence the sample size 
is smaller than the parent sample in this work. 

\begin{figure*}
\includegraphics[width=0.330\hsize]{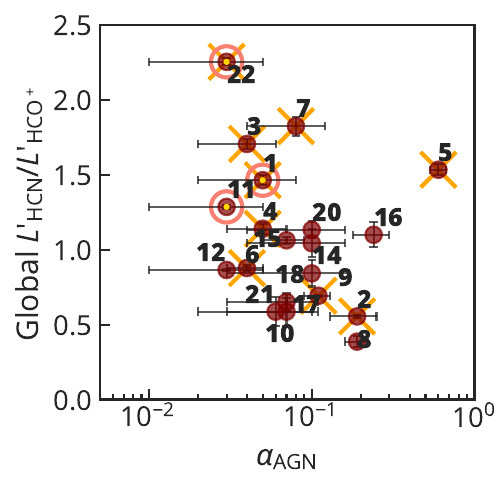}
\includegraphics[width=0.330\hsize]{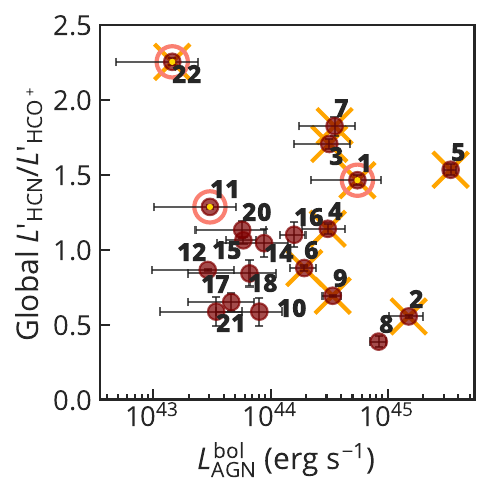}
\includegraphics[width=0.330\hsize]{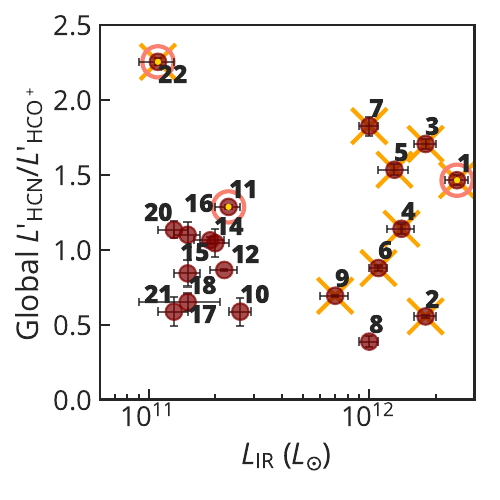}

\includegraphics[width=0.330\hsize]{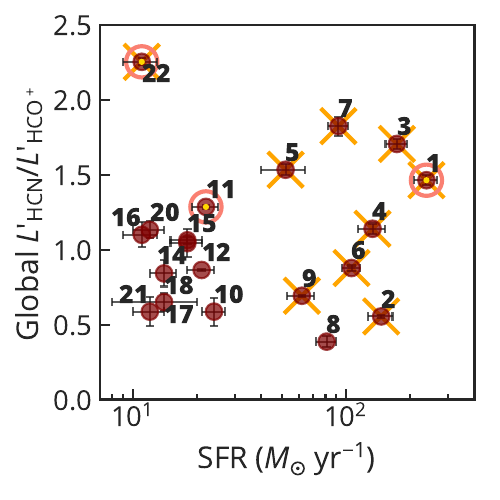}
\includegraphics[width=0.330\hsize]{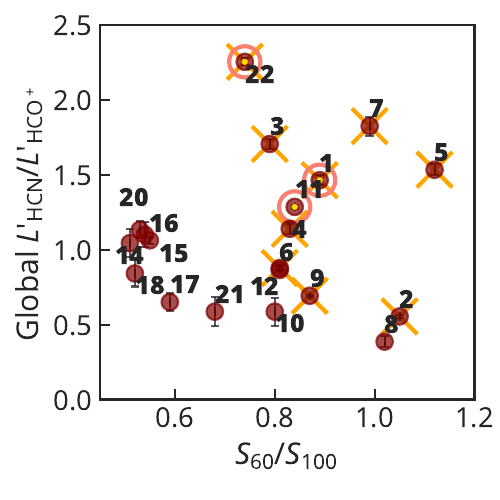}
\includegraphics[width=0.330\hsize]{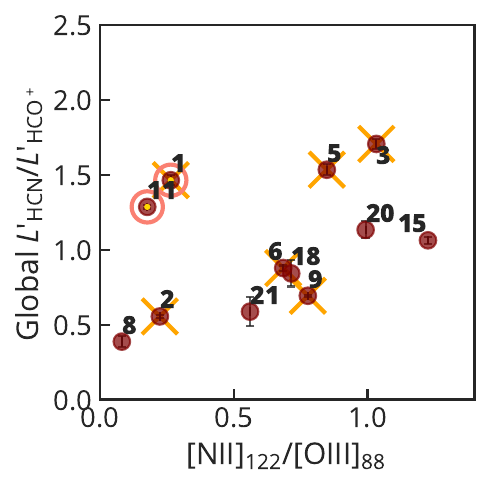}
\caption{Comparison of global $L'_\mathrm{HCN}/L'_\mathrm{HCO^+}$ 
with (\emph{top, from left to right}) AGN fraction, 
bolometric AGN luminosity, infrared luminosity, (\emph{bottom}) 
star formation rate, IRAS 25 $\mu$m/60 $\mu$m flux density ratio, 
and the [\ion{N}{II}]$_{122}$/[\ion{O}{III}]$_{88}$ line ratio. 
In all panels, the numbers refer to IDs as defined in Table \ref{table:sample}. 
Symbols are as in Fig.~\ref{figure:Gao-Solomon}. 
Uncertainties are $1\sigma$. 
For the [\ion{N}{II}]$_{122}$/[\ion{O}{III}]$_{88}$ line ratios, 
uncertainties are not available. 
Corresponding Spearman rank coefficients and p-values are 
listed in Table \ref{table:correlation}. 
}
\label{figure:correlation_global}
\end{figure*}

\begin{figure*}
\includegraphics[width=0.330\hsize]{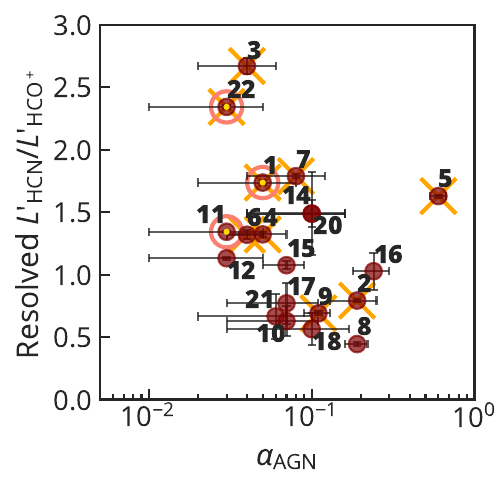}
\includegraphics[width=0.330\hsize]{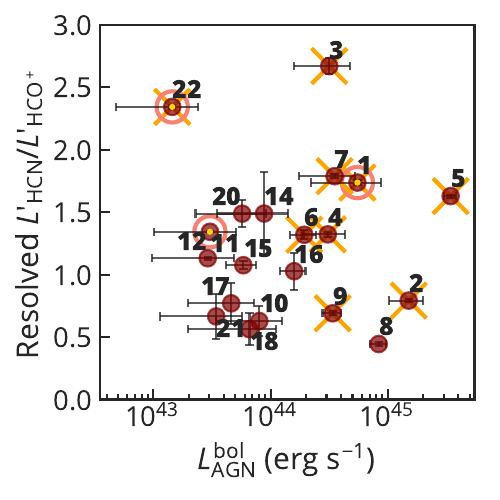}
\includegraphics[width=0.330\hsize]{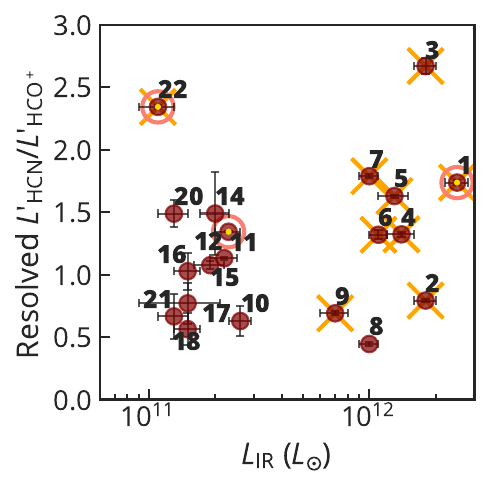}

\includegraphics[width=0.330\hsize]{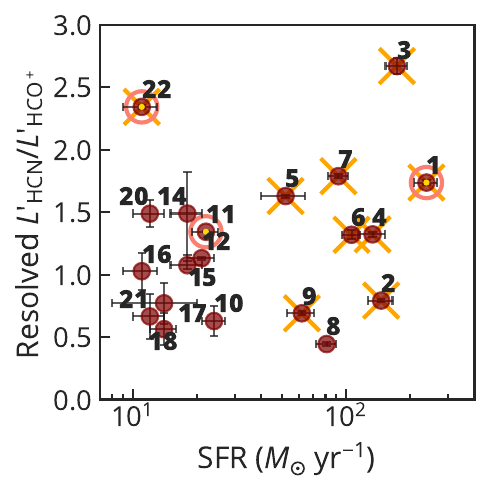}
\includegraphics[width=0.330\hsize]{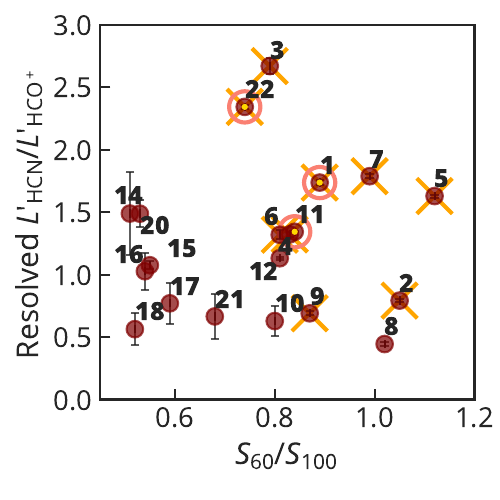}
\includegraphics[width=0.330\hsize]{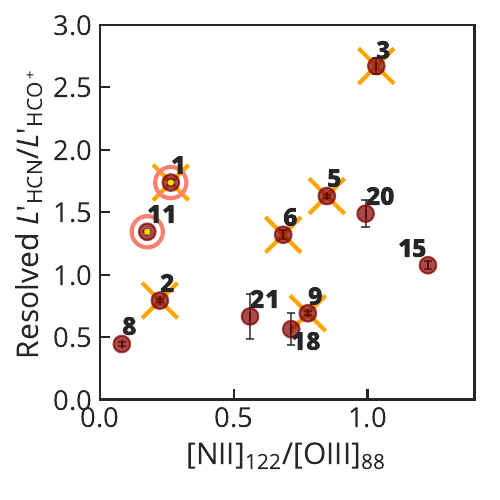}
\caption{Same as Fig.~\ref{figure:correlation_global} 
but for resolved $L'_\mathrm{HCN}/L'_\mathrm{HCO^+}$. 
}
\label{figure:correlation_resolved}
\end{figure*}

We performed a Spearman rank correlation analysis 
for the six above-mentioned quantities with each of the global 
and resolved $L'_\mathrm{HCN}/L'_\mathrm{HCO^+}$ 
using \texttt{pymccorrelation}\footnote{%
\url{https://github.com/privong/pymccorrelation}} 
\citep{Privon2020}, which implements Monte Carlo-based methods 
of uncertainty estimation introduced by \citet{Curran2014}. 
The galaxies in which either HCN or HCO$^+$ was not detected 
(i.e., UGC\,11763, UGC\,2982, and NGC\,5734) were excluded in the analysis. 
The coefficients and p-values are summarized in Table \ref{table:correlation}. 
Correlation plots are shown in Fig.~\ref{figure:correlation_global} 
and Fig.~\ref{figure:correlation_resolved} for the global and 
resolved $L'_\mathrm{HCN}/L'_\mathrm{HCO^+}$, respectively. 
We found no apparent correlations between $L'_\mathrm{HCN}/L'_\mathrm{HCO^+}$ 
and any of the galaxy properties. 

\subsubsection{Uncorrelated AGN strength}
The Spearman test suggests that AGN dominance ($\alpha_\mathrm{AGN}$) 
does not correlate with either the global $L'_\mathrm{HCN}/L'_\mathrm{HCO^+}$ 
or the resolved one (Table \ref{table:correlation}). 
The same analysis for AGN strength itself ($L_\mathrm{AGN}^\mathrm{bol}$) 
also showed no clear correlation with the global ratio nor the resolved one. 
Based on the p-values, we cannot reject the null hypothesis that 
the line ratio and AGN strength are uncorrelated in our sample. 

Whether $L'_\mathrm{HCN}/L'_\mathrm{HCO^+}$ 
correlates with AGN dominance has been a controversial issue
\citep[e.g.,][]{Gracia-Carpio2008,Privon2015,Imanishi2019}. 
Our results are qualitatively consistent with \citet{Privon2020}, for example, who found no correlation between the line ratio and X-ray measurements of AGN. 
The HCN enhancement by X-ray ionization from the AGN, if any, does not seem to
 significantly contribute to elevated $L'_\mathrm{HCN}/L'_\mathrm{HCO^+}$ 
in $\gtrsim$100\,pc apertures when multiple gas components with different 
physical and chemical conditions are laid along the line of sight. 

\subsubsection{Uncorrelated star formation activity}
Based on the Spearman test, we found no evidence for a correlation between 
$L_\mathrm{IR}$ and the global $L'_\mathrm{HCN}/L'_\mathrm{HCO^+}$ 
or between $L_\mathrm{IR}$ and the resolved one (Table \ref{table:correlation}). 
The result is the same as for the SFR corrected for the AGN contribution 
with the global ratio and with the resolved one. 
Although both $L'_\mathrm{HCN}$ and $L'_\mathrm{HCO^+}$ themselves are 
correlated with $L_\mathrm{IR}$ (Sect.~\ref{subsect:Gao-Solomon}), 
$L'_\mathrm{HCN}/L'_\mathrm{HCO^+}$ shows no clear trend 
with $L_\mathrm{IR}$ or the SFR. The variation of 
$L'_\mathrm{HCN}/L'_\mathrm{HCO^+}$ among galaxies on $\gtrsim$100\,pc scales 
cannot be accounted for by the difference in star formation activity. 
This result is consistent with the findings of 
\citet{Tan2018} and \citet{Israel2023}, 
who studied samples including galaxies with lower $L_\mathrm{IR}$. 

\subsubsection{Uncorrelated dust temperature}
The Spearman test showed that the $S_{60}/S_{100}$ color ratio 
does not correlate with the global $L'_\mathrm{HCN}/L'_\mathrm{HCO^+}$ 
or the resolved one (Table \ref{table:correlation}). 
The large p-values suggest that $L'_\mathrm{HCN}/L'_\mathrm{HCO^+}$ 
may be unrelated to the dust temperature on $\gtrsim$100\,pc scales. 
We note that \citet{Tan2018} also found a qualitatively similar result. 

\subsubsection{Marginally correlated N/O abundance ratio} 
According to the Spearman test, the [\ion{N}{ii}]$_{122}$/[\ion{O}{iii}]$_{88}$ 
line ratio may moderately correlate with the global 
$L'_\mathrm{HCN}/L'_\mathrm{HCO^+}$ ($\rho$=$+0.50^{+0.25}_{-0.32}$, 
p-value=$0.10^{+0.42}_{-0.09}$), although the p-value indicates that 
the two quantities may be uncorrelated with a probability of $\sim$10\%. 
The correlation appears less significant when tested with the resolved line 
ratio ($\rho$=$+0.38^{+0.26}_{-0.32}$, p-value=$0.21^{+0.46}_{-0.19}$). 
Considering that the [\ion{N}{ii}]$_{122}$/[\ion{O}{iii}]$_{88}$ line ratio 
is the galaxy-averaged value and both line ratios would be affected 
by the line optical depths, it would be reasonable that the correlation 
is weaker for the resolved $L'_\mathrm{HCN}/L'_\mathrm{HCO^+}$. 

The suggested possible correlation is consistent with the fact that 
in subsolar-metallicity dwarf galaxies, $L'_\mathrm{HCN}/L'_\mathrm{HCO^+}$ 
is smaller than that of solar-metallicity galaxies 
\citep[e.g.,][]{Nishimura2016b,Nishimura2016a,Braine2017}. 
Considering that $L'_\mathrm{HCN}/L'_\mathrm{HCO^+}$ can also be affected by 
the molecular chemistry, depending on the local physical conditions, 
this would imply that we should take into account the different 
elemental N/O abundances for the galaxies with different metallicities 
in order to highlight the peculiar molecular chemistry in specific regions. 
More accurate measurements of elemental abundances with a similar spatial 
resolution to molecular observations would be important for 
a robust and deeper understanding. 

\subsection{Spectrally resolved line ratio}
\label{subsect:resolved}

The challenges in studying the kinematics of galaxies 
using molecular lines lie in limited spatial resolution and 
the faintness of the line emission. The situation has greatly 
improved with the advent of ALMA, as demonstrated by earlier studies 
\citep[e.g.,][]{Garcia-Burillo2014, Martin2015, Saito2018}. 
These studies are, however, often restricted to individual sources 
and may lack sufficient spatial and/or spectral resolution 
to resolve the gas morphology. Our current study stands out 
due to its unique combination of homogeneously high sensitivity 
and a relatively large sample size. Leveraging this advantage, 
our aim is to investigate $L'_\mathrm{HCN}/L'_\mathrm{HCO^+}$ 
in a spatially and spectrally resolved manner. 

To explore the variation of $L'_\mathrm{HCN}/L'_\mathrm{HCO^+}$ 
across different positions and different velocities within each galaxy, 
we generated cubes of $L'_\mathrm{HCN}/L'_\mathrm{HCO^+}$ 
and conducted analyses in both a spectrally integrated manner (pixel-by-pixel analysis) and a spectrally resolved manner (spaxel-by-spaxel analysis). 
Each pixel and spaxel has dimensions of $0.05\arcsec\times0.05\arcsec$ 
and $0.05\arcsec\times0.05\arcsec\times20$ km s$^{-1}$, respectively, 
for all galaxies. 

For a pixel-by-pixel analysis, we initially performed spectral 
integration followed by applying a $3\sigma$ threshold clipping. 
In the spaxel-by-spaxel analysis, we first adopted a $3\sigma$ threshold 
clipping to both HCN and HCO$^+$ cubes 
and then calculated the ratio only for the spaxels where line emission 
from both species is detected with a greater than 3$\sigma$ significance. 
The adoption of the $3\sigma$ threshold clipping for the spaxels was primarily 
to reduce the contamination by the random Gaussian noise. 
We note, however, that faint but real emission contained in spaxels 
with a less than $3\sigma$ significance may be ignored by the clipping. 

\begin{figure*}
\includegraphics[width=0.99\hsize]{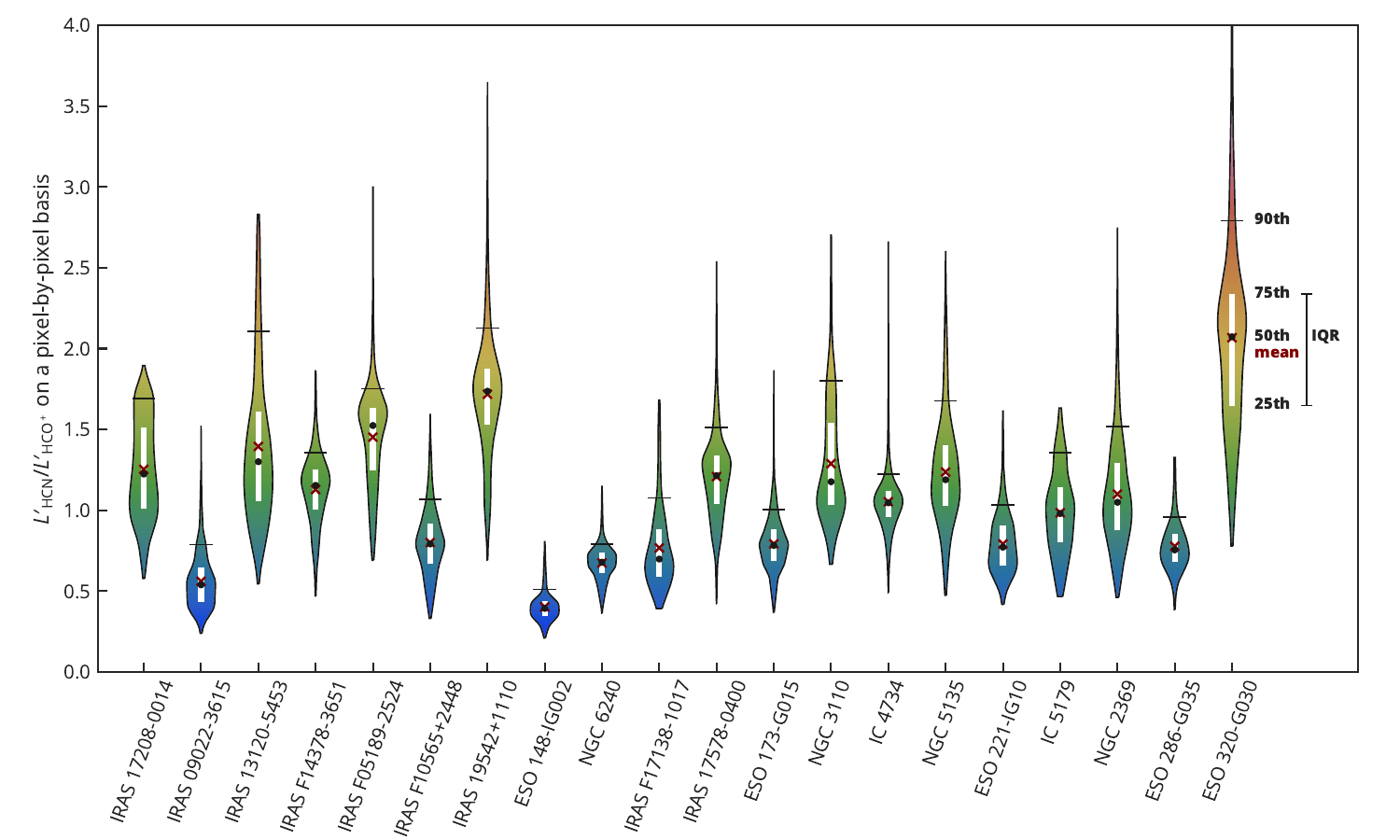}

\includegraphics[width=0.99\hsize]{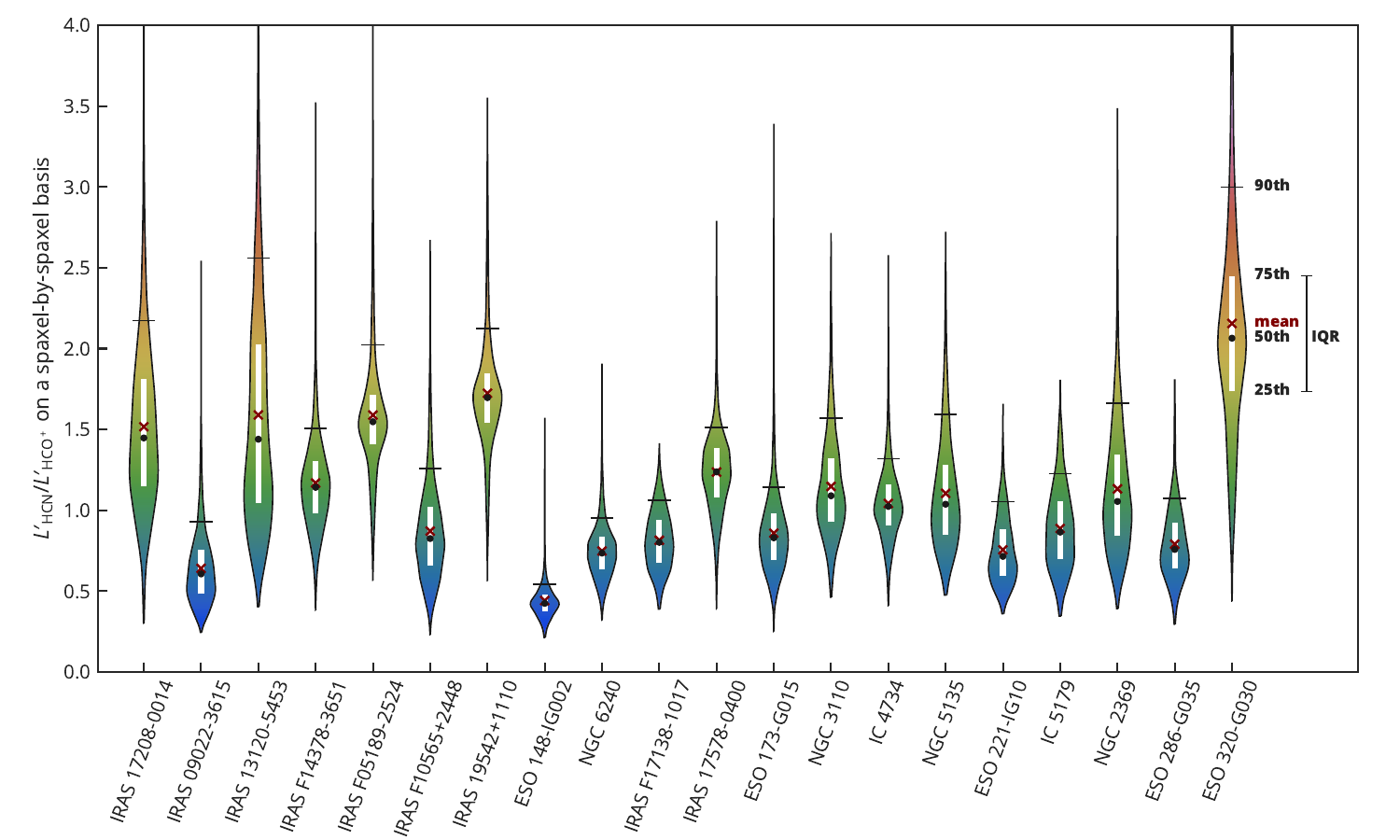}
\caption{Violin plots for $L'_\mathrm{HCN}/L'_\mathrm{HCO^+}$ 
on a pixel-by-pixel basis (\emph{top}) and 
on a spaxel-by-spaxel basis (\emph{bottom}). 
Mean, 50th, and 90th percentiles are denoted by a cross, dot, 
and a short horizontal bar, respectively, 
with the IQR displayed as a white bar. 
Corresponding numerical values are listed in Table~\ref{table:ratio}. 
See Sect.~\ref{subsect:resolved} for details. }
\label{figure:violin}
\end{figure*}

Table~\ref{table:ratio} summarizes the key statistical features, 
including the mean, the 25th-50th-75th and 90th percentiles, and 
the interquartile range for both the pixel-by-pixel and spaxel-by-spaxel analyses. 
Figure~\ref{figure:violin} visualizes the same quantities for 20 galaxies 
significantly detected in both HCN and HCO$^+$ emission 
(i.e., all the sample galaxies excluding UGC\,11763, UGC\,2982, and NGC\,5734). 
In general, the spectrally resolved (spaxel-by-spaxel) ratio 
tends to exhibit a wider range of values than the spectrally integrated 
(pixel-by-pixel) ratio within each galaxy. 
This broadening of the range is most significant and 
toward larger values in eight galaxies 
with known molecular outflows except for NGC\,6240 
(seven in the left-most side and one in the right-most side 
of Fig.\ref{figure:violin}). 

\begin{figure*}
\begin{minipage}{0.3\hsize}
\includegraphics[width=\hsize]{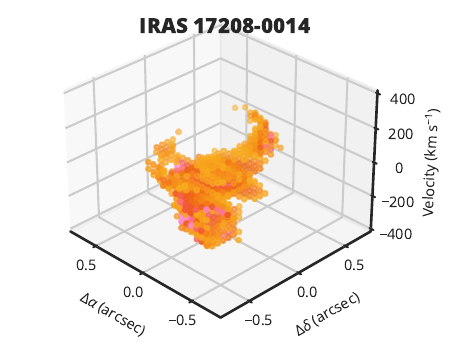}

\includegraphics[width=\hsize]{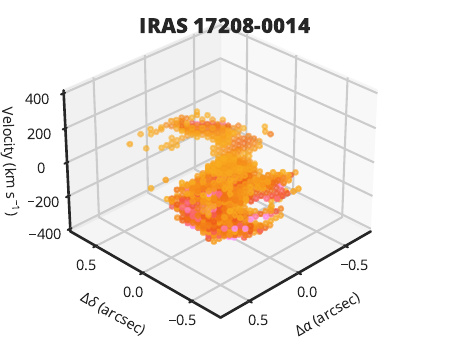}

\includegraphics[width=\hsize]{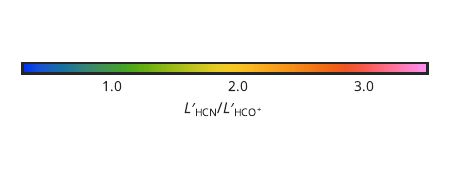}
\end{minipage}
\begin{minipage}{0.17\hsize}
\includegraphics[width=\hsize]{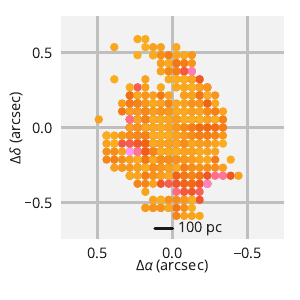}

\includegraphics[width=\hsize]{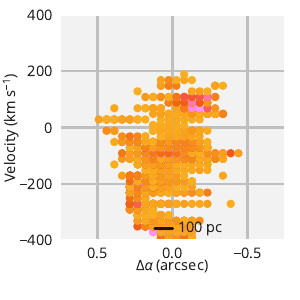}

\includegraphics[width=\hsize]{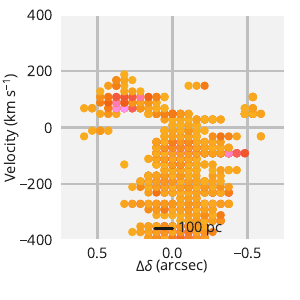}
\end{minipage}
\hspace{0.03\hsize}
\begin{minipage}{0.3\hsize}
\includegraphics[width=\hsize]{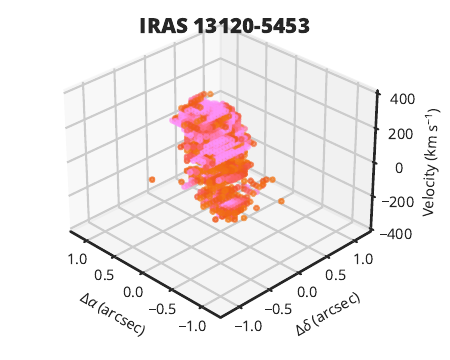}

\includegraphics[width=\hsize]{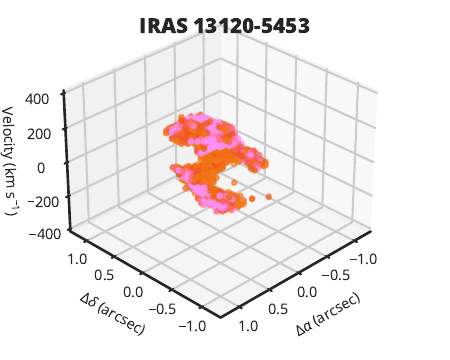}

\includegraphics[width=\hsize]{colorscale_mini.pdf}
\end{minipage}
\begin{minipage}{0.17\hsize}
\includegraphics[width=\hsize]{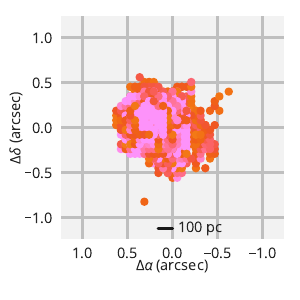}

\includegraphics[width=\hsize]{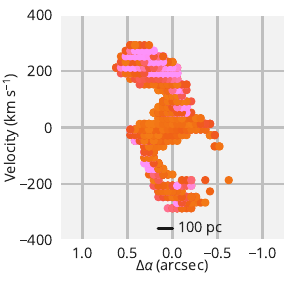}

\includegraphics[width=\hsize]{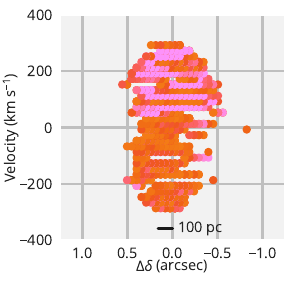}
\end{minipage}
\caption{Visualizations of $L'_\mathrm{HCN}/L'_\mathrm{HCO^+}$ 
exceeding the 90th percentile (see Table~\ref{table:ratio}) 
with projections onto the RA-Dec, RA-velocity, and Dec-velocity planes. 
The two 3D plots represent the same data from different angles. 
Displayed galaxies are IRAS\,17208-0014 and IRAS\,13120-5453. }
\label{figure:bicone}
\end{figure*}

\begin{figure*}
\begin{minipage}{0.3\hsize}
\includegraphics[width=\hsize]{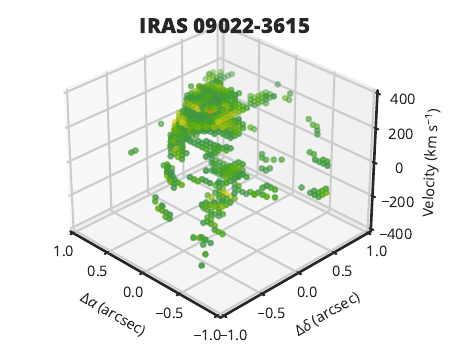}

\includegraphics[width=\hsize]{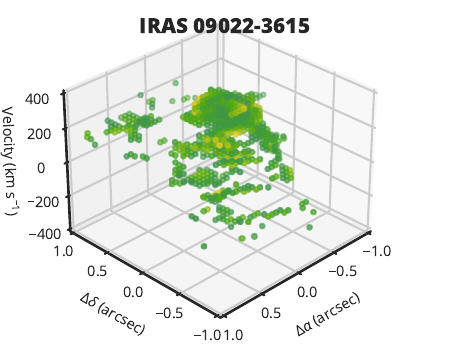}

\includegraphics[width=\hsize]{colorscale_mini.pdf}
\end{minipage}
\begin{minipage}{0.17\hsize}
\includegraphics[width=\hsize]{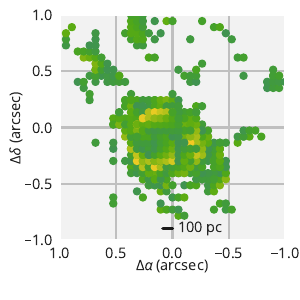}

\includegraphics[width=\hsize]{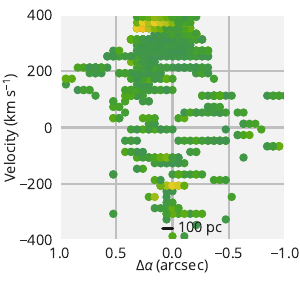}

\includegraphics[width=\hsize]{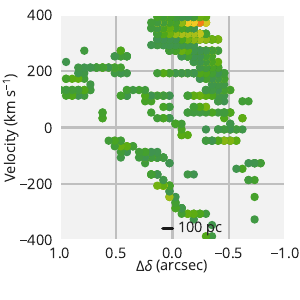}
\end{minipage}
\hspace{0.03\hsize}
\begin{minipage}{0.3\hsize}
\includegraphics[width=\hsize]{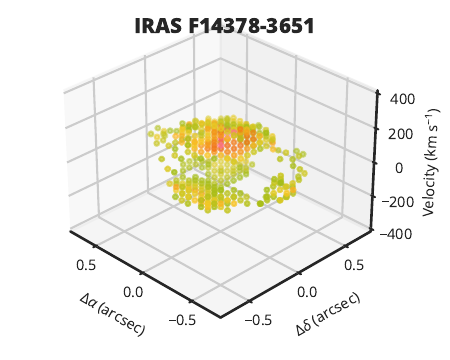}

\includegraphics[width=\hsize]{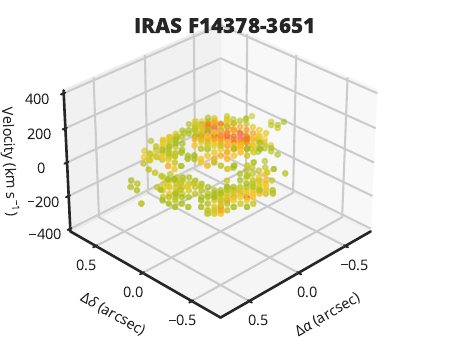}

\includegraphics[width=\hsize]{colorscale_mini.pdf}
\end{minipage}
\begin{minipage}{0.17\hsize}
\includegraphics[width=\hsize]{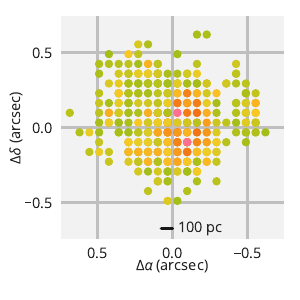}

\includegraphics[width=\hsize]{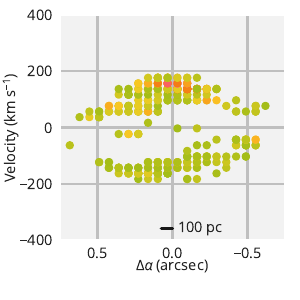}

\includegraphics[width=\hsize]{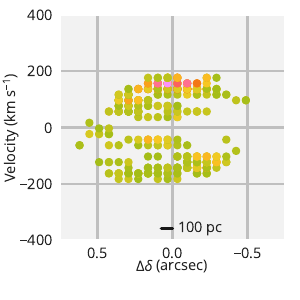}
\end{minipage}
\begin{minipage}{0.3\hsize}
\includegraphics[width=\hsize]{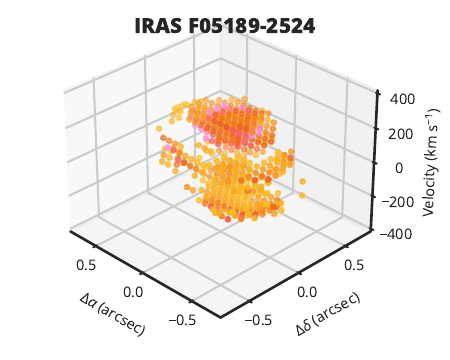}

\includegraphics[width=\hsize]{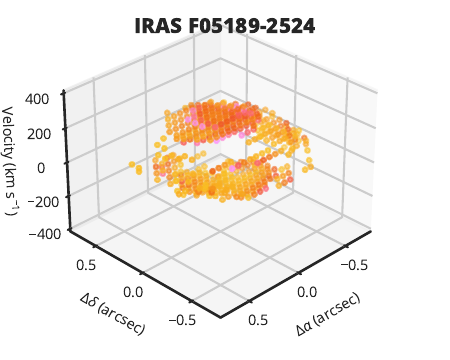}

\includegraphics[width=\hsize]{colorscale_mini.pdf}
\end{minipage}
\begin{minipage}{0.17\hsize}
\includegraphics[width=\hsize]{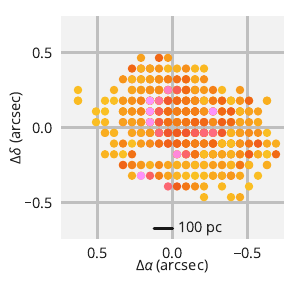}

\includegraphics[width=\hsize]{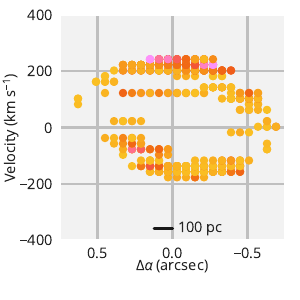}

\includegraphics[width=\hsize]{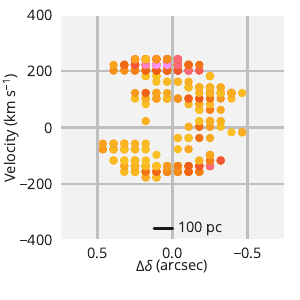}
\end{minipage}
\hspace{0.03\hsize}
\begin{minipage}{0.3\hsize}
\includegraphics[width=\hsize]{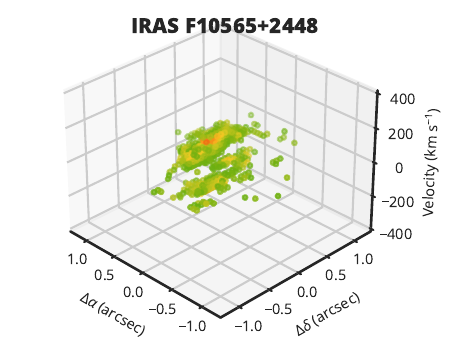}

\includegraphics[width=\hsize]{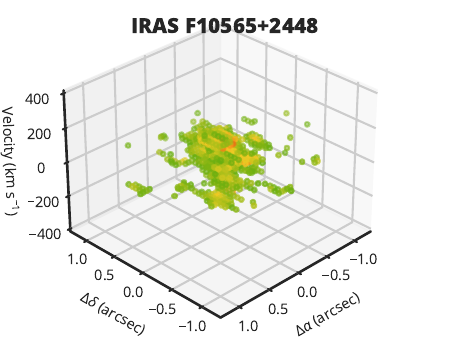}

\includegraphics[width=\hsize]{colorscale_mini.pdf}
\end{minipage}
\begin{minipage}{0.17\hsize}
\includegraphics[width=\hsize]{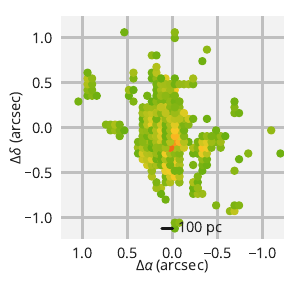}

\includegraphics[width=\hsize]{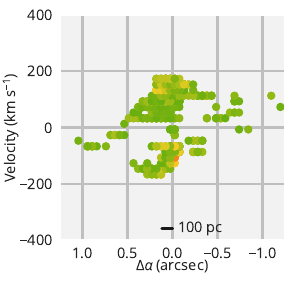}

\includegraphics[width=\hsize]{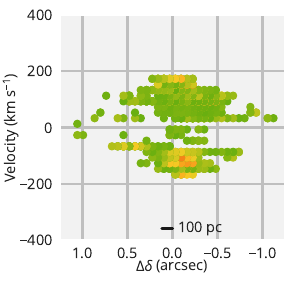}
\end{minipage}
\caption{Same as Fig.~\ref{figure:bicone} but for 
IRAS\,09022-3615, IRAS\,F14378-3651, IRAS\, F05189-2524, and IRAS\,F10565+2448. }
\label{figure:shell-1}
\end{figure*}

\begin{figure*}
\begin{minipage}{0.3\hsize}
\includegraphics[width=\hsize]{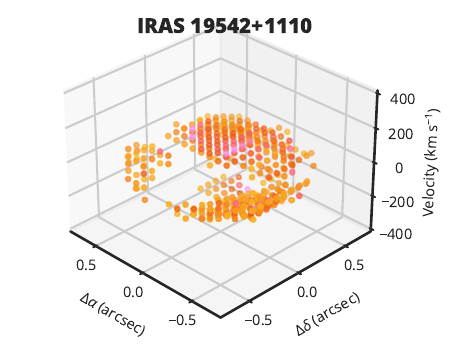}

\includegraphics[width=\hsize]{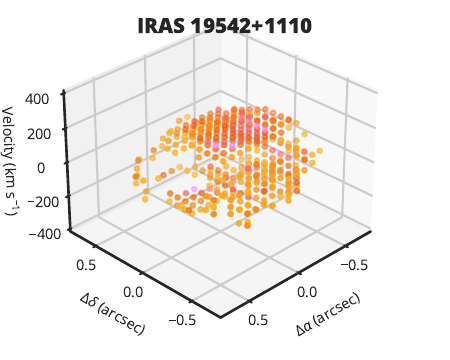}

\includegraphics[width=\hsize]{colorscale_mini.pdf}
\end{minipage}
\begin{minipage}{0.17\hsize}
\includegraphics[width=\hsize]{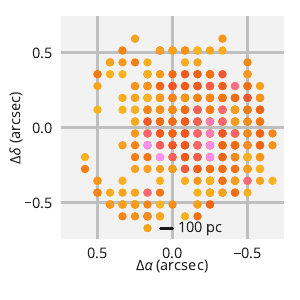}

\includegraphics[width=\hsize]{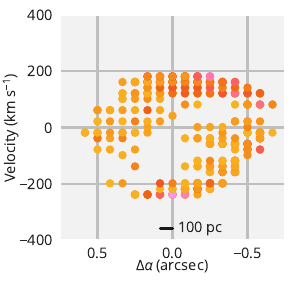}

\includegraphics[width=\hsize]{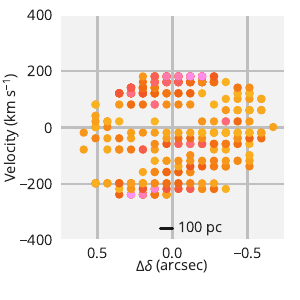}
\end{minipage}
\hspace{0.03\hsize}
\begin{minipage}{0.3\hsize}
\includegraphics[width=\hsize]{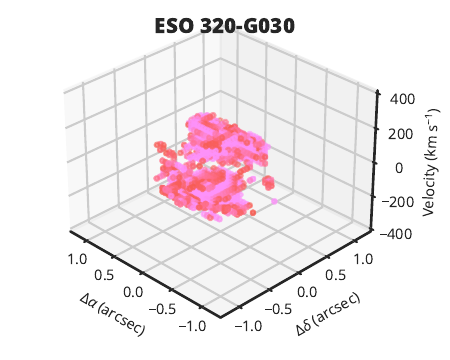}

\includegraphics[width=\hsize]{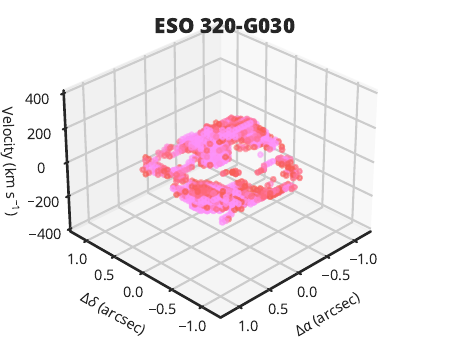}

\includegraphics[width=\hsize]{colorscale_mini.pdf}
\end{minipage}
\begin{minipage}{0.17\hsize}
\includegraphics[width=\hsize]{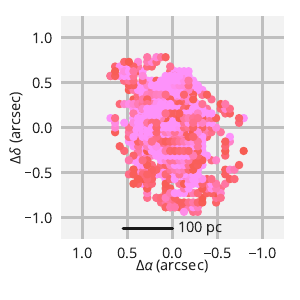}

\includegraphics[width=\hsize]{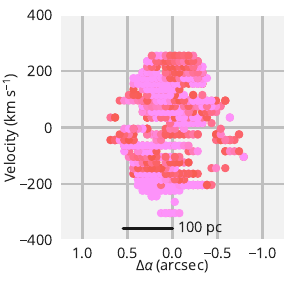}

\includegraphics[width=\hsize]{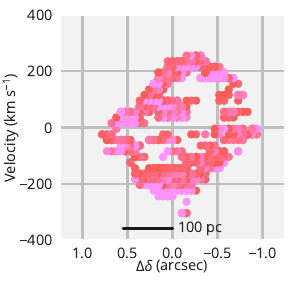}
\end{minipage}
\caption{Same as Fig.~\ref{figure:bicone} 
but for IRAS\,19542+1110 and ESO\,320-G030. }
\label{figure:shell-2}
\end{figure*}

\begin{figure*}
\begin{minipage}{0.3\hsize}
\includegraphics[width=\hsize]{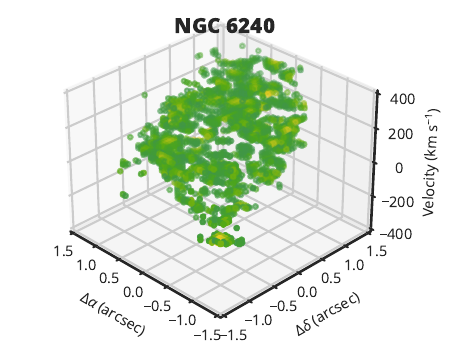}

\includegraphics[width=\hsize]{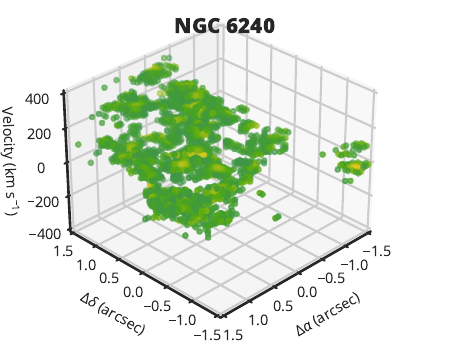}

\includegraphics[width=\hsize]{colorscale_mini.pdf}
\end{minipage}
\begin{minipage}{0.17\hsize}
\includegraphics[width=\hsize]{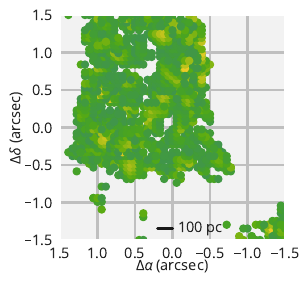}

\includegraphics[width=\hsize]{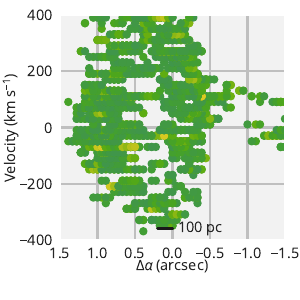}

\includegraphics[width=\hsize]{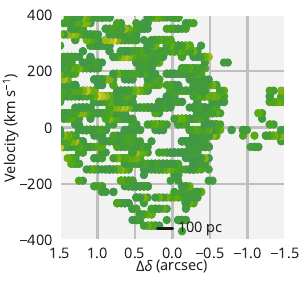}
\end{minipage}
\hspace{0.03\hsize}
\begin{minipage}{0.3\hsize}
\includegraphics[width=\hsize]{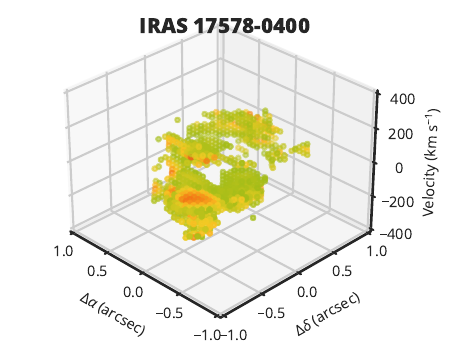}

\includegraphics[width=\hsize]{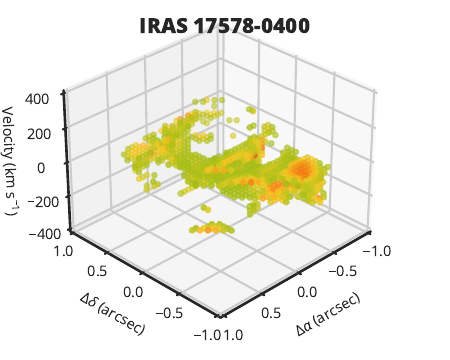}

\includegraphics[width=\hsize]{colorscale_mini.pdf}
\end{minipage}
\begin{minipage}{0.17\hsize}
\includegraphics[width=\hsize]{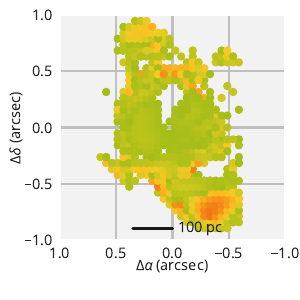}

\includegraphics[width=\hsize]{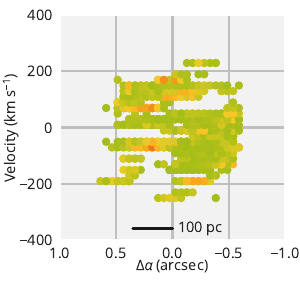}

\includegraphics[width=\hsize]{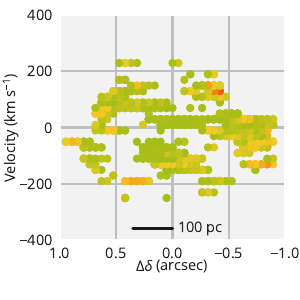}
\end{minipage}
\caption{Same as Fig.~\ref{figure:bicone} 
but for NGC\,6240 and IRAS\,17578-0400. 
We note that the continuum has not been subtracted for IRAS\,17578-0400.
}
\label{figure:random}
\end{figure*}

In some galaxies, the high $L'_\mathrm{HCN}/L'_\mathrm{HCO^+}$ 
regions show spatially and kinematically symmetric structures. 
The characteristic structure clearly emerged 
when we picked out the spaxels with $L'_\mathrm{HCN}/L'_\mathrm{HCO^+}$ 
higher than the 90th percentile from all spaxels 
in both HCN and HCO$^+$ in each galaxy. 
In Figs.~\ref{figure:bicone}--\ref{figure:random}, 
we present visualization of spaxels with $L'_\mathrm{HCN}/L'_\mathrm{HCO^+}$ 
exceeding the 90th percentile in the position-position-velocity (ppV) space 
for 10 galaxies with known molecular out- and/or inflows 
(See Table~\ref{table:sample}). 
Indeed, symmetric morphology can be found in eight galaxies 
shown in Figs.~\ref{figure:bicone}--\ref{figure:shell-2}. 
The morphology can be roughly categorized into two types: 
a filled bicone (IRAS\,17208$-$0014 and IRAS\,13120$-$5453; 
Fig.~\ref{figure:bicone}) and a thin spherical shell 
(IRAS\,09022$-$3615, IRAS\,F14378$-$3651, 
IRAS\,F05189$-$2524, IRAS\,F10565$+$2448, IRAS\,19542$+$1110, 
and ESO\,320$-$G030; Figs.~\ref{figure:shell-1} and \ref{figure:shell-2}). 
For NGC\,6240 and IRAS\,17578-0400 (Fig.~\ref{figure:random}), 
the spaxels with a high line ratio appear randomly in the ppV space, 
likely in part because of the unsubtracted continuum in IRAS\,17578-0400. 
For simplicity, we used the 90th percentile as a threshold 
for all galaxies. Through visual inspection, we found 
the 90th percentile generally produces suitable results to extract 
characteristic structures, as compared to the other neighboring values 
such as the 85th and 95th percentile. 
Understanding the underlying physics that make this threshold effective 
could be an interesting theme for future studies. 

For reference, the ppV plots for the entire set of spaxels 
as well as for those spaxels that exceed the 90th percentile 
within individual galaxies for all galaxies in our sample 
are presented in Appendix~\ref{appendix:ppV}. 
We note that the symmetry is rather distorted, 
but it is marginally seen in ESO\,148$-$IG002 and ESO\,173$-$G015. 
In the other galaxies, such symmetry is not noticeable, 
as we discuss in Appendix~\ref{appendix:notes}. 

\subsection{Relation to outflows and inflows}
\label{subsect:non-circular}

As mentioned in the previous section, symmetric structures 
of high $L'_\mathrm{HCN}/L'_\mathrm{HCO^+}$ are present 
in some of our sample (Figs.~\ref{figure:bicone}--\ref{figure:shell-2}). 
Notably, such symmetric structures are predominantly found in galaxies 
with molecular outflows and/or inflows previously 
found by CO and/or OH line observations. 
For a descriptive comparison with key parameters of out- and inflows 
found in the literature, we encourage readers to refer to 
Appendix~\ref{appendix:notes}. 
If we consider that the high $L'_\mathrm{HCN}/L'_\mathrm{HCO^+}$ regions 
are associated with the gas shocked by the out- and/or inflowing materials, 
plotting $L'_\mathrm{HCN}/L'_\mathrm{HCO^+}$ in the ppV space 
could be a useful method to look for out- and inflows 
and study their geometry independently from kinematic modeling. 

In our sample, symmetrically enhanced HCN is frequently seen in ULIRGs 
and is rare in LIRGs. Our results may suggest that nuclear feeding 
and feedback are predominantly taking place in IR-brighter galaxies. 
However, fast shocks ($v_\mathrm{shock}$\,$\gtrsim$\,20\,km\,s$^{-1}$)
may be required for the prominent enhancement of HCN  
(see Sect.~\ref{subsect:chemistry} for details), 
and thus we cannot rule out the presence of out- and inflows 
with slower shocks or no shock in our sample galaxies. 

\section{Discussion}
\label{sect:discussion}

As presented in Figs.~\ref{figure:violin} and \ref{figure:ppV_whole}, 
$L'_\mathrm{HCN}/L'_\mathrm{HCO^+}$ varies greatly 
from galaxy to galaxy and from position to position 
within each galaxy. In a galaxy-to-galaxy comparison, 
$L'_\mathrm{HCN}/L'_\mathrm{HCO^+}$ shows no clear 
correlation with galaxy properties such as AGN dominance, 
but it might be marginally scaled by the elemental N/O ratio 
(Sect.~\ref{subsect:galproperties}). 
On the other hand, the variation of 
$L'_\mathrm{HCN}/L'_\mathrm{HCO^+}$ in the ppV space of each galaxy 
seems to be related to out- and inflows. 
In this section, we try to figure out the most critical factor 
for the elevation of $L'_\mathrm{HCN}/L'_\mathrm{HCO^+}$ 
in out- and inflows. We discuss $L'_\mathrm{HCN}/L'_\mathrm{HCO^+}$ 
with regard to molecular abundances and excitation conditions, 
by non-LTE radiative transfer calculations 
(Sect.~\ref{subsect:excitation}). 
We also consider the excitation by collision with electrons, 
which could be effective for HCN in a moderately 
dense condition \citep{Goldsmith2017}. 
With constraints on the molecular abundances 
obtained from the non-LTE analysis, we ran chemical models 
and tested if shocks can reproduce the line ratio 
observed in out- and inflows (Sect.~\ref{subsect:chemistry}). 

\subsection{Excitation conditions}
\label{subsect:excitation}

To inspect the molecular abundances and excitation conditions 
for the observed $L'_\mathrm{HCN}/L'_\mathrm{HCO^+}$, 
we made calculations to study the line ratios 
as functions of gas density in various physical conditions. 
Given that both HCN and HCO$^+$ are likely to be 
subthermally excited (see Sect.~\ref{subsect:Gao-Solomon}), 
we adopted the non-local thermal equilibrium (non-LTE) 
radiative transfer model, assuming uniform sphere geometry 
with a large velocity gradient \citep[LVG, see e.g.,][]{Goldreich1974}. 
Practically, we used a publicly available code, 
RADEX\footnote{\url{https://home.strw.leidenuniv.nl/~moldata/radex.html}} 
\citep{vanderTak2007}, to predict the line intensity. 
RADEX requires five input parameters: the background temperature, 
the column density of the molecular species in question, 
the line width, the kinetic temperature, and the H$_2$ density. 
In all of our calculations, the background temperature was fixed 
to be the temperature of the cosmic microwave background (i.e., 2.7\,K).
We note that the background temperature could be higher in the nuclear region, 
but this fixed temperature assumption is acceptable for 
a large part of each outflow when considering that the continuum emission 
is much more compact \citep[e.g.,][]{Pereira-Santaella2021}. 
We implemented the molecular column density 
as a product of the H$_2$ column density ($N_\mathrm{H_2}$) 
and a fractional abundance of the species ($X_\mathrm{mol}$). 
For $N_\mathrm{H_2}$, we considered three plausible values 
 for U/LIRGs: $10^{22}$, $10^{23}$, and $10^{24}$\,cm$^{-2}$. 
We varied $X_\mathrm{HCN}$ to cover a wide range of values: 
$1\times10^{-8}$, $5\times10^{-8}$, $1\times10^{-7}$, 
$3\times10^{-7}$, and $5\times10^{-7}$. On the other hand, 
$X_\mathrm{HCO^+}$ was fixed to the reference value $10^{-8}$. 
The validity of these fractional abundances are discussed 
in Sect.~\ref{subsect:chemistry}. 
The line width ($\Delta v$) was set to be 50\,km\,s$^{-1}$. 
This value is somewhat smaller than the observed values, 
but we can consider the emission as arising 
from an ensemble of such clouds. 
We explored three values for the kinetic temperature 
($T_\mathrm{kin}$): 20, 50, and 100\,K. 
The H$_2$ density ($n_\mathrm{H_2}$) was set to 100 values 
logarithmically spaced in a range from $10^1$\,cm$^{-3}$ 
to $10^9$\,cm$^{-3}$. The collisional rate coefficients 
are taken from \citet{Dumouchel2010} for HCN 
and from \citet{Denis-Alpizar2020} for HCO$^+$. 

As partners for collisional excitation, 
we additionally took into account free electrons. 
This was motivated by the indication from \citet{Goldsmith2017} that 
electron excitation can be more significant for HCN compared to HCO$^+$ 
when the electron fractional abundance is $\gtrsim$$10^{-5}$ and 
the gas density is $\lesssim$$10^{5.5}$\,cm$^{-2}$ \citep{Goldsmith2017}. 
In addition to the case without collision with electrons, 
we explored cases with fractional abundances of electrons 
($X_{e^-}$) of $10^{-5}$ and $10^{-4}$, which could be achieved 
if the cosmic-ray ionization rate is high 
($\zeta$\,$\gtrsim$\,$10^{-15}$\,s$^{-1}$), 
such as in the vicinity of supernova remnants \citep{Ceccarelli2011}. 
This type of high cosmic-ray ionization rate is indeed reported for 
the nearby starburst galaxy NGC\,253 \citep{Harada2021,Holdship2022}. 
We note that the electron abundance was implemented into RADEX calculations 
in the form of a volume density of electrons 
($n_{e^-}$\,$=$\,$n_\mathrm{H_2}X_{e^-}$). 
The collisional rate coefficients for the HCN are taken from \citet{Faure2007}. 
For HCO$^+$, the published collisional rate coefficients are 
only available for $J$$\le$$3$ levels \citep{Faure2001,Singh2021}. 
As pointed out by \citet{Goldsmith2017}, these rate coefficients 
are comparable to those of HCN and consistent with scaling 
with the square of the dipole moment ($\mu^2$). 
We provisionally employed the rate coefficients for $J$$\ge$$3$ 
levels generated by scaling with $\mu^2$ based on the rate coefficients 
for HCN provided by \citet{Faure2007}. For a robust discussion in the future, 
more accurate rate coefficients for HCO$^+$ will be necessary. 

\begin{figure*}
\includegraphics[width=0.990\hsize]{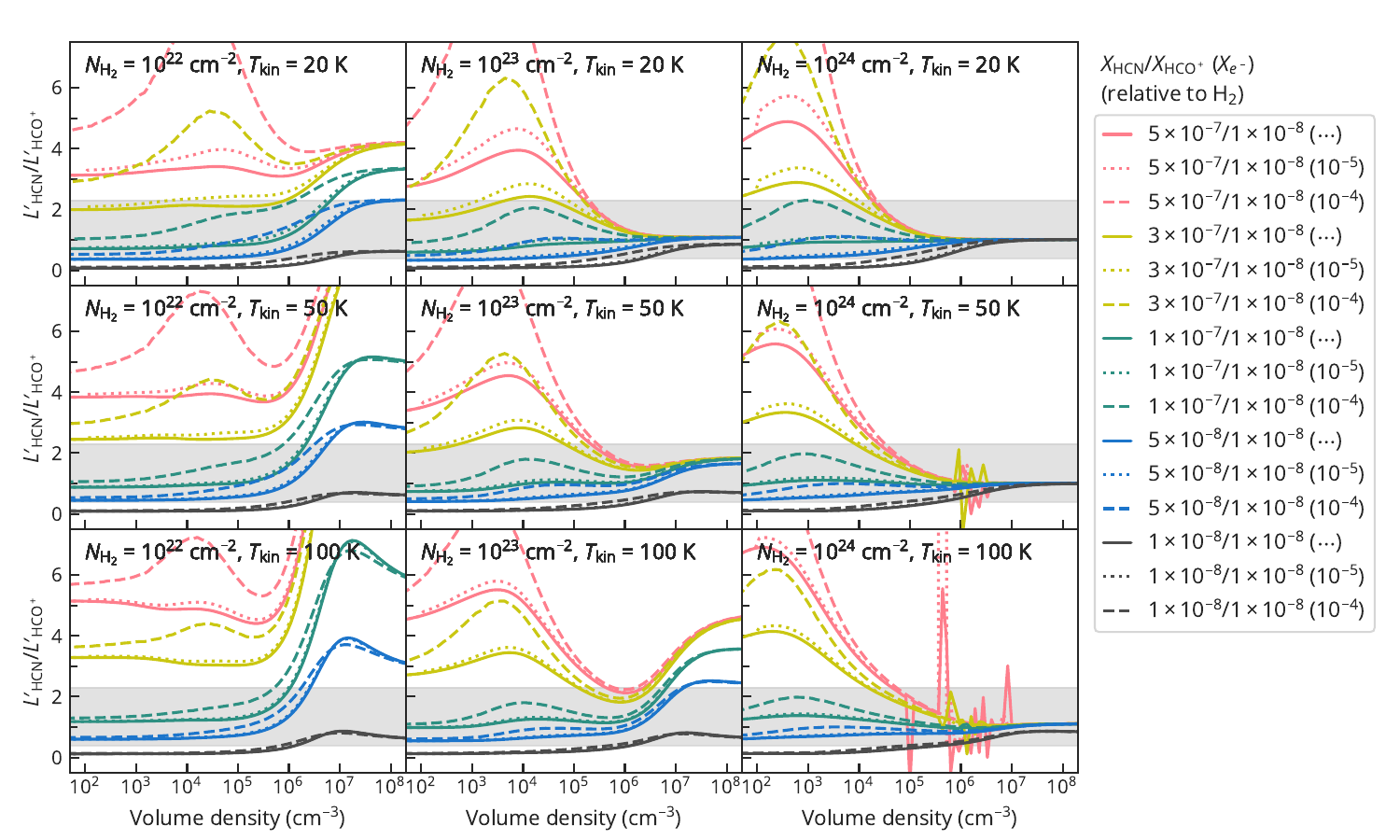}
\caption{One-zone non-LTE calculation of 
$L'_\mathrm{HCN}/L'_\mathrm{HCO^+}$ 
as a function of H$_2$ volume density. 
Gas kinetic temperatures of 20, 50, and 100 K 
and H$_2$ column densities of $10^{22}$, $10^{23}$, 
and $10^{24}$ cm$^{-2}$ were considered. 
The line width was fixed to be 50 km s$^{-1}$ for all models. 
Different color curves represent various 
fractional abundances of HCN and HCO$^+$ as indicated in the legend. 
Solid curves correspond to models without electron excitation, 
while dashed and dotted curves take into account excitation 
by electrons of different fractional abundances. 
Spikes seen in the lower right panels are 
caused by the population inversion of HCN. 
The gray shades indicate the range of 0.4--2.3, corresponding to 
the global $L'_\mathrm{HCN}/L'_\mathrm{HCO^+}$ found in the sample galaxies. 
See Sect.~\ref{subsect:excitation} for details. 
}
\label{figure:RADEX}
\end{figure*}

We ran RADEX with all combinations of parameters described above 
and calculated $L'_\mathrm{HCN}/L'_\mathrm{HCO^+}$ 
as a function of $n_\mathrm{H_2}$. 
The results are plotted in Fig.~\ref{figure:RADEX}. 
Each of the nine panels correspond to a different pairing 
of $N_\mathrm{H_2}$ and $T_\mathrm{kin}$. 
The different colors and line styles in the figure are employed to 
represent the varying values of $X_\mathrm{HCN}$, 
$X_\mathrm{HCO^+}$, and $X_{e^-}$. 
Consistent with such studies as \citet{Butterworth2022} 
and \citet{Imanishi2023}, Fig.~\ref{figure:RADEX} indicates 
that $L'_\mathrm{HCN}/L'_\mathrm{HCO^+}$ increases  
as $X_\mathrm{HCN}/X_\mathrm{HCO^+}$ increases 
for any pairings of $N_\mathrm{H_2}$ and $T_\mathrm{kin}$. 
As also noted in \citet{Yamada2007} and \citet{Izumi2016}, for example,
$X_\mathrm{HCN}/X_\mathrm{HCO^+}$\,$\gtrsim$\,$10$ 
is necessary to get $L'_\mathrm{HCN}/L'_\mathrm{HCO^+}$\,$\gtrsim$\,$1$ 
unless the density is very high ($\gtrsim$$10^6$\,cm$^{-3}$).

The H$_2$ density also plays an important role 
in regulating $L'_\mathrm{HCN}/L'_\mathrm{HCO^+}$. 
In the cases with low $N_\mathrm{H_2}$ 
(left panels in Fig.~\ref{figure:RADEX}), 
$L'_\mathrm{HCN}/L'_\mathrm{HCO^+}$ becomes larger at higher densities. 
This is because HCO$^+$ emission is brightest at densities 
slightly above its effective critical H$_2$ density and 
is a bit less bright at the highest densities, while HCN emission 
behaves similarly but with a larger intensity at a higher density. 
This trend is less pronounced as $N_\mathrm{H_2}$ increases 
because HCN and HCO$^+$ emission becomes similarly bright 
when they are thermalized at high density. 
In cases with $X_\mathrm{HCN}$\,$>$\,$10^{-7}$, 
bumps of $L'_\mathrm{HCN}/L'_\mathrm{HCO^+}$ 
are seen in the density range of $10^2$--$10^5$\,cm$^{-3}$. 
These bumps are attributed to the different effective critical 
density between HCN and HCO$^+$ \citep[for reference: %
$2.5\times10^4$\,cm$^{-3}$ for $N_\mathrm{HCN}$\,=\,$10^{14}$\,cm$^{-2}$ and 
$2.6\times10^3$\,cm$^{-3}$ for $N_\mathrm{HCO^+}$\,=\,$10^{14}$\,cm$^{-2}$, 
both for $T_\mathrm{kin}$\,$=$\,50\,K;][]{Shirley2015}. 
Because of radiative trapping, the effective critical density 
is roughly scaled by the inverse of $N_\mathrm{mol}$. 
As $N_\mathrm{HCN}$ increases, the effective critical density of 
HCN decreases, and hence the bump tends to appear at a lower density 
and becomes more pronounced. 
With $X_\mathrm{HCN}/X_\mathrm{HCO^+}$\,=\,$10$ 
(green curves in Fig.~\ref{figure:RADEX}), 
which results in $L'_\mathrm{HCN}/L'_\mathrm{HCO^+}$ of $\sim$1, 
the effective critical densities of HCN and HCO$^+$ become quite similar, 
and thus the $L'_\mathrm{HCN}/L'_\mathrm{HCO^+}$ curves are 
almost flat around the critical density. 
We also tested different reference abundances of HCO$^+$, 
such as $X_\mathrm{HCO^+}$\,=\,$10^{-7}$, and found qualitatively similar trends. 

Another notable feature is that a contribution from electron excitation 
significantly increases $L'_\mathrm{HCN}/L'_\mathrm{HCO^+}$ 
at moderate H$_2$ densities ($\sim$$10^3$--$10^5$\,cm$^{-3}$). 
The effect is especially noticeable when the electron abundance is highest 
($X_{e^-}$\,$=$\,$10^{-4}$) and the HCN abundance is also high 
($X_\mathrm{HCN}$\,$>$\,$1\times10^{-7}$), 
in which case $L'_\mathrm{HCN}/L'_\mathrm{HCO^+}$ can be more than twice 
as much as the ratio without electron excitation. 

In the case with $N_\mathrm{H_2}$\,$=$\,$10^{24}$\,cm$^{-2}$ and 
$T_\mathrm{kin}$\,$=$\,$50$ and $100$\,K 
(lower two panels on the right side of Fig.~\ref{figure:RADEX}), 
there are spikes at $n_\mathrm{H_2}$\,$\sim$\,$10^5$--$10^7$\,cm$^{-3}$, 
which are caused by population inversion between the $J$=2 and 3 levels of HCN. 
Because these variations are only seen for a very limited range 
of parameters, we consider the influence of such population inversion 
to be negligible in most observations. 

In summary, our one-zone models indicate that 
$L'_\mathrm{HCN}/L'_\mathrm{HCO^+}$ can attain high values ($>$1) 
when $X(\mathrm{HCN})/X(\mathrm{HCO^+})$ exceeds 10 
for a significant range of densities. 
$L'_\mathrm{HCN}/L'_\mathrm{HCO^+}$ will be further enhanced 
if electron abundance is considerably high ($X_{e^-}$\,$=$\,$10^{-4}$).

\subsection{Chemical pathways to the enhancement of HCN}
\label{subsect:chemistry}

Formation and destruction processes of molecules 
can be influenced by physical conditions. 
Galactic outflows and inflows can be distinguished from 
other parts of the galaxy by chemistry induced by shocks. 
Shock heating affects both gas-phase and grain-surface reactions and 
releases species in the icy grain mantles into the gas phase 
\citep[e.g.,][]{Bachiller2001}. Non-thermal desorption, such as 
sputtering, also helps in mantle release \citep[e.g.,][]{Bachiller2001}. 
In addition, outflows can induce turbulence at the interface between 
molecular clouds, enabling a continuous exposure of the gas to 
X-ray, UV, and/or cosmic-ray radiation at the surface. 
This process can lead to an effective ionization and, in turn, 
a refreshment of the molecular composition \citep{Garcia-Burillo2017}. 

The enhancement of HCN abundances in shocked regions is well known 
for protostellar outflows \citep[e.g., L1157;][]{Bachiller1997}, 
and many chemical models have been developed to reproduce 
the observed abundances \citep[e.g.,][]{Burkhardt2019}.  
Those chemical models consider the physical conditions as representative 
of protostellar outflows: the gas density is set to be 
as high as $\sim$$10^5$\,cm$^{-3}$. 
Under such density conditions, HCN abundances in the gas phase 
are enhanced immediately after the passage of shocks 
(post-shock time of $<$$10^2$\,yr) up to $X_\mathrm{HCN}$$\sim$$10^{-5}$ 
due to the release of the ice population by grain heating and sputtering 
\citep[see e.g.~Fig.~4 of][]{Burkhardt2019}. 
This enhancement mechanism of HCN could be applied to 
the case of galactic outflows and could be able to account, 
at least in part, for the observed line ratios in our sample. 
However, given that the majority of the gas is in a density range of 
$\sim$$10^3$--$10^4$\,cm$^{-3}$ in the beam of extragalactic observations, 
it is essential to consider the chemistry in the moderately-dense regime. 

To examine how shocks affect the chemistry in the moderately-dense gas 
($\sim$$10^3$--$10^4$\,cm$^{-3}$), we experimentally ran 
the time-dependent gas-grain chemical code, 
UCLCHEM\footnote{\url{https://uclchem.github.io}} \citep{Holdship2017}, 
which is publicly available. 
We adopted the ``C-shock'' model of UCLCHEM, which is based on 
the parameterization of C-type shocks by \citet{Jimenez-Serra2008}. 
For details, we refer to \citet{Viti2014}, \citet{Garcia-Burillo2017}, 
and \citet{Holdship2017}, where models similar to ours were computed. 
Here, we provide a concise overview of the most critical aspects. 
The model was run in two phases: the pre-shock and post-shock phases. 
As a pre-shock condition, we considered the typical condition of 
``standard'' molecular gas with Galactic elemental abundances. 
The pre-shock abundances follow the method employed 
by \citet{Harada2019}: all physical conditions were kept constant 
at standard Galactic values (H$_2$ density $10^3$\,cm$^{-3}$, 
temperature 10\,K, visual extinction 2\,mag, radiation field 1 Habing, 
cosmic ionization rate $1.3\times10^{-17}$\,s$^{-1}$; 
we note that these values are provisional and should be 
better calibrated for U/LIRGs in future works). 
The evolution of the molecular abundances in the gas phase and 
on the grain surfaces were calculated as a function of time using rate equations. 
The chemical evolution was followed for $10^5$\,yr in the pre-shock phase 
to reproduce realistic molecular abundances observed 
on cloud scales \citep[$>$10\,pc;][]{Harada2019}. 
In the post-shock phase, we tested four values of shock velocity: 
10, 20, 30, and 40\,km\,s$^{-1}$. During this phase, 
the density, temperature, and velocity profile 
were varied as a function of time, following the 
parameterizations of the shock model. 
Our model maintained constant values for the radiation field 
and the cosmic-ray ionization rate, but it is worth noting that 
these parameters would also fluctuate in a real molecular cloud.

\begin{figure}
\includegraphics[width=\hsize]{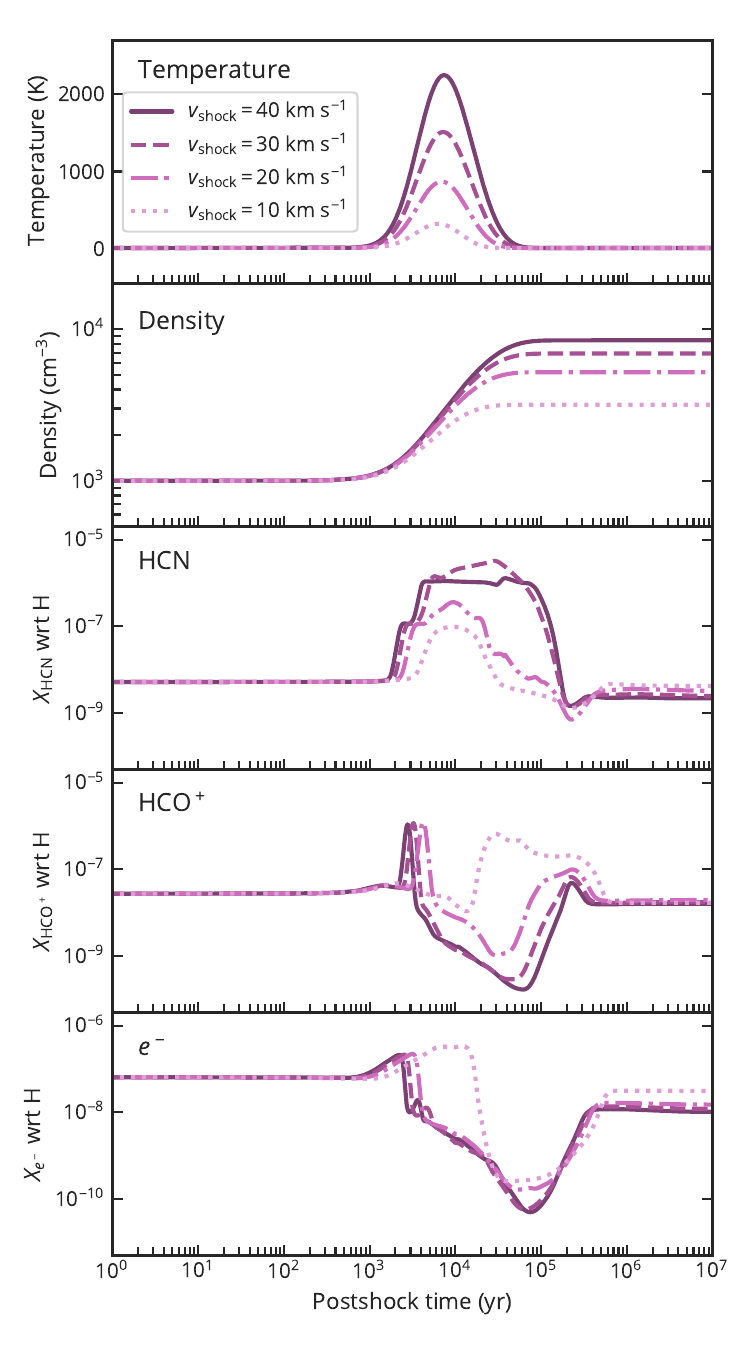}
\caption{Chemical model predictions to simulate 
gas temperature, density, and fractional abundances 
of HCN, HCO$^+$, and electrons under moderately-dense conditions 
($n_\mathrm{H_2}$$\sim$$10^3$--$10^4$\,cm$^{-3}$). 
C-type shocks with shock velocities of 
10, 20, 30, and 40\,km\,s$^{-1}$ are considered. 
The pre-shock H$_2$ density is assumed to be $10^3$\,cm$^{-2}$ for all cases. 
More details on the other assumed parameters are described 
in Sect.~\ref{subsect:chemistry}. 
}
\label{figure:UCLCHEM}
\end{figure}

In Fig.~\ref{figure:UCLCHEM}, the evolution of the gas temperature 
and density as well as the abundances of HCN, HCO$^+$, and the electrons
are shown as a function of post-shock time. 
As shown in Fig.~\ref{figure:UCLCHEM}, the HCN abundance is 
enhanced by a factor of $\sim$10--100 at $\sim$$10^3$--$10^5$\,yr 
after the shock, depending on the shock velocity. 
On the other hand, the HCO$^+$ and electron abundances are decreased 
during the HCN enhancement. It appears that the HCN abundance 
falls back to the pre-shock abundance at a post-shock time of $\sim$$10^5$\,yr. 
This timescale is roughly comparable to the dynamical time of outflows 
(calculated as $R_\mathrm{out}/v_\mathrm{out}$). 

In our model, unlike the chemistry under the high density conditions, 
thermal desorption or sputtering do not contribute to the HCN enhancement 
because HCN ice is almost absent in pre-shock conditions. 
Instead, the enhancement of HCN is mainly due to 
the gas-phase production via the high-temperature reaction: 
$\mathrm{CN} + \mathrm{H_2} \rightarrow \mathrm{HCN} + \mathrm{H}$. 
Because the rate coefficient of this reaction is a function of temperature 
\citep[$k = \alpha(T/300\,\mathrm{K})^{\beta} \exp(-\gamma/T)$ %
where $\alpha=4\times10^{-13}$, $\beta=2.87$, $\gamma=820$\,K, %
and $T$ is temperature;][]{Harada2010}, 
the degree of HCN enhancement depends on the heating. 
Consequently, the shock velocity affects the degree of HCN enhancement. 
As shown in Fig.~\ref{figure:UCLCHEM}, shock velocities of 
30 and 40\,km\,s$^{-1}$ can reproduce our HCN and HCO$^+$ abundances, 
accounting for the observed line ratio, while 10 and 20\,km\,s$^{-1}$ 
shocks seem to be insufficient. 
As mentioned in Appendix~\ref{appendix:notes}, the HCN enhancement 
was not found in NGC\,6240, and it might be because the shock velocity 
is too small to enhance the HCN abundance \citep{Meijerink2013}. 
However, considering the results of a high-resolution study 
in the CO $J$$=$2--1 line, which highlighted the existence of 
very high-velocity outflows \citep{Saito2018}, NGC\,6240 deserves 
further investigation. 

The key reaction 
($\mathrm{CN} + \mathrm{H_2} \rightarrow \mathrm{HCN} + \mathrm{H}$) 
suggests that HCN abundance increases at the expense of CN abundance. 
Considering that Mrk\,231 has an outflow that is notably bright 
in both CN \citep{Cicone2020} and HCN \citep{Aalto2015a,Lindberg2016}, 
it is likely that CN abundance is maintained by the high ionization 
rate in the outflow. In environments 
where C$^+$ is abundant, CN can be rapidly formed. A similar rationale can be applied to 
the CCH-bright outflow in NGC\,1068, 
as discussed in \citet{Garcia-Burillo2017}. 
We also note that CN emission is less affected by electron excitation 
than HCN \citep{Goldsmith2017}. This would provide an additional 
support for the enhancement of HCN abundance in outflows. 

\section{Summary and conclusions}
\label{sect:conclusions}

We conducted observations of the HCN and HCO$^+$ $J$$=$3--2 lines 
toward 23 U/LIRGs in the local Universe ($z<0.07$). 
The spatial resolution was $\sim$$0.3\arcsec$, which corresponds 
to $\sim$50--400\,pc at the distance of each target. 
The obtained dataset allowed us to explore 
the HCN/HCO$^+$ line ratio ($L'_\mathrm{HCN}/L'_\mathrm{HCO^+}$) 
of each galaxy in a spatially and spectrally resolved manner. 
We analyzed $L'_\mathrm{HCN}/L'_\mathrm{HCO^+}$ in relation to 
the galaxy properties such as AGN dominance and star formation activity, and 
we found the presence of outflows and/or inflows to have the most crucial influence. 
Non-LTE radiative transfer analysis and chemical models suggest that a
higher HCN abundance in shocked regions is likely to be responsible 
for a high $L'_\mathrm{HCN}/L'_\mathrm{HCO^+}$ in out- and inflows. 

The main results and key points from the analysis are the following: 
\begin{itemize}
\item \emph{Spectra and moment maps}. 
The line emission of HCN and HCO$^+$ was successfully detected 
with a signal-to-noise ratio of $>$5\,$\sigma$ in 21 out of the 23 galaxies. 
The line profiles of HCN and HCO$^+$ are often complex 
and cannot be fitted by a single Gaussian. 
The HCN and HCO$^+$ emission is mostly emitted 
from a central region of each galaxy. 
The apparent size of the line-emitting region is 
typically 1--3\,kpc in ULIRGs and 0.5--1\,kpc in LIRGs. 
In almost all the galaxies with known molecular outflows, 
the velocity dispersion of the HCN emission is higher than that of HCO$^+$. 

\item \emph{Correlation between line luminosity and infrared luminosity}. 
We found $L'_\mathrm{HCN}$ and $L'_\mathrm{HCO^+}$ correlate well with 
$L_\mathrm{IR}$ (correlation coefficients of 0.85 and 0.90, respectively). 
These relations are quantitatively consistent with 
previous observations and model predictions. 
The $L'_\mathrm{HCN}$--$L_\mathrm{IR}$ and 
$L'_\mathrm{HCO^+}$--$L_\mathrm{IR}$ slopes of 
$\sim$0.5--0.6 suggest that HCN and HCO$^+$ are ubiquitously 
distributed in the molecular gas with a wide range of densities, 
and most of them are subthermally excited. 

\item \emph{Line ratio and its correlation with galaxy properties}.
The velocity-integrated $L'_\mathrm{HCN}/L'_\mathrm{HCO^+}$ 
extracted from a large aperture ($\sim$$3\arcsec$) in each galaxy 
is in the range of 0.4--2.3. The unweighted mean is 1.1. 
Although the HCN/HCO$^+$ has been proposed as AGN/starburst 
diagnostics in the literature, we found no evidence for 
correlations of $L'_\mathrm{HCN}/L'_\mathrm{HCO^+}$ 
with AGN dominance, star formation activity, or dust temperature. 
The elemental N/O ratio potentially moderately correlates with 
$L'_\mathrm{HCN}/L'_\mathrm{HCO^+}$, 
although p-values suggest that those quantities may be 
uncorrelated, with a probability of $\sim$10\%. 

\item \emph{Spatially and spectrally resolved line ratios}. 
We explored the variation of $L'_\mathrm{HCN}/L'_\mathrm{HCO^+}$ 
in the ppV space of each galaxy. 
As a result, $L'_\mathrm{HCN}/L'_\mathrm{HCO^+}$ considerably 
varies from spaxel to spaxel, even within a galaxy. 
We found spatially and kinematically symmetric structures 
of high line ratio (the highest 10th percentile on a 
spaxel-by-spaxel basis) in some galaxies. 
Such structures are predominantly found in galaxies with 
molecular outflows previously found by CO and/or OH line observations. 

\item \emph{Non-LTE radiative transfer analysis}. 
One-zone non-LTE radiative transfer calculations were made 
to investigate the excitation conditions and molecular abundances 
in the gas phase in our sample galaxies. 
They revealed that $L'_\mathrm{HCN}/L'_\mathrm{HCO^+}$ 
depends on many parameters, such as the HCN/HCO$^+$ abundance ratio, 
H$_2$ volume density, and electron abundances. 
Taking all factors into consideration, the observed high 
$L'_\mathrm{HCN}/L'_\mathrm{HCO^+}$ ({$>$1}) in outflows and inflows 
are essentially caused by a high HCN/HCO$^+$ abundance ratio, 
and this can be further enhanced by electrons. 

\item \emph{Key chemical process to enhance HCN abundance}. 
Chemistry induced by shocks characterizes the molecular 
abundances in outflows and inflows. As reported by previous studies, 
thermal and non-thermal release of HCN ice mostly 
contributes to the enhancement of HCN in the gas phase 
under high density conditions ($\sim$$10^5$\,cm$^{-3}$). 
In the moderately-dense gas, which accounts for a larger fraction 
of the total gas, the HCN abundance is mainly enhanced 
by a high-temperature gas-phase reaction: 
$\mathrm{CN} + \mathrm{H_2} \rightarrow \mathrm{HCN} + \mathrm{H}$. 
The post-shock time of HCN enhancement is roughly comparable 
to the dynamical time of the outflow ($\sim$$10^5$\,yr). 

\end{itemize}

Finally, we emphasize the advantages of analyzing 
$L'_\mathrm{HCN}/L'_\mathrm{HCO^+}$ in a spatially and spectrally 
resolved manner. Considering that the mass of the regions 
influenced by outflows and/or inflows is smaller than the total mass 
of the galaxy, spatial or spectral integration may make it difficult 
to distinguish shocked regions from others. 
Our results imply that the spatially and spectrally resolved 
$L'_\mathrm{HCN}/L'_\mathrm{HCO^+}$ helps identify
the geometry of outflows and inflows, at least in some cases,
when the shock velocity of the outflows 
is sufficiently high to heat the gas. 
To develop line diagnostics for the characterization of physical 
properties of U/LIRGs, further exploration of molecular compositions 
and more detailed modeling of the kinematic structure of a galaxy 
should be investigated in future studies. 

\begin{acknowledgements}
We thank the anonymous referee for a constructive report 
that significantly contributed to the improvement of this paper. 

We thank Niklas Falstad for valuable discussions 
and for the earlier work on CON-quest.

YN gratefully acknowledges support by JSPS KAKENHI grant No.~JP23K13140, 
JP18K13577, and NAOJ ALMA Scientific Research Grant No.~2017-06B. 

SA and SK gratefully acknowledge funding from the European Research 
Council (ERC) under the European Union's Horizon 2020 research 
and innovation programme (grant agreement No 789410). 

SA and MG also acknowledge support from Swedish Research Council grant 621-2011-4143. 

J.S.G. thanks the University of Wisconsin College of Letters and Science for partial support of this research. 

E.G-A thanks the Spanish MICINN for support under projects PID2019-105552RB-C41 and PID2022-137779OB-C41. 

KK acknowledges the support by JSPS KAKENHI Grant Number JP17H06130 
and the NAOJ ALMA Scientific Research Grant No.~2017-06B. 

M.I. is supported by JSPS KAKENHI grant No.~JP21K03632. 

CR acknowledges support from Fondecyt Regular grant 1230345 
and ANID BASAL project FB210003. 

This paper makes use of the following ALMA data: 
ADS/JAO.ALMA\#2017.1.00759.S and ADS/JAO.ALMA\#2018.1.01344.S. 
ALMA is a partnership of ESO (representing its member states), 
NSF (USA) and NINS (Japan), together with NRC (Canada), 
MOST and ASIAA (Taiwan), and KASI (Republic of Korea), 
in cooperation with the Republic of Chile. 
The Joint ALMA Observatory is operated by ESO, AUI/NRAO and NAOJ. 
\end{acknowledgements}

\clearpage

\begin{table*}
\caption{Main properties of the sample galaxies.}
\label{table:sample}
\centering 
\begin{tabular}{clcccccccc}
\hline\hline
ID & Galaxy & $z$ & $D_L$ & $L_\mathrm{IR}$ & $\alpha_\mathrm{AGN}$ & Merger & CON & Outflow & Inflow \\
 & & & (Mpc) & ($10^{11}L_\sun$) & & stage & & & \\
(1) & (2) & (3) & (4) & (5) & (6) & (7) & (8) & (9) & (10) \\
\hline
 1 & IRAS\,17208$-$0014  & 0.0428 & $183 \pm12 $ & $25 \pm3  $ & $0.05\pm0.03$ & d & \Checkmark & i & $\alpha$ \\
 2 & IRAS\,09022$-$3615  & 0.0596 & $253 \pm17 $ & $18 \pm2  $ & $0.19\pm0.06$ & d & & vi & \\
 3 & IRAS\,13120$-$5453  & 0.0311 & $135 \pm9  $ & $18 \pm2  $ & $0.04\pm0.02$ & d & & ii, iii, vi, vii & \\
 4 & IRAS\,F14378$-$3651 & 0.0682 & $295 \pm20 $ & $14 \pm2  $ & $0.05\pm0.02$ & d & & vi, vii & \\
 5 & IRAS\,F05189$-$2524\tablefootmark{*} & 0.0427 & $176 \pm12 $ & $13 \pm2  $ & $0.60\pm0.07$ & d & & ii, vi, vii & \\
 6 & IRAS\,F10565$+$2448\tablefootmark{*} & 0.0431 & $184 \pm12 $ & $11 \pm1  $ & $0.04\pm0.01$ & d & & iv, vi, vii & \\
 7 & IRAS\,19542$+$1110\tablefootmark{*} & 0.0626 & $264 \pm18 $ & $10 \pm1  $ & $0.08\pm0.04$ & N & & vi & \\
 8 & ESO\,148$-$IG002\tablefootmark{*} & 0.0446 & $185 \pm12 $ & $10 \pm1  $ & $0.19\pm0.03$ & c & & & \\
 9 & NGC\,6240\tablefootmark{*} & 0.0243 & $106 \pm7  $ & $7  \pm1  $ & $0.11\pm0.02$ & d & & iv, vi & \\
10 & IRAS\,F17138$-$1017 & 0.0173 & $77.3\pm5.2$ & $2.6\pm0.3$ & $0.07\pm0.04$ & d & & & \\
11 & IRAS\,17578$-$0400  & 0.0134 & $60.0\pm4.1$ & $2.3\pm0.3$ & $0.03\pm0.02$ & b & \Checkmark & & $\beta$ \\
12 & ESO\,173$-$G015     & 0.0100 & $32.7\pm2.3$ & $2.2\pm0.3$ & $0.03\pm0.02$ & N & & & \\
13 & UGC\,11763\tablefootmark{*} & 0.0633 & $265 \pm18 $ & $2.2\pm0.5$ & $\cdots$      & N & & & \\
14 & NGC\,3110         & 0.0169 & $74.3\pm5.0$ & $2.0\pm0.3$ & $0.10\pm0.06$ & a & & & \\
15 & IC\,4734          & 0.0156 & $67.8\pm4.5$ & $1.9\pm0.3$ & $0.07\pm0.02$ & N & & & \\
16 & NGC\,5135         & 0.0137 & $52.6\pm3.5$ & $1.5\pm0.2$ & $0.24\pm0.06$ & N & & & \\
17 & ESO\,221$-$IG10     & 0.0105 & $59  \pm11 $ & $1.5\pm0.6$ & $0.07\pm0.04$ & N & & & \\
18 & IC\,5179          & 0.0113 & $47.3\pm3.2$ & $1.5\pm0.2$ & $0.10\pm0.07$ & N & & & \\
19 & UGC\,2982         & 0.0177 & $70.6\pm4.7$ & $1.4\pm0.2$ & $0.11\pm0.08$ & d & & & \\
20 & NGC\,2369         & 0.0108 & $44.6\pm3.0$ & $1.3\pm0.2$ & $0.10\pm0.06$ & N & & & \\
21 & ESO\,286$-$G035     & 0.0174 & $73.2\pm4.9$ & $1.3\pm0.2$ & $0.06\pm0.04$ & a & & & \\
22 & ESO\,320$-$G030     & 0.0103 & $36  \pm2.5$ & $1.1\pm0.2$ & $0.03\pm0.02$ & N & \Checkmark & v & $\gamma$ \\
23 & NGC\,5734         & 0.0136 & $60.2\pm4.0$ & $1.1\pm0.2$ & $0.14\pm0.07$ & a & & & \\
\hline
\end{tabular}
\tablefoot{
\tablefoottext{*}{Not included in the sample of \citetalias{Falstad2021}.}
(1) Identification number used in this paper. (2) Galaxy name. (3) Redshift. 
(4) Luminosity distance calculated from redshift 
following the same procedure as \citet{Sanders2003}. 
(5) Infrared luminosity based on IRAS fluxes 
taken from \citet{Sanders2003} and calculated with the same methods used by \citet{Sanders1996}. 
Only for UGC\,11763 are fluxes taken from \citet{Sanders1989}. 
(6) Bolometric AGN fraction ($\alpha_\mathrm{AGN}=L_\mathrm{AGN}/L_\mathrm{bol}$) 
calculated from mid-infrared diagnostics by \citet{Diaz-Santos2017}. See Sect.~\ref{sect:sample} for detail. 
(7) Merger stage visually determined by \citet{Stierwalt2013}. 
N = nonmerger, a = pre-merger, b = early-stage merger, c = mid-stage merger, and d = late-stage merger. 
(8) Presence of CON found by HCN-vib observations. Details are presented in \citetalias{Falstad2021}. 
(9) Presence of molecular outflow found by CO and/or OH observations. 
References are (i) \citet{Garcia-Burillo2015}; (ii) \citet{Fluetsch2019}; (iii) \citet{Lutz2020}; 
(iv) \citet{Cicone2014}; and (v) \citet{Pereira-Santaella2016} for CO observations 
and (vi) \citet{Veilleux2013} and (vii) \citet{Gonzalez-Alfonso2017} for OH observations. 
(10) Presence of molecular inflow found by CO and/or OH observations. 
References are ($\alpha$) \citet{Veilleux2013} and ($\beta$) \citet{Falstad2021} for OH observations 
and ($\gamma$) \citet{Gonzalez-Alfonso2021} for CO and OH observations. }
\end{table*}

\begin{table*}
\caption{Parameters of ALMA observations and adopted kinematics for analyses.}
\label{table:observations}
\centering 
\begin{tabular}{clcccccccc}
\hline\hline
 & & \multicolumn{3}{c}{ALMA observations} && \multicolumn{4}{c}{Adopted kinematics} \\
\cline{3-5}\cline{7-10}
ID & Galaxy & Beam (PA) & Resolution & Sensitivity && 
$\alpha^\mathrm{center}_\mathrm{J2000}$ & $\delta^\mathrm{center}_\mathrm{J2000}$ & PA & $v_\mathrm{sys}$ \\
 & & ($\arcsec\times\arcsec$ ($\degr$)) & (pc$\times$pc) & (mJy\,beam$^{-1}$) 
 && (h:m:s) & ($\degr$:$\arcmin$:$\arcsec$) & ($\degr$) & (km\,s$^{-1}$) \\
\hline
 1 & IRAS\,17208$-$0014  & $0.35\times0.32$ ($74 $) & $298\times272$ & 0.680 && 17:23:21.96 & $-$00:17:00.9 & 105 & 12321 \\
 2 & IRAS\,09022$-$3615  & $0.33\times0.28$ ($64 $) & $382\times324$ & 0.145 && 09:04:12.71 & $-$36:27:01.9 &  10 & 16847 \\
 3 & IRAS\,13120$-$5453  & $0.36\times0.32$ ($18 $) & $225\times200$ & 0.730 && 13:15:06.32 & $-$55:09:22.8 &  95 & 9037  \\
 4 & IRAS\,F14378$-$3651 & $0.29\times0.26$ ($66 $) & $388\times347$ & 0.194 && 14:40:59.01 & $-$37:04:31.9 &  30 & 19143 \\
 5 & IRAS\,F05189$-$2524 & $0.42\times0.27$ ($81 $) & $343\times220$ & 0.244 && 05:21:01.40 & $-$25:21:45.3 &  90 & 12257 \\
 6 & IRAS\,F10565$+$2448 & $0.33\times0.22$ ($-1 $) & $281\times187$ & 0.227 && 10:59:18.13 & $+$24:32:34.5 &  90 & 12387 \\
 7 & IRAS\,19542$+$1110  & $0.34\times0.29$ ($-53$) & $423\times360$ & 0.175 && 19:56:35.78 & $+$11:19:05.0 &  45 & 17678 \\
 8 & ESO\,148$-$IG002    & $0.40\times0.31$ ($31 $) & $344\times267$ & 0.236 && 23:15:46.75 & $-$59:03:15.5 &   0 & 12793 \\
 9 & NGC\,6240         & $0.32\times0.26$ ($58 $) & $161\times130$ & 0.281 && 16:52:58.89 & $+$02:24:03.4 &  40 & 7129  \\
10 & IRAS\,F17138-1017 & $0.38\times0.29$ ($-76$) & $139\times106$ & 0.576 && 17:16:35.81 & $-$10:20:39.0 &   5 & 5113  \\
11 & IRAS\,17578$-$0400  & $0.36\times0.29$ ($-65$) & $102\times 82$ & 0.512 && 18:00:31.84 & $-$04:00:53.5 & 110 & 3981  \\
12 & ESO\,173$-$G015     & $0.33\times0.30$ ($-7 $) & $ 51\times 47$ & 0.434 && 13:27:23.77 & $-$57:29:22.0 & 165 & 2963  \\
13 & UGC\,11763        & $0.35\times0.31$ ($51 $) & $397\times351$ & 0.206 && 21:32:27.82 & $+$10:08:19.2 &$\cdots$& 17859 \\
14 & NGC\,3110         & $0.40\times0.27$ ($56 $) & $142\times 96$ & 0.384 && 10:04:02.09 & $-$06:28:29.5 & 170 & 4973  \\
15 & IC\,4734          & $0.38\times0.29$ ($28 $) & $126\times 96$ & 0.525 && 18:38:25.70 & $-$57:29:25.2 & 105 & 4606  \\
16 & NGC\,5135         & $0.33\times0.27$ ($85 $) & $ 83\times 68$ & 0.630 && 13:25:43.99 & $-$29:49:60.0 &  15 & 4052  \\
17 & ESO\,221$-$IG10     & $0.34\times0.31$ ($-57$) & $ 95\times 87$ & 0.582 && 13:50:56.94 & $-$49:03:19.2 &  40 & 3113  \\
18 & IC\,5179          & $0.28\times0.23$ ($68 $) & $ 63\times 52$ & 0.391 && 22:16:09.13 & $-$36:50:37.0 &  50 & 3350  \\
19 & UGC\,2982         & $0.36\times0.30$ ($126$) & $117\times 98$ & 0.359 && 04:12:22.67 & $+$05:32:49.1 &$\cdots$& 5200  \\
20 & NGC\,2369         & $0.28\times0.25$ ($7  $) & $ 59\times 53$ & 0.511 && 07:16:37.68 & $-$62:20:36.6 &  50 & 3187  \\
21 & ESO\,286$-$G035     & $0.29\times0.26$ ($52 $) & $101\times 90$ & 0.409 && 21:04:11.12 & $-$43:35:35.8 &  20 & 5129  \\
22 & ESO\,320$-$G030     & $0.34\times0.29$ ($69 $) & $ 62\times 53$ & 0.453 && 11:53:11.72 & $-$39:07:49.1 & 130 & 3049  \\
23 & NGC\,5734         & $0.31\times0.30$ ($35 $) & $ 89\times 86$ & 0.417 && 14:45:09.04 & $-$20:52:13.5 &  40 & 4029  \\
\hline
\end{tabular}
\tablefoot{
``Beam'' refers to the synthesized beam obtained after cleaning with natural weighting. 
``Resolution'' is the projected physical size corresponding to the synthesized beam at the distance of the galaxy. 
``Sensitivity'' is 1$\sigma$ rms per beam in 20\,km\,s$^{-1}$ velocity channels calculated from flux-free parts. 
The adopted galaxy kinematics were roughly estimated 
from the velocity field seen outside the most nuclear region. 
The systemic velocity $v_\mathrm{sys}$ is given in radio convention with respect to the LSRK.
}
\end{table*}

\begin{table*}
\caption{Results of the HCN 3--2 and HCO$^+$ 3--2 line measurements: 
resolved and global line luminosities.}
\label{table:luminosities}
\centering 
\begin{tabular}{clcccccc}
\hline\hline
 & & & \multicolumn{2}{c}{Resolved} && \multicolumn{2}{c}{Global} \\
\cline{4-5}\cline{7-8}
ID & Galaxy & Range & $L'_\mathrm{HCN}$ & $L'_\mathrm{HCO^+}$ && $L'_\mathrm{HCN}$ & $L'_\mathrm{HCO^+}$ \\
 & & (km s$^{-1}$) & ($10^7$ K km s$^{-1}$ pc$^2$)& ($10^7$ K km s$^{-1}$ pc$^2$) %
 && ($10^7$ K km s$^{-1}$ pc$^2$)& ($10^7$ K km s$^{-1}$ pc$^2$) \\
\hline
 1 & IRAS\,17208$-$0014  & $\pm500$ &  $ 20.5  \pm 0.1  $ & $ 11.8  \pm0.1   $ && $94.5 \pm1.2 $ & $64.6 \pm1.2 $ \\
 2 & IRAS\,09022$-$3615  & $\pm500$ &  $ 5.12  \pm 0.04 $ & $ 6.46  \pm0.05  $ && $26.0 \pm0.5 $ & $46.9 \pm0.5 $ \\
 3 & IRAS\,13120$-$5453  & $\pm500$ &  $ 8.52  \pm 0.07 $ & $ 3.19  \pm0.07  $ && $77.2 \pm0.7 $ & $45.2 \pm0.7 $ \\
 4 & IRAS\,F14378$-$3651 & $\pm250$ &  $ 6.25  \pm 0.06 $ & $ 4.72  \pm0.06  $ && $29.4 \pm0.6 $ & $25.8 \pm0.6 $ \\
 5 & IRAS\,F05189$-$2524 & $\pm250$ &  $ 7.44  \pm 0.03 $ & $ 4.56  \pm0.03  $ && $20.6 \pm0.3 $ & $13.4 \pm0.3 $ \\
 6 & IRAS\,F10565$+$2448 & $\pm250$ &  $ 1.78  \pm 0.03 $ & $ 1.35  \pm0.03  $ && $19.1 \pm0.3 $ & $21.8 \pm0.3 $ \\
 7 & IRAS\,19542$+$1110  & $\pm300$ &  $ 9.39  \pm 0.05 $ & $ 5.25  \pm0.05  $ && $28.3 \pm0.5 $ & $15.5 \pm0.5 $ \\
 8 & ESO\,148$-$IG002    & $\pm200$ &  $ 0.87  \pm 0.03 $ & $ 1.95  \pm0.03  $ && $3.0  \pm0.3 $ & $7.8  \pm0.3 $ \\
 9 & NGC\,6240         & $\pm500$ &  $ 0.90  \pm 0.02 $ & $ 1.30  \pm0.02  $ && $21.1 \pm0.2 $ & $30.4 \pm0.2 $ \\
10 & IRAS\,F17138-1017 & $\pm150$ &  $ 0.06  \pm 0.01 $ & $ 0.10  \pm0.01  $ && $7.2  \pm0.1 $ & $1.2  \pm0.1 $ \\
11 & IRAS\,17578$-$0400%
\tablefootmark{a} & $\pm250$ &  $ 3.613 \pm 0.006$ & $ 2.697 \pm0.007 $ && $13.82\pm0.07$ & $10.74\pm0.07$ \\
12 & ESO\,173$-$G015     & $\pm250$ &  $ 0.297 \pm 0.002$ & $ 0.262 \pm0.002 $ && $3.38 \pm0.02$ & $3.91 \pm0.02$ \\
13 & UGC\,11763        & $\pm200$ &  $ 0.15  \pm 0.05 $ & $ 0.21  \pm0.04  $ && $<1.5        $ & $1.4  \pm0.4 $ \\
14 & NGC\,3110         & $\pm200$ &  $ 0.056 \pm 0.006$ & $ 0.036 \pm0.007 $ && $1.15 \pm0.07$ & $1.10 \pm0.08$ \\
15 & IC\,4734          & $\pm200$ &  $ 0.394 \pm 0.008$ & $ 0.367 \pm0.009 $ && $5.79 \pm0.08$ & $5.45 \pm0.09$ \\
16 & NGC\,5135         & $\pm150$ &  $ 0.048 \pm 0.005$ & $ 0.047 \pm0.005 $ && $1.00 \pm0.05$ & $0.91 \pm0.05$ \\
17 & ESO\,221$-$IG10     & $\pm150$ &  $ 0.035 \pm 0.006$ & $ 0.046 \pm0.006 $ && $0.80 \pm0.06$ & $1.23 \pm0.06$ \\
18 & IC\,5179          & $\pm200$ &  $ 0.015 \pm 0.003$ & $ 0.026 \pm0.003 $ && $0.37 \pm0.03$ & $0.44 \pm0.03$ \\
19 & UGC\,2982         & $\pm150$ &  $<0.014          $ & $<0.016          $ && $<0.14       $ & $<0.16       $ \\
20 & NGC\,2369         & $\pm300$ &  $ 0.111 \pm 0.004$ & $ 0.074 \pm0.005 $ && $1.27 \pm0.04$ & $1.12 \pm0.04$ \\
21 & ESO\,286$-$G035     & $\pm200$ &  $ 0.033 \pm 0.007$ & $ 0.050 \pm0.008 $ && $0.49 \pm0.07$ & $0.84 \pm0.08$ \\
22 & ESO\,320$-$G030     & $\pm300$ &  $ 0.798 \pm 0.002$ & $ 0.341 \pm0.002 $ && $5.67 \pm0.02$ & $2.52 \pm0.02$ \\
23 & NGC\,5734         & $\pm150$ &  $ 0.032 \pm 0.004$ & $<0.013          $ && $0.31 \pm0.04$ & $<0.13       $ \\
\hline
\end{tabular}
\tablefoot{
The spectra were extracted from the beam-sized aperture and 
the elliptical aperture of $10\times10$ times the beam 
for the resolved and global luminosity, respectively, 
and both were centered at the position listed in Table \ref{table:observations}. 
Line flux densities were integrated over the range with respect to 
the systemic velocity given in Table \ref{table:observations}. 
Uncertainties and upper limits correspond to 1$\sigma$ and 
3$\sigma$ statistical errors, respectively. \\
\tablefoottext{a}{Continuum subtraction was done after spectral extraction 
assuming continuum levels of 0.02\,Jy and 0.07\,Jy for the resolved and global spectra, respectively. }
}
\end{table*}

\begin{table}
\caption{Measured $L'_\mathrm{HCN}/L'_\mathrm{HCO^+}$ 
for the resolved and global apertures.}
\label{table:integratedratio}
\centering 
\begin{tabular}{clcc}
\hline\hline
ID & Galaxy & Resolved & Global \\
\hline
1 & IRAS\,17208$-$0014  & $1.738\pm0.019$ & $1.464\pm0.032$ \\
2 & IRAS\,09022$-$3615  & $0.793\pm0.009$ & $0.555\pm0.011$ \\
3 & IRAS\,13120$-$5453  & $2.671\pm0.064$ & $1.706\pm0.031$ \\
4 & IRAS\,F14378$-$3651 & $1.323\pm0.022$ & $1.140\pm0.035$ \\
5 & IRAS\,F05189$-$2524 & $1.630\pm0.012$ & $1.534\pm0.038$ \\
6 & IRAS\,F10565$+$2448 & $1.320\pm0.035$ & $0.878\pm0.017$ \\
7 & IRAS\,19542$+$1110  & $1.789\pm0.018$ & $1.825\pm0.063$ \\
8 & ESO\,148$-$IG002    & $0.445\pm0.015$ & $0.387\pm0.035$ \\
9 & NGC\,6240           & $0.691\pm0.015$ & $0.693\pm0.007$ \\
10 & IRAS\,F17138-1017  & $0.629\pm0.123$ & $0.587\pm0.093$ \\
11 & IRAS\,17578$-$0400\tablefootmark{a} & $1.340\pm0.004$ & $1.283\pm0.011$ \\
12 & ESO\,173$-$G015    & $1.132\pm0.010$ & $0.865\pm0.006$ \\
13 & UGC\,11763         & $0.728\pm0.283$ & $<1.1$ \\
14 & NGC\,3110          & $1.490\pm0.330$ & $1.044\pm0.093$ \\
15 & IC\,4734           & $1.076\pm0.033$ & $1.063\pm0.022$ \\
16 & NGC\,5135          & $1.028\pm0.147$ & $1.100\pm0.083$ \\
17 & ESO\,221$-$IG10    & $0.771\pm0.164$ & $0.651\pm0.057$ \\
18 & IC\,5179           & $0.564\pm0.127$ & $0.842\pm0.087$ \\
19 & UGC\,2982          & $\cdots$        & $\cdots$ \\
20 & NGC\,2369          & $1.489\pm0.110$ & $1.133\pm0.057$ \\
21 & ESO\,286$-$G035    & $0.666\pm0.178$ & $0.587\pm0.098$ \\
22 & ESO\,320$-$G030    & $2.344\pm0.018$ & $2.254\pm0.023$ \\
23 & NGC\,5734          & $>2.5$          & $>2.4$ \\
\hline
\multicolumn{2}{c}{average (excl.~13, 19, 23)} & 1.216 & 1.080 \\
\hline
\end{tabular}
\tablefoot{
\tablefoottext{a}{Continuum was not subtracted. }
}
\end{table}

\begin{table}
\caption{Spearman rank coefficients and p-values.}
\label{table:correlation}
\centering 
\begin{tabular}{lcc}
\hline\hline
Quantities & $\rho$ & p-value \\
\hline
Global $L'_\mathrm{HCN}/L'_\mathrm{HCO^+}$ vs. \\
\hspace{2eM}$\alpha_\mathrm{AGN}$ 
& $-0.17^{+0.24}_{-0.23}$ & $0.38^{+0.41}_{-0.31}$ \\ \noalign {\smallskip}
\hspace{2eM}$\log(L_\mathrm{AGN}^\mathrm{bol}/\mathrm{erg\ s}^{-1})$
& $+0.00^{+0.27}_{-0.26}$ & $0.44^{+0.38}_{-0.34}$ \\ \noalign {\smallskip}
\hspace{2eM}$\log(L_\mathrm{IR}/L_\sun)$ 
& $+0.11^{+0.25}_{-0.26}$ & $0.42^{+0.39}_{-0.33}$ \\ \noalign {\smallskip}
\hspace{2eM}$\log(\mathrm{SFR}/M_\sun\ \mathrm{yr}^{-1})$
& $+0.09^{+0.24}_{-0.26}$ & $0.44^{+0.38}_{-0.33}$ \\ \noalign {\smallskip}
\hspace{2eM}$S_{60}/S_{100}$
& $+0.02^{+0.25}_{-0.25}$ & $0.47^{+0.36}_{-0.34}$ \\ \noalign {\smallskip}
\hspace{2eM}[\ion{N}{ii}]$_{122}$/[\ion{O}{iii}]$_{88}$ 
& $+0.50^{+0.25}_{-0.32}$ & $0.10^{+0.42}_{-0.09}$ \\ \noalign {\smallskip}
\hline
Resolved $L'_\mathrm{HCN}/L'_\mathrm{HCO^+}$ vs. \\
\hspace{2eM}$\alpha_\mathrm{AGN}$ 
& $-0.22^{+0.23}_{-0.22}$ & $0.31^{+0.44}_{-0.26}$ \\ \noalign {\smallskip}
\hspace{2eM}$\log(L_\mathrm{AGN}^\mathrm{bol}/\mathrm{erg\ s}^{-1})$
& $+0.05^{+0.24}_{-0.25}$ & $0.47^{+0.36}_{-0.34}$ \\ \noalign {\smallskip}
\hspace{2eM}$\log(L_\mathrm{IR}/L_\sun)$ 
& $+0.23^{+0.23}_{-0.25}$ & $0.29^{+0.45}_{-0.25}$ \\ \noalign {\smallskip}
\hspace{2eM}$\log(\mathrm{SFR}/M_\sun\ \mathrm{yr}^{-1})$
& $+0.23^{+0.22}_{-0.25}$ & $0.29^{+0.45}_{-0.25}$ \\ \noalign {\smallskip}
\hspace{2eM}$S_{60}/S_{100}$
& $+0.07^{+0.24}_{-0.24}$ & $0.47^{+0.35}_{-0.34}$ \\ \noalign {\smallskip}
\hspace{2eM}[\ion{N}{ii}]$_{122}$/[\ion{O}{iii}]$_{88}$ 
& $+0.38^{+0.26}_{-0.32}$ & $0.21^{+0.46}_{-0.19}$ \\ \noalign {\smallskip}
\hline
\end{tabular}
\tablefoot{We performed Spearman rank correlation analyses 
using the Monte Carlo perturbation plus bootstrapping method 
as implemented by \texttt{pymccorrelation} 
\citep[][see Sect. \ref{subsect:galproperties} for detail]{Privon2020}.
Correlation coefficients $\rho$ and p-values are 
the median and range of the 16 and 84 percentiles. 
Corresponding plots are shown in Figs.~\ref{figure:correlation_global} 
and \ref{figure:correlation_resolved}.
}
\end{table}

\begin{table*}
\caption{$L'_\mathrm{HCN}/L'_\mathrm{HCO^+}$ 
on a pixel-by-pixel basis and on a spaxel-by-spaxel basis.}
\label{table:ratio}
\centering 
\begin{tabular}{clcccccccc}
\hline\hline
& && \multicolumn{3}{c}{Spectrally integrated (pixel-by-pixel basis)} 
&& \multicolumn{3}{c}{Spectrally resolved (spaxel-by-spaxel basis)} \\
\cline{4-6}\cline{8-10}
ID & Galaxy && Mean & 25th--50th--75th (IQR) & 90th 
&& Mean & 25th--50th--75th (IQR) & 90th \\
\hline
1 & IRAS\,17208$-$0014 && 1.253 & 1.012--1.224--1.513 (0.501) & 1.689 && 1.517 & 1.147--1.447--1.812 (0.664) & 2.173 \\
2 & IRAS\,09022$-$3615 && 0.561 & 0.433--0.539--0.648 (0.215) & 0.788 && 0.641 & 0.482--0.607--0.754 (0.272) & 0.928 \\
3 & IRAS\,13120$-$5453 && 1.395 & 1.056--1.301--1.609 (0.553) & 2.109 && 1.590 & 1.046--1.440--2.024 (0.977) & 2.560 \\
4 & IRAS\,F14378$-$365 && 1.129 & 1.004--1.154--1.255 (0.251) & 1.355 && 1.169 & 0.980--1.142--1.305 (0.325) & 1.507 \\
5 & IRAS\,F05189$-$252 && 1.452 & 1.246--1.524--1.633 (0.387) & 1.752 && 1.589 & 1.409--1.547--1.717 (0.308) & 2.022 \\
6 & IRAS\,F10565$+$244 && 0.801 & 0.667--0.792--0.918 (0.251) & 1.068 && 0.871 & 0.660--0.825--1.023 (0.364) & 1.258 \\
7 & IRAS\,19542$+$1110 && 1.718 & 1.532--1.736--1.873 (0.342) & 2.127 && 1.724 & 1.544--1.697--1.848 (0.304) & 2.124 \\
8 & ESO\,148$-$IG002   && 0.402 & 0.345--0.394--0.437 (0.093) & 0.510 && 0.442 & 0.377--0.424--0.478 (0.101) & 0.544 \\
9 & NGC\,6240          && 0.674 & 0.613--0.679--0.736 (0.123) & 0.792 && 0.748 & 0.635--0.736--0.835 (0.200) & 0.954 \\
10 & IRAS\,F17138-101  && 0.767 & 0.591--0.698--0.886 (0.295) & 1.076 && 0.815 & 0.675--0.802--0.939 (0.264) & 1.060 \\
11 & IRAS\,17578$-$0400
     \tablefootmark{a} && 1.208 & 1.038--1.215--1.336 (0.298) & 1.514 && 1.237 & 1.081--1.238--1.383 (0.302) & 1.513 \\
12 & ESO\,173$-$G015   && 0.792 & 0.685--0.783--0.884 (0.199) & 1.004 && 0.858 & 0.694--0.830--0.979 (0.285) & 1.143 \\
13 & UGC\,11763        && $\cdots$ & $\cdots$ & $\cdots$ && $\cdots$ & $\cdots$ & $\cdots$ \\
14 & NGC\,3110         && 1.289 & 1.032--1.176--1.538 (0.506) & 1.802 && 1.147 & 0.932--1.090--1.324 (0.392) & 1.572 \\
15 & IC\,4734          && 1.053 & 0.957--1.046--1.121 (0.164) & 1.222 && 1.042 & 0.905--1.024--1.159 (0.254) & 1.319 \\
16 & NGC\,5135         && 1.237 & 1.029--1.189--1.405 (0.376) & 1.677 && 1.105 & 0.850--1.038--1.280 (0.430) & 1.594 \\
17 & ESO\,221$-$IG10   && 0.792 & 0.656--0.772--0.906 (0.250) & 1.034 && 0.755 & 0.596--0.715--0.886 (0.290) & 1.054 \\
18 & IC\,5179          && 0.985 & 0.804--0.980--1.145 (0.341) & 1.357 && 0.887 & 0.698--0.865--1.054 (0.357) & 1.227 \\
19 & UGC\,2982         && $\cdots$ & $\cdots$ & $\cdots$ && $\cdots$ & $\cdots$ & $\cdots$ \\
20 & NGC\,2369         && 1.100 & 0.876--1.049--1.294 (0.418) & 1.517 && 1.133 & 0.841--1.055--1.343 (0.502) & 1.663 \\
21 & ESO\,286$-$G035   && 0.778 & 0.680--0.755--0.854 (0.174) & 0.960 && 0.790 & 0.638--0.758--0.922 (0.284) & 1.072 \\
22 & ESO\,320$-$G030   && 2.068 & 1.647--2.073--2.336 (0.689) & 2.791 && 2.156 & 1.735--2.064--2.450 (0.716) & 2.998 \\
23 & NGC\,5734         && $\cdots$ & $\cdots$ & $\cdots$ && $\cdots$ & $\cdots$ & $\cdots$ \\
\hline
\multicolumn{2}{c}{average (excl.~13, 19, 23)} 
&& 1.073 & 0.895--1.054--1.216 (0.321) & 1.407 
&& 1.111 & 0.896--1.067--1.276 (0.380) & 1.514 \\
\hline
\end{tabular}
\tablefoot{
We examined pixels and spaxels within the global aperture 
(see Sect.~\ref{subsect:spectra}) and within the velocity range 
listed in Table~\ref{table:luminosities}. 
For pixel-by-pixel analysis, we first performed a spectral integration 
followed by a $3\sigma$ threshold clipping. 
Each pixel has a size of $0.05\arcsec\times0.05\arcsec$.
In the spaxel-by-spaxel analysis, we focused solely on spaxels 
where HCN and HCO$^+$ were both detected with $>$$3\sigma$ significance. 
Each spaxel has a size of $0.05\arcsec\times0.05\arcsec\times20$ km s$^{-1}$. 
We note that pixels and spaxels within a beam are correlated. 
IQR stands for the interquartile range. \\
\tablefoottext{a}{Continuum was not subtracted. }
}
\end{table*}

\clearpage


\bibliographystyle{aa}
\bibliography{ref}

\begin{thebibliography}{120}
\expandafter\ifx\csname natexlab\endcsname\relax\def\natexlab#1{#1}\fi

\bibitem[{{Aalto} {et~al.}(2015{\natexlab{a}}){Aalto}, {Garcia-Burillo},
  {Muller}, {Winters}, {Gonzalez-Alfonso}, {van der Werf}, {Henkel},
  {Costagliola}, \& {Neri}}]{Aalto2015a}
{Aalto}, S., {Garcia-Burillo}, S., {Muller}, S., {et~al.} 2015{\natexlab{a}},
  \aap, 574, A85

\bibitem[{{Aalto} {et~al.}(2015{\natexlab{b}}){Aalto}, {Mart{\'\i}n},
  {Costagliola}, {Gonz{\'a}lez-Alfonso}, {Muller}, {Sakamoto}, {Fuller},
  {Garc{\'\i}a-Burillo}, {van der Werf}, {Neri}, {Spaans}, {Combes}, {Viti},
  {M{\"u}hle}, {Armus}, {Evans}, {Sturm}, {Cernicharo}, {Henkel}, \&
  {Greve}}]{Aalto2015b}
{Aalto}, S., {Mart{\'\i}n}, S., {Costagliola}, F., {et~al.} 2015{\natexlab{b}},
  \aap, 584, A42

\bibitem[{{Aalto} {et~al.}(2019){Aalto}, {Muller}, {K{\"o}nig}, {Falstad},
  {Mangum}, {Sakamoto}, {Privon}, {Gallagher}, {Combes}, {Garc{\'\i}a-Burillo},
  {Mart{\'\i}n}, {Viti}, {van der Werf}, {Evans}, {Black}, {Varenius},
  {Beswick}, {Fuller}, {Henkel}, {Kohno}, {Alatalo}, \&
  {M{\"u}hle}}]{Aalto2019}
{Aalto}, S., {Muller}, S., {K{\"o}nig}, S., {et~al.} 2019, \aap, 627, A147

\bibitem[{{Aalto} {et~al.}(2007){Aalto}, {Spaans}, {Wiedner}, \&
  {H{\"u}ttemeister}}]{Aalto2007}
{Aalto}, S., {Spaans}, M., {Wiedner}, M.~C., \& {H{\"u}ttemeister}, S. 2007,
  \aap, 464, 193

\bibitem[{{Astropy Collaboration} {et~al.}(2018){Astropy Collaboration},
  {Price-Whelan}, {Sip{\H{o}}cz}, {G{\"u}nther}, {Lim}, {Crawford}, {Conseil},
  {Shupe}, {Craig}, {Dencheva}, {Ginsburg}, {VanderPlas}, {Bradley},
  {P{\'e}rez-Su{\'a}rez}, {de Val-Borro}, {Aldcroft}, {Cruz}, {Robitaille},
  {Tollerud}, {Ardelean}, {Babej}, {Bach}, {Bachetti}, {Bakanov}, {Bamford},
  {Barentsen}, {Barmby}, {Baumbach}, {Berry}, {Biscani}, {Boquien}, {Bostroem},
  {Bouma}, {Brammer}, {Bray}, {Breytenbach}, {Buddelmeijer}, {Burke},
  {Calderone}, {Cano Rodr{\'\i}guez}, {Cara}, {Cardoso}, {Cheedella}, {Copin},
  {Corrales}, {Crichton}, {D'Avella}, {Deil}, {Depagne}, {Dietrich}, {Donath},
  {Droettboom}, {Earl}, {Erben}, {Fabbro}, {Ferreira}, {Finethy}, {Fox},
  {Garrison}, {Gibbons}, {Goldstein}, {Gommers}, {Greco}, {Greenfield},
  {Groener}, {Grollier}, {Hagen}, {Hirst}, {Homeier}, {Horton}, {Hosseinzadeh},
  {Hu}, {Hunkeler}, {Ivezi{\'c}}, {Jain}, {Jenness}, {Kanarek}, {Kendrew},
  {Kern}, {Kerzendorf}, {Khvalko}, {King}, {Kirkby}, {Kulkarni}, {Kumar},
  {Lee}, {Lenz}, {Littlefair}, {Ma}, {Macleod}, {Mastropietro}, {McCully},
  {Montagnac}, {Morris}, {Mueller}, {Mumford}, {Muna}, {Murphy}, {Nelson},
  {Nguyen}, {Ninan}, {N{\"o}the}, {Ogaz}, {Oh}, {Parejko}, {Parley}, {Pascual},
  {Patil}, {Patil}, {Plunkett}, {Prochaska}, {Rastogi}, {Reddy Janga},
  {Sabater}, {Sakurikar}, {Seifert}, {Sherbert}, {Sherwood-Taylor}, {Shih},
  {Sick}, {Silbiger}, {Singanamalla}, {Singer}, {Sladen}, {Sooley},
  {Sornarajah}, {Streicher}, {Teuben}, {Thomas}, {Tremblay}, {Turner},
  {Terr{\'o}n}, {van Kerkwijk}, {de la Vega}, {Watkins}, {Weaver}, {Whitmore},
  {Woillez}, {Zabalza}, \& {Astropy Contributors}}]{Astropy2018}
{Astropy Collaboration}, {Price-Whelan}, A.~M., {Sip{\H{o}}cz}, B.~M., {et~al.}
  2018, \aj, 156, 123

\bibitem[{{Astropy Collaboration} {et~al.}(2013){Astropy Collaboration},
  {Robitaille}, {Tollerud}, {Greenfield}, {Droettboom}, {Bray}, {Aldcroft},
  {Davis}, {Ginsburg}, {Price-Whelan}, {Kerzendorf}, {Conley}, {Crighton},
  {Barbary}, {Muna}, {Ferguson}, {Grollier}, {Parikh}, {Nair}, {Unther},
  {Deil}, {Woillez}, {Conseil}, {Kramer}, {Turner}, {Singer}, {Fox}, {Weaver},
  {Zabalza}, {Edwards}, {Azalee Bostroem}, {Burke}, {Casey}, {Crawford},
  {Dencheva}, {Ely}, {Jenness}, {Labrie}, {Lim}, {Pierfederici}, {Pontzen},
  {Ptak}, {Refsdal}, {Servillat}, \& {Streicher}}]{Astropy2013}
{Astropy Collaboration}, {Robitaille}, T.~P., {Tollerud}, E.~J., {et~al.} 2013,
  \aap, 558, A33

\bibitem[{{Bachiller} \& {P{\'e}rez Guti{\'e}rrez}(1997)}]{Bachiller1997}
{Bachiller}, R. \& {P{\'e}rez Guti{\'e}rrez}, M. 1997, \apjl, 487, L93

\bibitem[{{Bachiller} {et~al.}(2001){Bachiller}, {P{\'e}rez Guti{\'e}rrez},
  {Kumar}, \& {Tafalla}}]{Bachiller2001}
{Bachiller}, R., {P{\'e}rez Guti{\'e}rrez}, M., {Kumar}, M.~S.~N., \&
  {Tafalla}, M. 2001, \aap, 372, 899

\bibitem[{{Barcos-Mu{\~n}oz} {et~al.}(2018){Barcos-Mu{\~n}oz}, {Aalto},
  {Thompson}, {Sakamoto}, {Mart{\'\i}n}, {Leroy}, {Privon}, {Evans}, \&
  {Kepley}}]{Barcos-Munos2018}
{Barcos-Mu{\~n}oz}, L., {Aalto}, S., {Thompson}, T.~A., {et~al.} 2018, \apjl,
  853, L28

\bibitem[{{Bayet} {et~al.}(2008){Bayet}, {Viti}, {Williams}, \&
  {Rawlings}}]{Bayet2008}
{Bayet}, E., {Viti}, S., {Williams}, D.~A., \& {Rawlings}, J.~M.~C. 2008, \apj,
  676, 978

\bibitem[{{Braine} {et~al.}(2017){Braine}, {Shimajiri}, {Andr{\'e}},
  {Bontemps}, {Gao}, {Chen}, \& {Kramer}}]{Braine2017}
{Braine}, J., {Shimajiri}, Y., {Andr{\'e}}, P., {et~al.} 2017, \aap, 597, A44

\bibitem[{{Burkhardt} {et~al.}(2019){Burkhardt}, {Shingledecker}, {Le Gal},
  {McGuire}, {Remijan}, \& {Herbst}}]{Burkhardt2019}
{Burkhardt}, A.~M., {Shingledecker}, C.~N., {Le Gal}, R., {et~al.} 2019, \apj,
  881, 32

\bibitem[{{Bussmann} {et~al.}(2008){Bussmann}, {Narayanan}, {Shirley},
  {Juneau}, {Wu}, {Solomon}, {Vanden Bout}, {Moustakas}, \&
  {Walker}}]{Bussmann2008}
{Bussmann}, R.~S., {Narayanan}, D., {Shirley}, Y.~L., {et~al.} 2008, \apjl,
  681, L73

\bibitem[{{Butterworth} {et~al.}(2022){Butterworth}, {Holdship}, {Viti}, \&
  {Garc{\'\i}a-Burillo}}]{Butterworth2022}
{Butterworth}, J., {Holdship}, J., {Viti}, S., \& {Garc{\'\i}a-Burillo}, S.
  2022, \aap, 667, A131

\bibitem[{{CASA Team} {et~al.}(2022){CASA Team}, {Bean}, {Bhatnagar}, {Castro},
  {Donovan Meyer}, {Emonts}, {Garcia}, {Garwood}, {Golap}, {Gonzalez Villalba},
  {Harris}, {Hayashi}, {Hoskins}, {Hsieh}, {Jagannathan}, {Kawasaki},
  {Keimpema}, {Kettenis}, {Lopez}, {Marvil}, {Masters}, {McNichols},
  {Mehringer}, {Miel}, {Moellenbrock}, {Montesino}, {Nakazato}, {Ott}, {Petry},
  {Pokorny}, {Raba}, {Rau}, {Schiebel}, {Schweighart}, {Sekhar}, {Shimada},
  {Small}, {Steeb}, {Sugimoto}, {Suoranta}, {Tsutsumi}, {van Bemmel},
  {Verkouter}, {Wells}, {Xiong}, {Szomoru}, {Griffith}, {Glendenning}, \&
  {Kern}}]{CASAteam2022}
{CASA Team}, {Bean}, B., {Bhatnagar}, S., {et~al.} 2022, \pasp, 134, 114501

\bibitem[{{Ceccarelli} {et~al.}(2011){Ceccarelli}, {Hily-Blant}, {Montmerle},
  {Dubus}, {Gallant}, \& {Fiasson}}]{Ceccarelli2011}
{Ceccarelli}, C., {Hily-Blant}, P., {Montmerle}, T., {et~al.} 2011, \apjl, 740,
  L4

\bibitem[{{Cicone} {et~al.}(2020){Cicone}, {Maiolino}, {Aalto}, {Muller}, \&
  {Feruglio}}]{Cicone2020}
{Cicone}, C., {Maiolino}, R., {Aalto}, S., {Muller}, S., \& {Feruglio}, C.
  2020, \aap, 633, A163

\bibitem[{{Cicone} {et~al.}(2014){Cicone}, {Maiolino}, {Sturm},
  {Graci{\'a}-Carpio}, {Feruglio}, {Neri}, {Aalto}, {Davies}, {Fiore},
  {Fischer}, {Garc{\'\i}a-Burillo}, {Gonz{\'a}lez-Alfonso}, {Hailey-Dunsheath},
  {Piconcelli}, \& {Veilleux}}]{Cicone2014}
{Cicone}, C., {Maiolino}, R., {Sturm}, E., {et~al.} 2014, \aap, 562, A21

\bibitem[{{Cicone} {et~al.}(2018){Cicone}, {Severgnini}, {Papadopoulos},
  {Maiolino}, {Feruglio}, {Treister}, {Privon}, {Zhang}, {Della Ceca}, {Fiore},
  {Schawinski}, \& {Wagg}}]{Cicone2018}
{Cicone}, C., {Severgnini}, P., {Papadopoulos}, P.~P., {et~al.} 2018, \apj,
  863, 143

\bibitem[{{Costagliola} {et~al.}(2011){Costagliola}, {Aalto}, {Rodriguez},
  {Muller}, {Spoon}, {Mart{\'\i}n}, {Per{\'e}z-Torres}, {Alberdi}, {Lindberg},
  {Batejat}, {J{\"u}tte}, {van der Werf}, \& {Lahuis}}]{Costagliola2011}
{Costagliola}, F., {Aalto}, S., {Rodriguez}, M.~I., {et~al.} 2011, \aap, 528,
  A30

\bibitem[{{Costagliola} {et~al.}(2013){Costagliola}, {Aalto}, {Sakamoto},
  {Mart{\'\i}n}, {Beswick}, {Muller}, \& {Kl{\"o}ckner}}]{Costagliola2013}
{Costagliola}, F., {Aalto}, S., {Sakamoto}, K., {et~al.} 2013, \aap, 556, A66

\bibitem[{Curran(2014)}]{Curran2014}
Curran, P.~A. 2014, Monte Carlo error analyses of Spearman's rank test

\bibitem[{{Denis-Alpizar} {et~al.}(2020){Denis-Alpizar}, {Stoecklin}, {Dutrey},
  \& {Guilloteau}}]{Denis-Alpizar2020}
{Denis-Alpizar}, O., {Stoecklin}, T., {Dutrey}, A., \& {Guilloteau}, S. 2020,
  \mnras, 497, 4276

\bibitem[{{D{\'\i}az-Santos} {et~al.}(2017){D{\'\i}az-Santos}, {Armus},
  {Charmandaris}, {Lu}, {Stierwalt}, {Stacey}, {Malhotra}, {van der Werf},
  {Howell}, {Privon}, {Mazzarella}, {Goldsmith}, {Murphy}, {Barcos-Mu{\~n}oz},
  {Linden}, {Inami}, {Larson}, {Evans}, {Appleton}, {Iwasawa}, {Lord},
  {Sanders}, \& {Surace}}]{Diaz-Santos2017}
{D{\'\i}az-Santos}, T., {Armus}, L., {Charmandaris}, V., {et~al.} 2017, \apj,
  846, 32

\bibitem[{{Dumouchel} {et~al.}(2010){Dumouchel}, {Faure}, \&
  {Lique}}]{Dumouchel2010}
{Dumouchel}, F., {Faure}, A., \& {Lique}, F. 2010, \mnras, 406, 2488

\bibitem[{{Falstad} {et~al.}(2021){Falstad}, {Aalto}, {K{\"o}nig}, {Onishi},
  {Muller}, {Gorski}, {Sato}, {Stanley}, {Combes}, {Gonz{\'a}lez-Alfonso},
  {Mangum}, {Evans}, {Barcos-Mu{\~n}oz}, {Privon}, {Linden},
  {D{\'\i}az-Santos}, {Mart{\'\i}n}, {Sakamoto}, {Harada}, {Fuller},
  {Gallagher}, {van der Werf}, {Viti}, {Greve}, {Garc{\'\i}a-Burillo},
  {Henkel}, {Imanishi}, {Izumi}, {Nishimura}, {Ricci}, \&
  {M{\"u}hle}}]{Falstad2021}
{Falstad}, N., {Aalto}, S., {K{\"o}nig}, S., {et~al.} 2021, \aap, 649, A105

\bibitem[{{Falstad} {et~al.}(2019){Falstad}, {Hallqvist}, {Aalto}, {K{\"o}nig},
  {Muller}, {Aladro}, {Combes}, {Evans}, {Fuller}, {Gallagher},
  {Garc{\'\i}a-Burillo}, {Gonz{\'a}lez-Alfonso}, {Greve}, {Henkel}, {Imanishi},
  {Izumi}, {Mangum}, {Mart{\'\i}n}, {Privon}, {Sakamoto}, {Veilleux}, \& {van
  der Werf}}]{Falstad2019}
{Falstad}, N., {Hallqvist}, F., {Aalto}, S., {et~al.} 2019, \aap, 623, A29

\bibitem[{{Faure} \& {Tennyson}(2001)}]{Faure2001}
{Faure}, A. \& {Tennyson}, J. 2001, \mnras, 325, 443

\bibitem[{{Faure} {et~al.}(2007){Faure}, {Varambhia}, {Stoecklin}, \&
  {Tennyson}}]{Faure2007}
{Faure}, A., {Varambhia}, H.~N., {Stoecklin}, T., \& {Tennyson}, J. 2007,
  \mnras, 382, 840

\bibitem[{{Fluetsch} {et~al.}(2019){Fluetsch}, {Maiolino}, {Carniani},
  {Marconi}, {Cicone}, {Bourne}, {Costa}, {Fabian}, {Ishibashi}, \&
  {Venturi}}]{Fluetsch2019}
{Fluetsch}, A., {Maiolino}, R., {Carniani}, S., {et~al.} 2019, \mnras, 483,
  4586

\bibitem[{{Gao} \& {Solomon}(2004)}]{GaoSolomon2004}
{Gao}, Y. \& {Solomon}, P.~M. 2004, \apjs, 152, 63

\bibitem[{{Garc{\'\i}a-Burillo} {et~al.}(2015){Garc{\'\i}a-Burillo}, {Combes},
  {Usero}, {Aalto}, {Colina}, {Alonso-Herrero}, {Hunt}, {Arribas},
  {Costagliola}, {Labiano}, {Neri}, {Pereira-Santaella}, {Tacconi}, \& {van der
  Werf}}]{Garcia-Burillo2015}
{Garc{\'\i}a-Burillo}, S., {Combes}, F., {Usero}, A., {et~al.} 2015, \aap, 580,
  A35

\bibitem[{{Garc{\'\i}a-Burillo} {et~al.}(2014){Garc{\'\i}a-Burillo}, {Combes},
  {Usero}, {Aalto}, {Krips}, {Viti}, {Alonso-Herrero}, {Hunt}, {Schinnerer},
  {Baker}, {Boone}, {Casasola}, {Colina}, {Costagliola}, {Eckart}, {Fuente},
  {Henkel}, {Labiano}, {Mart{\'\i}n}, {M{\'a}rquez}, {Muller}, {Planesas},
  {Ramos Almeida}, {Spaans}, {Tacconi}, \& {van der Werf}}]{Garcia-Burillo2014}
{Garc{\'\i}a-Burillo}, S., {Combes}, F., {Usero}, A., {et~al.} 2014, \aap, 567,
  A125

\bibitem[{{Garc{\'\i}a-Burillo} {et~al.}(2017){Garc{\'\i}a-Burillo}, {Viti},
  {Combes}, {Fuente}, {Usero}, {Hunt}, {Mart{\'\i}n}, {Krips}, {Aalto},
  {Aladro}, {Ramos Almeida}, {Alonso-Herrero}, {Casasola}, {Henkel},
  {Querejeta}, {Neri}, {Costagliola}, {Tacconi}, \& {van der
  Werf}}]{Garcia-Burillo2017}
{Garc{\'\i}a-Burillo}, S., {Viti}, S., {Combes}, F., {et~al.} 2017, \aap, 608,
  A56

\bibitem[{{Goldreich} \& {Kwan}(1974)}]{Goldreich1974}
{Goldreich}, P. \& {Kwan}, J. 1974, \apj, 189, 441

\bibitem[{{Goldsmith} \& {Kauffmann}(2017)}]{Goldsmith2017}
{Goldsmith}, P.~F. \& {Kauffmann}, J. 2017, \apj, 841, 25

\bibitem[{{Gonz{\'a}lez-Alfonso} {et~al.}(2017){Gonz{\'a}lez-Alfonso},
  {Fischer}, {Spoon}, {Stewart}, {Ashby}, {Veilleux}, {Smith}, {Sturm},
  {Farrah}, {Falstad}, {Mel{\'e}ndez}, {Graci{\'a}-Carpio}, {Janssen}, \&
  {Lebouteiller}}]{Gonzalez-Alfonso2017}
{Gonz{\'a}lez-Alfonso}, E., {Fischer}, J., {Spoon}, H.~W.~W., {et~al.} 2017,
  \apj, 836, 11

\bibitem[{{Gonz{\'a}lez-Alfonso} {et~al.}(2021){Gonz{\'a}lez-Alfonso},
  {Pereira-Santaella}, {Fischer}, {Garc{\'\i}a-Burillo}, {Yang},
  {Alonso-Herrero}, {Colina}, {Ashby}, {Smith}, {Rico-Villas},
  {Mart{\'\i}n-Pintado}, {Cazzoli}, \& {Stewart}}]{Gonzalez-Alfonso2021}
{Gonz{\'a}lez-Alfonso}, E., {Pereira-Santaella}, M., {Fischer}, J., {et~al.}
  2021, \aap, 645, A49

\bibitem[{{Gorski} {et~al.}(2023){Gorski}, {Aalto}, {K{\"o}nig}, {Wethers},
  {Yang}, {Muller}, {Viti}, {Black}, {Onishi}, \& {Sato}}]{Gorski2023}
{Gorski}, M.~D., {Aalto}, S., {K{\"o}nig}, S., {et~al.} 2023, \aap, 670, A70

\bibitem[{{Graci{\'a}-Carpio} {et~al.}(2008){Graci{\'a}-Carpio},
  {Garc{\'\i}a-Burillo}, {Planesas}, {Fuente}, \& {Usero}}]{Gracia-Carpio2008}
{Graci{\'a}-Carpio}, J., {Garc{\'\i}a-Burillo}, S., {Planesas}, P., {Fuente},
  A., \& {Usero}, A. 2008, \aap, 479, 703

\bibitem[{{Hagiwara} {et~al.}(2011){Hagiwara}, {Baan}, \&
  {Kl{\"o}ckner}}]{Hagiwara2011}
{Hagiwara}, Y., {Baan}, W.~A., \& {Kl{\"o}ckner}, H.-R. 2011, \aj, 142, 17

\bibitem[{{Harada} {et~al.}(2010){Harada}, {Herbst}, \& {Wakelam}}]{Harada2010}
{Harada}, N., {Herbst}, E., \& {Wakelam}, V. 2010, \apj, 721, 1570

\bibitem[{{Harada} {et~al.}(2021){Harada}, {Mart{\'\i}n}, {Mangum}, {Sakamoto},
  {Muller}, {Tanaka}, {Nakanishi}, {Herrero-Illana}, {Yoshimura}, {M{\"u}hle},
  {Aladro}, {Colzi}, {Rivilla}, {Aalto}, {Behrens}, {Henkel}, {Holdship},
  {Humire}, {Meier}, {Nishimura}, {van der Werf}, \& {Viti}}]{Harada2021}
{Harada}, N., {Mart{\'\i}n}, S., {Mangum}, J.~G., {et~al.} 2021, \apj, 923, 24

\bibitem[{{Harada} {et~al.}(2019){Harada}, {Nishimura}, {Watanabe}, {Yamamoto},
  {Aikawa}, {Sakai}, \& {Shimonishi}}]{Harada2019}
{Harada}, N., {Nishimura}, Y., {Watanabe}, Y., {et~al.} 2019, \apj, 871, 238

\bibitem[{{Harada} {et~al.}(2018){Harada}, {Sakamoto}, {Mart{\'\i}n}, {Aalto},
  {Aladro}, \& {Sliwa}}]{Harada2018}
{Harada}, N., {Sakamoto}, K., {Mart{\'\i}n}, S., {et~al.} 2018, \apj, 855, 49

\bibitem[{{Harada} {et~al.}(2013){Harada}, {Thompson}, \&
  {Herbst}}]{Harada2013}
{Harada}, N., {Thompson}, T.~A., \& {Herbst}, E. 2013, \apj, 765, 108

\bibitem[{{Helou} {et~al.}(1988){Helou}, {Khan}, {Malek}, \&
  {Boehmer}}]{Helou1988}
{Helou}, G., {Khan}, I.~R., {Malek}, L., \& {Boehmer}, L. 1988, \apjs, 68, 151

\bibitem[{{Herrera-Camus} {et~al.}(2018){Herrera-Camus}, {Sturm},
  {Graci{\'a}-Carpio}, {Lutz}, {Contursi}, {Veilleux}, {Fischer},
  {Gonz{\'a}lez-Alfonso}, {Poglitsch}, {Tacconi}, {Genzel}, {Maiolino},
  {Sternberg}, {Davies}, \& {Verma}}]{Herrera-Camus2018}
{Herrera-Camus}, R., {Sturm}, E., {Graci{\'a}-Carpio}, J., {et~al.} 2018, \apj,
  861, 94

\bibitem[{{Holdship} {et~al.}(2022){Holdship}, {Mangum}, {Viti}, {Behrens},
  {Harada}, {Mart{\'\i}n}, {Sakamoto}, {Muller}, {Tanaka}, {Nakanishi},
  {Herrero-Illana}, {Yoshimura}, {Aladro}, {Colzi}, {Emig}, {Henkel},
  {Nishimura}, {Rivilla}, {van der Werf}, \& {Alma Comprehensive
  High-Resolution Extragalactic Molecular Inventory (Alchemi)
  Collaboration}}]{Holdship2022}
{Holdship}, J., {Mangum}, J.~G., {Viti}, S., {et~al.} 2022, \apj, 931, 89

\bibitem[{{Holdship} {et~al.}(2017){Holdship}, {Viti}, {Jim{\'e}nez-Serra},
  {Makrymallis}, \& {Priestley}}]{Holdship2017}
{Holdship}, J., {Viti}, S., {Jim{\'e}nez-Serra}, I., {Makrymallis}, A., \&
  {Priestley}, F. 2017, \aj, 154, 38

\bibitem[{{Hopkins} {et~al.}(2006){Hopkins}, {Somerville}, {Hernquist}, {Cox},
  {Robertson}, \& {Li}}]{Hopkins2006}
{Hopkins}, P.~F., {Somerville}, R.~S., {Hernquist}, L., {et~al.} 2006, \apj,
  652, 864

\bibitem[{{Imanishi} {et~al.}(2023){Imanishi}, {Baba}, {Nakanishi}, \&
  {Izumi}}]{Imanishi2023}
{Imanishi}, M., {Baba}, S., {Nakanishi}, K., \& {Izumi}, T. 2023, \apj, 950, 75

\bibitem[{{Imanishi} \& {Nakanishi}(2013)}]{Imanishi2013}
{Imanishi}, M. \& {Nakanishi}, K. 2013, \aj, 146, 91

\bibitem[{{Imanishi} {et~al.}(2019){Imanishi}, {Nakanishi}, \&
  {Izumi}}]{Imanishi2019}
{Imanishi}, M., {Nakanishi}, K., \& {Izumi}, T. 2019, \apjs, 241, 19

\bibitem[{{Imanishi} {et~al.}(2009){Imanishi}, {Nakanishi}, {Tamura}, \&
  {Peng}}]{Imanishi2009}
{Imanishi}, M., {Nakanishi}, K., {Tamura}, Y., \& {Peng}, C.-H. 2009, \aj, 137,
  3581

\bibitem[{{Israel}(2023)}]{Israel2023}
{Israel}, F.~P. 2023, \aap, 671, A59

\bibitem[{{Izumi} {et~al.}(2016){Izumi}, {Kohno}, {Aalto}, {Espada}, {Fathi},
  {Harada}, {Hatsukade}, {Hsieh}, {Imanishi}, {Krips}, {Mart{\'\i}n},
  {Matsushita}, {Meier}, {Nakai}, {Nakanishi}, {Schinnerer}, {Sheth},
  {Terashima}, \& {Turner}}]{Izumi2016}
{Izumi}, T., {Kohno}, K., {Aalto}, S., {et~al.} 2016, \apj, 818, 42

\bibitem[{{Jim{\'e}nez-Donaire} {et~al.}(2019){Jim{\'e}nez-Donaire}, {Bigiel},
  {Leroy}, {Usero}, {Cormier}, {Puschnig}, {Gallagher}, {Kepley}, {Bolatto},
  {Garc{\'\i}a-Burillo}, {Hughes}, {Kramer}, {Pety}, {Schinnerer}, {Schruba},
  {Schuster}, \& {Walter}}]{Jimenez-Donaire2019}
{Jim{\'e}nez-Donaire}, M.~J., {Bigiel}, F., {Leroy}, A.~K., {et~al.} 2019,
  \apj, 880, 127

\bibitem[{{Jim{\'e}nez-Serra} {et~al.}(2008){Jim{\'e}nez-Serra}, {Caselli},
  {Mart{\'\i}n-Pintado}, \& {Hartquist}}]{Jimenez-Serra2008}
{Jim{\'e}nez-Serra}, I., {Caselli}, P., {Mart{\'\i}n-Pintado}, J., \&
  {Hartquist}, T.~W. 2008, \aap, 482, 549

\bibitem[{{Juneau} {et~al.}(2009){Juneau}, {Narayanan}, {Moustakas}, {Shirley},
  {Bussmann}, {Kennicutt}, \& {Vanden Bout}}]{Juneau2009}
{Juneau}, S., {Narayanan}, D.~T., {Moustakas}, J., {et~al.} 2009, \apj, 707,
  1217

\bibitem[{{Kauffmann} {et~al.}(2017){Kauffmann}, {Goldsmith}, {Melnick},
  {Tolls}, {Guzman}, \& {Menten}}]{Kauffmann2017}
{Kauffmann}, J., {Goldsmith}, P.~F., {Melnick}, G., {et~al.} 2017, \aap, 605,
  L5

\bibitem[{{Kawana} {et~al.}(2022){Kawana}, {Saito}, {Okumura}, {Kawabe},
  {Espada}, {Iono}, {Kaneko}, {Lee}, {Michiyama}, {Motohara}, {Nakanishi},
  {Pettitt}, {Randriamanakoto}, {Ueda}, \& {Yamashita}}]{Kawana2022}
{Kawana}, Y., {Saito}, T., {Okumura}, S.~K., {et~al.} 2022, \apj, 929, 100

\bibitem[{{Kennicutt}(1998)}]{Kennicutt1998}
{Kennicutt}, Robert~C., J. 1998, \apj, 498, 541

\bibitem[{{Kohno} {et~al.}(2001){Kohno}, {Matsushita}, {Vila-Vilar{\'o}},
  {Okumura}, {Shibatsuka}, {Okiura}, {Ishizuki}, \& {Kawabe}}]{Kohno2001}
{Kohno}, K., {Matsushita}, S., {Vila-Vilar{\'o}}, B., {et~al.} 2001, in
  Astronomical Society of the Pacific Conference Series, Vol. 249, The Central
  Kiloparsec of Starbursts and AGN: The La Palma Connection, ed. J.~H.
  {Knapen}, J.~E. {Beckman}, I.~{Shlosman}, \& T.~J. {Mahoney}, 672

\bibitem[{{Komossa} {et~al.}(2003){Komossa}, {Burwitz}, {Hasinger}, {Predehl},
  {Kaastra}, \& {Ikebe}}]{Komossa2003}
{Komossa}, S., {Burwitz}, V., {Hasinger}, G., {et~al.} 2003, \apjl, 582, L15

\bibitem[{{K{\"o}nig} {et~al.}(2018){K{\"o}nig}, {Aalto}, {Muller},
  {Gallagher}, {Beswick}, {Varenius}, {J{\"u}tte}, {Krips}, \&
  {Adamo}}]{Koenig2018}
{K{\"o}nig}, S., {Aalto}, S., {Muller}, S., {et~al.} 2018, \aap, 615, A122

\bibitem[{{Krips} {et~al.}(2008){Krips}, {Neri}, {Garc{\'\i}a-Burillo},
  {Mart{\'\i}n}, {Combes}, {Graci{\'a}-Carpio}, \& {Eckart}}]{Krips2008}
{Krips}, M., {Neri}, R., {Garc{\'\i}a-Burillo}, S., {et~al.} 2008, \apj, 677,
  262

\bibitem[{{Krumholz} \& {Thompson}(2007)}]{Krumholz2007}
{Krumholz}, M.~R. \& {Thompson}, T.~A. 2007, \apj, 669, 289

\bibitem[{{Laurent} {et~al.}(2000){Laurent}, {Mirabel}, {Charmandaris},
  {Gallais}, {Madden}, {Sauvage}, {Vigroux}, \& {Cesarsky}}]{Laurent2000}
{Laurent}, O., {Mirabel}, I.~F., {Charmandaris}, V., {et~al.} 2000, \aap, 359,
  887

\bibitem[{{Lepp} \& {Dalgarno}(1996)}]{Lepp1996}
{Lepp}, S. \& {Dalgarno}, A. 1996, \aap, 306, L21

\bibitem[{{Lindberg} {et~al.}(2016){Lindberg}, {Aalto}, {Muller},
  {Mart{\'\i}-Vidal}, {Falstad}, {Costagliola}, {Henkel}, {van der Werf},
  {Garc{\'\i}a-Burillo}, \& {Gonz{\'a}lez-Alfonso}}]{Lindberg2016}
{Lindberg}, J.~E., {Aalto}, S., {Muller}, S., {et~al.} 2016, \aap, 587, A15

\bibitem[{{Lutz} {et~al.}(1996){Lutz}, {Genzel}, {Sternberg}, {Netzer},
  {Kunze}, {Rigopoulou}, {Sturm}, {Egami}, {Feuchtgruber}, {Moorwood}, \& {de
  Graauw}}]{Lutz1996}
{Lutz}, D., {Genzel}, R., {Sternberg}, A., {et~al.} 1996, \aap, 315, L137

\bibitem[{{Lutz} {et~al.}(2020){Lutz}, {Sturm}, {Janssen}, {Veilleux}, {Aalto},
  {Cicone}, {Contursi}, {Davies}, {Feruglio}, {Fischer}, {Fluetsch},
  {Garcia-Burillo}, {Genzel}, {Gonz{\'a}lez-Alfonso}, {Graci{\'a}-Carpio},
  {Herrera-Camus}, {Maiolino}, {Schruba}, {Shimizu}, {Sternberg}, {Tacconi}, \&
  {Wei{\ss}}}]{Lutz2020}
{Lutz}, D., {Sturm}, E., {Janssen}, A., {et~al.} 2020, \aap, 633, A134

\bibitem[{{Maloney} {et~al.}(1996){Maloney}, {Hollenbach}, \&
  {Tielens}}]{Maloney1996}
{Maloney}, P.~R., {Hollenbach}, D.~J., \& {Tielens}, A.~G.~G.~M. 1996, \apj,
  466, 561

\bibitem[{{Mart{\'\i}n} {et~al.}(2016){Mart{\'\i}n}, {Aalto}, {Sakamoto},
  {Gonz{\'a}lez-Alfonso}, {Muller}, {Henkel}, {Garc{\'\i}a-Burillo}, {Aladro},
  {Costagliola}, {Harada}, {Krips}, {Mart{\'\i}n-Pintado}, {M{\"u}hle}, {van
  der Werf}, \& {Viti}}]{Martin2016}
{Mart{\'\i}n}, S., {Aalto}, S., {Sakamoto}, K., {et~al.} 2016, \aap, 590, A25

\bibitem[{{Mart{\'\i}n} {et~al.}(2015){Mart{\'\i}n}, {Kohno}, {Izumi}, {Krips},
  {Meier}, {Aladro}, {Matsushita}, {Takano}, {Turner}, {Espada}, {Nakajima},
  {Terashima}, {Fathi}, {Hsieh}, {Imanishi}, {Lundgren}, {Nakai}, {Schinnerer},
  {Sheth}, \& {Wiklind}}]{Martin2015}
{Mart{\'\i}n}, S., {Kohno}, K., {Izumi}, T., {et~al.} 2015, \aap, 573, A116

\bibitem[{{Meijerink} {et~al.}(2013){Meijerink}, {Kristensen}, {Wei{\ss}}, {van
  der Werf}, {Walter}, {Spaans}, {Loenen}, {Fischer}, {Israel}, {Isaak},
  {Papadopoulos}, {Aalto}, {Armus}, {Charmandaris}, {Dasyra}, {Diaz-Santos},
  {Evans}, {Gao}, {Gonz{\'a}lez-Alfonso}, {G{\"u}sten}, {Henkel}, {Kramer},
  {Lord}, {Mart{\'\i}n-Pintado}, {Naylor}, {Sanders}, {Smith}, {Spinoglio},
  {Stacey}, {Veilleux}, \& {Wiedner}}]{Meijerink2013}
{Meijerink}, R., {Kristensen}, L.~E., {Wei{\ss}}, A., {et~al.} 2013, \apjl,
  762, L16

\bibitem[{{Meijerink} \& {Spaans}(2005)}]{Meijerink2005}
{Meijerink}, R. \& {Spaans}, M. 2005, \aap, 436, 397

\bibitem[{{Meijerink} {et~al.}(2007){Meijerink}, {Spaans}, \&
  {Israel}}]{Meijerink2007}
{Meijerink}, R., {Spaans}, M., \& {Israel}, F.~P. 2007, \aap, 461, 793

\bibitem[{{Montoya Arroyave} {et~al.}(2023){Montoya Arroyave}, {Cicone},
  {Makroleivaditi}, {Weiss}, {Lundgren}, {Severgnini}, {De Breuck},
  {Baumschlager}, {Schimek}, {Shen}, \& {Aravena}}]{MontoyaArroyave2023}
{Montoya Arroyave}, I., {Cicone}, C., {Makroleivaditi}, E., {et~al.} 2023,
  arXiv e-prints, arXiv:2302.06629

\bibitem[{{Narayanan} {et~al.}(2008){Narayanan}, {Cox}, {Shirley}, {Dav{\'e}},
  {Hernquist}, \& {Walker}}]{Narayanan2008}
{Narayanan}, D., {Cox}, T.~J., {Shirley}, Y., {et~al.} 2008, \apj, 684, 996

\bibitem[{{Nishimura} {et~al.}(2016{\natexlab{a}}){Nishimura}, {Shimonishi},
  {Watanabe}, {Sakai}, {Aikawa}, {Kawamura}, \& {Yamamoto}}]{Nishimura2016a}
{Nishimura}, Y., {Shimonishi}, T., {Watanabe}, Y., {et~al.} 2016{\natexlab{a}},
  \apj, 829, 94

\bibitem[{{Nishimura} {et~al.}(2016{\natexlab{b}}){Nishimura}, {Shimonishi},
  {Watanabe}, {Sakai}, {Aikawa}, {Kawamura}, \& {Yamamoto}}]{Nishimura2016b}
{Nishimura}, Y., {Shimonishi}, T., {Watanabe}, Y., {et~al.} 2016{\natexlab{b}},
  \apj, 818, 161

\bibitem[{{Nishimura} {et~al.}(2017){Nishimura}, {Watanabe}, {Harada},
  {Shimonishi}, {Sakai}, {Aikawa}, {Kawamura}, \& {Yamamoto}}]{Nishimura2017}
{Nishimura}, Y., {Watanabe}, Y., {Harada}, N., {et~al.} 2017, \apj, 848, 17

\bibitem[{{Oteo} {et~al.}(2017){Oteo}, {Zhang}, {Yang}, {Ivison}, {Omont},
  {Bremer}, {Bussmann}, {Cooray}, {Cox}, {Dannerbauer}, {Dunne}, {Eales},
  {Furlanetto}, {Gavazzi}, {Gao}, {Greve}, {Nayyeri}, {Negrello}, {Neri},
  {Riechers}, {Tunnard}, {Wagg}, \& {Van der Werf}}]{Oteo2017}
{Oteo}, I., {Zhang}, Z.~Y., {Yang}, C., {et~al.} 2017, \apj, 850, 170

\bibitem[{{Pereira-Santaella} {et~al.}(2016){Pereira-Santaella}, {Colina},
  {Garc{\'\i}a-Burillo}, {Alonso-Herrero}, {Arribas}, {Cazzoli}, {Emonts},
  {Piqueras L{\'o}pez}, {Planesas}, {Storchi Bergmann}, {Usero}, \&
  {Villar-Mart{\'\i}n}}]{Pereira-Santaella2016}
{Pereira-Santaella}, M., {Colina}, L., {Garc{\'\i}a-Burillo}, S., {et~al.}
  2016, \aap, 594, A81

\bibitem[{{Pereira-Santaella} {et~al.}(2021){Pereira-Santaella}, {Colina},
  {Garc{\'\i}a-Burillo}, {Lamperti}, {Gonz{\'a}lez-Alfonso}, {Perna},
  {Arribas}, {Alonso-Herrero}, {Aalto}, {Combes}, {Labiano},
  {Piqueras-L{\'o}pez}, {Rigopoulou}, \& {van der
  Werf}}]{Pereira-Santaella2021}
{Pereira-Santaella}, M., {Colina}, L., {Garc{\'\i}a-Burillo}, S., {et~al.}
  2021, \aap, 651, A42

\bibitem[{{Pereira-Santaella} {et~al.}(2017){Pereira-Santaella}, {Rigopoulou},
  {Farrah}, {Lebouteiller}, \& {Li}}]{Pereira-Santaella2017}
{Pereira-Santaella}, M., {Rigopoulou}, D., {Farrah}, D., {Lebouteiller}, V., \&
  {Li}, J. 2017, \mnras, 470, 1218

\bibitem[{{P{\'e}rez-Torres} {et~al.}(2021){P{\'e}rez-Torres}, {Mattila},
  {Alonso-Herrero}, {Aalto}, \& {Efstathiou}}]{Perez-Torres2021}
{P{\'e}rez-Torres}, M., {Mattila}, S., {Alonso-Herrero}, A., {Aalto}, S., \&
  {Efstathiou}, A. 2021, \aapr, 29, 2

\bibitem[{{Pety} {et~al.}(2017){Pety}, {Guzm{\'a}n}, {Orkisz}, {Liszt},
  {Gerin}, {Bron}, {Bardeau}, {Goicoechea}, {Gratier}, {Le Petit}, {Levrier},
  {{\"O}berg}, {Roueff}, \& {Sievers}}]{Pety2017}
{Pety}, J., {Guzm{\'a}n}, V.~V., {Orkisz}, J.~H., {et~al.} 2017, \aap, 599, A98

\bibitem[{{Privon} {et~al.}(2017){Privon}, {Aalto}, {Falstad}, {Muller},
  {Gonz{\'a}lez-Alfonso}, {Sliwa}, {Treister}, {Costagliola}, {Armus}, {Evans},
  {Garcia-Burillo}, {Izumi}, {Sakamoto}, {van der Werf}, \& {Chu}}]{Privon2017}
{Privon}, G.~C., {Aalto}, S., {Falstad}, N., {et~al.} 2017, \apj, 835, 213

\bibitem[{{Privon} {et~al.}(2015){Privon}, {Herrero-Illana}, {Evans},
  {Iwasawa}, {Perez-Torres}, {Armus}, {D{\'\i}az-Santos}, {Murphy},
  {Stierwalt}, {Aalto}, {Mazzarella}, {Barcos-Mu{\~n}oz}, {Borish}, {Inami},
  {Kim}, {Treister}, {Surace}, {Lord}, {Conway}, {Frayer}, \&
  {Alberdi}}]{Privon2015}
{Privon}, G.~C., {Herrero-Illana}, R., {Evans}, A.~S., {et~al.} 2015, \apj,
  814, 39

\bibitem[{{Privon} {et~al.}(2020){Privon}, {Ricci}, {Aalto}, {Viti}, {Armus},
  {D{\'\i}az-Santos}, {Gonz{\'a}lez-Alfonso}, {Iwasawa}, {Jeff}, {Treister},
  {Bauer}, {Evans}, {Garg}, {Herrero-Illana}, {Mazzarella}, {Larson}, {Blecha},
  {Barcos-Mu{\~n}oz}, {Charmandaris}, {Stierwalt}, \&
  {P{\'e}rez-Torres}}]{Privon2020}
{Privon}, G.~C., {Ricci}, C., {Aalto}, S., {et~al.} 2020, \apj, 893, 149

\bibitem[{{Ricci} {et~al.}(2021){Ricci}, {Privon}, {Pfeifle}, {Armus},
  {Iwasawa}, {Torres-Alb{\`a}}, {Satyapal}, {Bauer}, {Treister}, {Ho}, {Aalto},
  {Ar{\'e}valo}, {Barcos-Mu{\~n}oz}, {Charmandaris}, {Diaz-Santos}, {Evans},
  {Gao}, {Inami}, {Koss}, {Lansbury}, {Linden}, {Medling}, {Sanders}, {Song},
  {Stern}, {U}, {Ueda}, \& {Yamada}}]{Ricci2021}
{Ricci}, C., {Privon}, G.~C., {Pfeifle}, R.~W., {et~al.} 2021, \mnras, 506,
  5935

\bibitem[{{Romano} {et~al.}(2021){Romano}, {Cassata}, {Morselli}, {Jones},
  {Ginolfi}, {Zanella}, {B{\'e}thermin}, {Capak}, {Faisst}, {Le F{\`e}vre},
  {Schaerer}, {Silverman}, {Yan}, {Bardelli}, {Boquien}, {Cimatti},
  {Dessauges-Zavadsky}, {Enia}, {Fujimoto}, {Gruppioni}, {Hathi}, {Ibar},
  {Koekemoer}, {Lemaux}, {Rodighiero}, {Vergani}, {Zamorani}, \&
  {Zucca}}]{Romano2021}
{Romano}, M., {Cassata}, P., {Morselli}, L., {et~al.} 2021, \aap, 653, A111

\bibitem[{{Saito} {et~al.}(2018){Saito}, {Iono}, {Ueda}, {Espada}, {Sliwa},
  {Nakanishi}, {Lu}, {Xu}, {Michiyama}, {Kaneko}, {Yamashita}, {Ando}, {Yun},
  {Motohara}, \& {Kawabe}}]{Saito2018}
{Saito}, T., {Iono}, D., {Ueda}, J., {et~al.} 2018, \mnras, 475, L52

\bibitem[{{Sakamoto} {et~al.}(2013){Sakamoto}, {Aalto}, {Costagliola},
  {Mart{\'\i}n}, {Ohyama}, {Wiedner}, \& {Wilner}}]{Sakamoto2013}
{Sakamoto}, K., {Aalto}, S., {Costagliola}, F., {et~al.} 2013, \apj, 764, 42

\bibitem[{{Sakamoto} {et~al.}(2010){Sakamoto}, {Aalto}, {Evans}, {Wiedner}, \&
  {Wilner}}]{Sakamoto2010}
{Sakamoto}, K., {Aalto}, S., {Evans}, A.~S., {Wiedner}, M.~C., \& {Wilner},
  D.~J. 2010, \apjl, 725, L228

\bibitem[{{Sanders} {et~al.}(2003){Sanders}, {Mazzarella}, {Kim}, {Surace}, \&
  {Soifer}}]{Sanders2003}
{Sanders}, D.~B., {Mazzarella}, J.~M., {Kim}, D.~C., {Surace}, J.~A., \&
  {Soifer}, B.~T. 2003, \aj, 126, 1607

\bibitem[{{Sanders} \& {Mirabel}(1996)}]{Sanders1996}
{Sanders}, D.~B. \& {Mirabel}, I.~F. 1996, \araa, 34, 749

\bibitem[{{Sanders} {et~al.}(1989){Sanders}, {Phinney}, {Neugebauer}, {Soifer},
  \& {Matthews}}]{Sanders1989}
{Sanders}, D.~B., {Phinney}, E.~S., {Neugebauer}, G., {Soifer}, B.~T., \&
  {Matthews}, K. 1989, \apj, 347, 29

\bibitem[{{Sanders} {et~al.}(1988){Sanders}, {Soifer}, {Elias}, {Madore},
  {Matthews}, {Neugebauer}, \& {Scoville}}]{Sanders1988}
{Sanders}, D.~B., {Soifer}, B.~T., {Elias}, J.~H., {et~al.} 1988, \apj, 325, 74

\bibitem[{{Schmidt}(1959)}]{Schmidt1959}
{Schmidt}, M. 1959, \apj, 129, 243

\bibitem[{{Shirley}(2015)}]{Shirley2015}
{Shirley}, Y.~L. 2015, \pasp, 127, 299

\bibitem[{{Singh}(2021)}]{Singh2021}
{Singh}, J. 2021, \mnras, 504, 1531

\bibitem[{{Solomon} \& {Vanden Bout}(2005)}]{Solomon2005}
{Solomon}, P.~M. \& {Vanden Bout}, P.~A. 2005, \araa, 43, 677

\bibitem[{{Stierwalt} {et~al.}(2013){Stierwalt}, {Armus}, {Surace}, {Inami},
  {Petric}, {Diaz-Santos}, {Haan}, {Charmandaris}, {Howell}, {Kim}, {Marshall},
  {Mazzarella}, {Spoon}, {Veilleux}, {Evans}, {Sanders}, {Appleton}, {Bothun},
  {Bridge}, {Chan}, {Frayer}, {Iwasawa}, {Kewley}, {Lord}, {Madore},
  {Melbourne}, {Murphy}, {Rich}, {Schulz}, {Sturm}, {Vavilkin}, \&
  {Xu}}]{Stierwalt2013}
{Stierwalt}, S., {Armus}, L., {Surace}, J.~A., {et~al.} 2013, \apjs, 206, 1

\bibitem[{{Sturm} {et~al.}(2011){Sturm}, {Gonz{\'a}lez-Alfonso}, {Veilleux},
  {Fischer}, {Graci{\'a}-Carpio}, {Hailey-Dunsheath}, {Contursi}, {Poglitsch},
  {Sternberg}, {Davies}, {Genzel}, {Lutz}, {Tacconi}, {Verma}, {Maiolino}, \&
  {de Jong}}]{Sturm2011}
{Sturm}, E., {Gonz{\'a}lez-Alfonso}, E., {Veilleux}, S., {et~al.} 2011, \apjl,
  733, L16

\bibitem[{{Tan} {et~al.}(2018){Tan}, {Gao}, {Zhang}, {Greve}, {Jiang},
  {Wilson}, {Yang}, {Bemis}, {Chung}, {Matsushita}, {Shi}, {Ao}, {Brinks},
  {Currie}, {Davis}, {de Grijs}, {Ho}, {Imanishi}, {Kohno}, {Lee}, {Parsons},
  {Rawlings}, {Rigopoulou}, {Rosolowsky}, {Bulger}, {Chen}, {Chapman}, {Eden},
  {Gear}, {Gu}, {He}, {Jiao}, {Liu}, {Liu}, {Li}, {Micha{\l}owski},
  {Nguyen-Luong}, {Qiu}, {Smith}, {Violino}, {Wang}, {Wang}, {Wang}, {Yeh},
  {Zhao}, \& {Zhu}}]{Tan2018}
{Tan}, Q.-H., {Gao}, Y., {Zhang}, Z.-Y., {et~al.} 2018, \apj, 860, 165

\bibitem[{{Tielens} \& {Hollenbach}(1985)}]{Tielens1985}
{Tielens}, A.~G.~G.~M. \& {Hollenbach}, D. 1985, \apj, 291, 722

\bibitem[{{Treister} {et~al.}(2020){Treister}, {Messias}, {Privon}, {Nagar},
  {Medling}, {U}, {Bauer}, {Cicone}, {Mu{\~n}oz}, {Evans}, {Muller-Sanchez},
  {Comerford}, {Armus}, {Chang}, {Koss}, {Venturi}, {Schawinski}, {Casey},
  {Urry}, {Sanders}, {Scoville}, \& {Sheth}}]{Treister2020}
{Treister}, E., {Messias}, H., {Privon}, G.~C., {et~al.} 2020, \apj, 890, 149

\bibitem[{{van der Tak} {et~al.}(2007){van der Tak}, {Black}, {Sch{\"o}ier},
  {Jansen}, \& {van Dishoeck}}]{vanderTak2007}
{van der Tak}, F.~F.~S., {Black}, J.~H., {Sch{\"o}ier}, F.~L., {Jansen}, D.~J.,
  \& {van Dishoeck}, E.~F. 2007, \aap, 468, 627

\bibitem[{{Veilleux} {et~al.}(2013){Veilleux}, {Mel{\'e}ndez}, {Sturm},
  {Gracia-Carpio}, {Fischer}, {Gonz{\'a}lez-Alfonso}, {Contursi}, {Lutz},
  {Poglitsch}, {Davies}, {Genzel}, {Tacconi}, {de Jong}, {Sternberg}, {Netzer},
  {Hailey-Dunsheath}, {Verma}, {Rupke}, {Maiolino}, {Teng}, \&
  {Polisensky}}]{Veilleux2013}
{Veilleux}, S., {Mel{\'e}ndez}, M., {Sturm}, E., {et~al.} 2013, \apj, 776, 27

\bibitem[{{Veilleux} {et~al.}(2009){Veilleux}, {Rupke}, {Kim}, {Genzel},
  {Sturm}, {Lutz}, {Contursi}, {Schweitzer}, {Tacconi}, {Netzer}, {Sternberg},
  {Mihos}, {Baker}, {Mazzarella}, {Lord}, {Sanders}, {Stockton}, {Joseph}, \&
  {Barnes}}]{Veilleux2009}
{Veilleux}, S., {Rupke}, D.~S.~N., {Kim}, D.~C., {et~al.} 2009, \apjs, 182, 628

\bibitem[{{Vincenzo} {et~al.}(2016){Vincenzo}, {Belfiore}, {Maiolino},
  {Matteucci}, \& {Ventura}}]{Vincenzo2016}
{Vincenzo}, F., {Belfiore}, F., {Maiolino}, R., {Matteucci}, F., \& {Ventura},
  P. 2016, \mnras, 458, 3466

\bibitem[{{Viti} {et~al.}(2014){Viti}, {Garc{\'\i}a-Burillo}, {Fuente}, {Hunt},
  {Usero}, {Henkel}, {Eckart}, {Martin}, {Spaans}, {Muller}, {Combes}, {Krips},
  {Schinnerer}, {Casasola}, {Costagliola}, {Marquez}, {Planesas}, {van der
  Werf}, {Aalto}, {Baker}, {Boone}, \& {Tacconi}}]{Viti2014}
{Viti}, S., {Garc{\'\i}a-Burillo}, S., {Fuente}, A., {et~al.} 2014, \aap, 570,
  A28

\bibitem[{{Yamada} {et~al.}(2007){Yamada}, {Wada}, \& {Tomisaka}}]{Yamada2007}
{Yamada}, M., {Wada}, K., \& {Tomisaka}, K. 2007, \apj, 671, 73

\bibitem[{{Yamada} {et~al.}(2019){Yamada}, {Ueda}, {Tanimoto}, {Kawamuro},
  {Imanishi}, \& {Toba}}]{Yamada2019}
{Yamada}, S., {Ueda}, Y., {Tanimoto}, A., {et~al.} 2019, \apj, 876, 96

\bibitem[{{Yang} {et~al.}(2023){Yang}, {Aalto}, {K{\"o}nig}, {Del Palacio},
  {Gorski}, {Linden}, {Muller}, {Onishi}, {Sato}, \& {Wethers}}]{Yang2023}
{Yang}, C., {Aalto}, S., {K{\"o}nig}, S., {et~al.} 2023, arXiv e-prints,
  arXiv:2307.07641

\bibitem[{{Zhang} {et~al.}(2014){Zhang}, {Gao}, {Henkel}, {Zhao}, {Wang},
  {Menten}, \& {G{\"u}sten}}]{Zhang2014}
{Zhang}, Z.-Y., {Gao}, Y., {Henkel}, C., {et~al.} 2014, \apjl, 784, L31

\end{thebibliography}

\begin{appendix}
\label{sect:appendix}

\section{Spectra and moment maps}
\label{appendix:figures}

\begin{figure*}
\includegraphics[width=0.4\hsize]{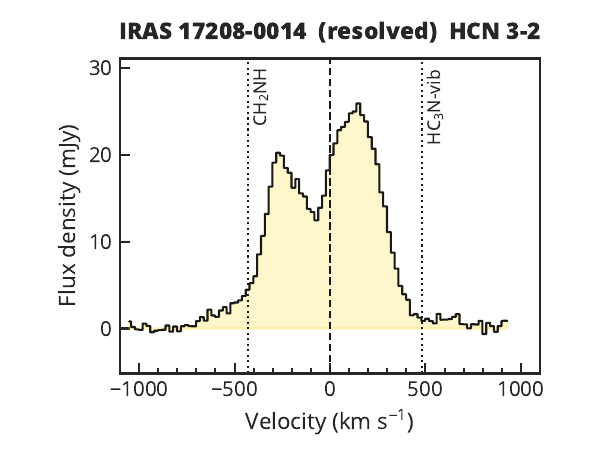}
\includegraphics[width=0.6\hsize]{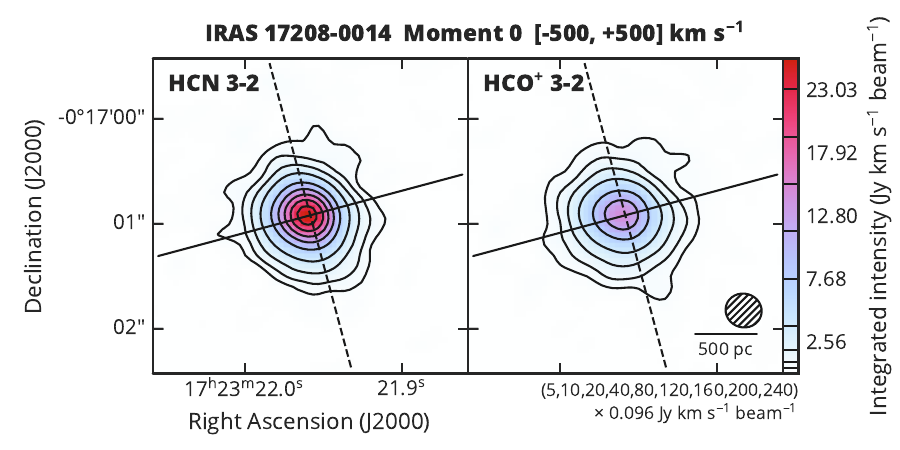}

\includegraphics[width=0.4\hsize]{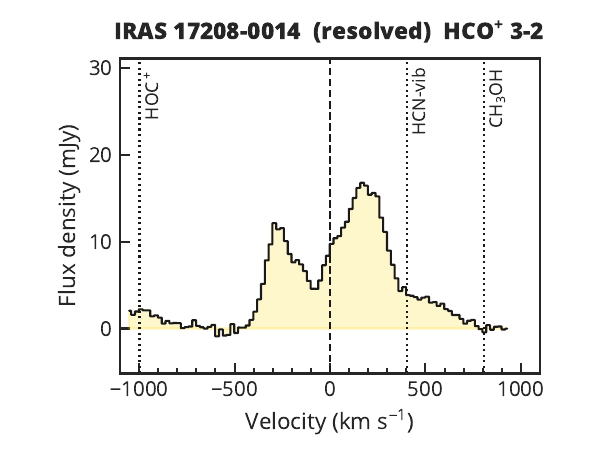}
\includegraphics[width=0.6\hsize]{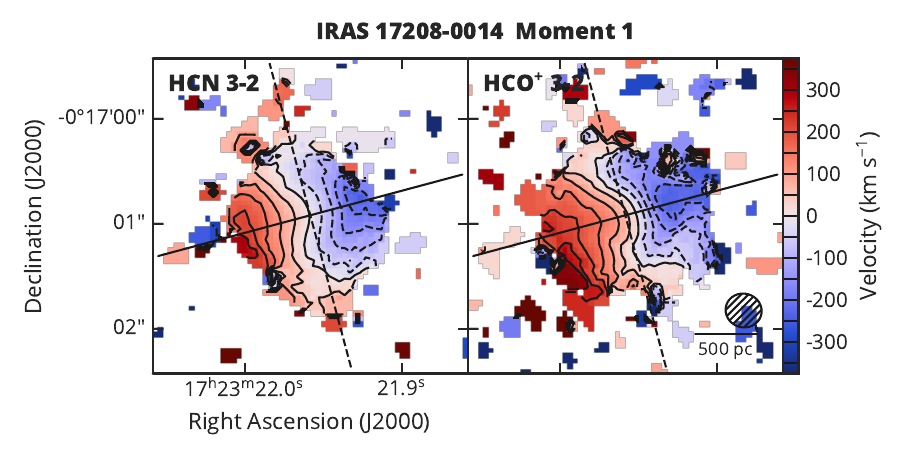}

\includegraphics[width=0.4\hsize]{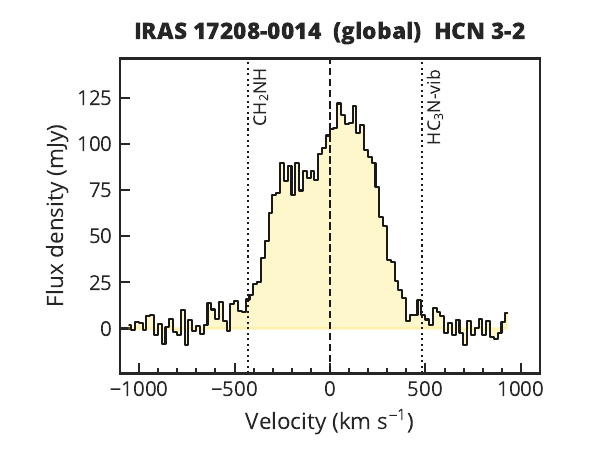}
\includegraphics[width=0.6\hsize]{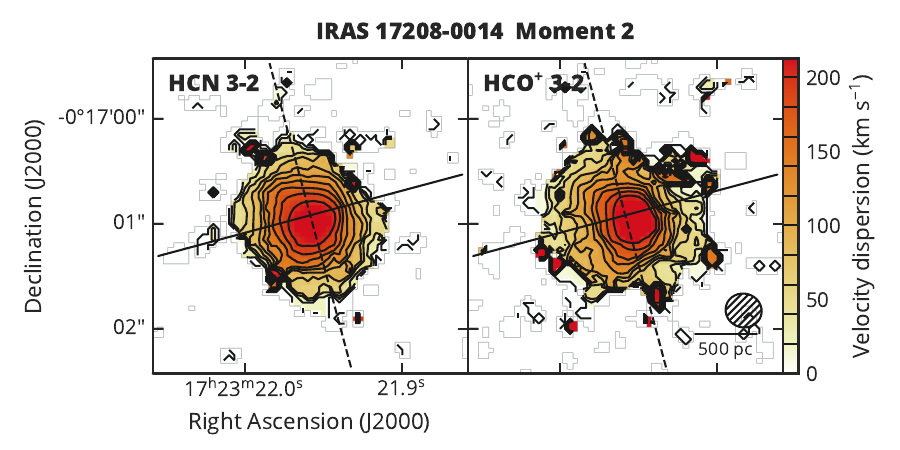}

\includegraphics[width=0.4\hsize]{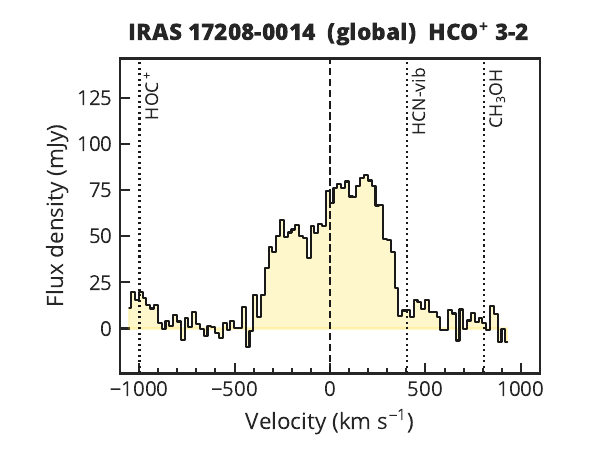}
\includegraphics[width=0.4\hsize]{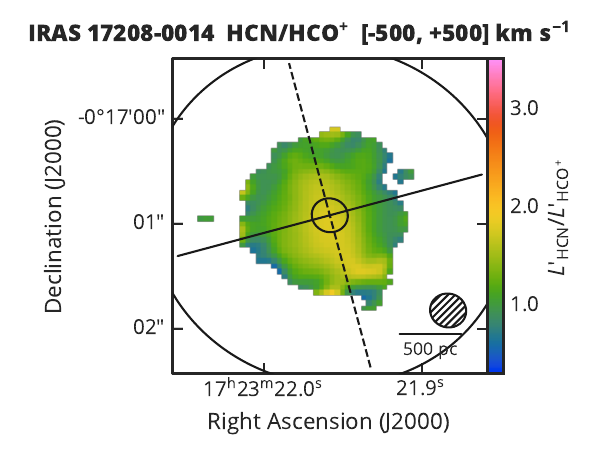}
\caption{HCN 3--2 and HCO$^+$ 3--2 for IRAS\,17208$-$0014. 
\emph{Left panels}: (\emph{Top two panels}) 
Spectra extracted from the resolved aperture. 
(\emph{Lower two panels}) Spectra extracted from the global aperture. 
Velocities are relative to the systemic velocity. 
The corresponding velocities of potentially detected species 
are indicated by vertical dotted lines.
\emph{Right panels}: 
(\emph{Top}) Integrated intensity over $\pm500$ km s$^{-1}$ (moment 0). 
Contours are (5, 10, 20, 40, 80, 120, 160, 200, 240) $\times\sigma$, 
where $\sigma$ is 0.096 Jy km s$^{-1}$ beam$^{-1}$. 
(\emph{Second from top}) Velocity field (moment 1). 
Contours are in steps of $\pm50$ km s$^{-1}$. 
(\emph{Third from top}) Velocity dispersion (moment 2). 
Contours are in steps of 20 km s$^{-1}$. 
Moment 1 and 2 were derived with $3\sigma$ clipping. 
(\emph{Bottom}) $L'_\mathrm{HCN}/L'_\mathrm{HCO^+}$. 
Color scale is from 0.285 to 3.5. 
Overlaid ellipses represent the apertures used for spectral extraction. 
Solid and dashed lines represent the kinematic major and minor axes, respectively. 
The synthesized beam is indicated by hatched ellipses in the lower right corners.
}
\label{figure:17208-0014}
\end{figure*}

\begin{figure*}
\includegraphics[width=0.4\hsize]{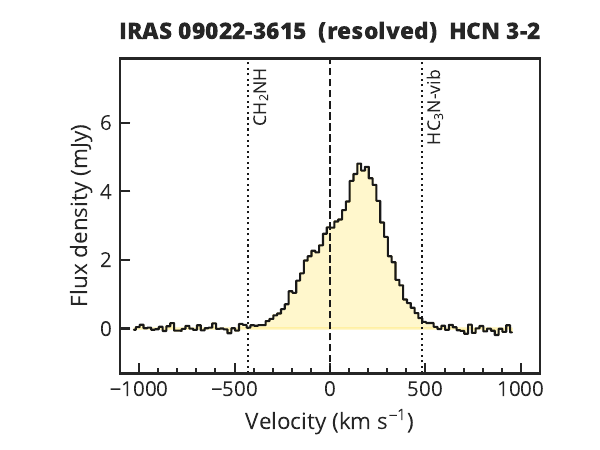}
\includegraphics[width=0.6\hsize]{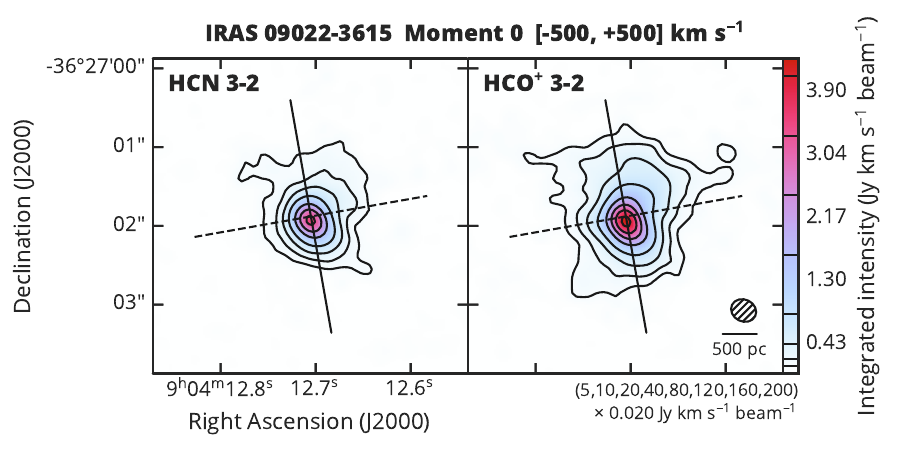}

\includegraphics[width=0.4\hsize]{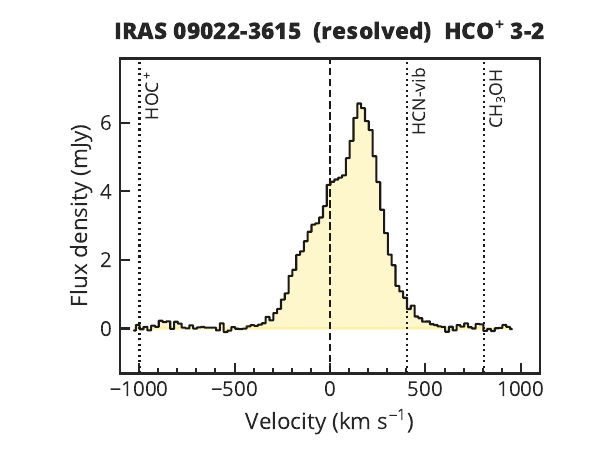}
\includegraphics[width=0.6\hsize]{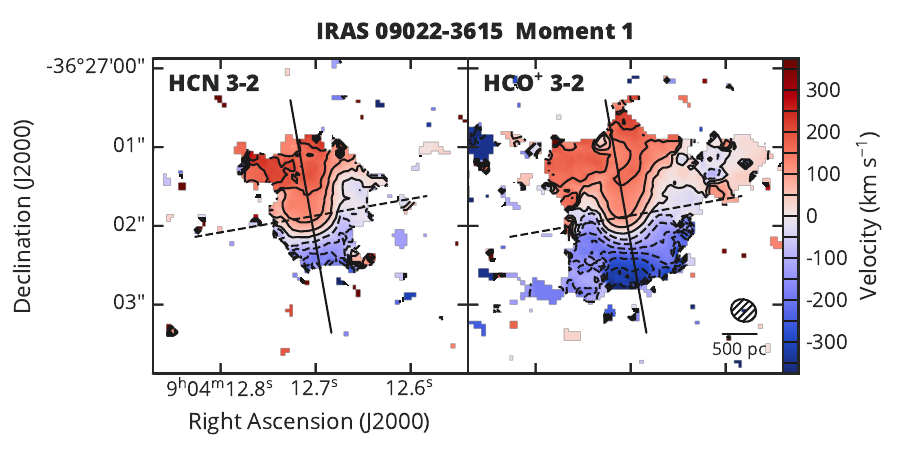}

\includegraphics[width=0.4\hsize]{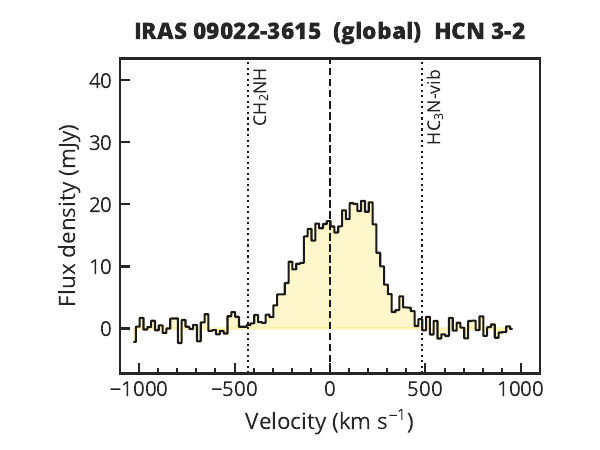}
\includegraphics[width=0.6\hsize]{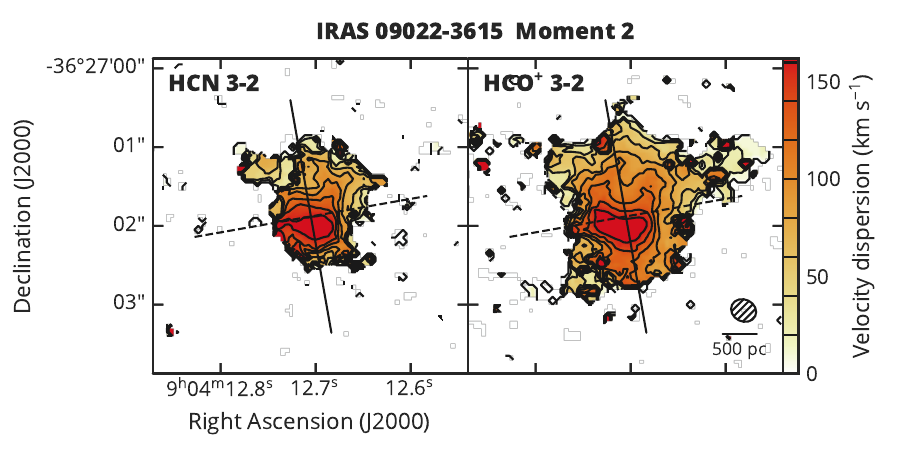}

\includegraphics[width=0.4\hsize]{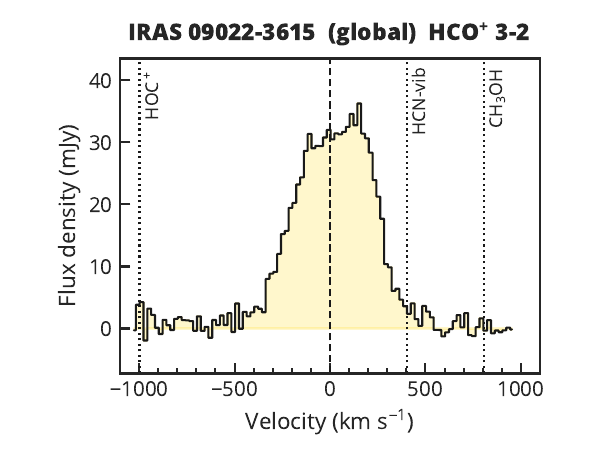}
\includegraphics[width=0.4\hsize]{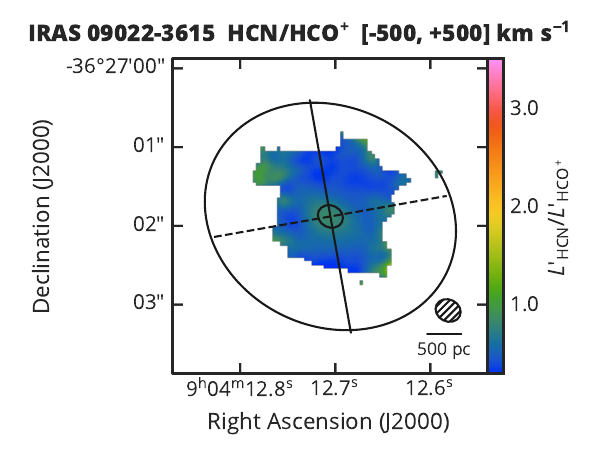}
\caption{HCN 3--2 and HCO$^+$ 3--2 for IRAS\,09022$-$3615. 
\emph{Left panels}: (\emph{Top two panels}) 
Spectra extracted from the resolved aperture. 
(\emph{Lower two panels}) Spectra extracted from the global aperture. 
Velocities are relative to the systemic velocity. 
The corresponding velocities of potentially detected species 
are indicated by vertical dotted lines.
\emph{Right panels}: 
(\emph{Top}) Integrated intensity over $\pm500$ km s$^{-1}$ (moment 0). 
Contours are (5, 10, 20, 40, 80, 120, 160, 200) $\times\sigma$, 
where $\sigma$ is 0.020 Jy km s$^{-1}$ beam$^{-1}$. 
(\emph{Second from top}) Velocity field (moment 1). 
Contours are in steps of $\pm50$ km s$^{-1}$. 
(\emph{Third from top}) Velocity dispersion (moment 2). 
Contours are in steps of 20 km s$^{-1}$. 
Moment 1 and 2 were derived with $3\sigma$ clipping. 
(\emph{Bottom}) $L'_\mathrm{HCN}/L'_\mathrm{HCO^+}$. 
Color scale is from 0.285 to 3.5. 
Overlaid ellipses represent the apertures used for spectral extraction. 
Solid and dashed lines represent the kinematic major and minor axes, respectively. 
The synthesized beam is indicated by hatched ellipses in the lower right corners.
}
\label{figure:09022-3615}
\end{figure*}

\begin{figure*}
\includegraphics[width=0.4\hsize]{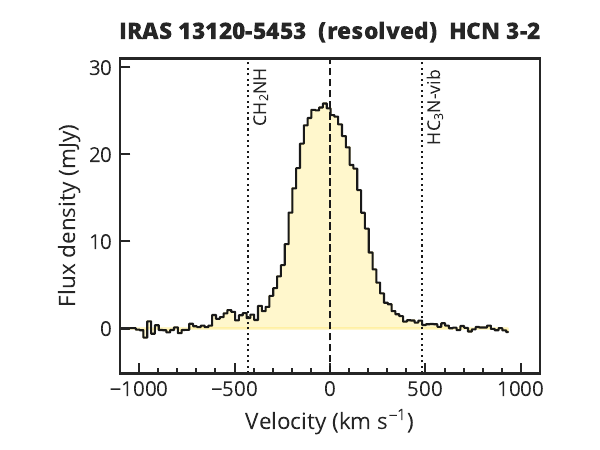}
\includegraphics[width=0.6\hsize]{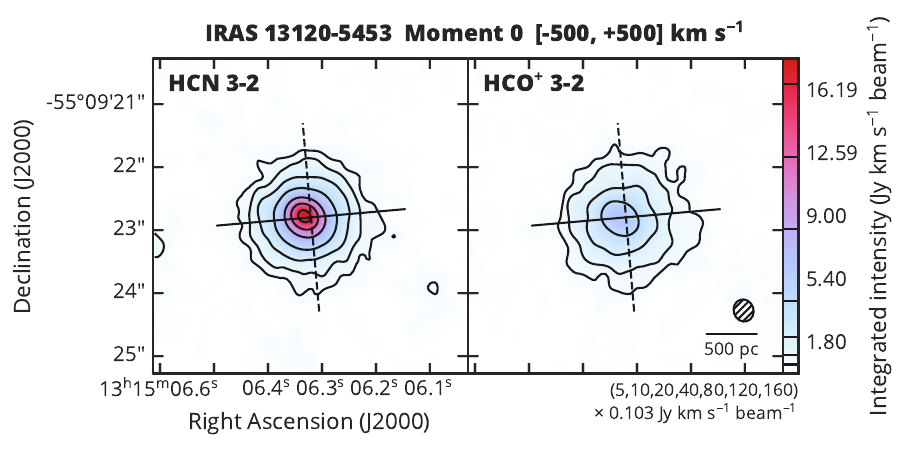}

\includegraphics[width=0.4\hsize]{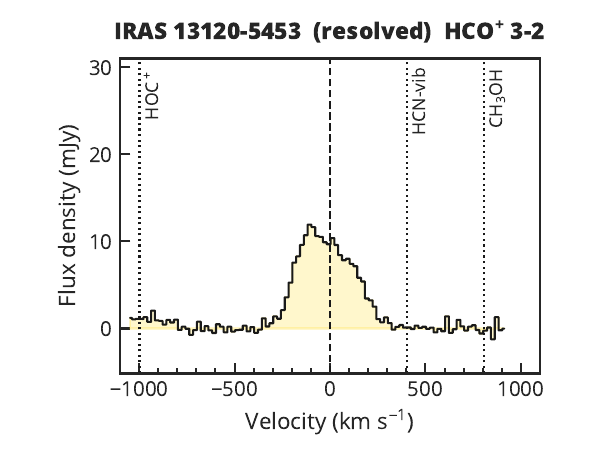}
\includegraphics[width=0.6\hsize]{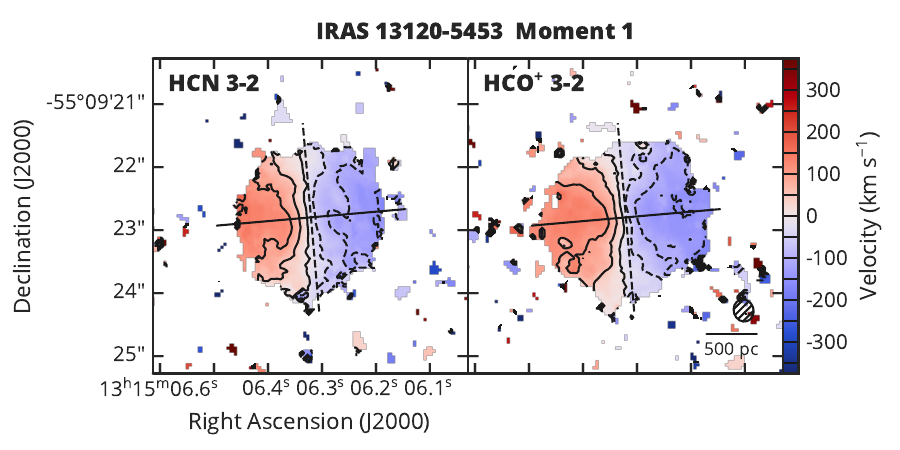}

\includegraphics[width=0.4\hsize]{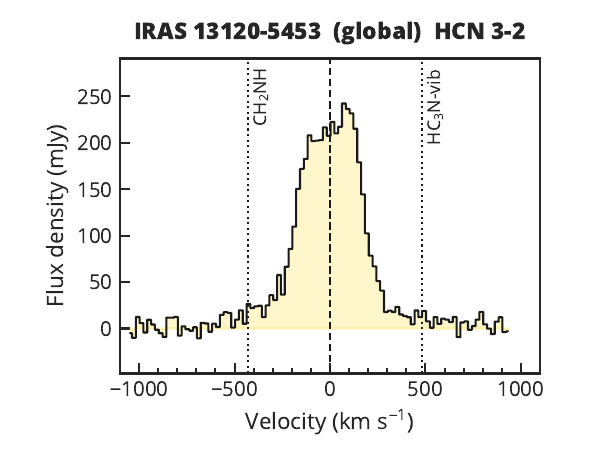}
\includegraphics[width=0.6\hsize]{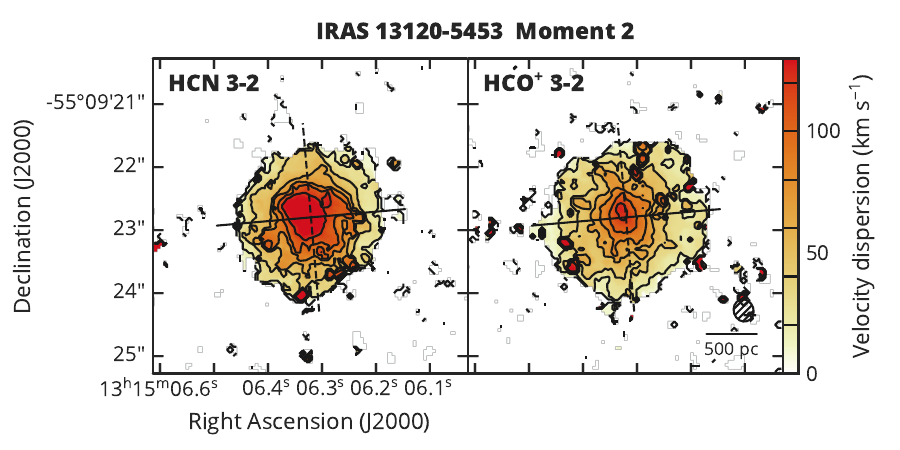}

\includegraphics[width=0.4\hsize]{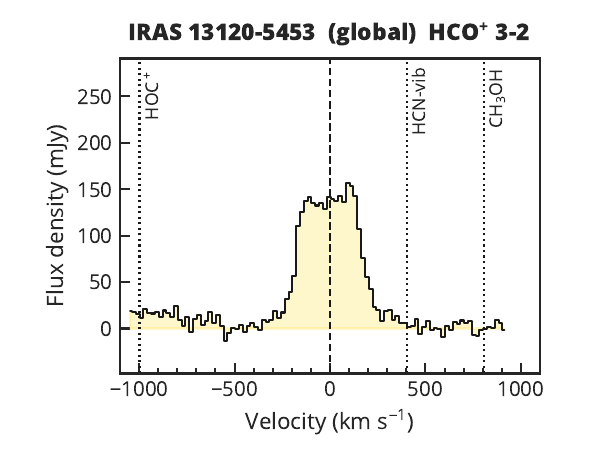}
\includegraphics[width=0.4\hsize]{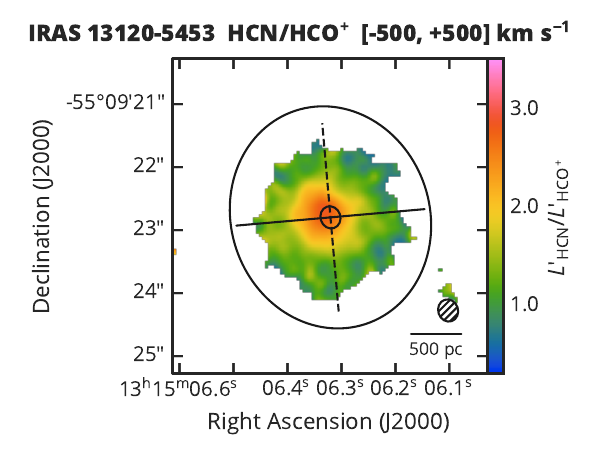}
\caption{HCN 3--2 and HCO$^+$ 3--2 for IRAS\,13120$-$5453. 
\emph{Left panels}: (\emph{Top two panels}) 
Spectra extracted from the resolved aperture. 
(\emph{Lower two panels}) Spectra extracted from the global aperture. 
Velocities are relative to the systemic velocity. 
The corresponding velocities of potentially detected species 
are indicated by vertical dotted lines.
\emph{Right panels}: 
(\emph{Top}) Integrated intensity over $\pm500$ km s$^{-1}$ (moment 0). 
Contours are (5, 10, 20, 40, 80, 120, 160) $\times\sigma$, 
where $\sigma$ is 0.103 Jy km s$^{-1}$ beam$^{-1}$. 
(\emph{Second from top}) Velocity field (moment 1). 
Contours are in steps of $\pm50$ km s$^{-1}$. 
(\emph{Third from top}) Velocity dispersion (moment 2). 
Contours are in steps of 20 km s$^{-1}$. 
Moment 1 and 2 were derived with $3\sigma$ clipping. 
(\emph{Bottom}) $L'_\mathrm{HCN}/L'_\mathrm{HCO^+}$. 
Color scale is from 0.285 to 3.5. 
Overlaid ellipses represent the apertures used for spectral extraction. 
Solid and dashed lines represent the kinematic major and minor axes, respectively. 
The synthesized beam is indicated by hatched ellipses in the lower right corners.
}
\label{figure:13120-5453}
\end{figure*}

\begin{figure*}
\includegraphics[width=0.4\hsize]{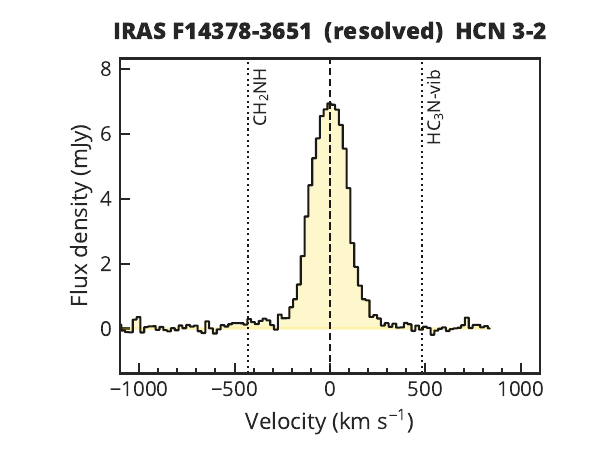}
\includegraphics[width=0.6\hsize]{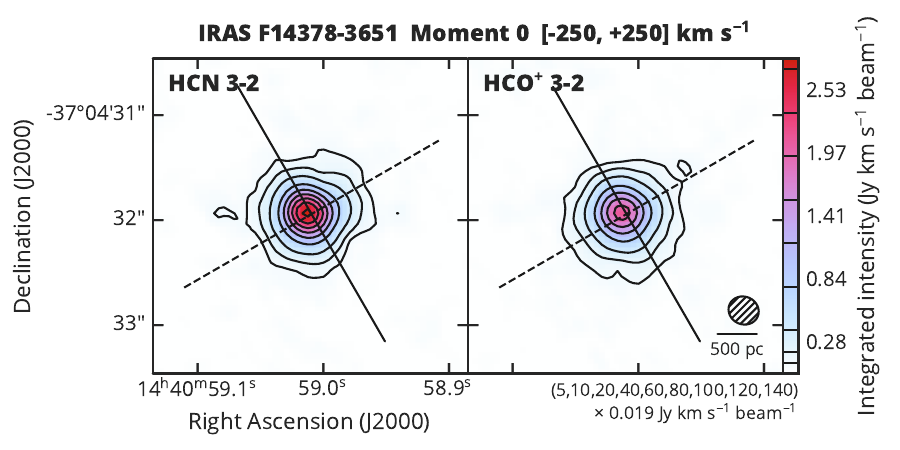}

\includegraphics[width=0.4\hsize]{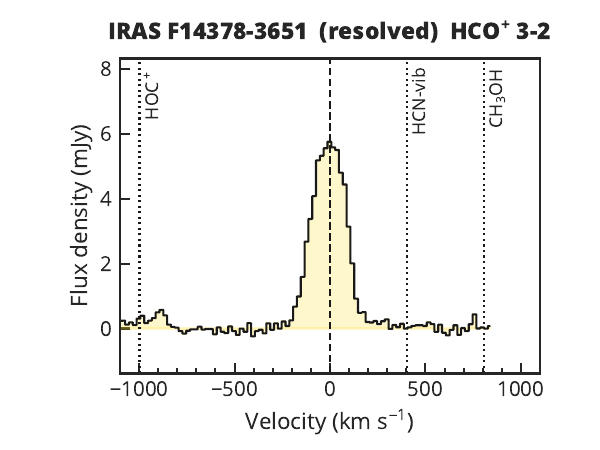}
\includegraphics[width=0.6\hsize]{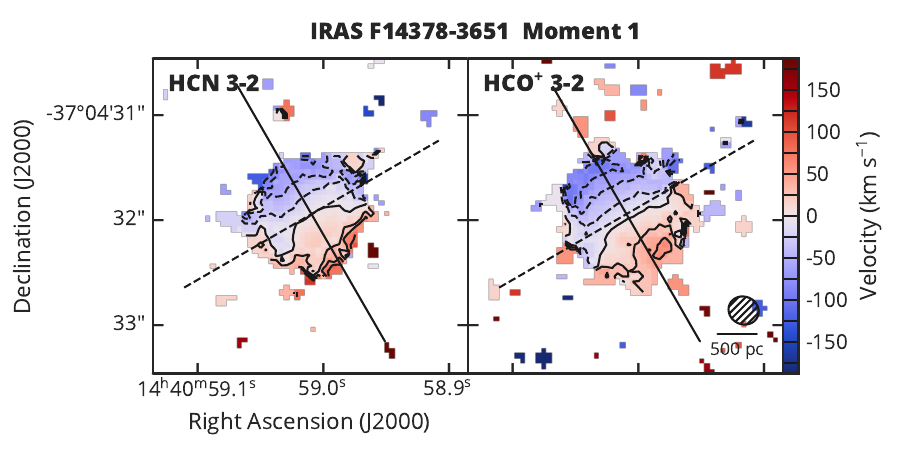}

\includegraphics[width=0.4\hsize]{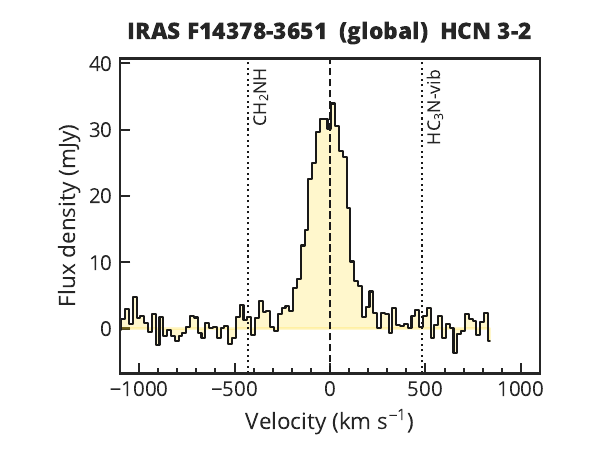}
\includegraphics[width=0.6\hsize]{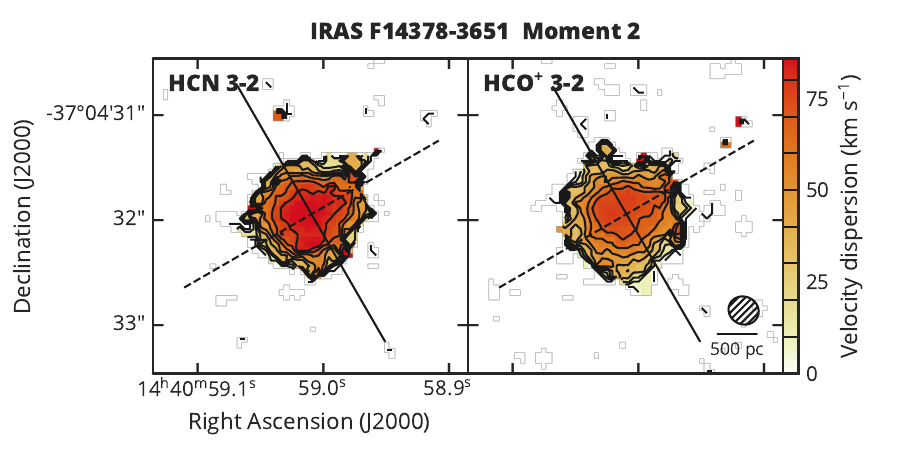}

\includegraphics[width=0.4\hsize]{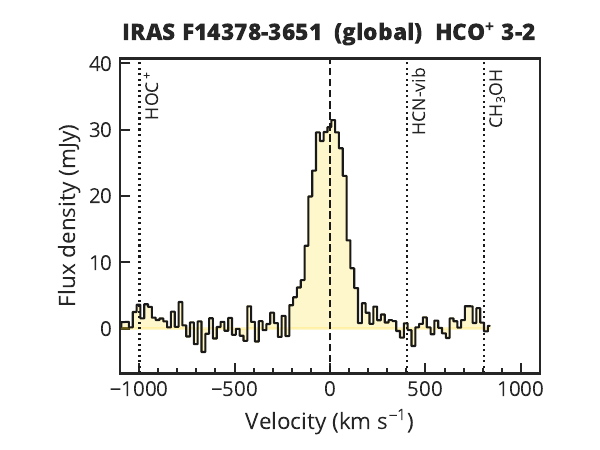}
\includegraphics[width=0.4\hsize]{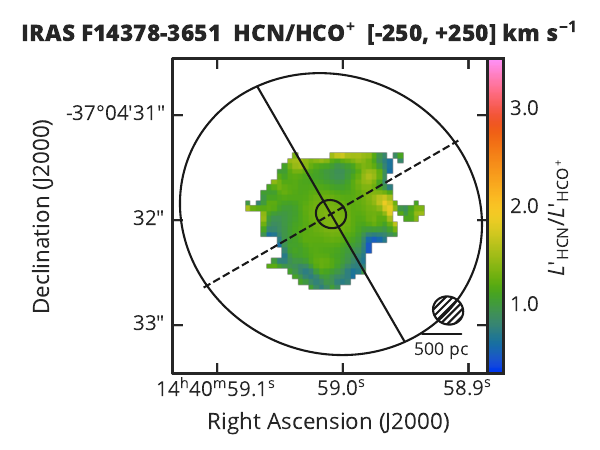}
\caption{HCN 3--2 and HCO$^+$ 3--2 for IRAS\,F14378$-$3651. 
\emph{Left panels}: (\emph{Top two panels}) 
Spectra extracted from the resolved aperture. 
(\emph{Lower two panels}) Spectra extracted from the global aperture. 
Velocities are relative to the systemic velocity. 
The corresponding velocities of potentially detected species 
are indicated by vertical dotted lines.
\emph{Right panels}: 
(\emph{Top}) Integrated intensity over $\pm250$ km s$^{-1}$ (moment 0). 
Contours are (5, 10, 20, 40, 60, 80, 100, 120, 140) $\times\sigma$, 
where $\sigma$ is 0.019 Jy km s$^{-1}$ beam$^{-1}$. 
(\emph{Second from top}) Velocity field (moment 1). 
Contours are in steps of $\pm25$ km s$^{-1}$. 
(\emph{Third from top}) Velocity dispersion (moment 2). 
Contours are in steps of 10 km s$^{-1}$. 
Moment 1 and 2 were derived with $3\sigma$ clipping. 
(\emph{Bottom}) $L'_\mathrm{HCN}/L'_\mathrm{HCO^+}$. 
Color scale is from 0.285 to 3.5. 
Overlaid ellipses represent the apertures used for spectral extraction. 
Solid and dashed lines represent the kinematic major and minor axes, respectively. 
The synthesized beam is indicated by hatched ellipses in the lower right corners.
}
\label{figure:F14378-3651}
\end{figure*}

\begin{figure*}
\includegraphics[width=0.4\hsize]{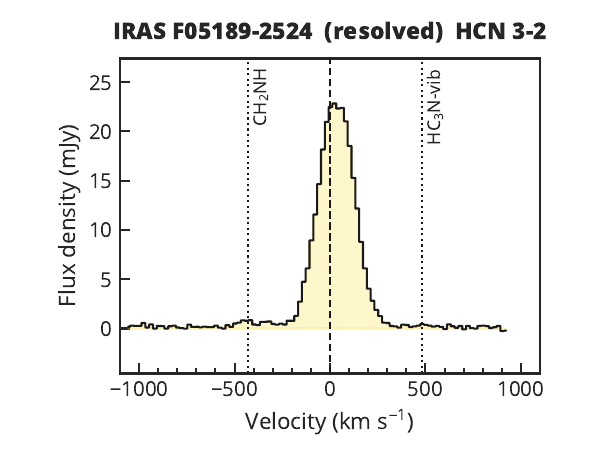}
\includegraphics[width=0.6\hsize]{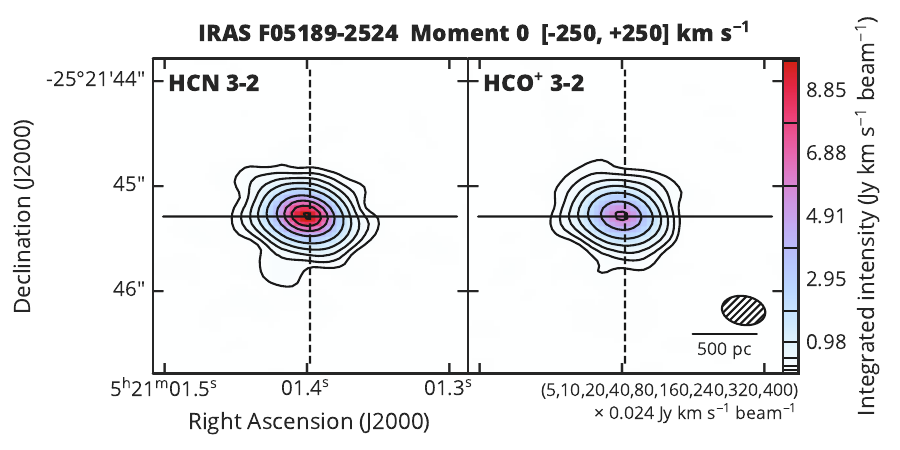}

\includegraphics[width=0.4\hsize]{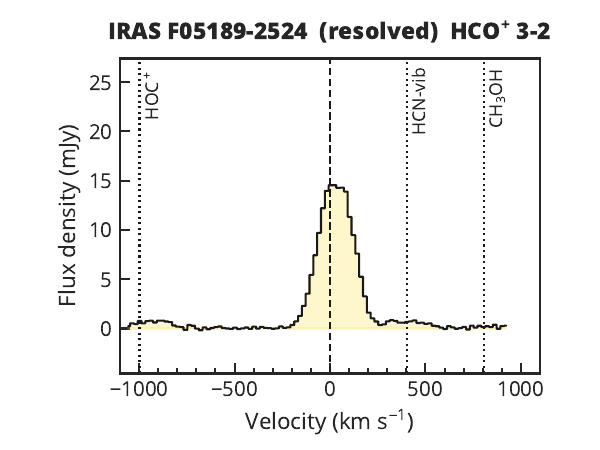}
\includegraphics[width=0.6\hsize]{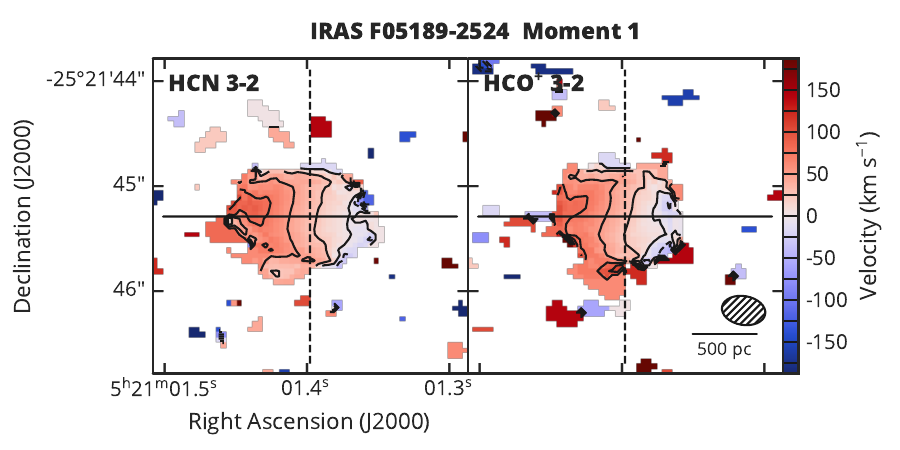}

\includegraphics[width=0.4\hsize]{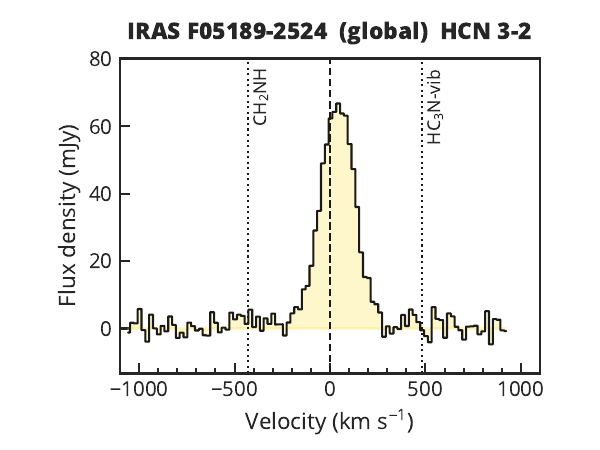}
\includegraphics[width=0.6\hsize]{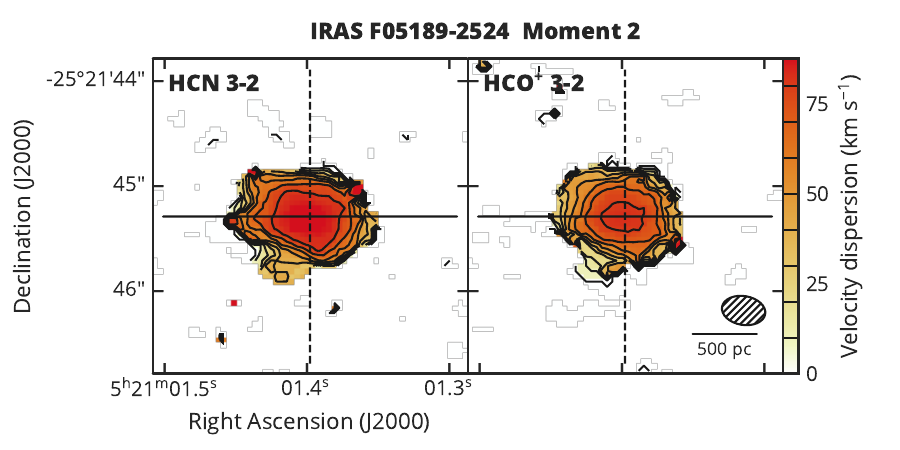}

\includegraphics[width=0.4\hsize]{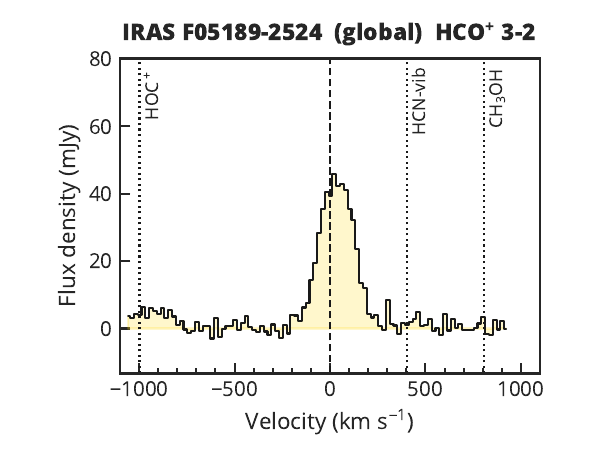}
\includegraphics[width=0.4\hsize]{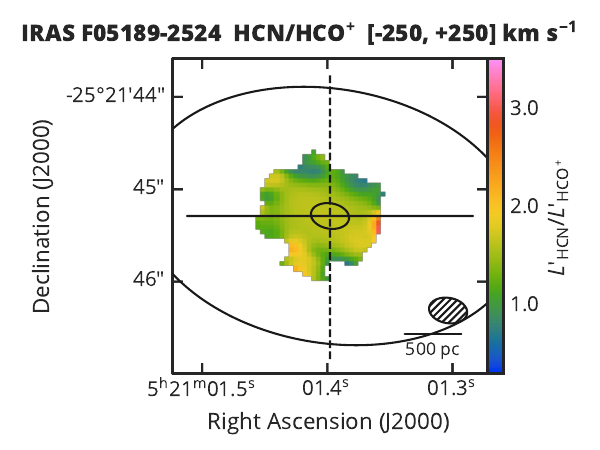}
\caption{HCN 3--2 and HCO$^+$ 3--2 for IRAS\,F05189$-$2524.
\emph{Left panels}: (\emph{Top two panels}) 
Spectra extracted from the resolved aperture. 
(\emph{Lower two panels}) Spectra extracted from the global aperture. 
Velocities are relative to the systemic velocity. 
The corresponding velocities of potentially detected species 
are indicated by vertical dotted lines.
\emph{Right panels}: 
(\emph{Top}) Integrated intensity over $\pm250$ km s$^{-1}$ (moment 0). 
Contours are (5, 10, 20, 40, 80, 160, 240, 320, 400) $\times\sigma$, 
where $\sigma$ is 0.024 Jy km s$^{-1}$ beam$^{-1}$. 
(\emph{Second from top}) Velocity field (moment 1). 
Contours are in steps of $\pm25$ km s$^{-1}$. 
(\emph{Third from top}) Velocity dispersion (moment 2). 
Contours are in steps of 10 km s$^{-1}$. 
Moment 1 and 2 were derived with $3\sigma$ clipping. 
(\emph{Bottom}) $L'_\mathrm{HCN}/L'_\mathrm{HCO^+}$. 
Color scale is from 0.285 to 3.5. 
Overlaid ellipses represent the apertures used for spectral extraction. 
Solid and dashed lines represent the kinematic major and minor axes, respectively. 
The synthesized beam is indicated by hatched ellipses in the lower right corners.
}
\label{figure:F05189-2524}
\end{figure*}

\begin{figure*}
\includegraphics[width=0.4\hsize]{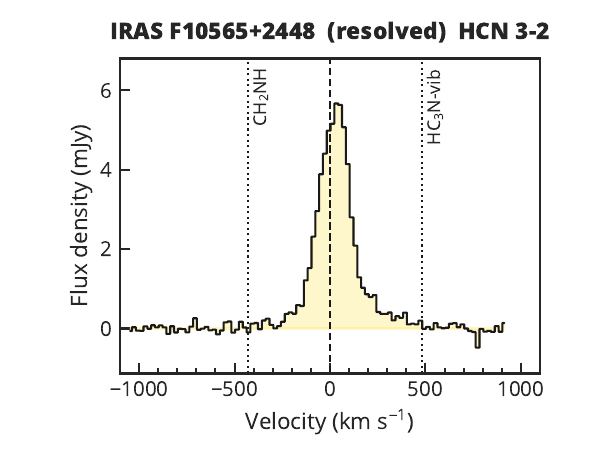}
\includegraphics[width=0.6\hsize]{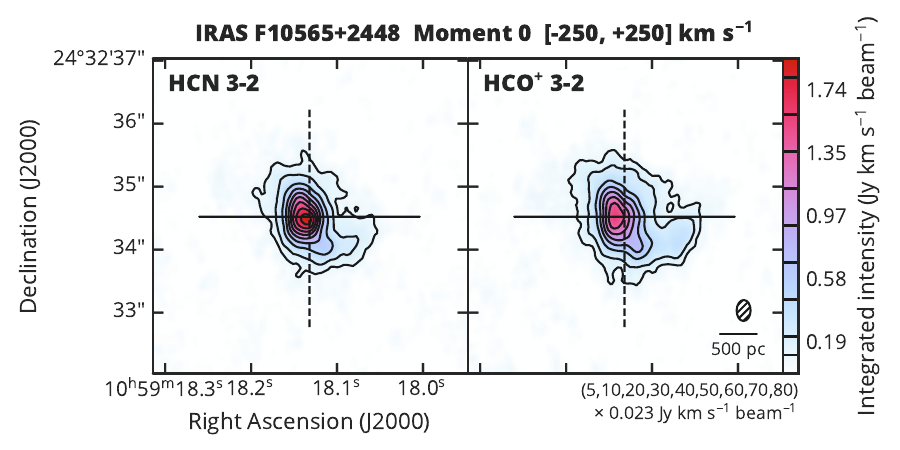}

\includegraphics[width=0.4\hsize]{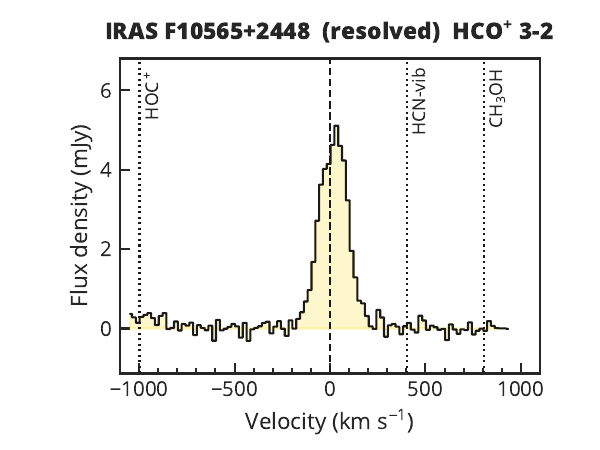}
\includegraphics[width=0.6\hsize]{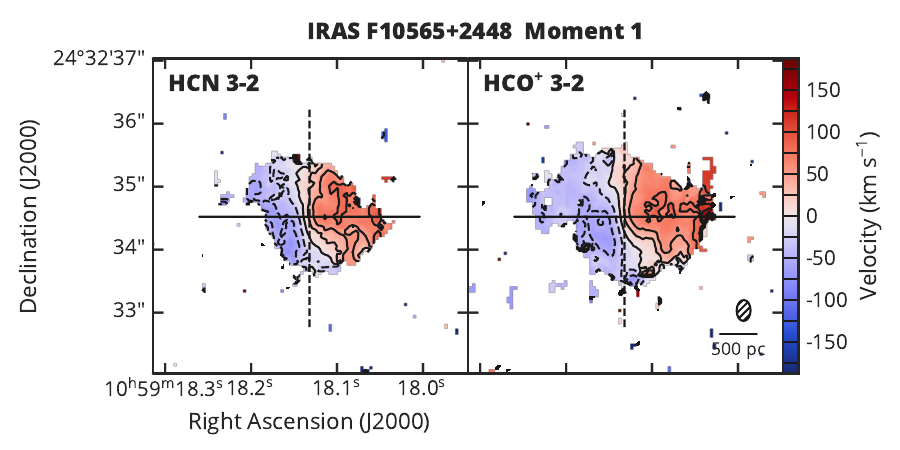}

\includegraphics[width=0.4\hsize]{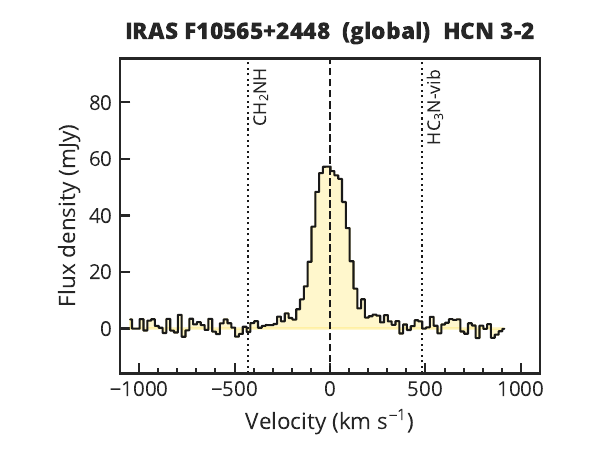}
\includegraphics[width=0.6\hsize]{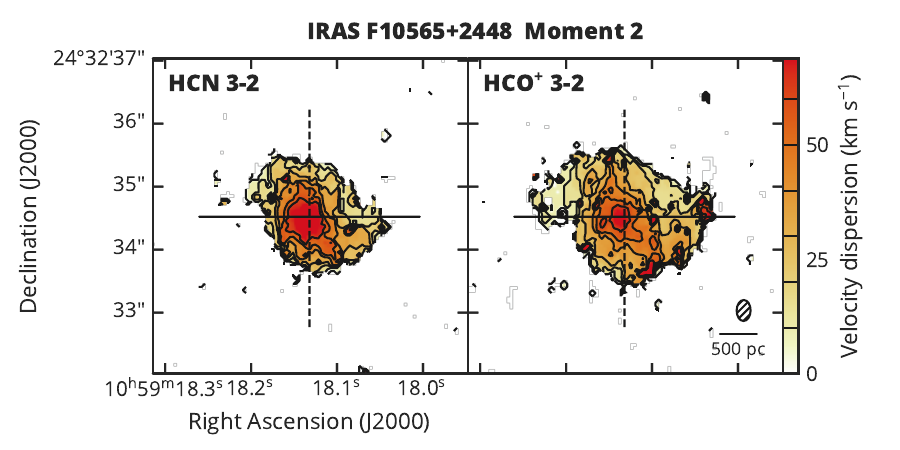}

\includegraphics[width=0.4\hsize]{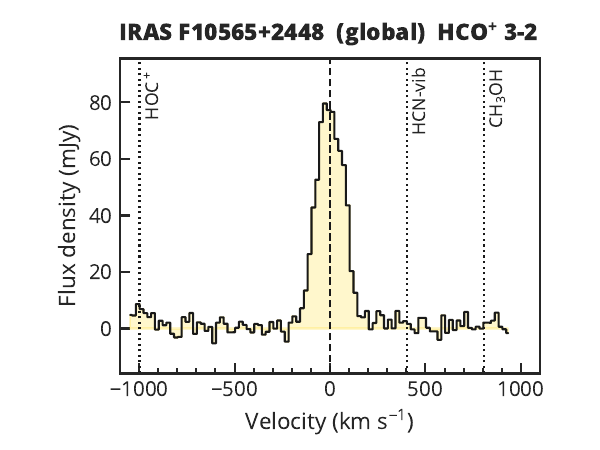}
\includegraphics[width=0.4\hsize]{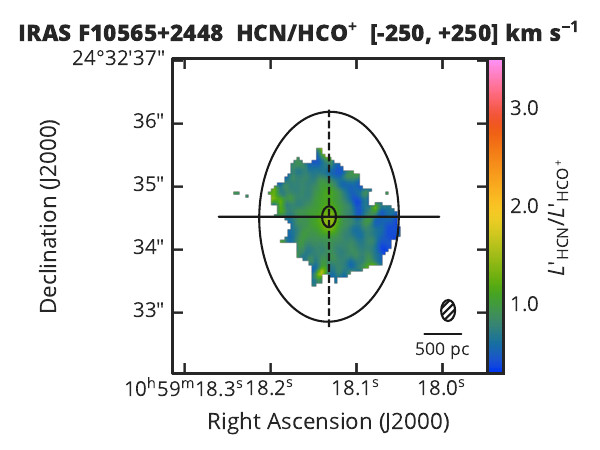}
\caption{HCN 3--2 and HCO$^+$ 3--2 for IRAS\,F10565$+$2448. 
\emph{Left panels}: (\emph{Top two panels}) 
Spectra extracted from the resolved aperture. 
(\emph{Lower two panels}) Spectra extracted from the global aperture. 
Velocities are relative to the systemic velocity. 
The corresponding velocities of potentially detected species 
are indicated by vertical dotted lines.
\emph{Right panels}: 
(\emph{Top}) Integrated intensity over $\pm250$ km s$^{-1}$ (moment 0). 
Contours are (5, 10, 20, 30, 40, 50, 60, 70, 80) $\times\sigma$, 
where $\sigma$ is 0.023 Jy km s$^{-1}$ beam$^{-1}$. 
(\emph{Second from top}) Velocity field (moment 1). 
Contours are in steps of $\pm25$ km s$^{-1}$. 
(\emph{Third from top}) Velocity dispersion (moment 2). 
Contours are in steps of 10 km s$^{-1}$. 
Moment 1 and 2 were derived with $3\sigma$ clipping. 
(\emph{Bottom}) $L'_\mathrm{HCN}/L'_\mathrm{HCO^+}$. 
Color scale is from 0.285 to 3.5. 
Overlaid ellipses represent the apertures used for spectral extraction. 
Solid and dashed lines represent the kinematic major and minor axes, respectively. 
The synthesized beam is indicated by hatched ellipses in the lower right corners.
}
\label{figure:F10565+2448}
\end{figure*}

\begin{figure*}
\includegraphics[width=0.4\hsize]{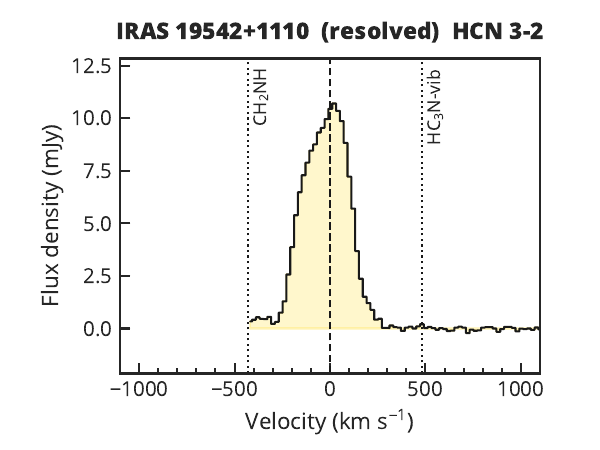}
\includegraphics[width=0.6\hsize]{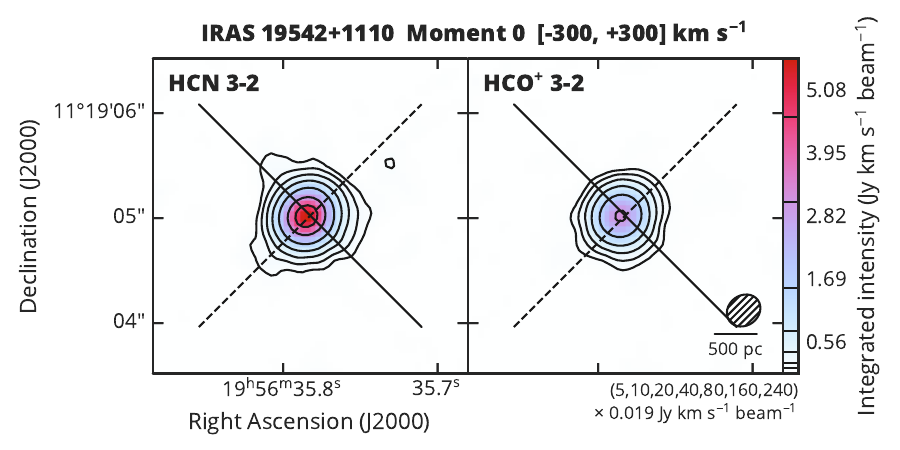}

\includegraphics[width=0.4\hsize]{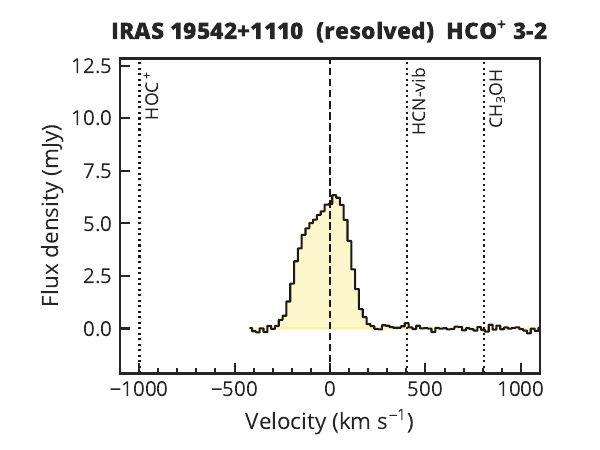}
\includegraphics[width=0.6\hsize]{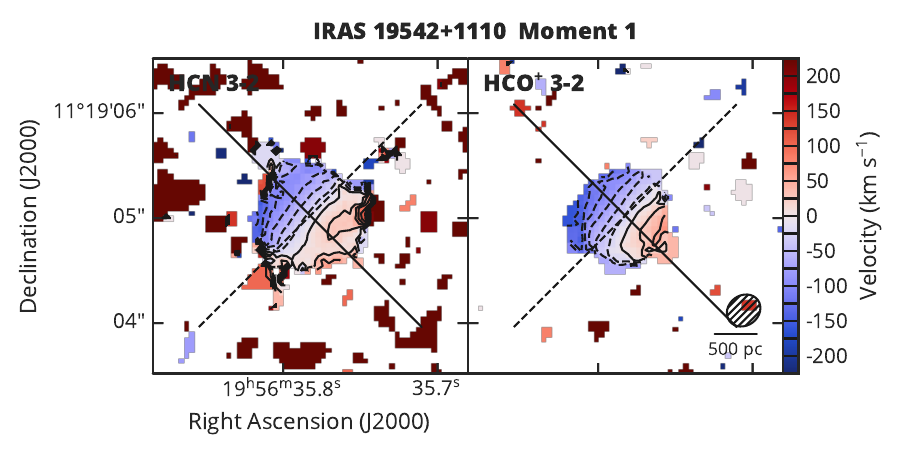}

\includegraphics[width=0.4\hsize]{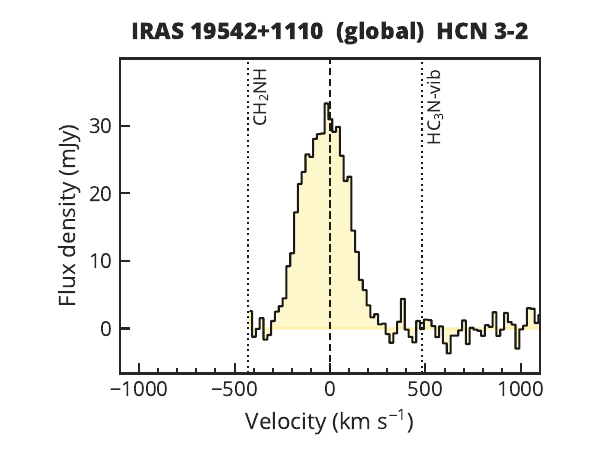}
\includegraphics[width=0.6\hsize]{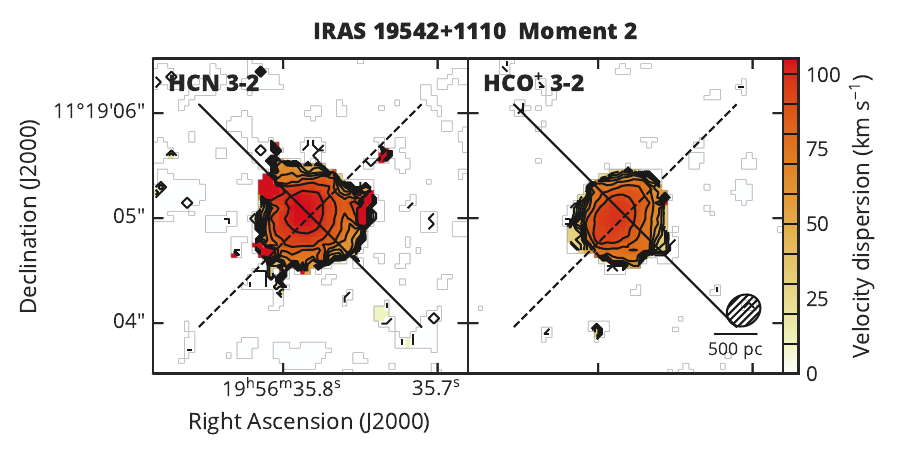}

\includegraphics[width=0.4\hsize]{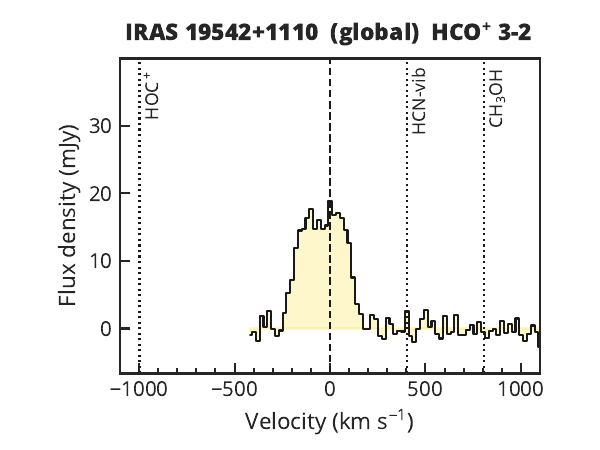}
\includegraphics[width=0.4\hsize]{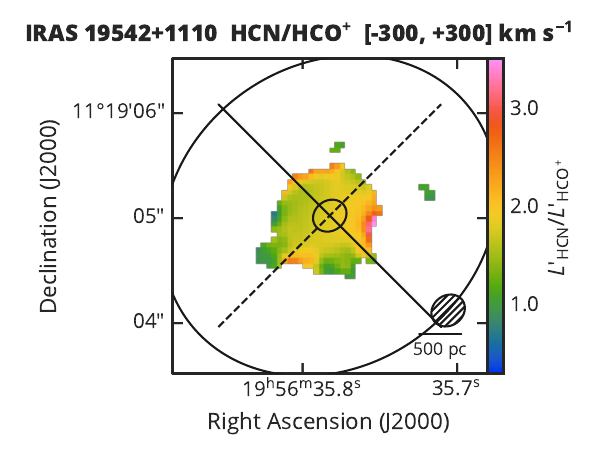}
\caption{HCN 3--2 and HCO$^+$ 3--2 for IRAS\,19542$+$1110. 
\emph{Left panels}: (\emph{Top two panels}) 
Spectra extracted from the resolved aperture. 
(\emph{Lower two panels}) Spectra extracted from the global aperture. 
Velocities are relative to the systemic velocity. 
The corresponding velocities of potentially detected species 
are indicated by vertical dotted lines.
\emph{Right panels}: 
(\emph{Top}) Integrated intensity over $\pm300$ km s$^{-1}$ (moment 0). 
Contours are (5, 10, 20, 40, 80, 160, 240) $\times\sigma$, 
where $\sigma$ is 0.019 Jy km s$^{-1}$ beam$^{-1}$. 
(\emph{Second from top}) Velocity field (moment 1). 
Contours are in steps of $\pm25$ km s$^{-1}$. 
(\emph{Third from top}) Velocity dispersion (moment 2). 
Contours are in steps of 10 km s$^{-1}$. 
Moment 1 and 2 were derived with $3\sigma$ clipping. 
(\emph{Bottom}) $L'_\mathrm{HCN}/L'_\mathrm{HCO^+}$. 
Color scale is from 0.285 to 3.5. 
Overlaid ellipses represent the apertures used for spectral extraction. 
Solid and dashed lines represent the kinematic major and minor axes, respectively. 
The synthesized beam is indicated by hatched ellipses in the lower right corners.
}
\label{figure:19542+1110}
\end{figure*}

\begin{figure*}
\includegraphics[width=0.4\hsize]{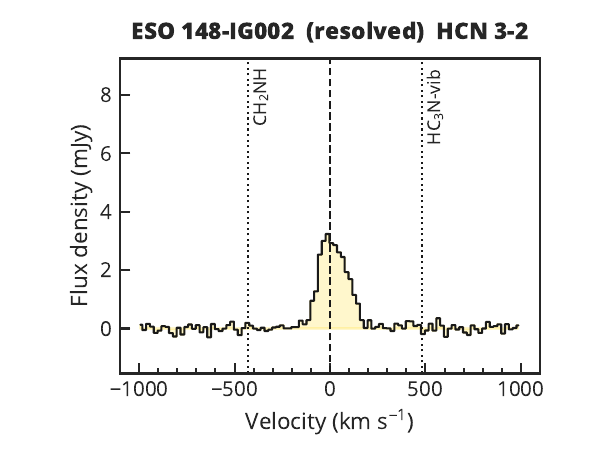}
\includegraphics[width=0.6\hsize]{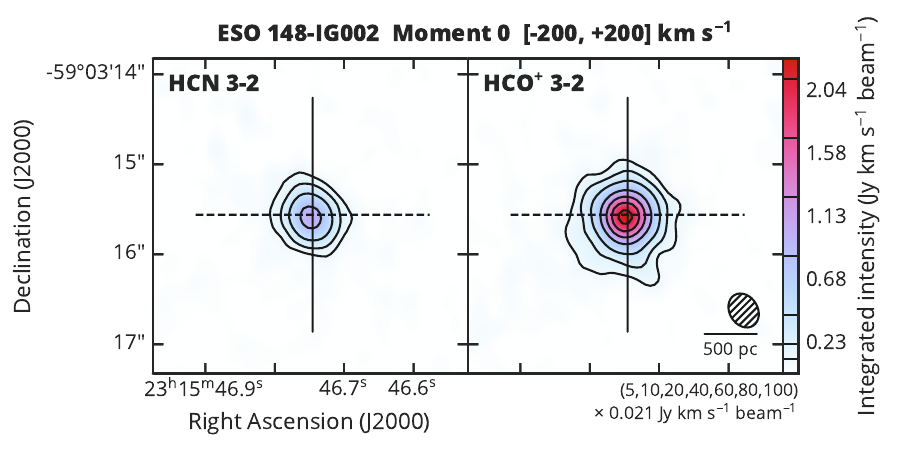}

\includegraphics[width=0.4\hsize]{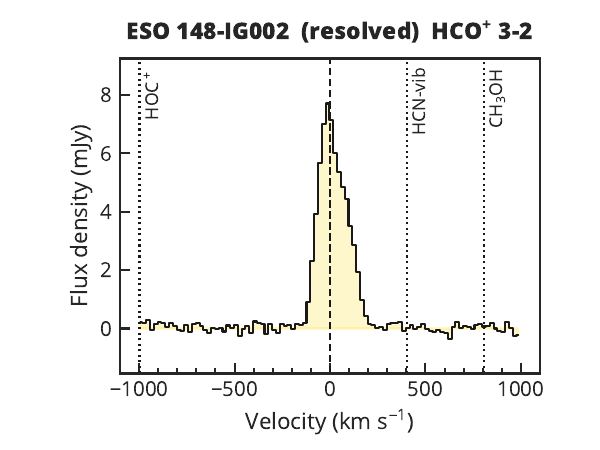}
\includegraphics[width=0.6\hsize]{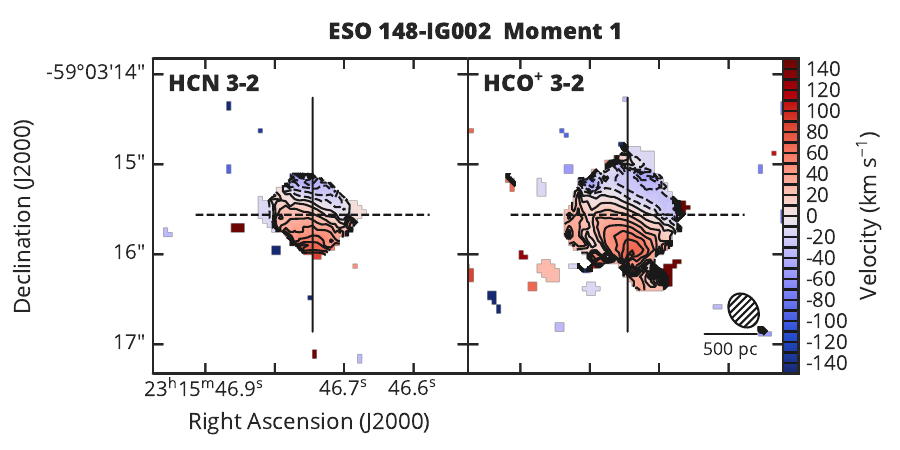}

\includegraphics[width=0.4\hsize]{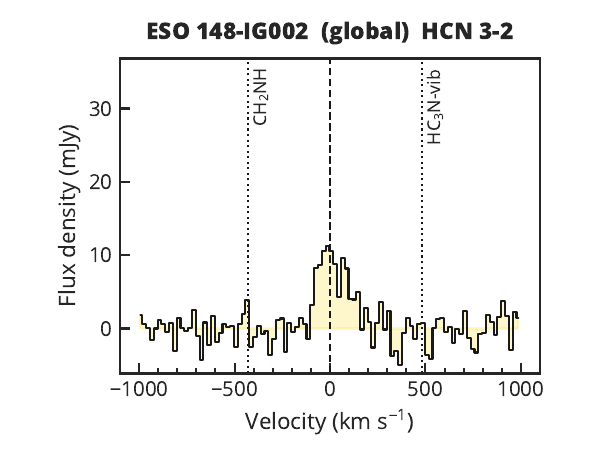}
\includegraphics[width=0.6\hsize]{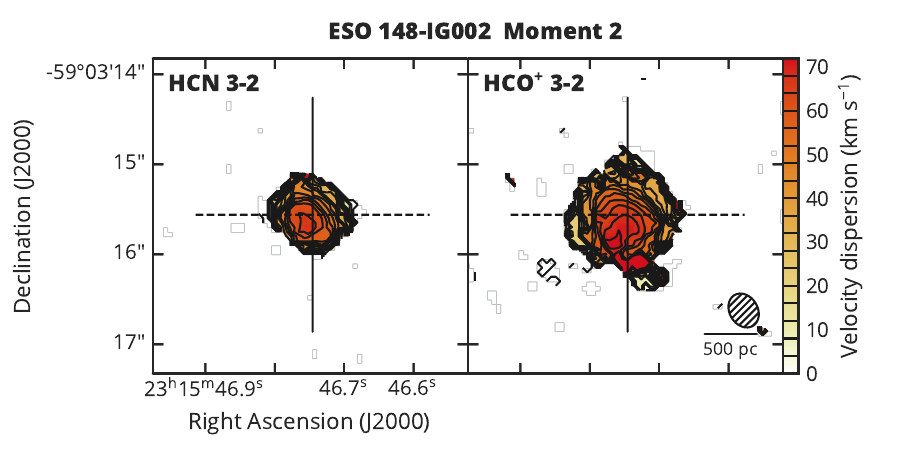}

\includegraphics[width=0.4\hsize]{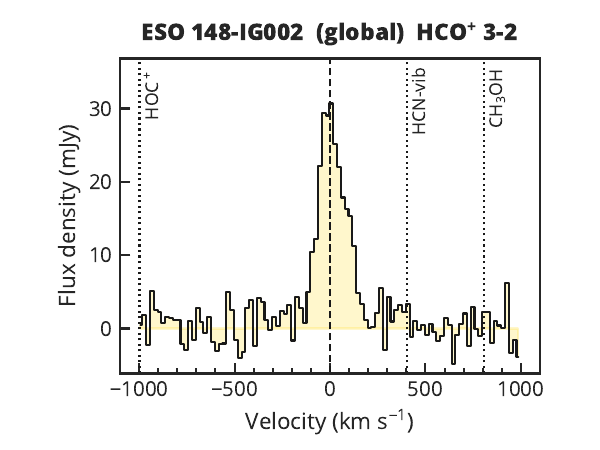}
\includegraphics[width=0.4\hsize]{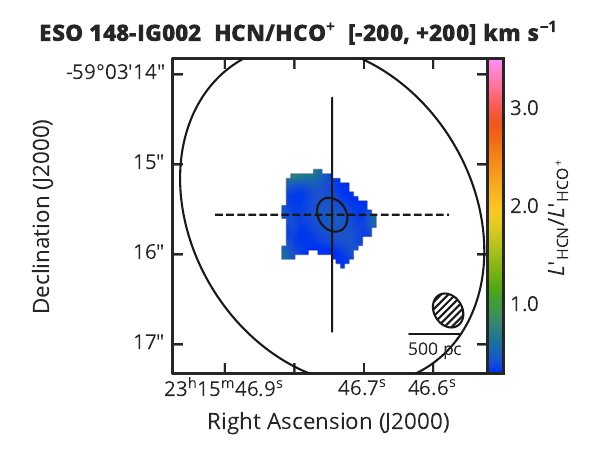}
\caption{HCN 3--2 and HCO$^+$ 3--2 for ESO\,148$-$IG002. 
\emph{Left panels}: (\emph{Top two panels}) 
Spectra extracted from the resolved aperture. 
(\emph{Lower two panels}) Spectra extracted from the global aperture. 
Velocities are relative to the systemic velocity. 
The corresponding velocities of potentially detected species 
are indicated by vertical dotted lines.
\emph{Right panels}: 
(\emph{Top}) Integrated intensity over $\pm200$ km s$^{-1}$ (moment 0). 
Contours are (5, 10, 20, 40, 60, 80, 100) $\times\sigma$, 
where $\sigma$ is 0.019 Jy km s$^{-1}$ beam$^{-1}$. 
(\emph{Second from top}) Velocity field (moment 1). 
Contours are in steps of $\pm10$ km s$^{-1}$. 
(\emph{Third from top}) Velocity dispersion (moment 2). 
Contours are in steps of 4 km s$^{-1}$. 
Moment 1 and 2 were derived with $3\sigma$ clipping. 
(\emph{Bottom}) $L'_\mathrm{HCN}/L'_\mathrm{HCO^+}$. 
Color scale is from 0.285 to 3.5. 
Overlaid ellipses represent the apertures used for spectral extraction. 
Solid and dashed lines represent the kinematic major and minor axes, respectively. 
The synthesized beam is indicated by hatched ellipses in the lower right corners.
}
\label{figure:148-IG002}
\end{figure*}

\begin{figure*}
\includegraphics[width=0.4\hsize]{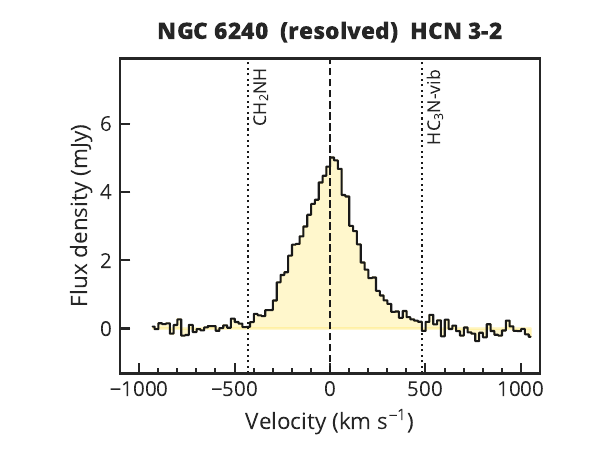}
\includegraphics[width=0.6\hsize]{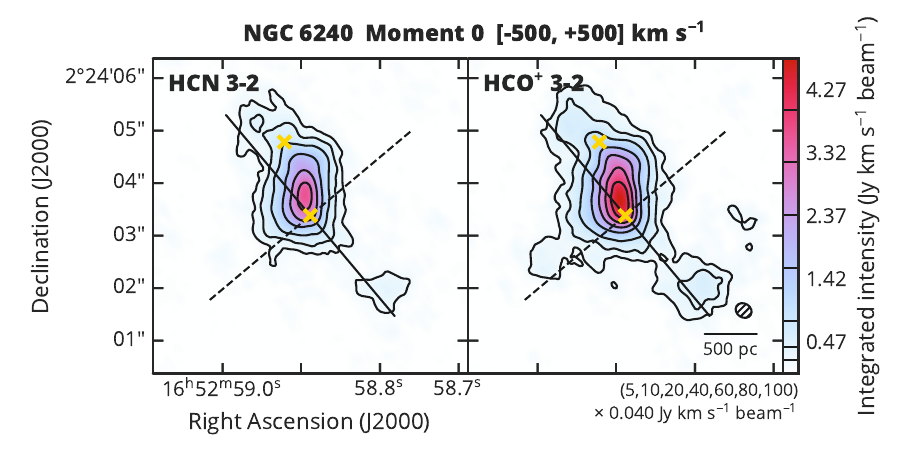}

\includegraphics[width=0.4\hsize]{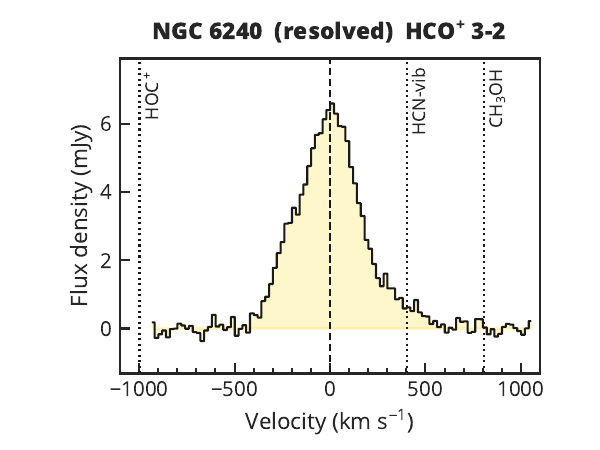}
\includegraphics[width=0.6\hsize]{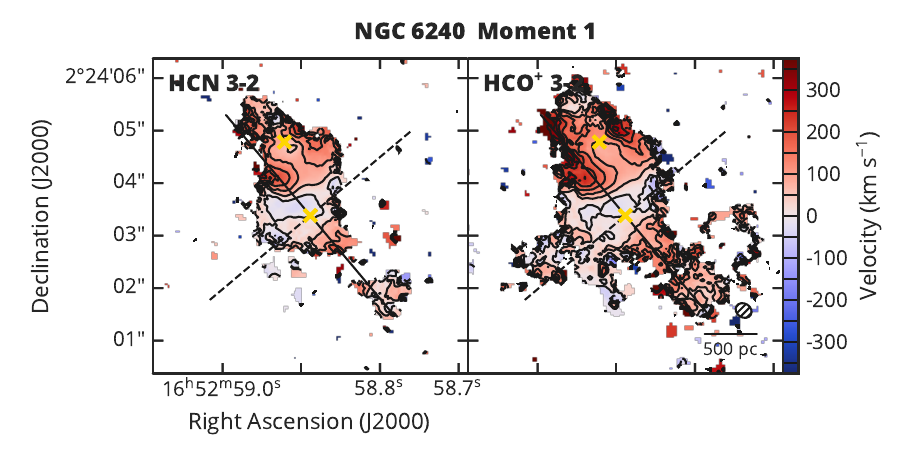}

\includegraphics[width=0.4\hsize]{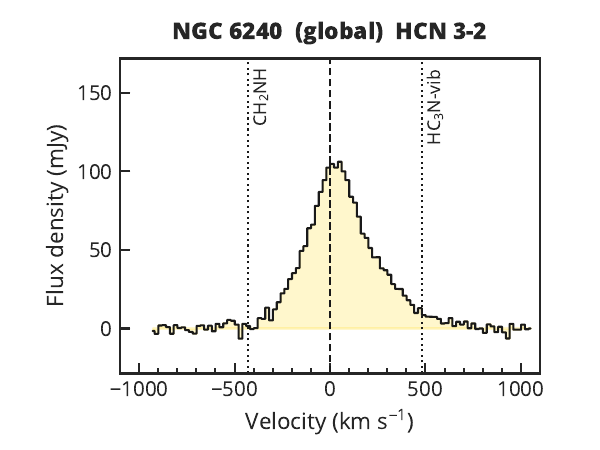}
\includegraphics[width=0.6\hsize]{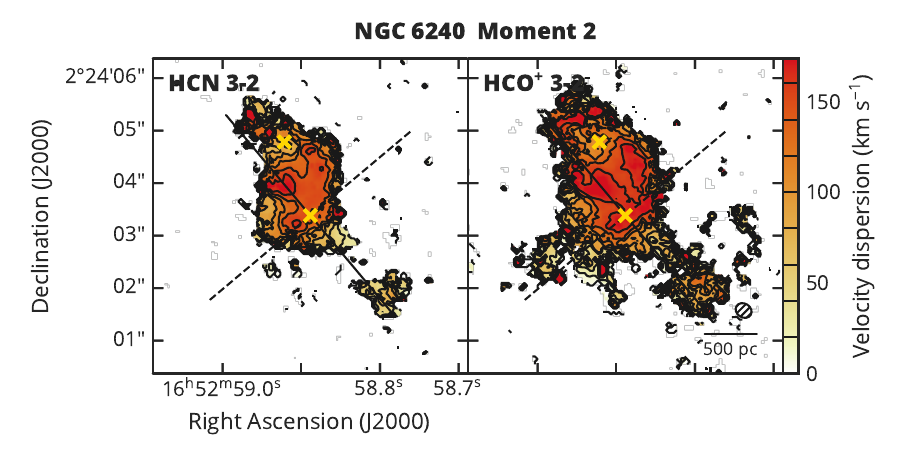}

\includegraphics[width=0.4\hsize]{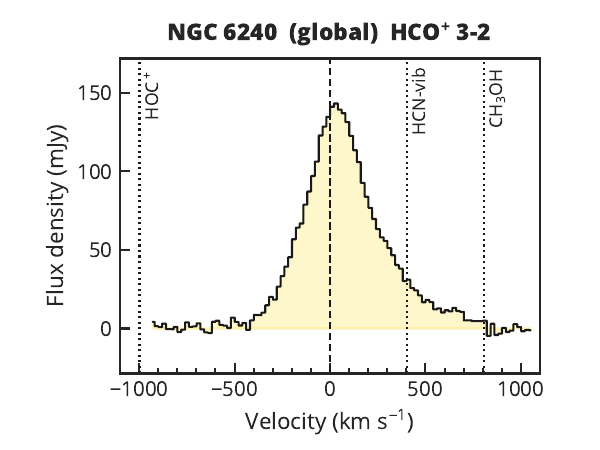}
\includegraphics[width=0.4\hsize]{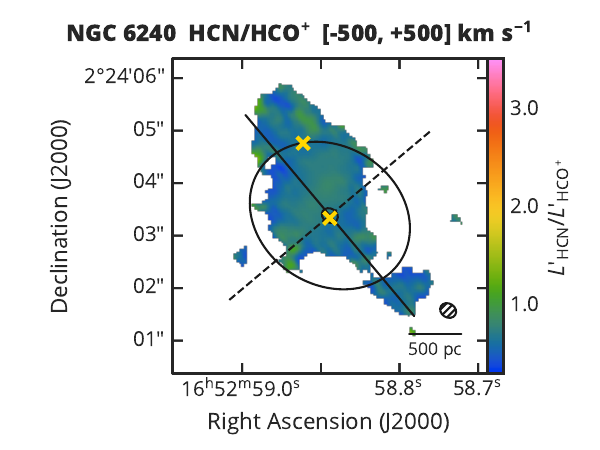}
\caption{HCN 3--2 and HCO$^+$ 3--2 for NGC\,6240. 
\emph{Left panels}: (\emph{Top two panels}) 
Spectra extracted from the resolved aperture. 
(\emph{Lower two panels}) Spectra extracted from the global aperture. 
Velocities are relative to the systemic velocity. 
The corresponding velocities of potentially detected species 
are indicated by vertical dotted lines.
\emph{Right panels}: 
(\emph{Top}) Integrated intensity over $\pm500$ km s$^{-1}$ (moment 0). 
Contours are (5, 10, 20, 40, 60, 80, 100) $\times\sigma$, 
where $\sigma$ is 0.040 Jy km s$^{-1}$ beam$^{-1}$. 
(\emph{Second from top}) Velocity field (moment 1). 
Contours are in steps of $\pm50$ km s$^{-1}$. 
(\emph{Third from top}) Velocity dispersion (moment 2). 
Contours are in steps of 20 km s$^{-1}$. 
Moment 1 and 2 were derived with $3\sigma$ clipping. 
(\emph{Bottom}) $L'_\mathrm{HCN}/L'_\mathrm{HCO^+}$. 
Color scale is from 0.285 to 3.5. 
Overlaid ellipses represent the apertures used for spectral extraction. 
Solid and dashed lines represent the kinematic major and minor axes, respectively. 
The synthesized beam is indicated by hatched ellipses in the lower right corners.
Yellow crosses show the positions of two nuclei taken from \citet{Hagiwara2011}. 
}
\label{figure:6240}
\end{figure*}

\begin{figure*}
\includegraphics[width=0.4\hsize]{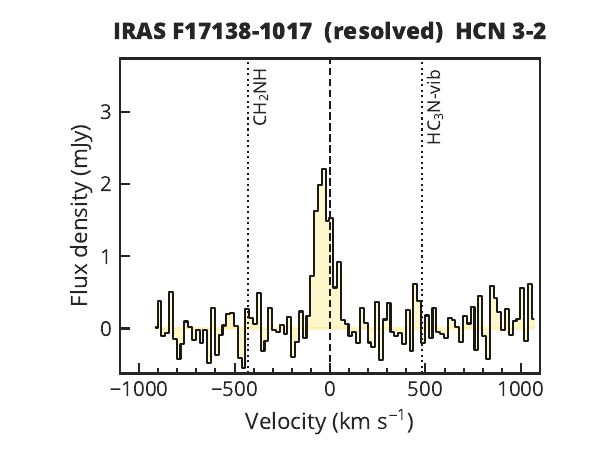}
\includegraphics[width=0.6\hsize]{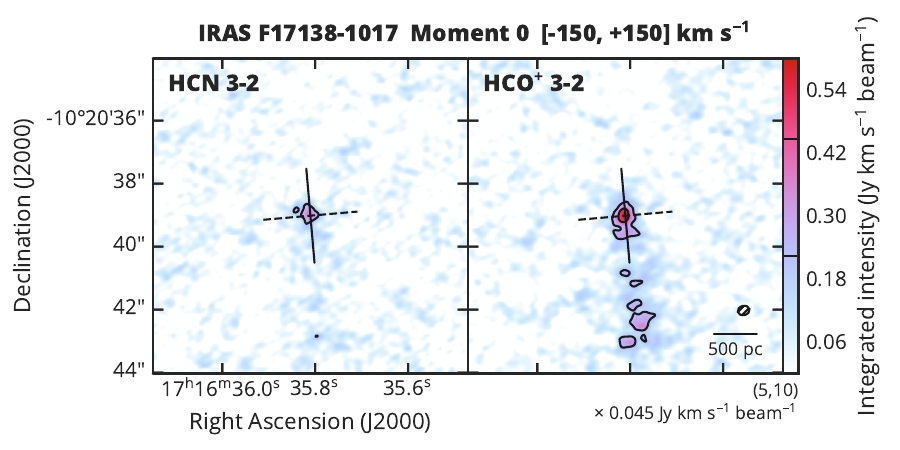}

\includegraphics[width=0.4\hsize]{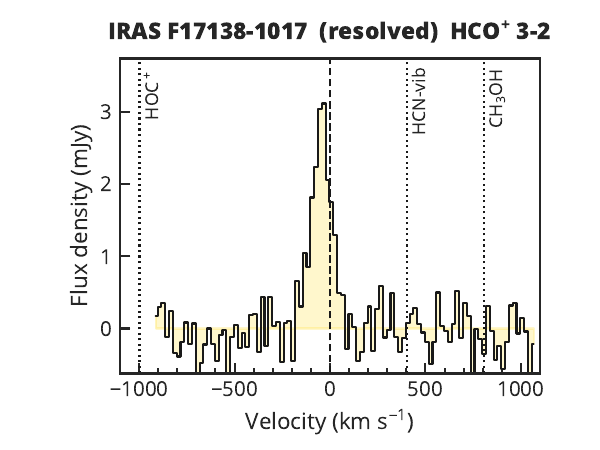}
\includegraphics[width=0.6\hsize]{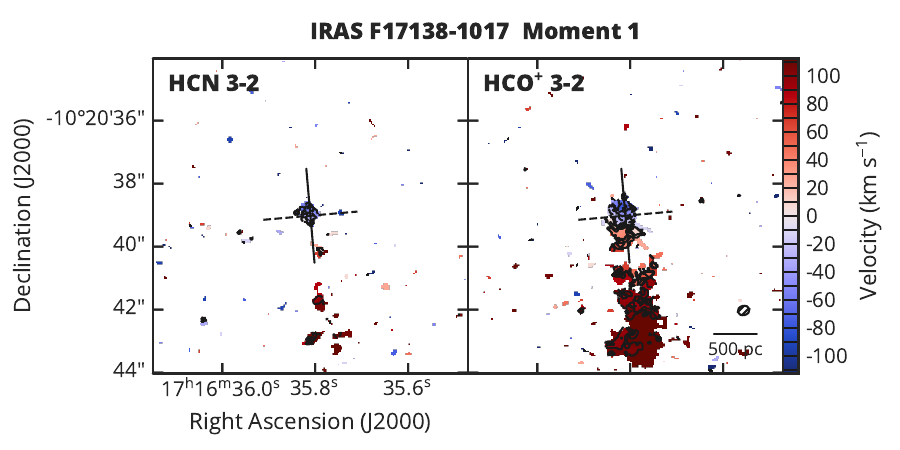}

\includegraphics[width=0.4\hsize]{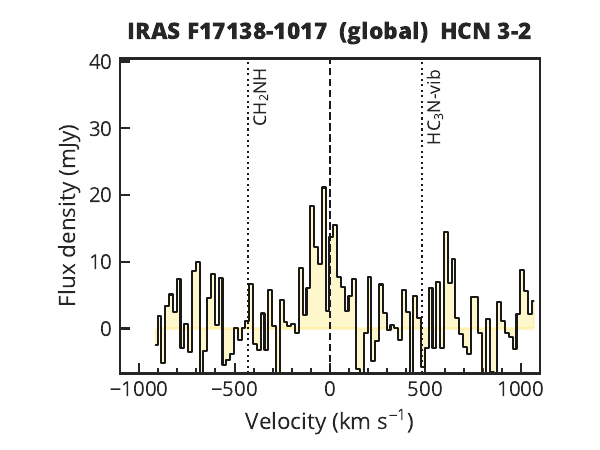}
\includegraphics[width=0.6\hsize]{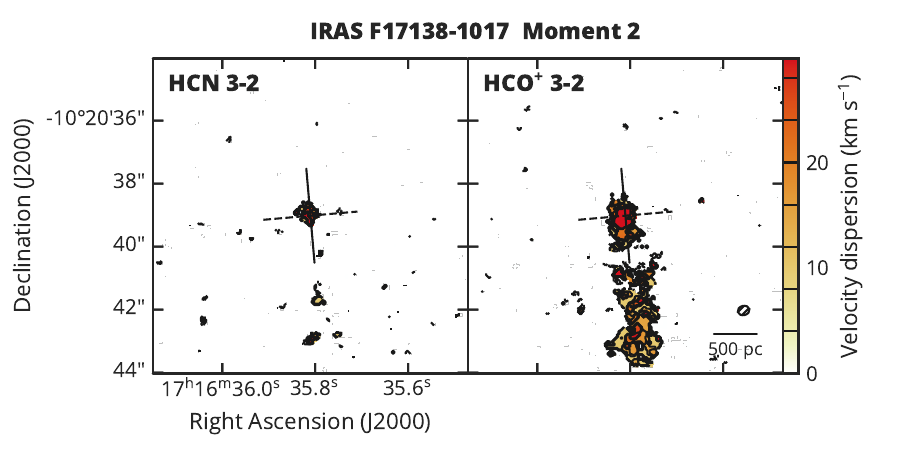}

\includegraphics[width=0.4\hsize]{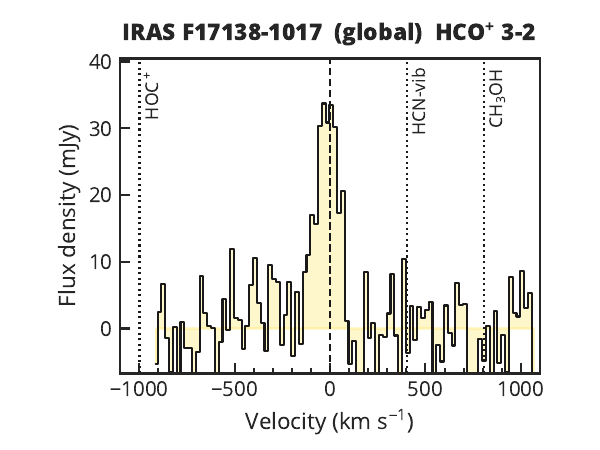}
\includegraphics[width=0.4\hsize]{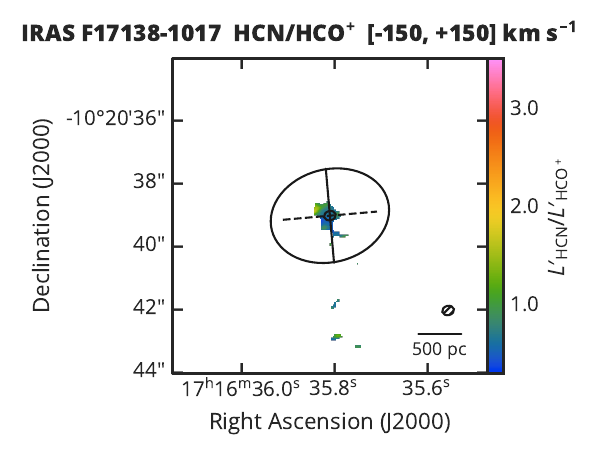}
\caption{HCN 3--2 and HCO$^+$ 3--2 for IRAS\,F17138$-$1017. 
\emph{Left panels}: (\emph{Top two panels}) 
Spectra extracted from the resolved aperture. 
(\emph{Lower two panels}) Spectra extracted from the global aperture. 
Velocities are relative to the systemic velocity. 
The corresponding velocities of potentially detected species 
are indicated by vertical dotted lines.
\emph{Right panels}: 
(\emph{Top}) Integrated intensity over $\pm150$ km s$^{-1}$ (moment 0). 
Contours are (5, 10) $\times\sigma$, 
where $\sigma$ is 0.045 Jy km s$^{-1}$ beam$^{-1}$. 
(\emph{Second from top}) Velocity field (moment 1). 
Contours are in steps of $\pm10$ km s$^{-1}$. 
(\emph{Third from top}) Velocity dispersion (moment 2). 
Contours are in steps of 4 km s$^{-1}$. 
Moment 1 and 2 were derived with $3\sigma$ clipping. 
(\emph{Bottom}) $L'_\mathrm{HCN}/L'_\mathrm{HCO^+}$. 
Color scale is from 0.285 to 3.5. 
Overlaid ellipses represent the apertures used for spectral extraction. 
Solid and dashed lines represent the kinematic major and minor axes, respectively. 
The synthesized beam is indicated by hatched ellipses in the lower right corners.
}
\label{figure:F17138-1017}
\end{figure*}

\begin{figure*}
\includegraphics[width=0.4\hsize]{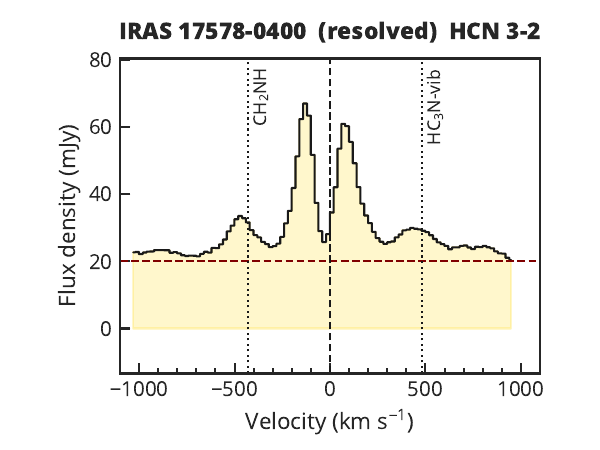}
\includegraphics[width=0.6\hsize]{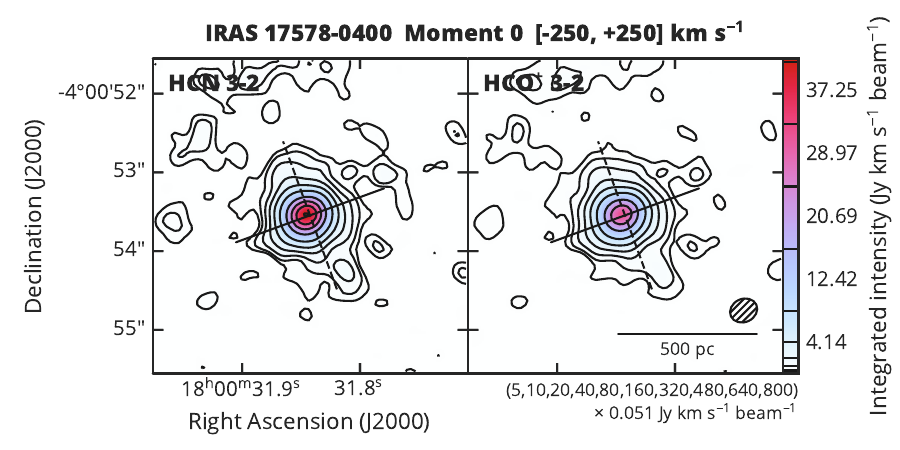}

\includegraphics[width=0.4\hsize]{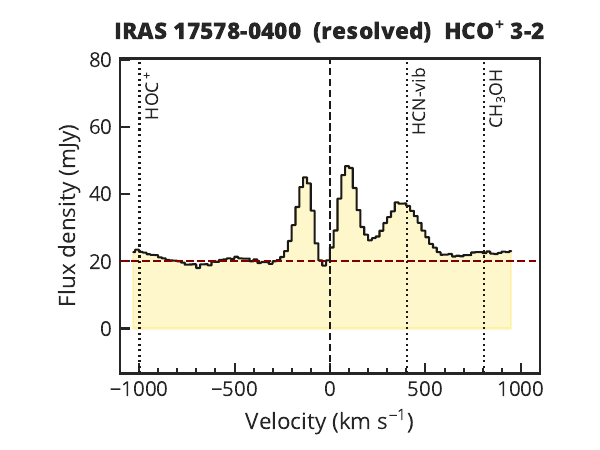}
\includegraphics[width=0.6\hsize]{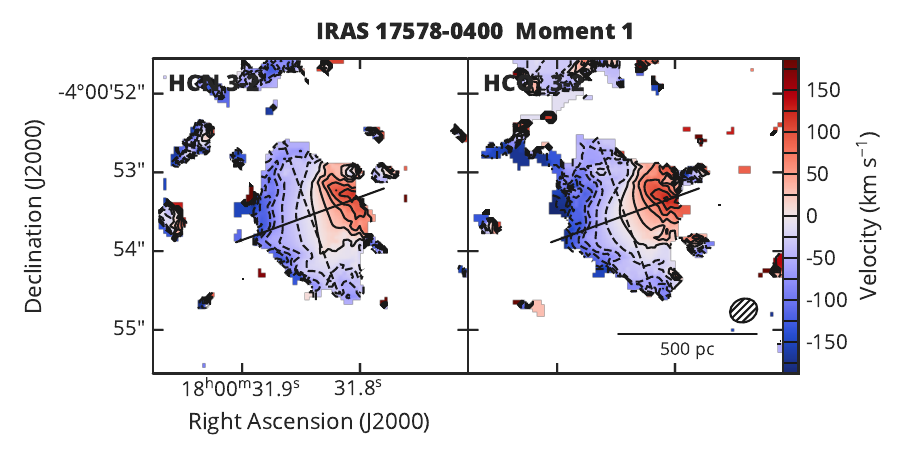}

\includegraphics[width=0.4\hsize]{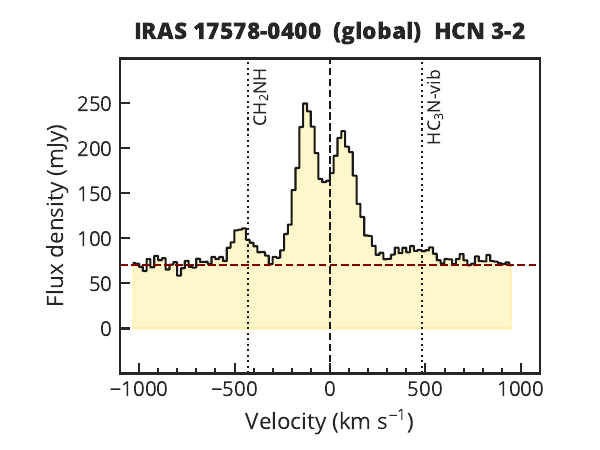}
\includegraphics[width=0.6\hsize]{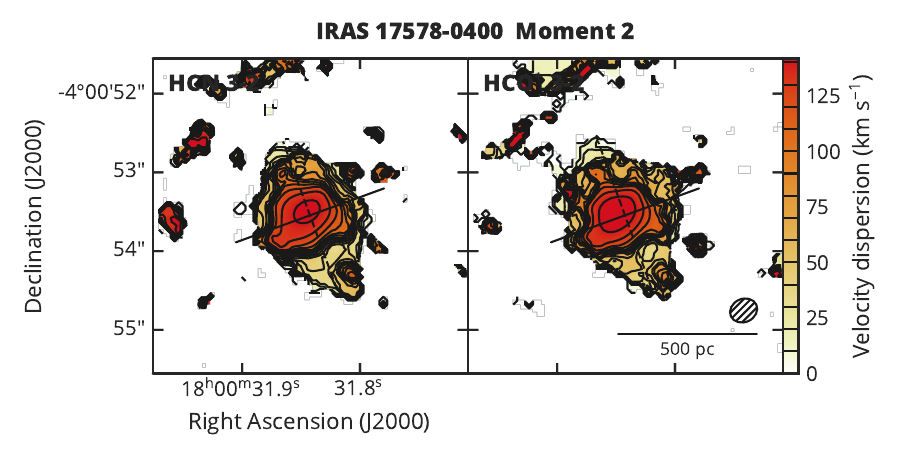}

\includegraphics[width=0.4\hsize]{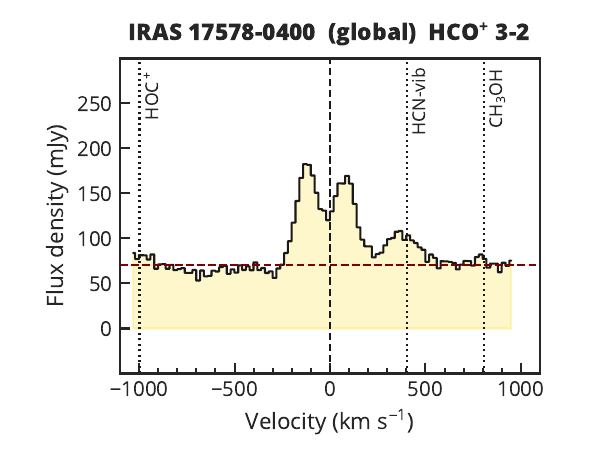}
\includegraphics[width=0.4\hsize]{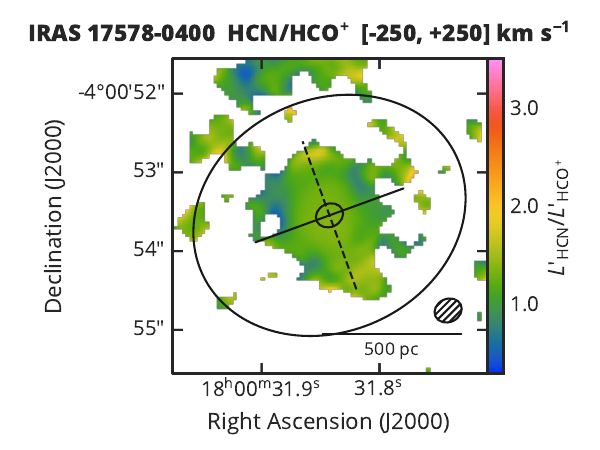}
\caption{HCN 3--2 and HCO$^+$ 3--2 for IRAS\,17578$-$0400. 
\emph{Left panels}: (\emph{Top two panels}) 
Spectra extracted from the resolved aperture. 
(\emph{Lower two panels}) Spectra extracted from the global aperture. 
Velocities are relative to the systemic velocity. 
The corresponding velocities of potentially detected species 
are indicated by vertical dotted lines. 
Horizontal dashed lines are the assumed continuum level. 
\emph{Right panels}: 
(\emph{Top}) Integrated intensity over $\pm250$ km s$^{-1}$ (moment 0). 
Contours are (5, 10, 20, 40, 80, 160, 320, 480, 640, 800) $\times\sigma$, 
where $\sigma$ is 0.051 Jy km s$^{-1}$ beam$^{-1}$. 
(\emph{Second from top}) Velocity field (moment 1). 
Contours are in steps of $\pm25$ km s$^{-1}$. 
(\emph{Third from top}) Velocity dispersion (moment 2). 
Contours are in steps of 10 km s$^{-1}$. 
Moment 1 and 2 were derived with $3\sigma$ clipping. 
(\emph{Bottom}) $L'_\mathrm{HCN}/L'_\mathrm{HCO^+}$. 
Color scale is from 0.285 to 3.5. 
Overlaid ellipses represent the apertures used for spectral extraction. 
Solid and dashed lines represent the kinematic major and minor axes, respectively. 
The synthesized beam is indicated by hatched ellipses in the lower right corners. 
Note that continuum is not subtracted. 
}
\label{figure:17578-0400}
\end{figure*}

\begin{figure*}
\includegraphics[width=0.4\hsize]{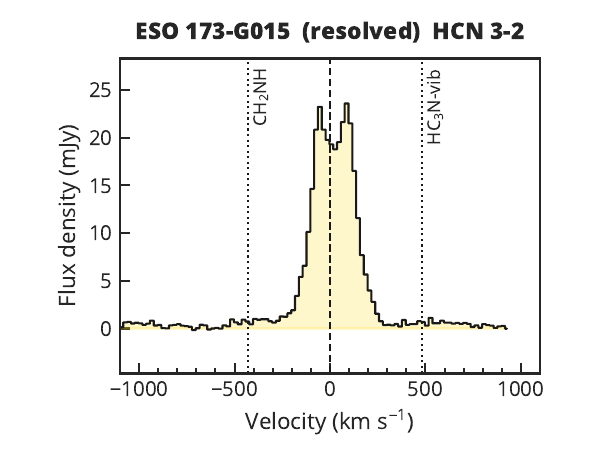}
\includegraphics[width=0.6\hsize]{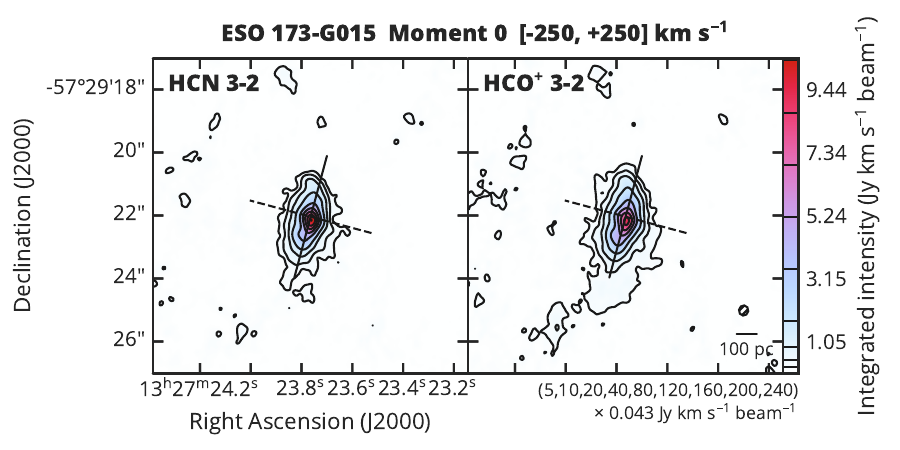}

\includegraphics[width=0.4\hsize]{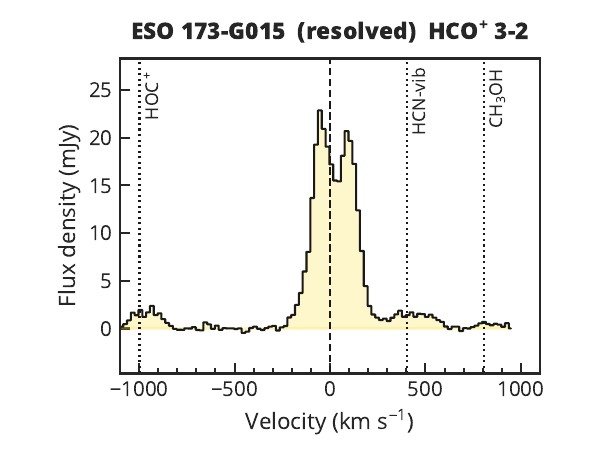}
\includegraphics[width=0.6\hsize]{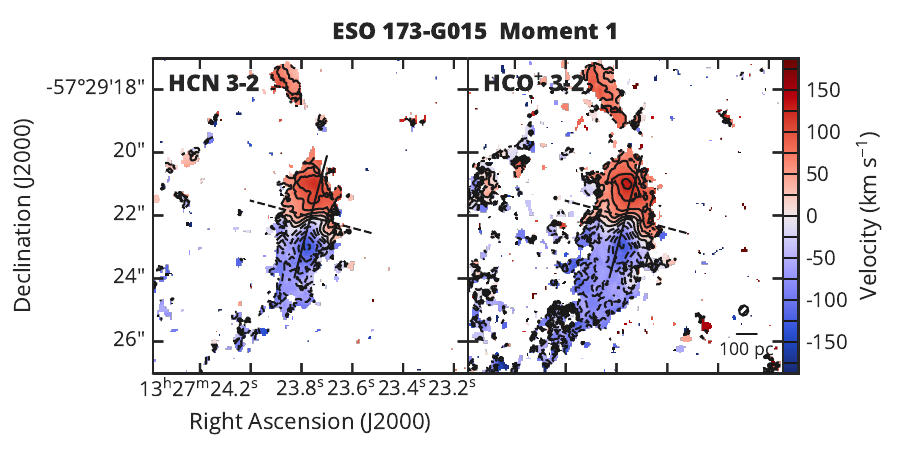}

\includegraphics[width=0.4\hsize]{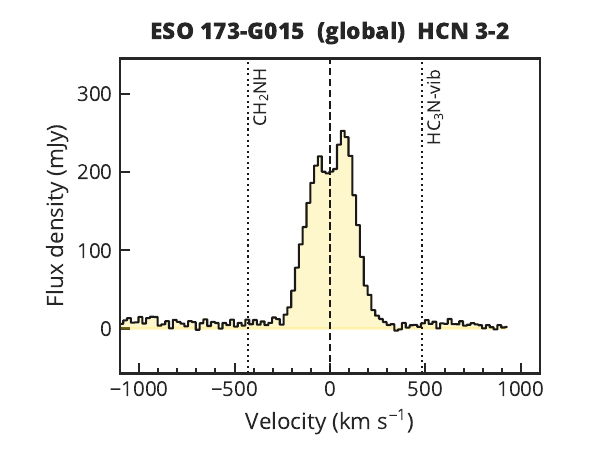}
\includegraphics[width=0.6\hsize]{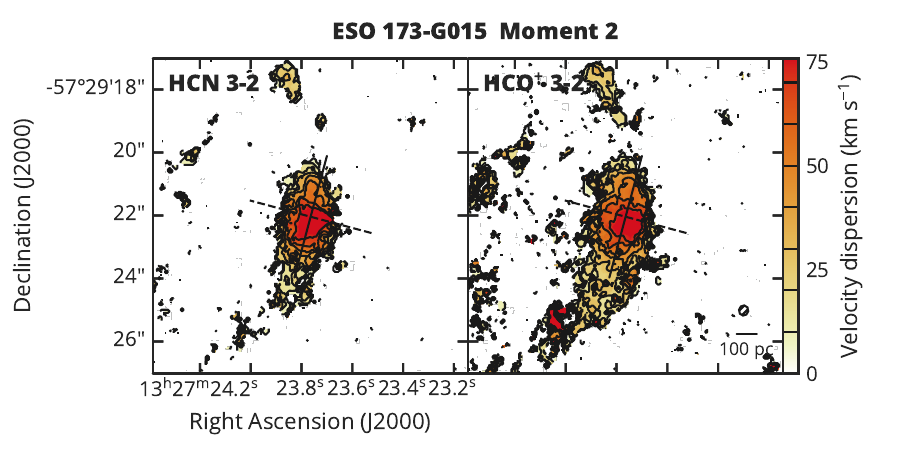}

\includegraphics[width=0.4\hsize]{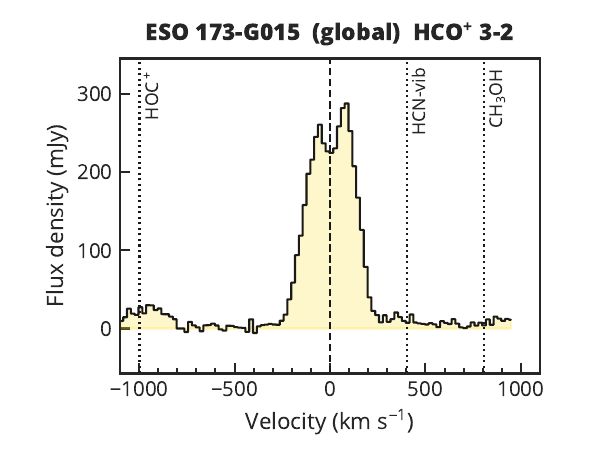}
\includegraphics[width=0.4\hsize]{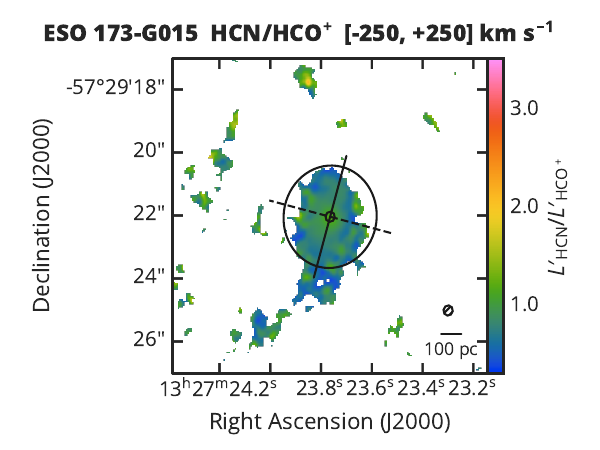}
\caption{HCN 3--2 and HCO$^+$ 3--2 for ESO\,173$-$G015. 
\emph{Left panels}: (\emph{Top two panels}) 
Spectra extracted from the resolved aperture. 
(\emph{Lower two panels}) Spectra extracted from the global aperture. 
Velocities are relative to the systemic velocity. 
The corresponding velocities of potentially detected species 
are indicated by vertical dotted lines.
\emph{Right panels}: 
(\emph{Top}) Integrated intensity over $\pm250$ km s$^{-1}$ (moment 0). 
Contours are (5, 10, 20, 40, 80, 120, 160, 200, 240) $\times\sigma$, 
where $\sigma$ is 0.043 Jy km s$^{-1}$ beam$^{-1}$. 
(\emph{Second from top}) Velocity field (moment 1). 
Contours are in steps of $\pm25$ km s$^{-1}$. 
(\emph{Third from top}) Velocity dispersion (moment 2). 
Contours are in steps of 10 km s$^{-1}$. 
Moment 1 and 2 were derived with $3\sigma$ clipping. 
(\emph{Bottom}) $L'_\mathrm{HCN}/L'_\mathrm{HCO^+}$. 
Color scale is from 0.285 to 3.5. 
Overlaid ellipses represent the apertures used for spectral extraction. 
Solid and dashed lines represent the kinematic major and minor axes, respectively. 
The synthesized beam is indicated by hatched ellipses in the lower right corners.
}
\label{figure:173-G015}
\end{figure*}

\begin{figure*}
\includegraphics[width=0.4\hsize]{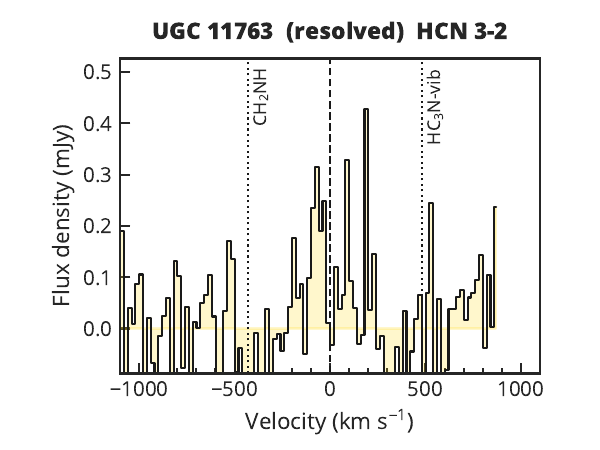}
\includegraphics[width=0.6\hsize]{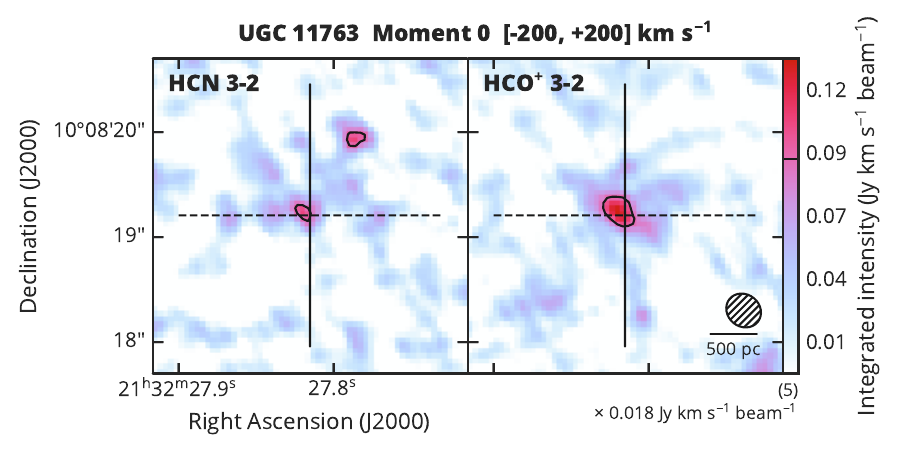}

\includegraphics[width=0.4\hsize]{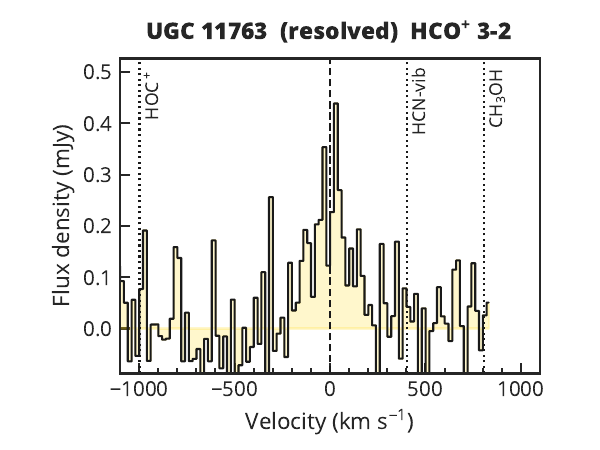}
\includegraphics[width=0.6\hsize]{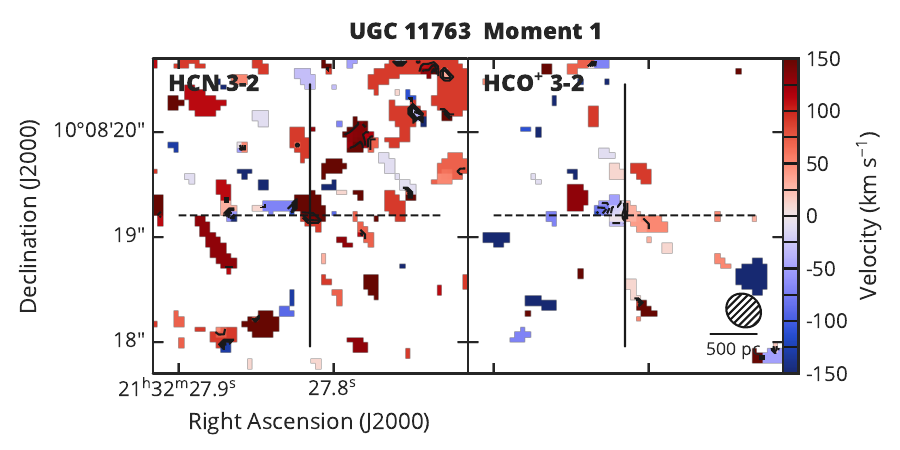}

\includegraphics[width=0.4\hsize]{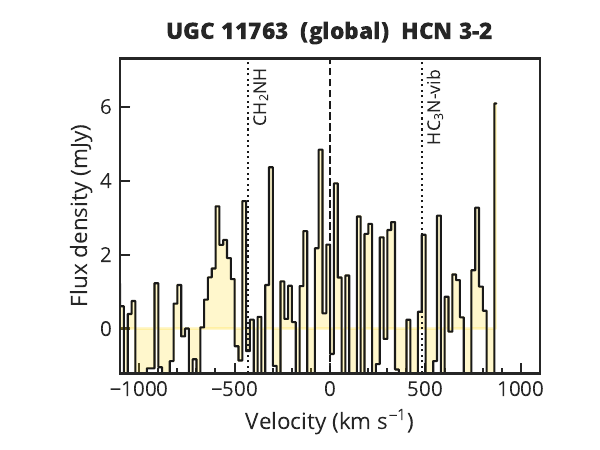}
\includegraphics[width=0.6\hsize]{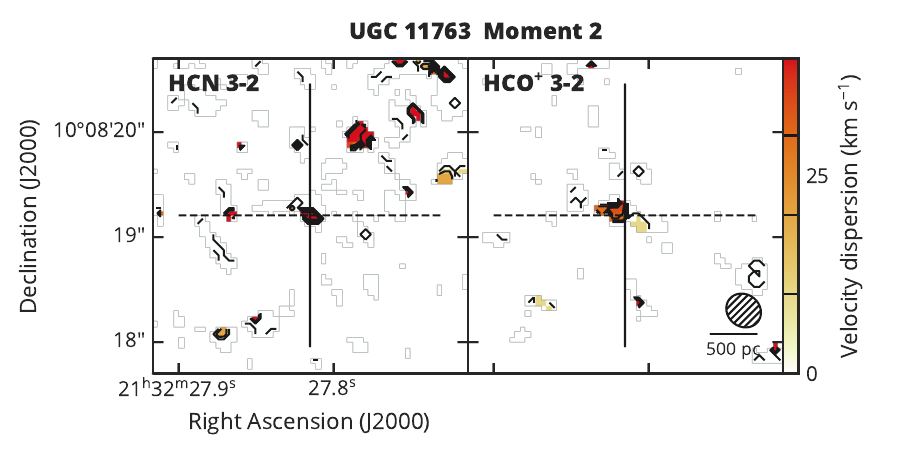}

\includegraphics[width=0.4\hsize]{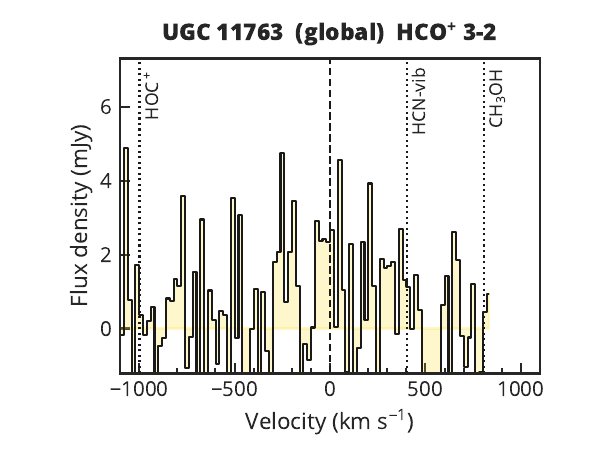}
\includegraphics[width=0.4\hsize]{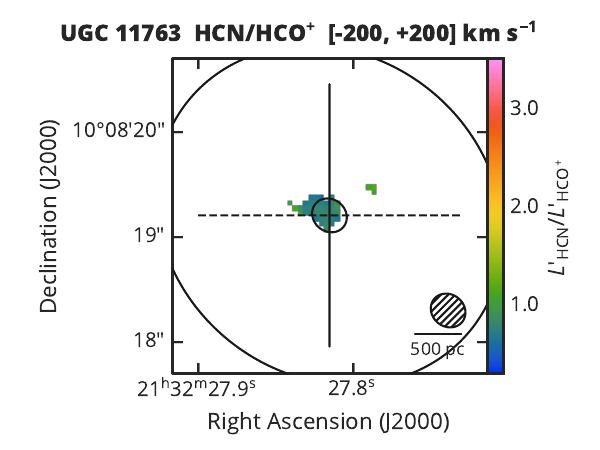}
\caption{HCN 3--2 and HCO$^+$ 3--2 for UGC\,11763. 
\emph{Left panels}: (\emph{Top two panels}) 
Spectra extracted from the resolved aperture. 
(\emph{Lower two panels}) Spectra extracted from the global aperture. 
Velocities are relative to the systemic velocity. 
The corresponding velocities of potentially detected species 
are indicated by vertical dotted lines.
\emph{Right panels}: 
(\emph{Top}) Integrated intensity over $\pm200$ km s$^{-1}$ (moment 0). 
Contours are (5) $\times\sigma$, 
where $\sigma$ is 0.018 Jy km s$^{-1}$ beam$^{-1}$. 
(\emph{Second from top}) Velocity field (moment 1). 
Contours are in steps of $\pm25$ km s$^{-1}$. 
(\emph{Third from top}) Velocity dispersion (moment 2). 
Contours are in steps of 10 km s$^{-1}$. 
Moment 1 and 2 were derived with $3\sigma$ clipping. 
(\emph{Bottom}) $L'_\mathrm{HCN}/L'_\mathrm{HCO^+}$. 
Color scale is from 0.285 to 3.5. 
Overlaid ellipses represent the apertures used for spectral extraction. 
Solid and dashed lines represent the kinematic major and minor axes, respectively. 
The synthesized beam is indicated by hatched ellipses in the lower right corners.
}
\label{figure:11763}
\end{figure*}

\begin{figure*}
\includegraphics[width=0.4\hsize]{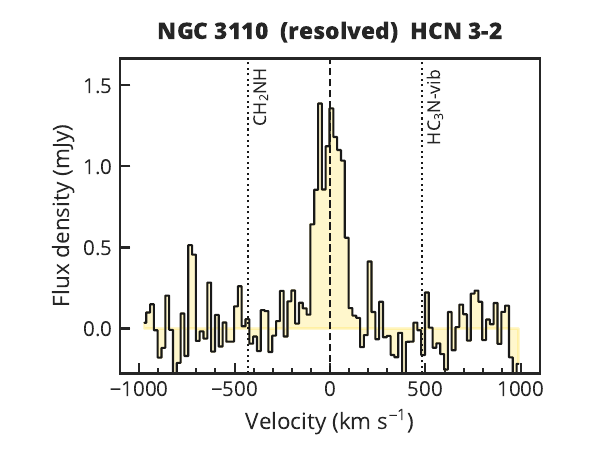}
\includegraphics[width=0.6\hsize]{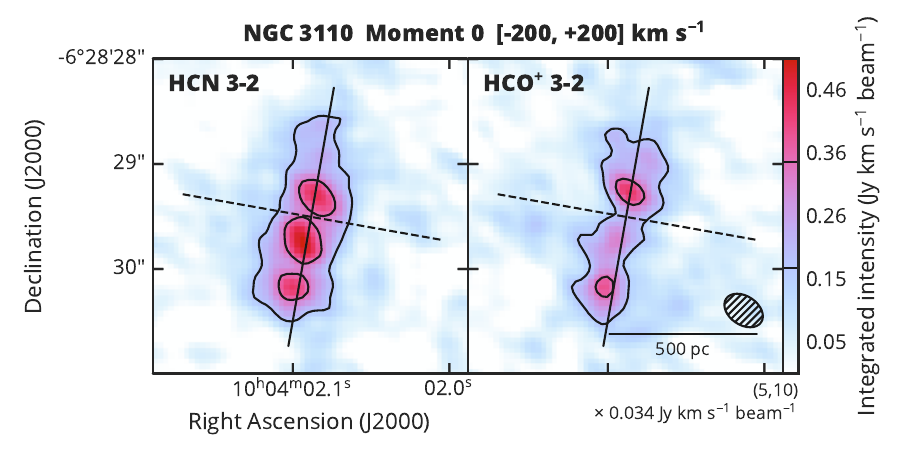}

\includegraphics[width=0.4\hsize]{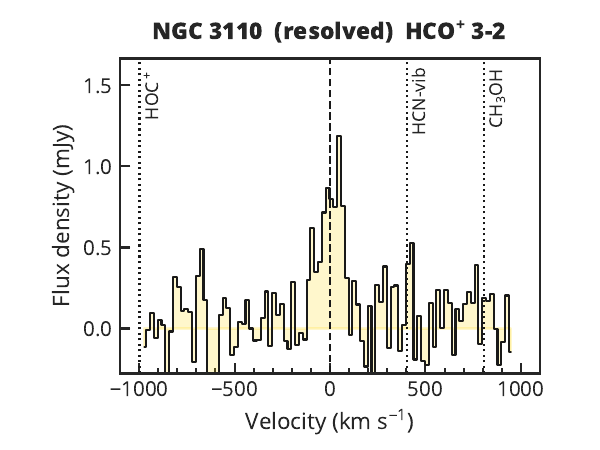}
\includegraphics[width=0.6\hsize]{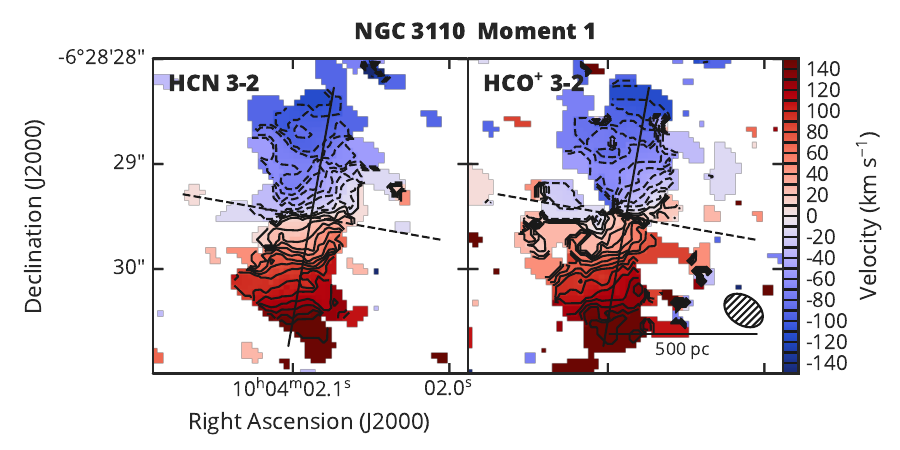}

\includegraphics[width=0.4\hsize]{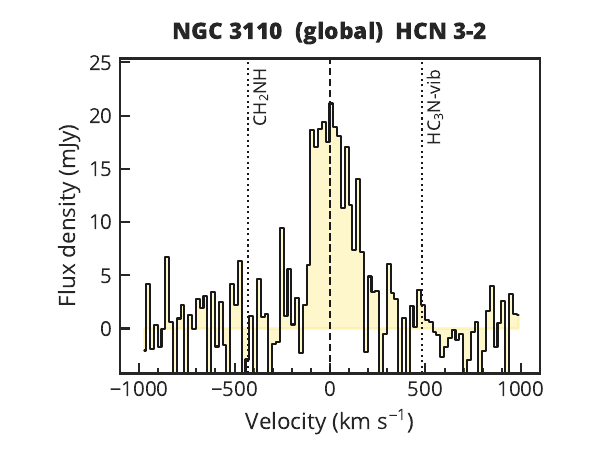}
\includegraphics[width=0.6\hsize]{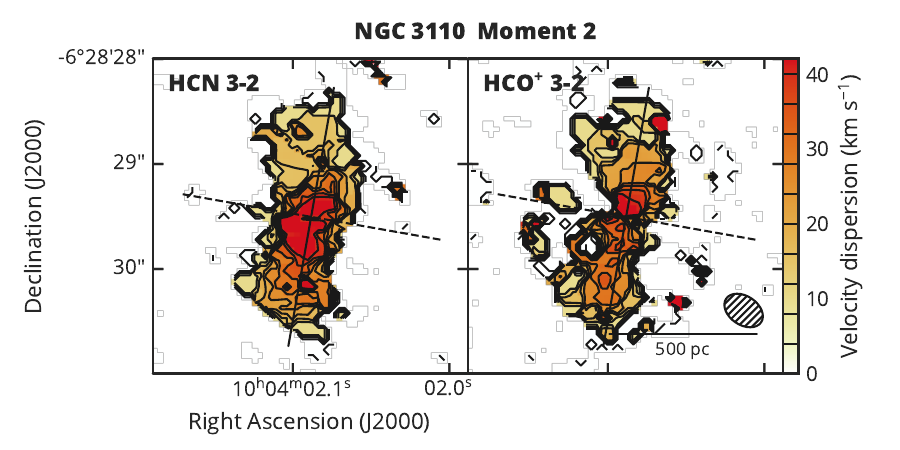}

\includegraphics[width=0.4\hsize]{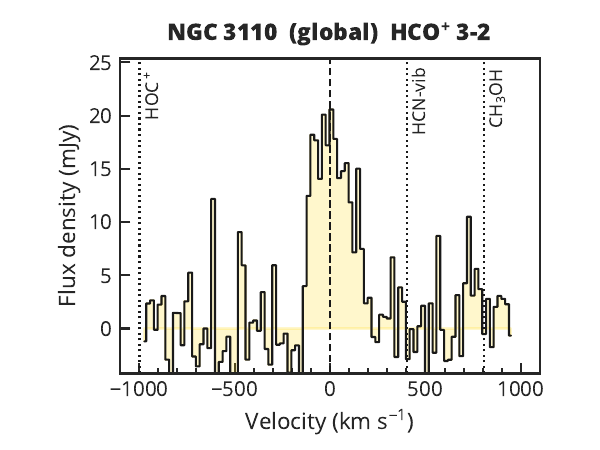}
\includegraphics[width=0.4\hsize]{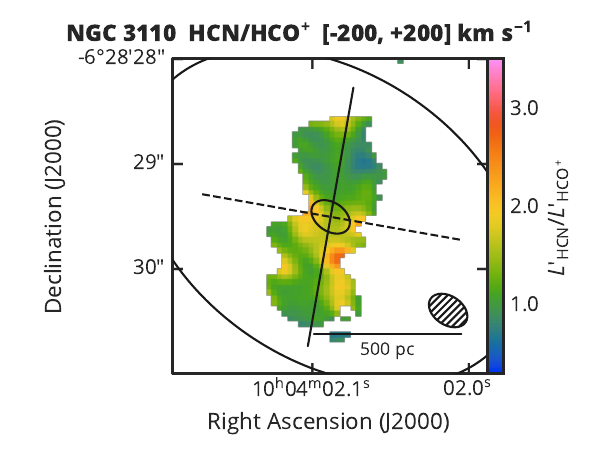}
\caption{HCN 3--2 and HCO$^+$ 3--2 for NGC\,3110. 
\emph{Left panels}: (\emph{Top two panels}) 
Spectra extracted from the resolved aperture. 
(\emph{Lower two panels}) Spectra extracted from the global aperture. 
Velocities are relative to the systemic velocity. 
The corresponding velocities of potentially detected species 
are indicated by vertical dotted lines.
\emph{Right panels}: 
(\emph{Top}) Integrated intensity over $\pm200$ km s$^{-1}$ (moment 0). 
Contours are (5, 10) $\times\sigma$, 
where $\sigma$ is 0.034 Jy km s$^{-1}$ beam$^{-1}$. 
(\emph{Second from top}) Velocity field (moment 1). 
Contours are in steps of $\pm10$ km s$^{-1}$. 
(\emph{Third from top}) Velocity dispersion (moment 2). 
Contours are in steps of 4 km s$^{-1}$. 
Moment 1 and 2 were derived with $3\sigma$ clipping. 
(\emph{Bottom}) $L'_\mathrm{HCN}/L'_\mathrm{HCO^+}$. 
Color scale is from 0.285 to 3.5. 
Overlaid ellipses represent the apertures used for spectral extraction. 
Solid and dashed lines represent the kinematic major and minor axes, respectively. 
The synthesized beam is indicated by hatched ellipses in the lower right corners.
}
\label{figure:3110}
\end{figure*}

\begin{figure*}
\includegraphics[width=0.4\hsize]{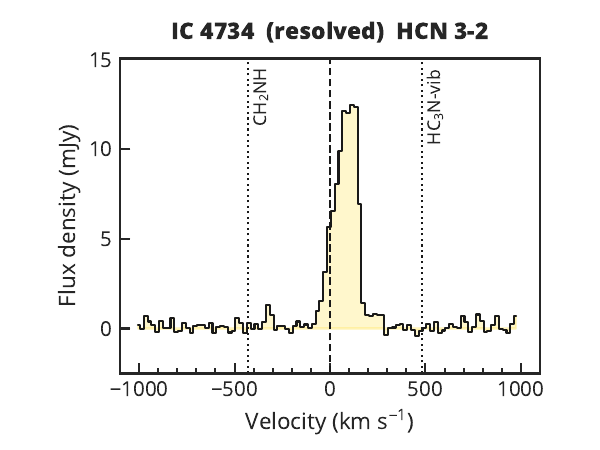}
\includegraphics[width=0.6\hsize]{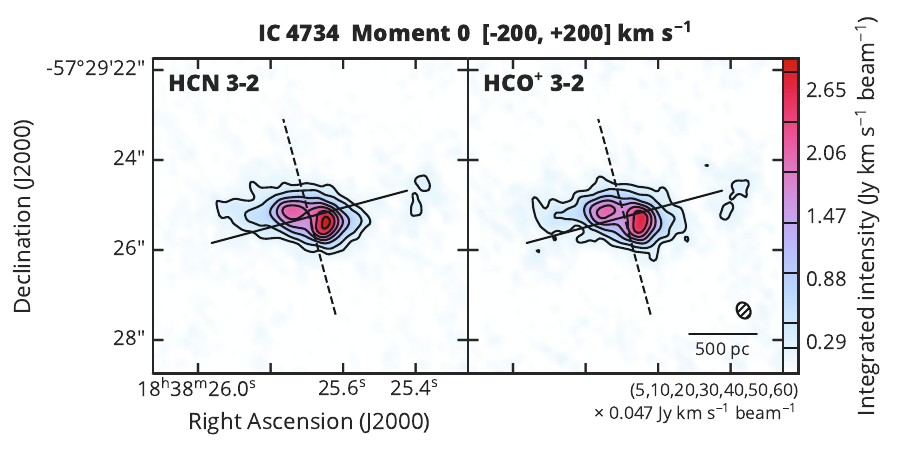}

\includegraphics[width=0.4\hsize]{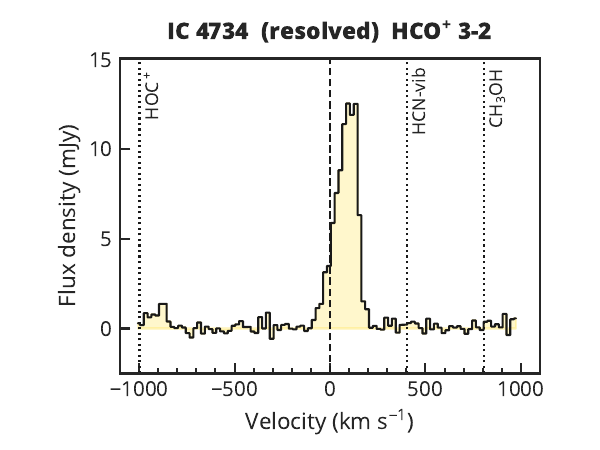}
\includegraphics[width=0.6\hsize]{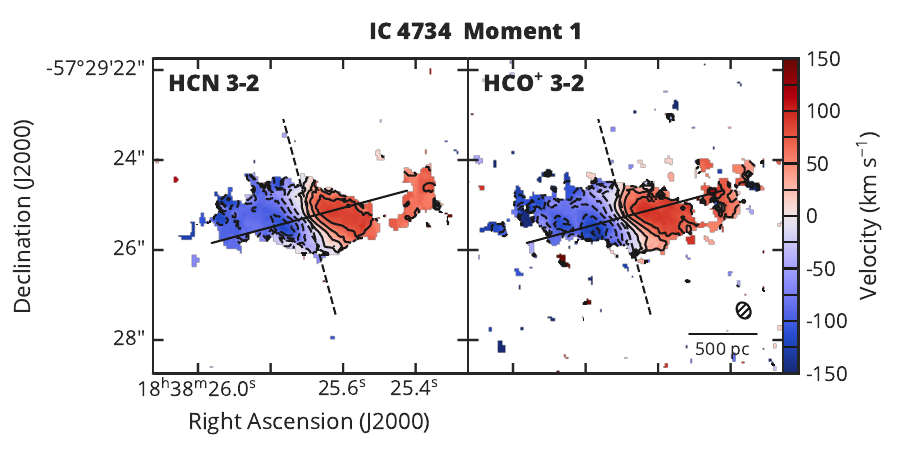}

\includegraphics[width=0.4\hsize]{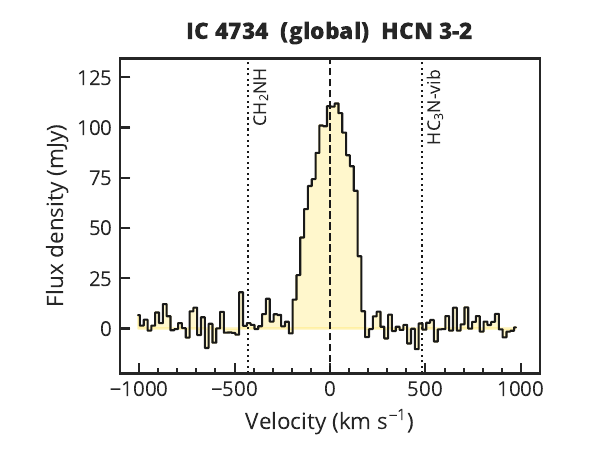}
\includegraphics[width=0.6\hsize]{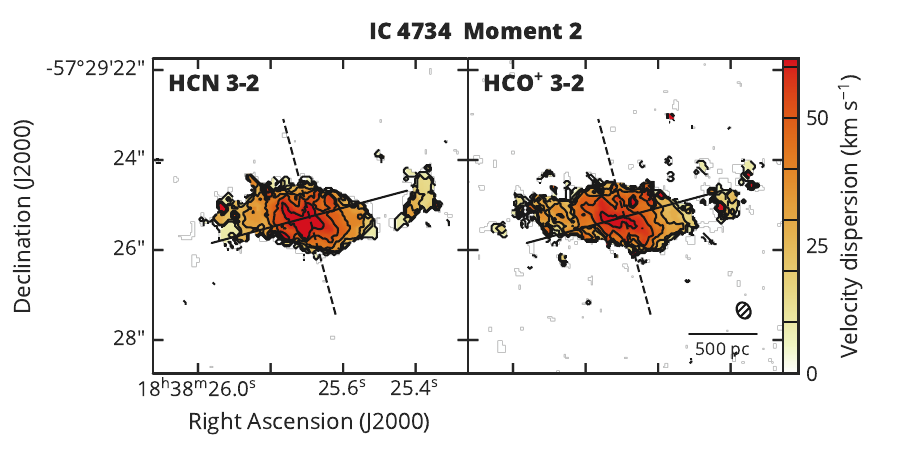}

\includegraphics[width=0.4\hsize]{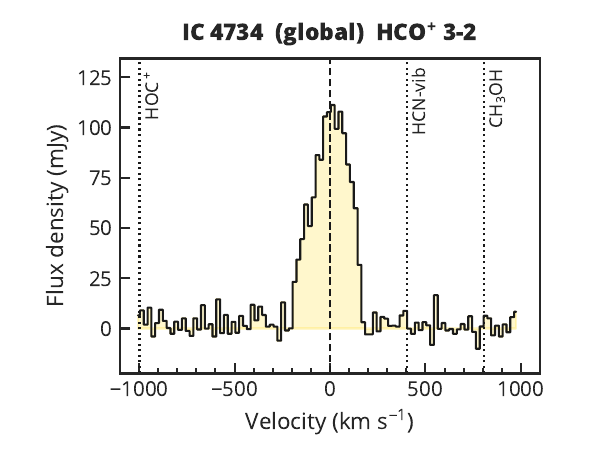}
\includegraphics[width=0.4\hsize]{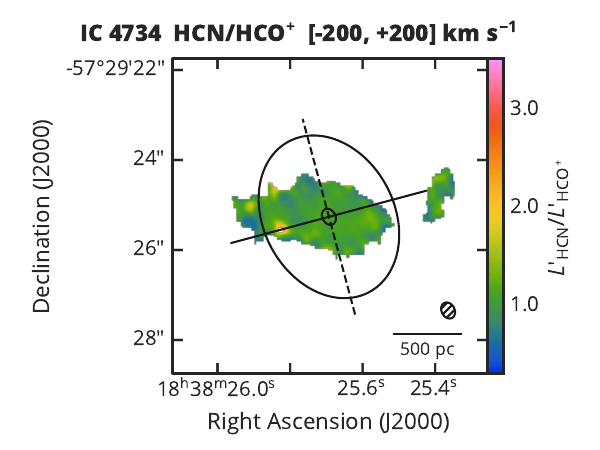}
\caption{HCN 3--2 and HCO$^+$ 3--2 for IC\,4734. 
\emph{Left panels}: (\emph{Top two panels}) 
Spectra extracted from the resolved aperture. 
(\emph{Lower two panels}) Spectra extracted from the global aperture. 
Velocities are relative to the systemic velocity. 
The corresponding velocities of potentially detected species 
are indicated by vertical dotted lines.
\emph{Right panels}: 
(\emph{Top}) Integrated intensity over $\pm200$ km s$^{-1}$ (moment 0). 
Contours are (5, 10, 20, 30, 40, 50, 60) $\times\sigma$, 
where $\sigma$ is 0.047 Jy km s$^{-1}$ beam$^{-1}$. 
(\emph{Second from top}) Velocity field (moment 1). 
Contours are in steps of $\pm25$ km s$^{-1}$. 
(\emph{Third from top}) Velocity dispersion (moment 2). 
Contours are in steps of 10 km s$^{-1}$. 
Moment 1 and 2 were derived with $3\sigma$ clipping. 
(\emph{Bottom}) $L'_\mathrm{HCN}/L'_\mathrm{HCO^+}$. 
Color scale is from 0.285 to 3.5. 
Overlaid ellipses represent the apertures used for spectral extraction. 
Solid and dashed lines represent the kinematic major and minor axes, respectively. 
The synthesized beam is indicated by hatched ellipses in the lower right corners.
}
\label{figure:4734}
\end{figure*}

\begin{figure*}
\includegraphics[width=0.4\hsize]{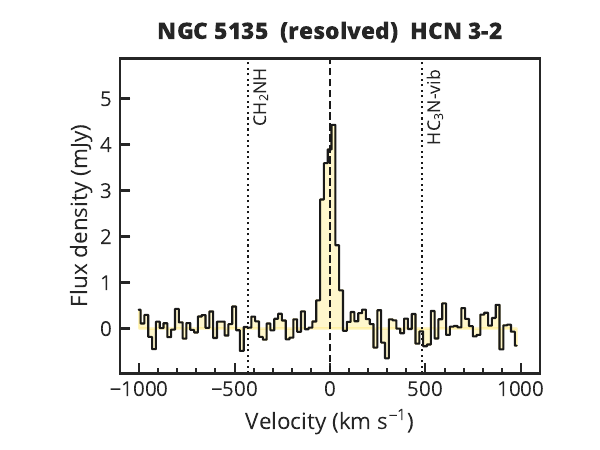}
\includegraphics[width=0.6\hsize]{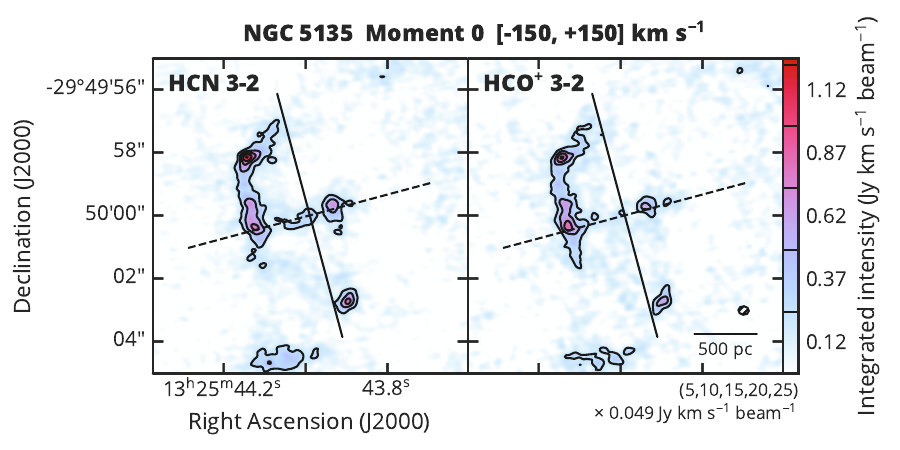}

\includegraphics[width=0.4\hsize]{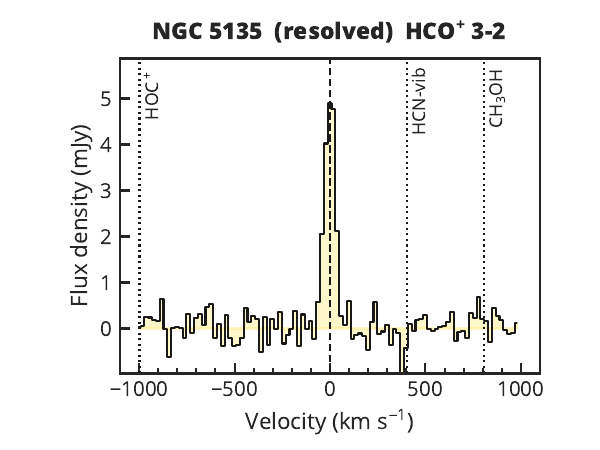}
\includegraphics[width=0.6\hsize]{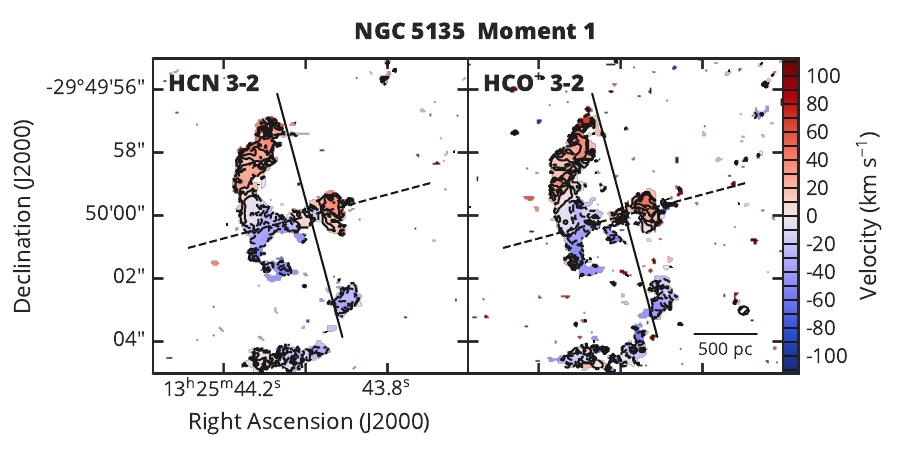}

\includegraphics[width=0.4\hsize]{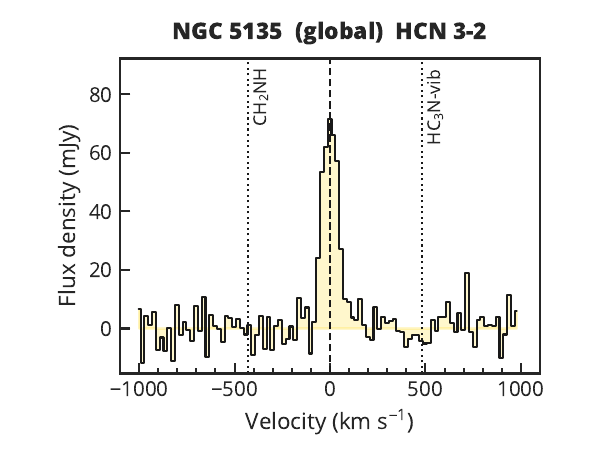}
\includegraphics[width=0.6\hsize]{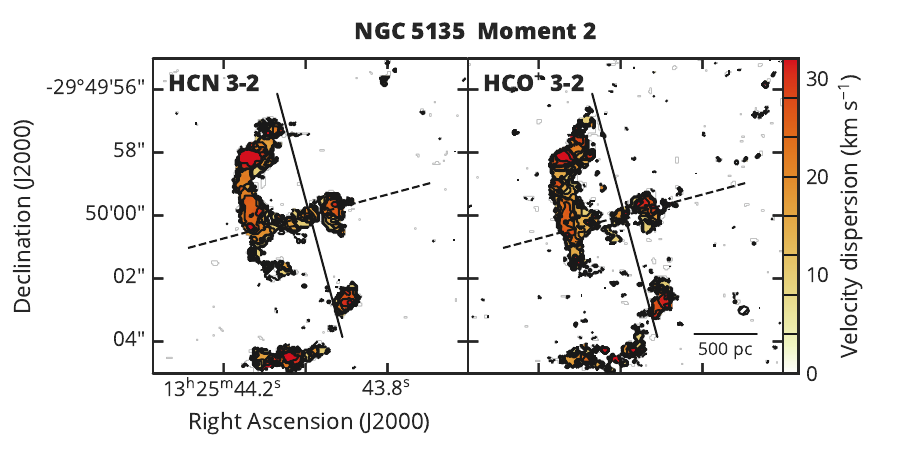}

\includegraphics[width=0.4\hsize]{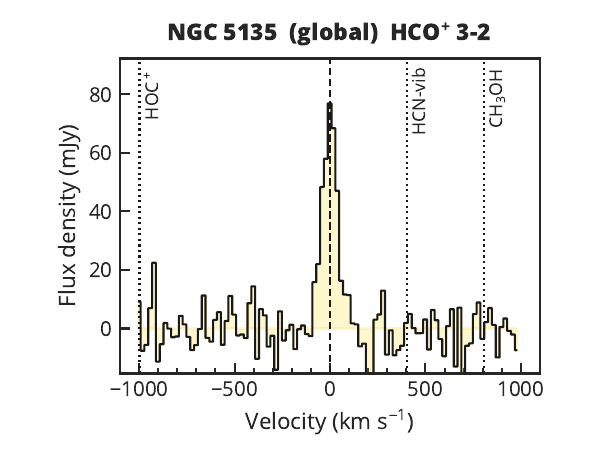}
\includegraphics[width=0.4\hsize]{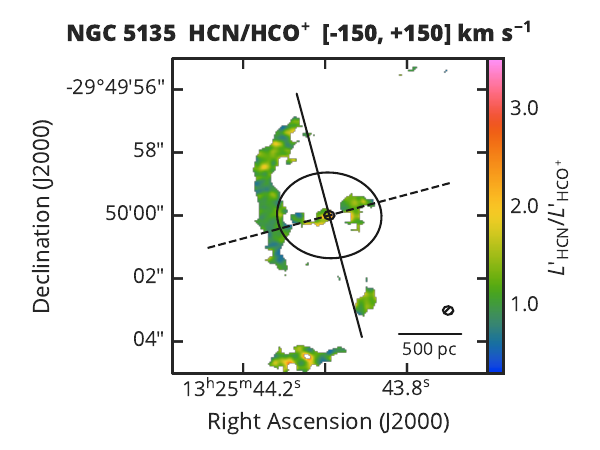}
\caption{HCN 3--2 and HCO$^+$ 3--2 for NGC\,5135. 
\emph{Left panels}: (\emph{Top two panels}) 
Spectra extracted from the resolved aperture. 
(\emph{Lower two panels}) Spectra extracted from the global aperture. 
Velocities are relative to the systemic velocity. 
The corresponding velocities of potentially detected species 
are indicated by vertical dotted lines.
\emph{Right panels}: 
(\emph{Top}) Integrated intensity over $\pm150$ km s$^{-1}$ (moment 0). 
Contours are (5, 10, 15, 20, 25) $\times\sigma$, 
where $\sigma$ is 0.049 Jy km s$^{-1}$ beam$^{-1}$. 
(\emph{Second from top}) Velocity field (moment 1). 
Contours are in steps of $\pm10$ km s$^{-1}$. 
(\emph{Third from top}) Velocity dispersion (moment 2). 
Contours are in steps of 4 km s$^{-1}$. 
Moment 1 and 2 were derived with $3\sigma$ clipping. 
(\emph{Bottom}) $L'_\mathrm{HCN}/L'_\mathrm{HCO^+}$. 
Color scale is from 0.285 to 3.5. 
Overlaid ellipses represent the apertures used for spectral extraction. 
Solid and dashed lines represent the kinematic major and minor axes, respectively. 
The synthesized beam is indicated by hatched ellipses in the lower right corners.
}
\label{figure:5135}
\end{figure*}

\begin{figure*}
\includegraphics[width=0.4\hsize]{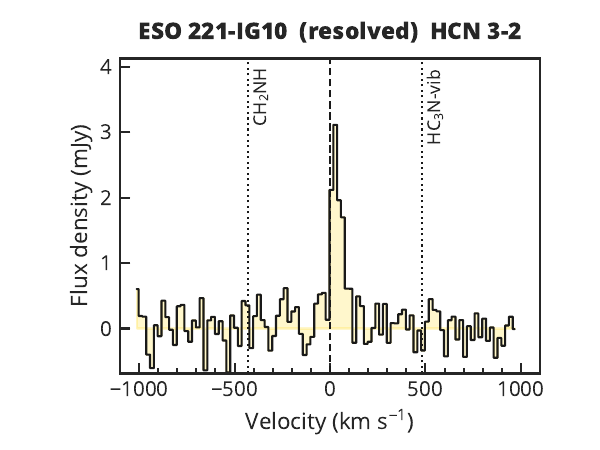}
\includegraphics[width=0.6\hsize]{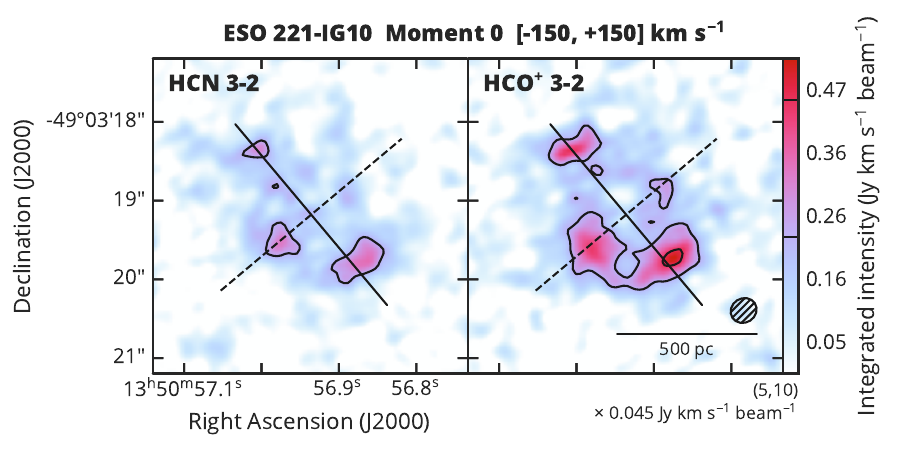}

\includegraphics[width=0.4\hsize]{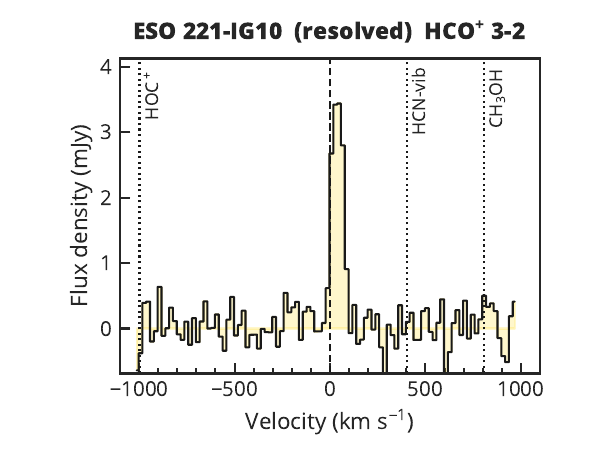}
\includegraphics[width=0.6\hsize]{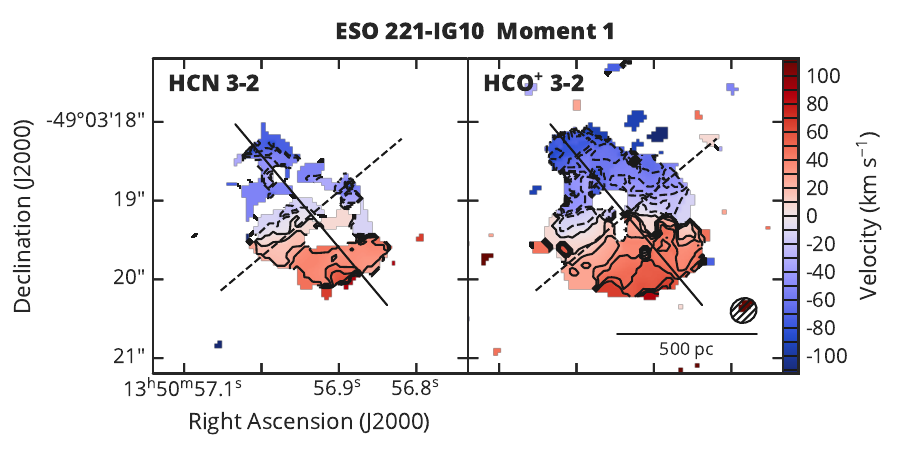}

\includegraphics[width=0.4\hsize]{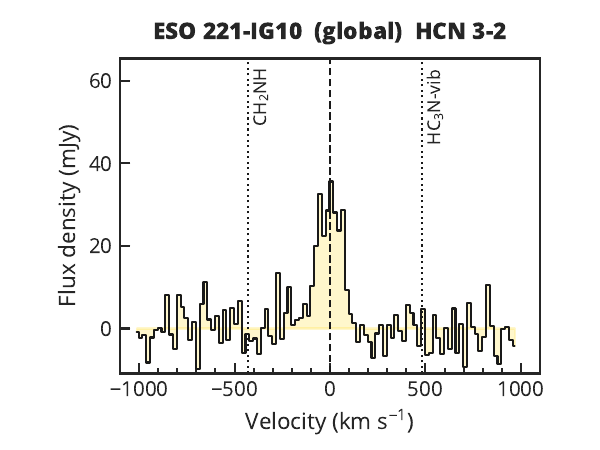}
\includegraphics[width=0.6\hsize]{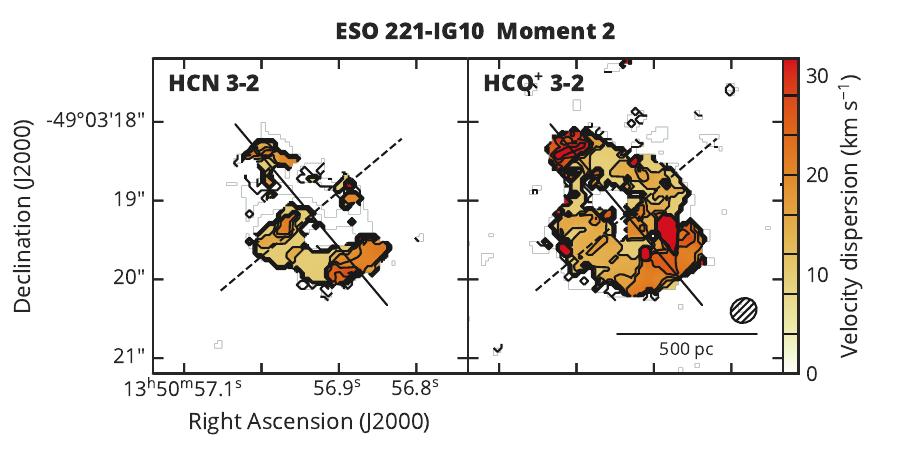}

\includegraphics[width=0.4\hsize]{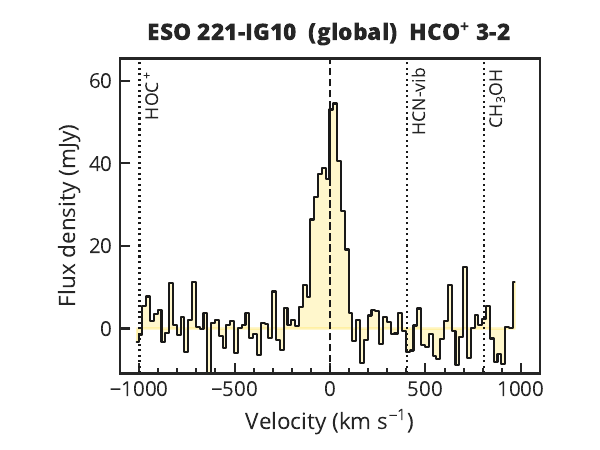}
\includegraphics[width=0.4\hsize]{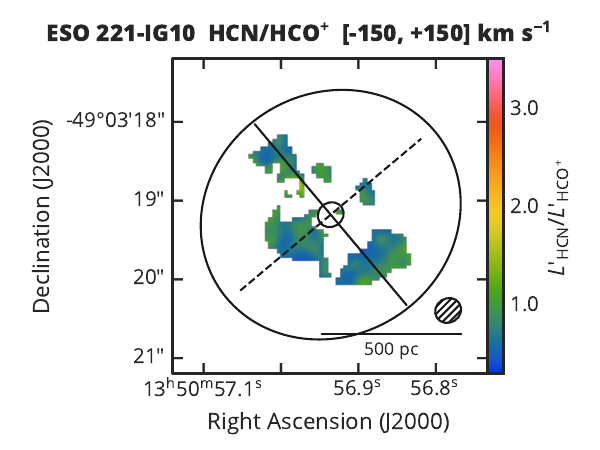}
\caption{HCN 3--2 and HCO$^+$ 3--2 for ESO\,221$-$IG10. 
\emph{Left panels}: (\emph{Top two panels}) 
Spectra extracted from the resolved aperture. 
(\emph{Lower two panels}) Spectra extracted from the global aperture. 
Velocities are relative to the systemic velocity. 
The corresponding velocities of potentially detected species 
are indicated by vertical dotted lines.
\emph{Right panels}: 
(\emph{Top}) Integrated intensity over $\pm150$ km s$^{-1}$ (moment 0). 
Contours are (5, 10) $\times\sigma$, 
where $\sigma$ is 0.045 Jy km s$^{-1}$ beam$^{-1}$. 
(\emph{Second from top}) Velocity field (moment 1). 
Contours are in steps of $\pm10$ km s$^{-1}$. 
(\emph{Third from top}) Velocity dispersion (moment 2). 
Contours are in steps of 4 km s$^{-1}$. 
Moment 1 and 2 were derived with $3\sigma$ clipping. 
(\emph{Bottom}) $L'_\mathrm{HCN}/L'_\mathrm{HCO^+}$. 
Color scale is from 0.285 to 3.5. 
Overlaid ellipses represent the apertures used for spectral extraction. 
Solid and dashed lines represent the kinematic major and minor axes, respectively. 
The synthesized beam is indicated by hatched ellipses in the lower right corners.
}
\label{figure:221-IG10}
\end{figure*}

\begin{figure*}
\includegraphics[width=0.4\hsize]{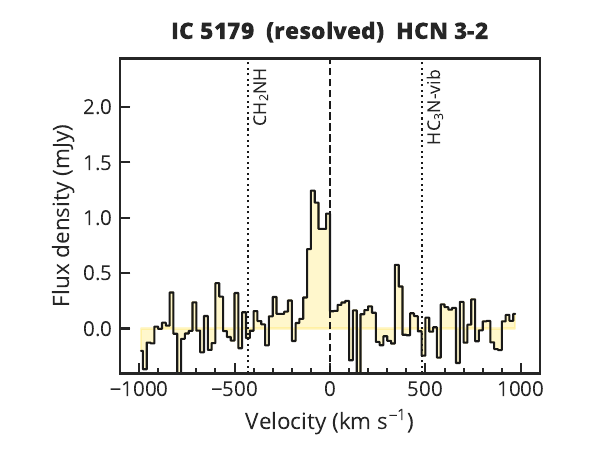}
\includegraphics[width=0.6\hsize]{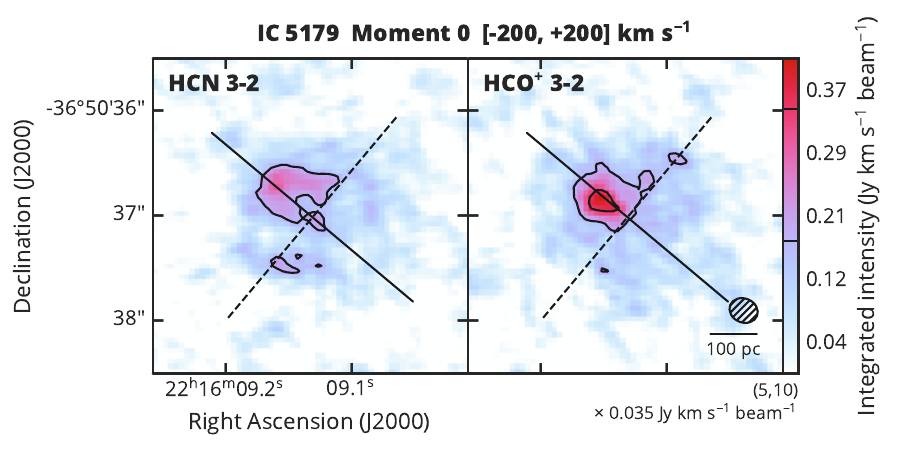}

\includegraphics[width=0.4\hsize]{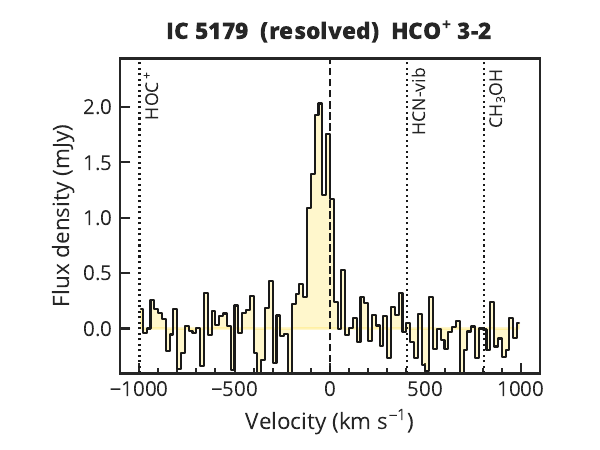}
\includegraphics[width=0.6\hsize]{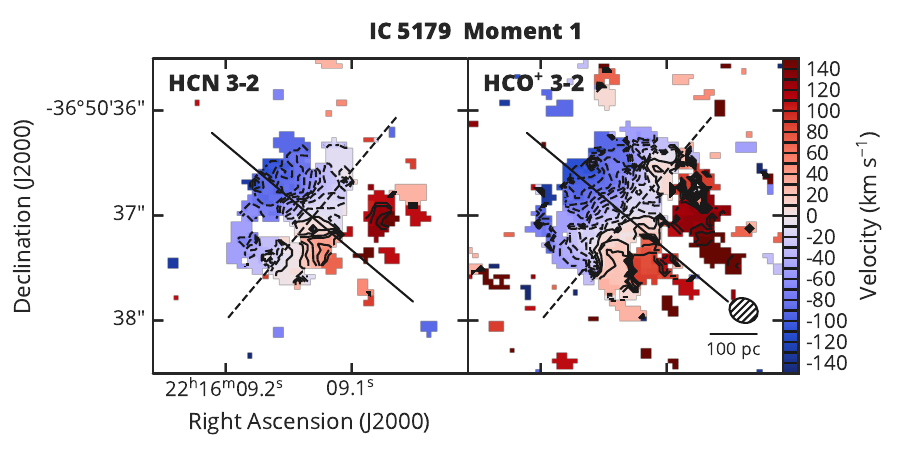}

\includegraphics[width=0.4\hsize]{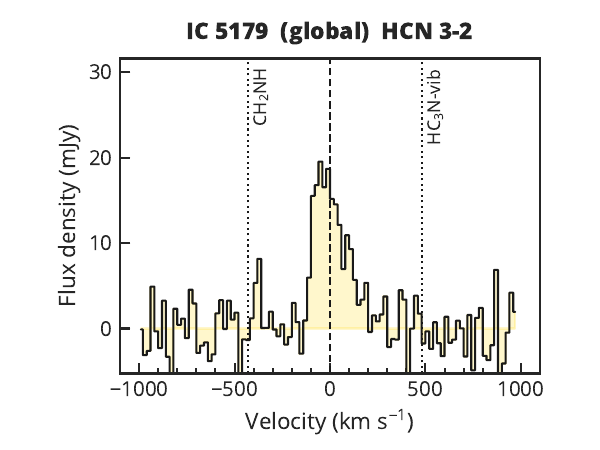}
\includegraphics[width=0.6\hsize]{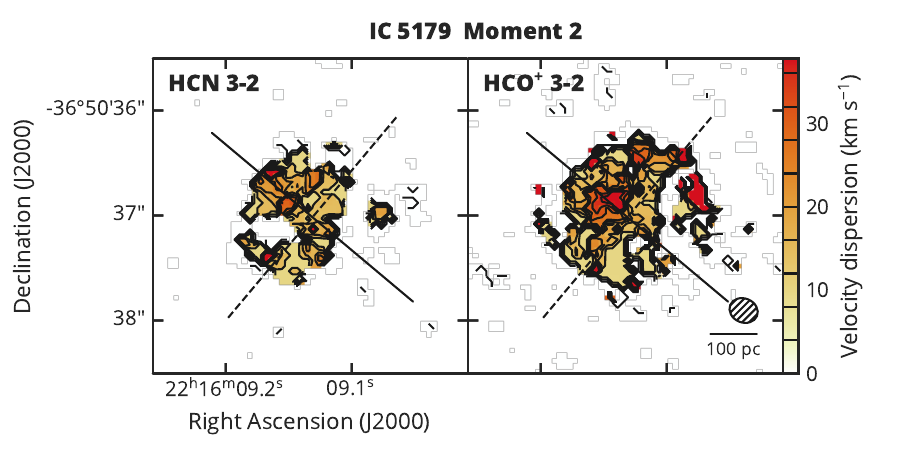}

\includegraphics[width=0.4\hsize]{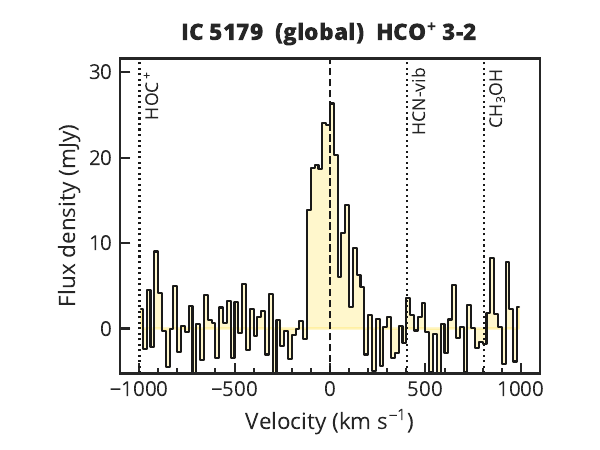}
\includegraphics[width=0.4\hsize]{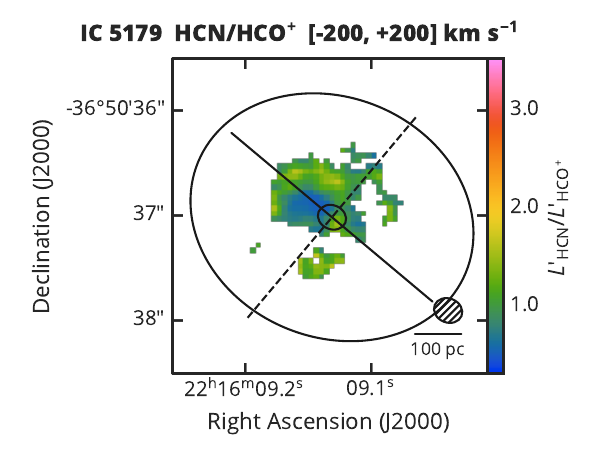}
\caption{HCN 3--2 and HCO$^+$ 3--2 for IC\,5179. 
\emph{Left panels}: (\emph{Top two panels}) 
Spectra extracted from the resolved aperture. 
(\emph{Lower two panels}) Spectra extracted from the global aperture. 
Velocities are relative to the systemic velocity. 
The corresponding velocities of potentially detected species 
are indicated by vertical dotted lines.
\emph{Right panels}: 
(\emph{Top}) Integrated intensity over $\pm200$ km s$^{-1}$ (moment 0). 
Contours are (5, 10) $\times\sigma$, 
where $\sigma$ is 0.035 Jy km s$^{-1}$ beam$^{-1}$. 
(\emph{Second from top}) Velocity field (moment 1). 
Contours are in steps of $\pm10$ km s$^{-1}$. 
(\emph{Third from top}) Velocity dispersion (moment 2). 
Contours are in steps of 4 km s$^{-1}$. 
Moment 1 and 2 were derived with $3\sigma$ clipping. 
(\emph{Bottom}) $L'_\mathrm{HCN}/L'_\mathrm{HCO^+}$. 
Color scale is from 0.285 to 3.5. 
Overlaid ellipses represent the apertures used for spectral extraction. 
Solid and dashed lines represent the kinematic major and minor axes, respectively. 
The synthesized beam is indicated by hatched ellipses in the lower right corners.
}
\label{figure:5179}
\end{figure*}

\begin{figure*}
\includegraphics[width=0.4\hsize]{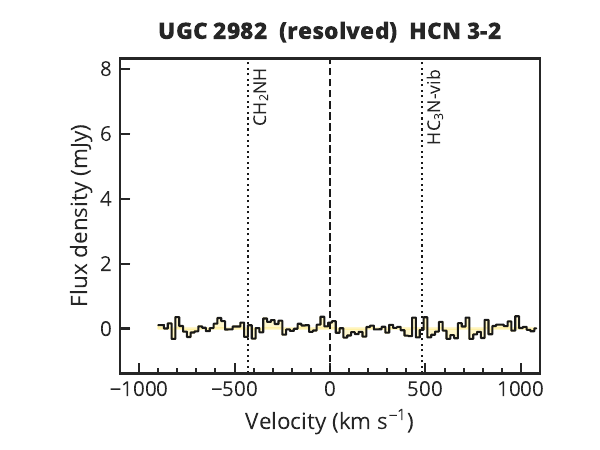}
\includegraphics[width=0.6\hsize]{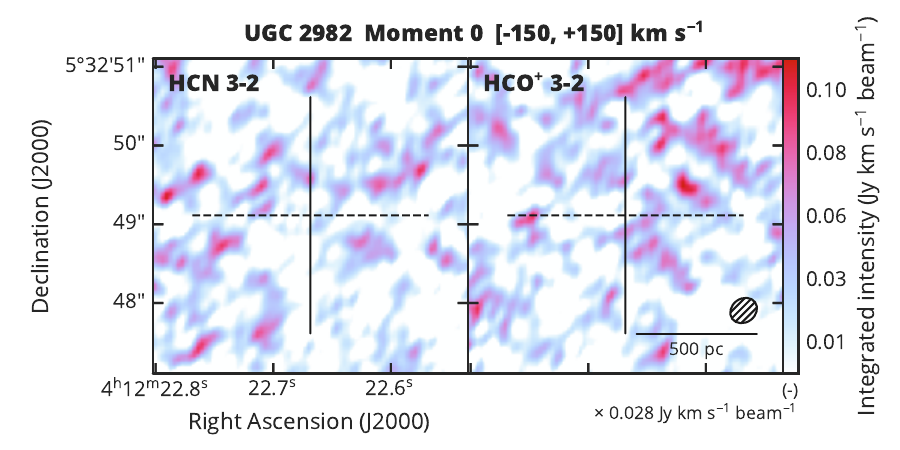}

\includegraphics[width=0.4\hsize]{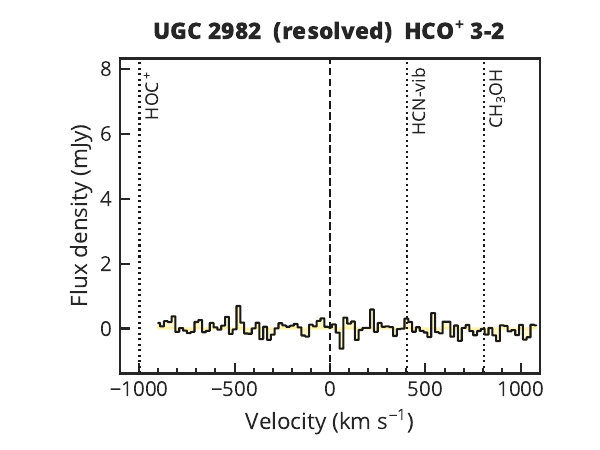}
\includegraphics[width=0.6\hsize]{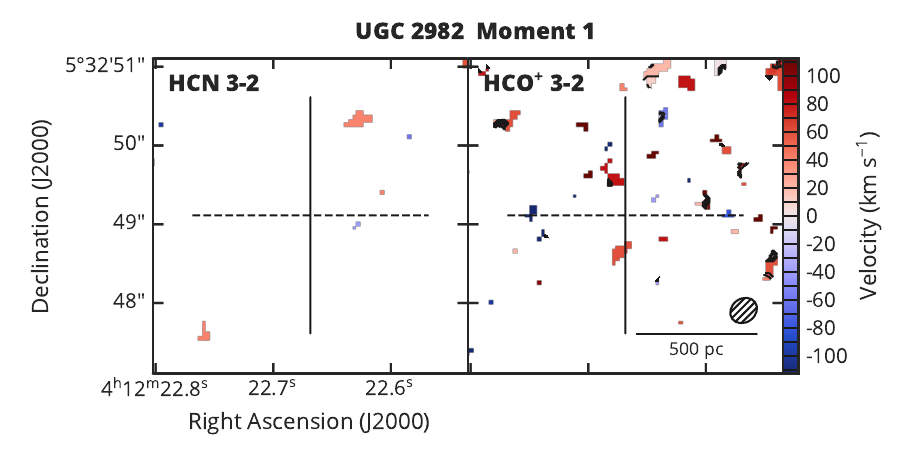}

\includegraphics[width=0.4\hsize]{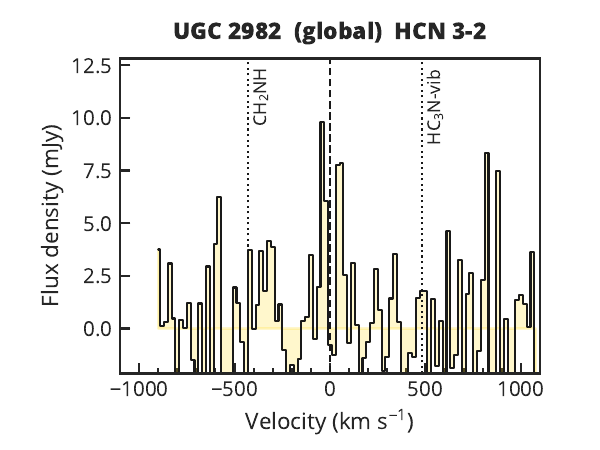}
\includegraphics[width=0.6\hsize]{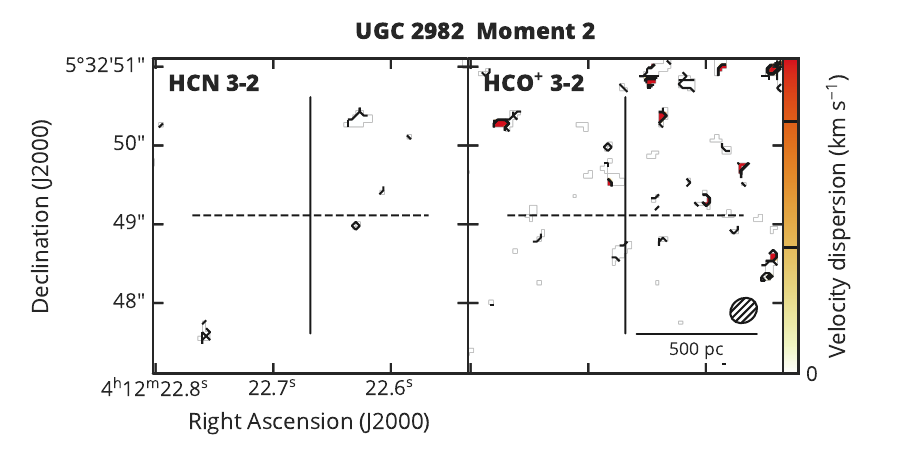}

\includegraphics[width=0.4\hsize]{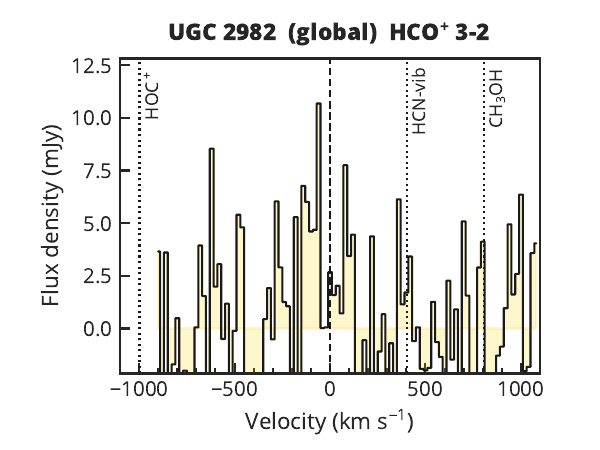}
\includegraphics[width=0.4\hsize]{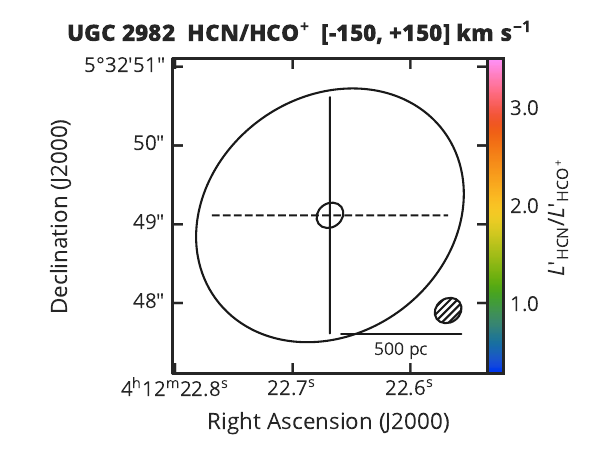}
\caption{HCN 3--2 and HCO$^+$ 3--2 for UGC\,2982. 
\emph{Left panels}: (\emph{Top two panels}) 
Spectra extracted from the resolved aperture. 
(\emph{Lower two panels}) Spectra extracted from the global aperture. 
Velocities are relative to the systemic velocity. 
The corresponding velocities of potentially detected species 
are indicated by vertical dotted lines.
\emph{Right panels}: 
(\emph{Top}) Integrated intensity over $\pm150$ km s$^{-1}$ (moment 0). 
No emission is found above $5\sigma$ threshold. 
(\emph{Second from top}) Velocity field (moment 1). 
Contours are in steps of $\pm10$ km s$^{-1}$. 
(\emph{Third from top}) Velocity dispersion (moment 2). 
Contours are in steps of 4 km s$^{-1}$. 
Moment 1 and 2 were derived with $3\sigma$ clipping. 
(\emph{Bottom}) $L'_\mathrm{HCN}/L'_\mathrm{HCO^+}$. 
Color scale is from 0.285 to 3.5. 
Overlaid ellipses represent the apertures used for spectral extraction. 
Solid and dashed lines represent the kinematic major and minor axes, respectively. 
The synthesized beam is indicated by hatched ellipses in the lower right corners.
}
\label{figure:2982}
\end{figure*}

\begin{figure*}
\includegraphics[width=0.4\hsize]{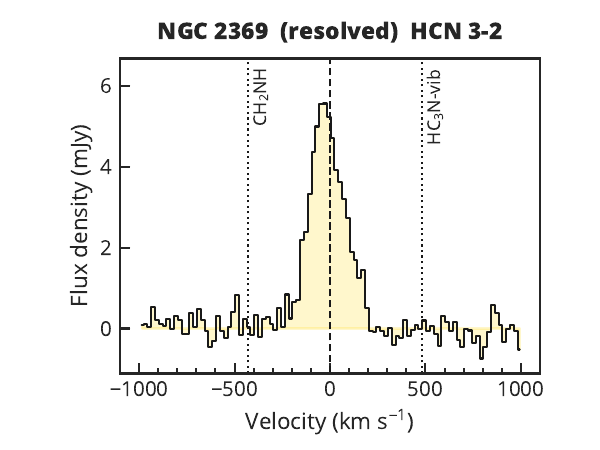}
\includegraphics[width=0.6\hsize]{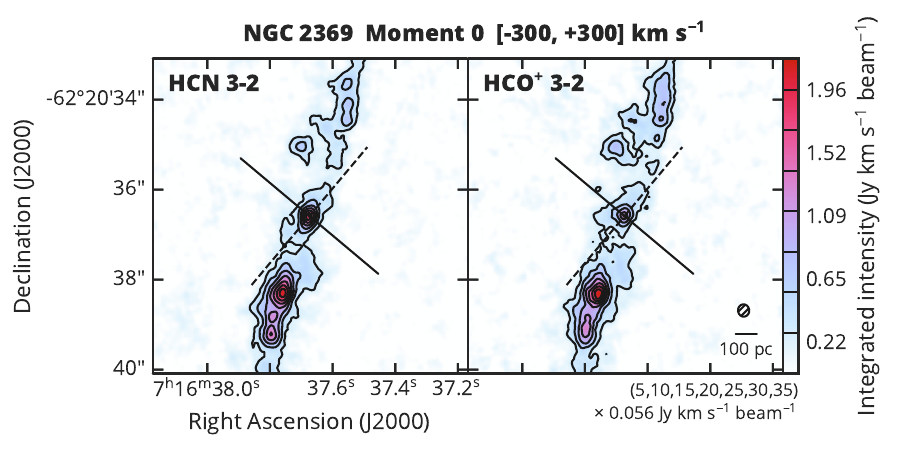}

\includegraphics[width=0.4\hsize]{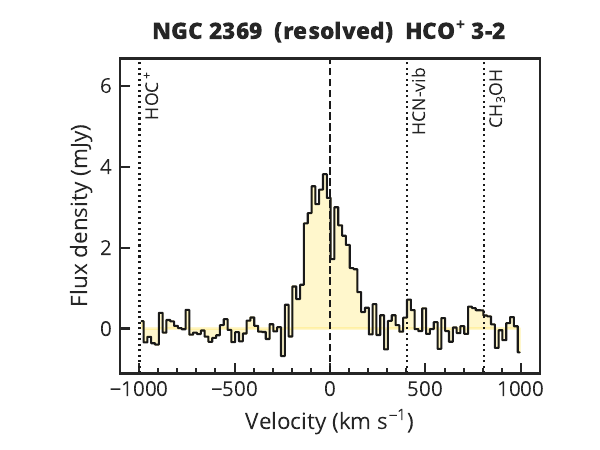}
\includegraphics[width=0.6\hsize]{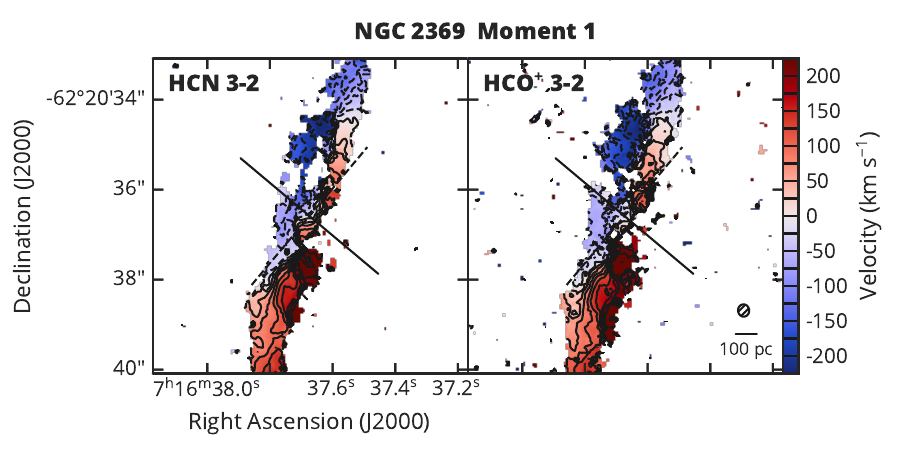}

\includegraphics[width=0.4\hsize]{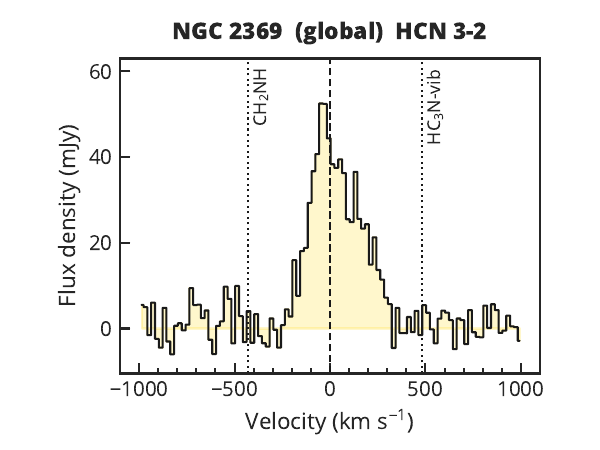}
\includegraphics[width=0.6\hsize]{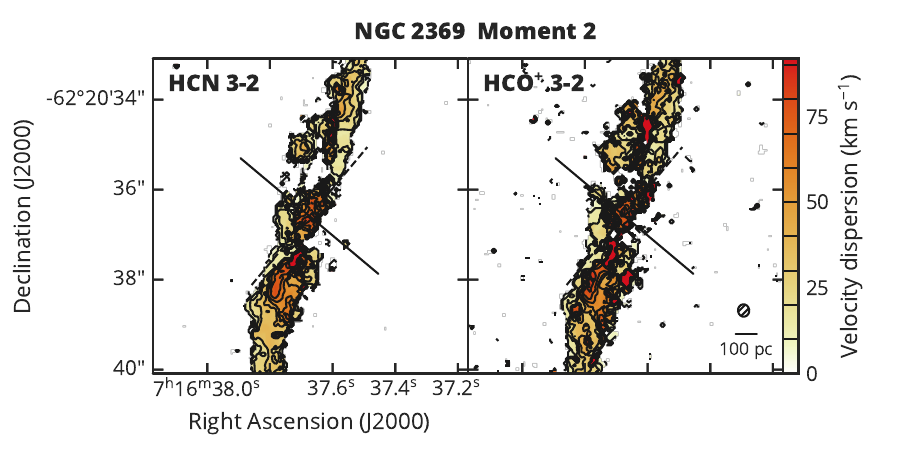}

\includegraphics[width=0.4\hsize]{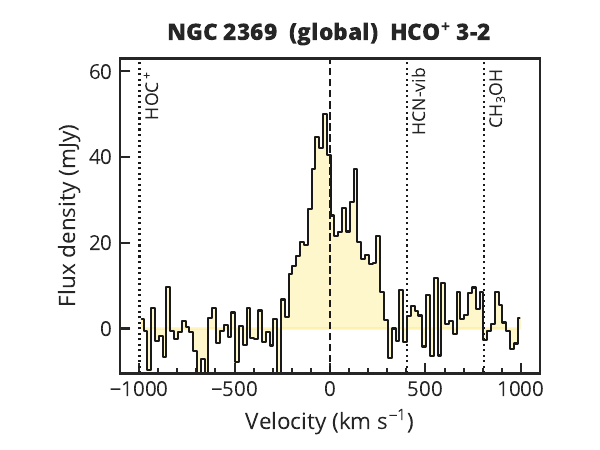}
\includegraphics[width=0.4\hsize]{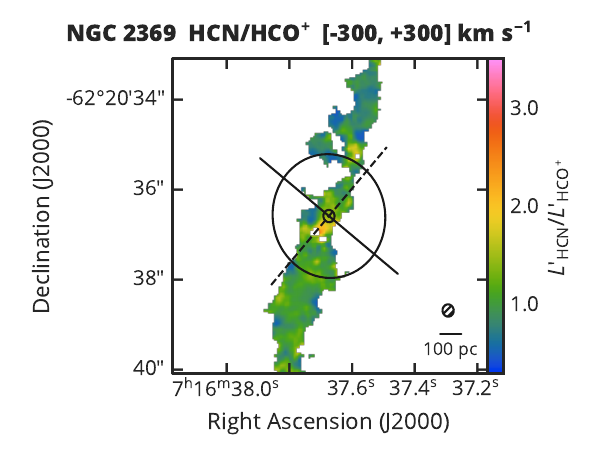}
\caption{HCN 3--2 and HCO$^+$ 3--2 for NGC\,2369. 
\emph{Left panels}: (\emph{Top two panels}) 
Spectra extracted from the resolved aperture. 
(\emph{Lower two panels}) Spectra extracted from the global aperture. 
Velocities are relative to the systemic velocity. 
The corresponding velocities of potentially detected species 
are indicated by vertical dotted lines.
\emph{Right panels}: 
(\emph{Top}) Integrated intensity over $\pm300$ km s$^{-1}$ (moment 0). 
Contours are (5, 10, 15, 20, 25, 30, 35) $\times\sigma$, 
where $\sigma$ is 0.056 Jy km s$^{-1}$ beam$^{-1}$. 
(\emph{Second from top}) Velocity field (moment 1). 
Contours are in steps of $\pm25$ km s$^{-1}$. 
(\emph{Third from top}) Velocity dispersion (moment 2). 
Contours are in steps of 10 km s$^{-1}$. 
Moment 1 and 2 were derived with $3\sigma$ clipping. 
(\emph{Bottom}) $L'_\mathrm{HCN}/L'_\mathrm{HCO^+}$. 
Color scale is from 0.285 to 3.5. 
Overlaid ellipses represent the apertures used for spectral extraction. 
Solid and dashed lines represent the kinematic major and minor axes, respectively. 
The synthesized beam is indicated by hatched ellipses in the lower right corners.
}
\label{figure:2369}
\end{figure*}

\begin{figure*}
\includegraphics[width=0.4\hsize]{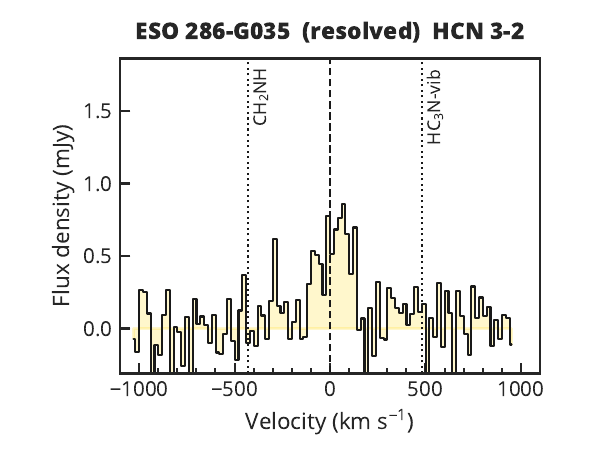}
\includegraphics[width=0.6\hsize]{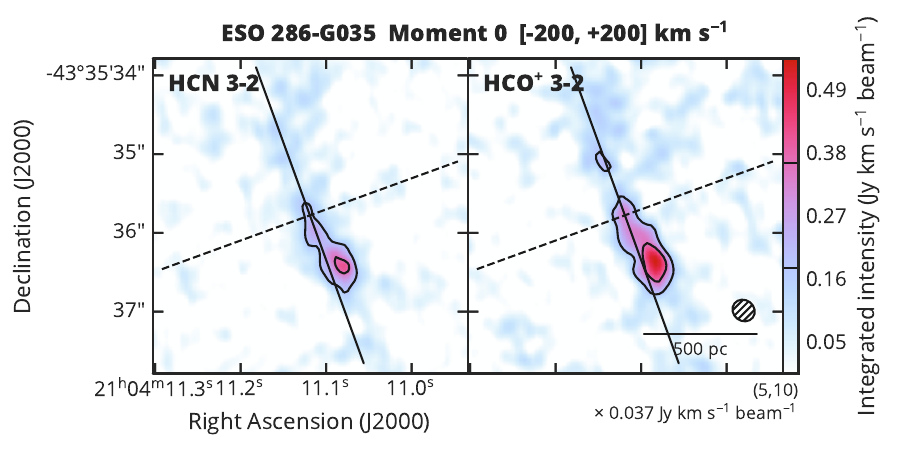}

\includegraphics[width=0.4\hsize]{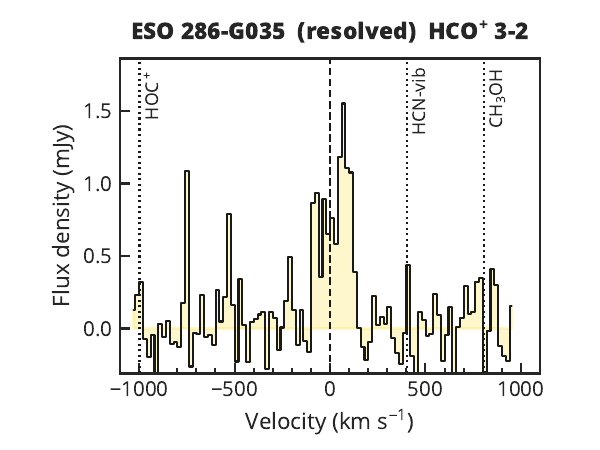}
\includegraphics[width=0.6\hsize]{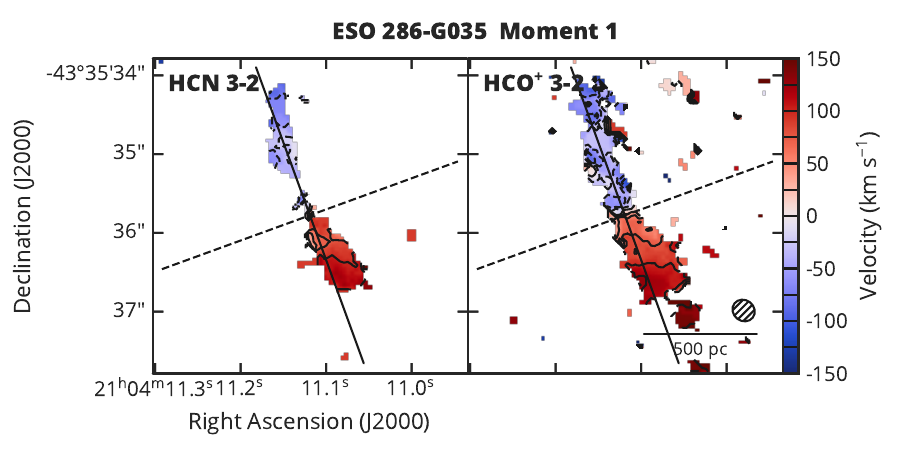}

\includegraphics[width=0.4\hsize]{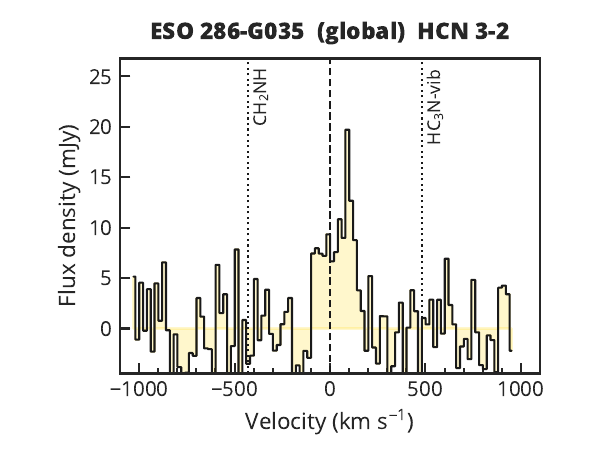}
\includegraphics[width=0.6\hsize]{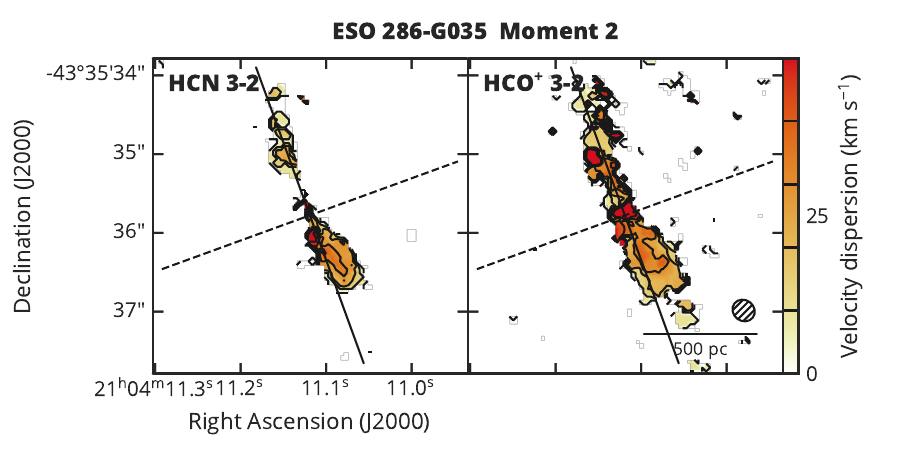}

\includegraphics[width=0.4\hsize]{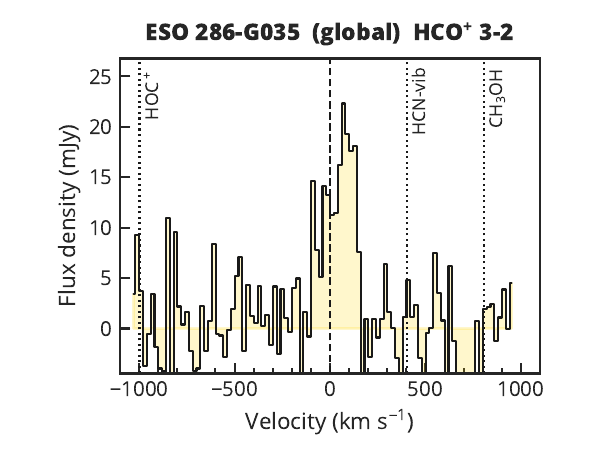}
\includegraphics[width=0.4\hsize]{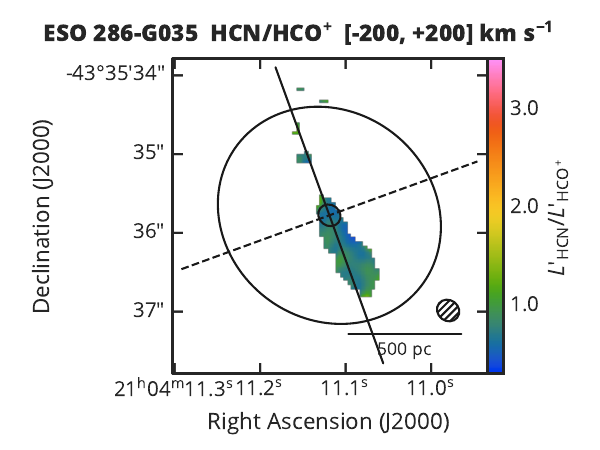}
\caption{HCN 3--2 and HCO$^+$ 3--2 for ESO\,286$-$G035. 
\emph{Left panels}: (\emph{Top two panels}) 
Spectra extracted from the resolved aperture. 
(\emph{Lower two panels}) Spectra extracted from the global aperture. 
Velocities are relative to the systemic velocity. 
The corresponding velocities of potentially detected species 
are indicated by vertical dotted lines.
\emph{Right panels}: 
(\emph{Top}) Integrated intensity over $\pm200$ km s$^{-1}$ (moment 0). 
Contours are (5, 10) $\times\sigma$, 
where $\sigma$ is 0.037 Jy km s$^{-1}$ beam$^{-1}$. 
(\emph{Second from top}) Velocity field (moment 1). 
Contours are in steps of $\pm25$ km s$^{-1}$. 
(\emph{Third from top}) Velocity dispersion (moment 2). 
Contours are in steps of 10 km s$^{-1}$. 
Moment 1 and 2 were derived with $3\sigma$ clipping. 
(\emph{Bottom}) $L'_\mathrm{HCN}/L'_\mathrm{HCO^+}$. 
Color scale is from 0.285 to 3.5. 
Overlaid ellipses represent the apertures used for spectral extraction. 
Solid and dashed lines represent the kinematic major and minor axes, respectively. 
The synthesized beam is indicated by hatched ellipses in the lower right corners.
}
\label{figure:286-G035}
\end{figure*}

\begin{figure*}
\includegraphics[width=0.4\hsize]{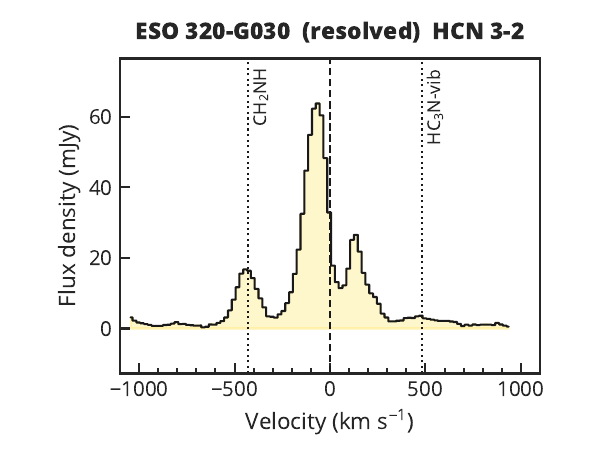}
\includegraphics[width=0.6\hsize]{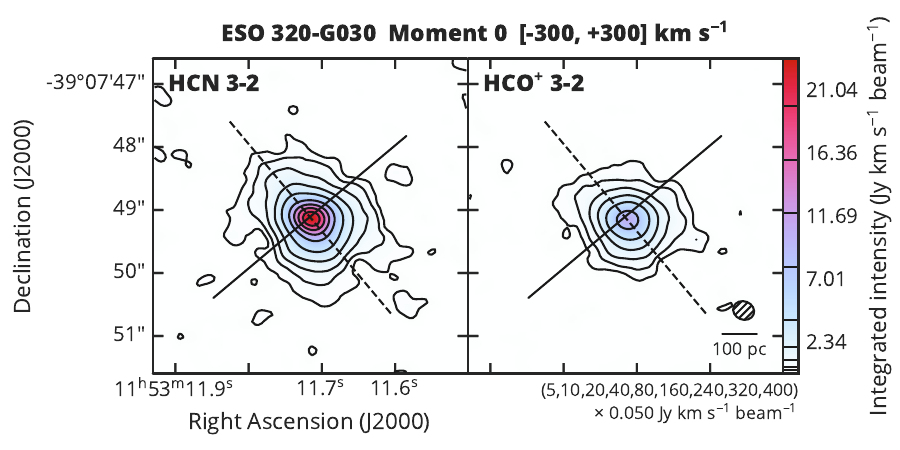}

\includegraphics[width=0.4\hsize]{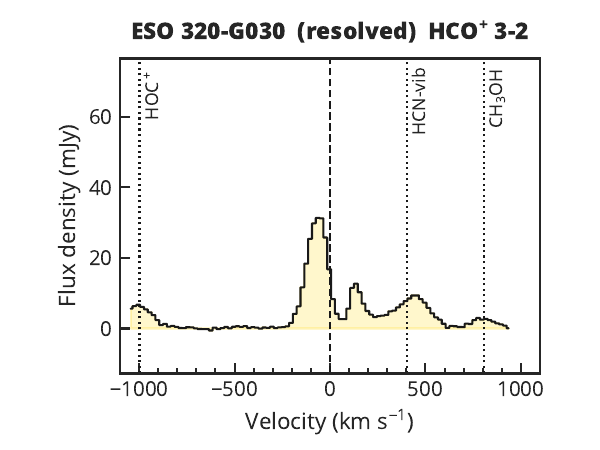}
\includegraphics[width=0.6\hsize]{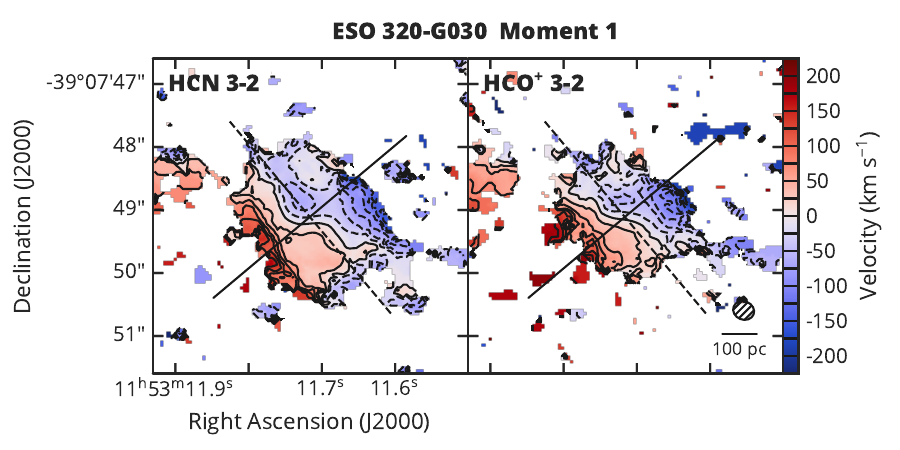}

\includegraphics[width=0.4\hsize]{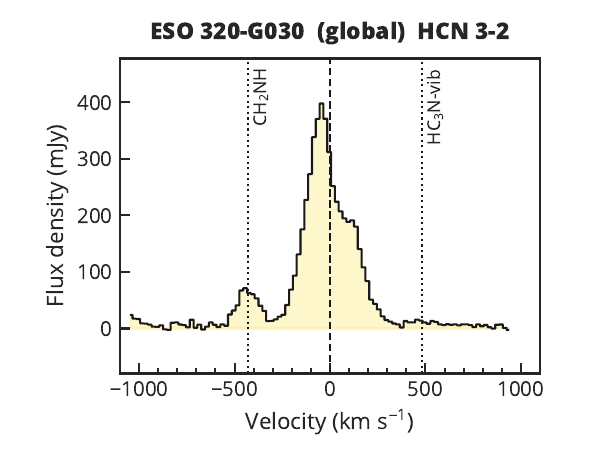}
\includegraphics[width=0.6\hsize]{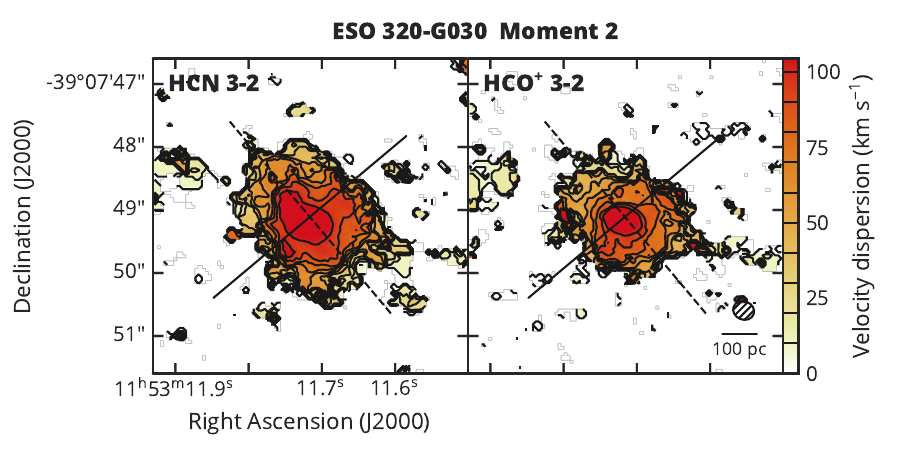}

\includegraphics[width=0.4\hsize]{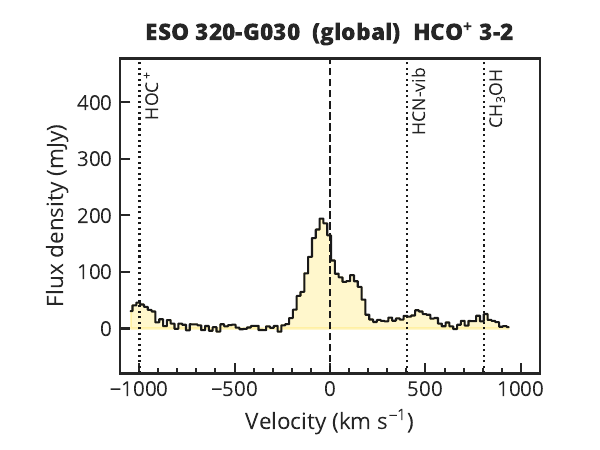}
\includegraphics[width=0.4\hsize]{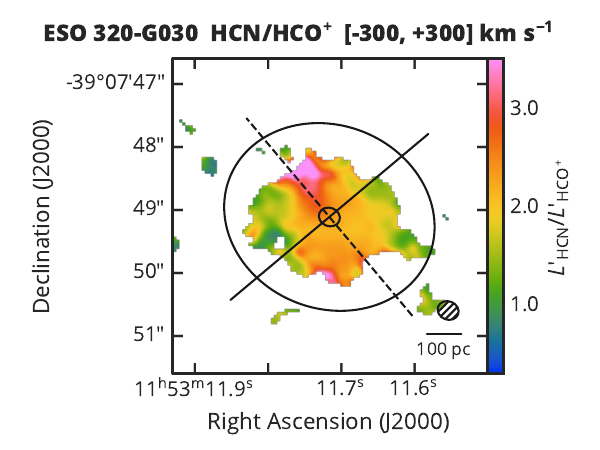}
\caption{HCN 3--2 and HCO$^+$ 3--2 for ESO\,320$-$G030. 
\emph{Left panels}: (\emph{Top two panels}) 
Spectra extracted from the resolved aperture. 
(\emph{Lower two panels}) Spectra extracted from the global aperture. 
Velocities are relative to the systemic velocity. 
The corresponding velocities of potentially detected species 
are indicated by vertical dotted lines.
\emph{Right panels}: 
(\emph{Top}) Integrated intensity over $\pm300$ km s$^{-1}$ (moment 0). 
Contours are (5, 10, 20, 40, 80, 160, 240, 320, 400) $\times\sigma$, 
where $\sigma$ is 0.050 Jy km s$^{-1}$ beam$^{-1}$. 
(\emph{Second from top}) Velocity field (moment 1). 
Contours are in steps of $\pm25$ km s$^{-1}$. 
(\emph{Third from top}) Velocity dispersion (moment 2). 
Contours are in steps of 10 km s$^{-1}$. 
Moment 1 and 2 were derived with $3\sigma$ clipping. 
(\emph{Bottom}) $L'_\mathrm{HCN}/L'_\mathrm{HCO^+}$. 
Color scale is from 0.285 to 3.5. 
Overlaid ellipses represent the apertures used for spectral extraction. 
Solid and dashed lines represent the kinematic major and minor axes, respectively. 
The synthesized beam is indicated by hatched ellipses in the lower right corners.
}
\label{figure:320-G030}
\end{figure*}

\begin{figure*}
\includegraphics[width=0.4\hsize]{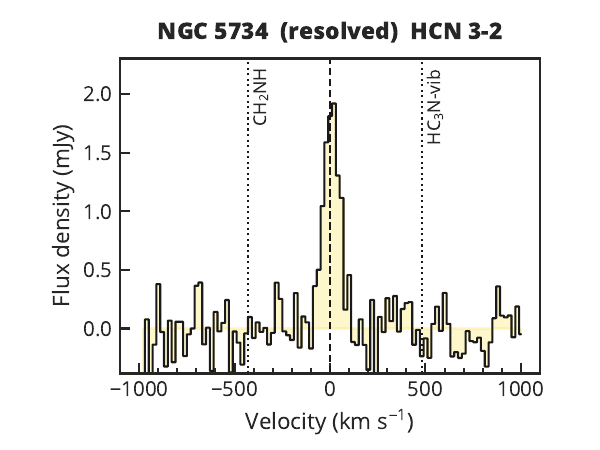}
\includegraphics[width=0.6\hsize]{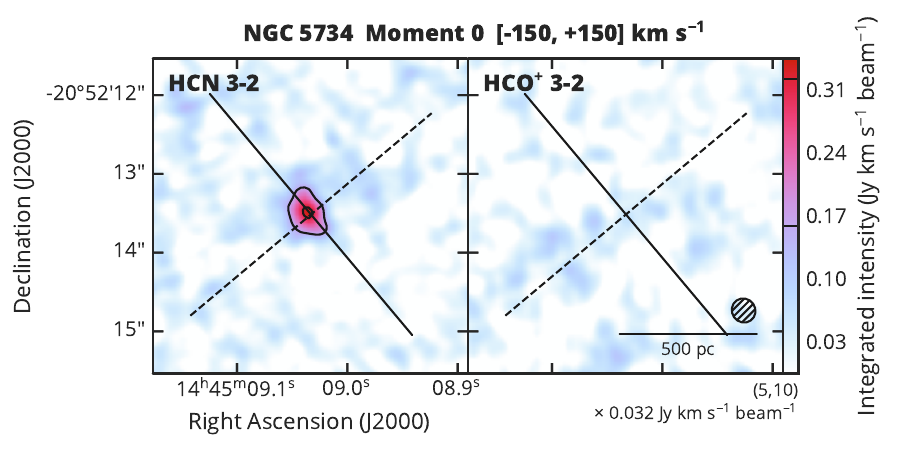}

\includegraphics[width=0.4\hsize]{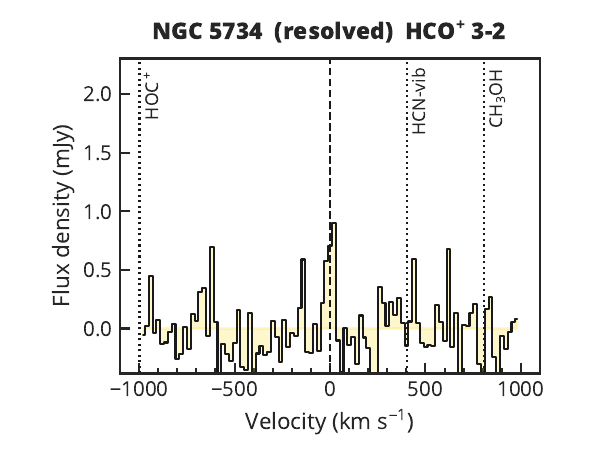}
\includegraphics[width=0.6\hsize]{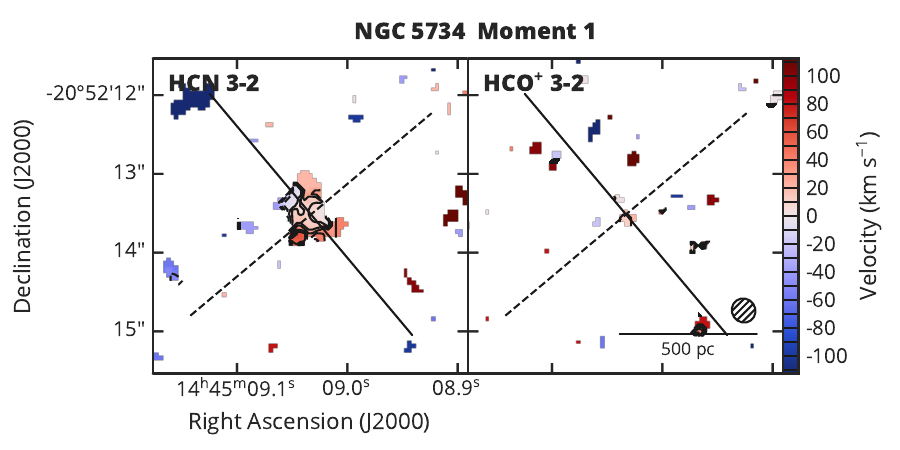}

\includegraphics[width=0.4\hsize]{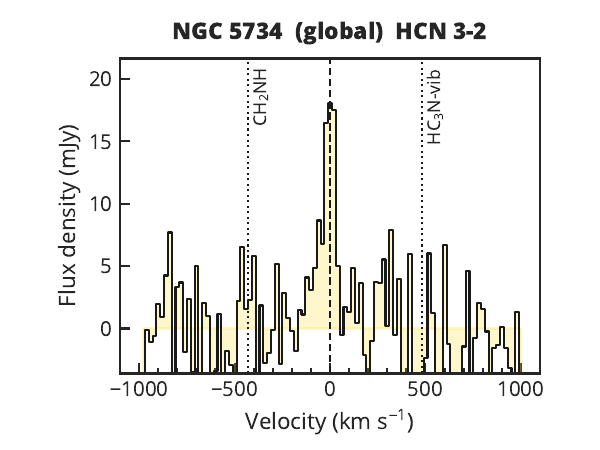}
\includegraphics[width=0.6\hsize]{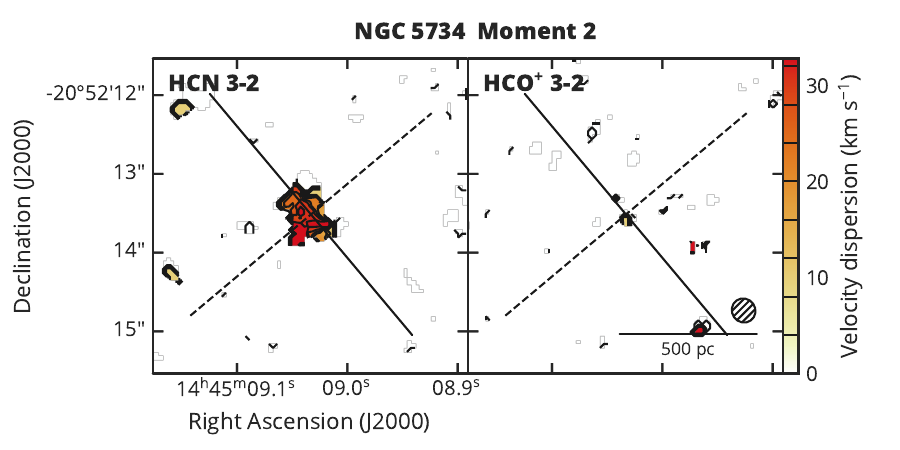}

\includegraphics[width=0.4\hsize]{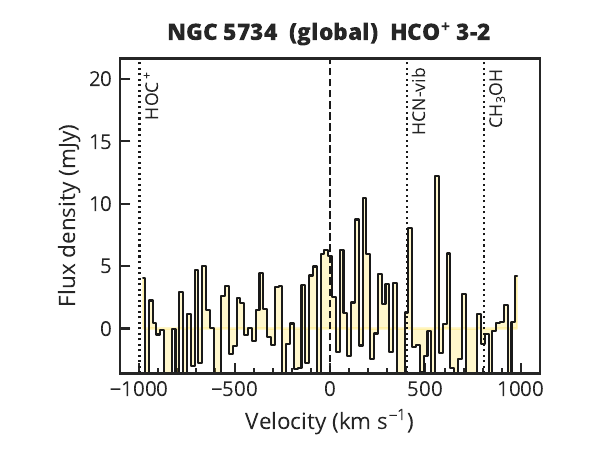}
\includegraphics[width=0.4\hsize]{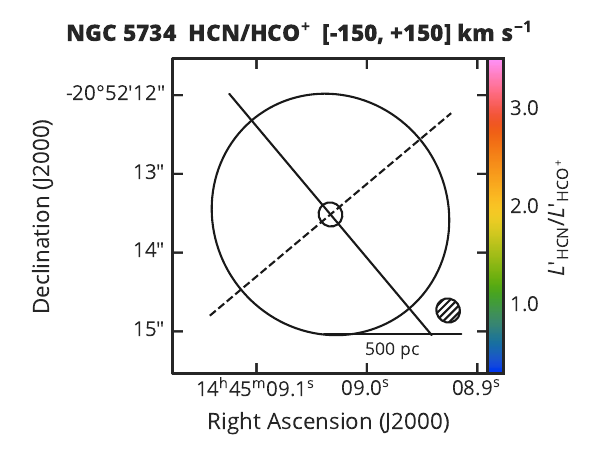}
\caption{HCN 3--2 and HCO$^+$ 3--2 for NGC\,5734. 
\emph{Left panels}: (\emph{Top two panels}) 
Spectra extracted from the resolved aperture. 
(\emph{Lower two panels}) Spectra extracted from the global aperture. 
Velocities are relative to the systemic velocity. 
The corresponding velocities of potentially detected species 
are indicated by vertical dotted lines.
\emph{Right panels}: 
(\emph{Top}) Integrated intensity over $\pm150$ km s$^{-1}$ (moment 0). 
Contours are (5, 10) $\times\sigma$, 
where $\sigma$ is 0.026 Jy km s$^{-1}$ beam$^{-1}$. 
(\emph{Second from top}) Velocity field (moment 1). 
Contours are in steps of $\pm10$ km s$^{-1}$. 
(\emph{Third from top}) Velocity dispersion (moment 2). 
Contours are in steps of 4 km s$^{-1}$. 
Moment 1 and 2 were derived with $3\sigma$ clipping. 
(\emph{Bottom}) $L'_\mathrm{HCN}/L'_\mathrm{HCO^+}$. 
Color scale is from 0.285 to 3.5. 
Overlaid ellipses represent the apertures used for spectral extraction. 
Solid and dashed lines represent the kinematic major and minor axes, respectively. 
The synthesized beam is indicated by hatched ellipses in the lower right corners.
}
\label{figure:5734}
\end{figure*}

In Figs.~\ref{figure:17208-0014}--\ref{figure:5734}, 
we present the basic figures, 
the resolved and global spectra and moment maps 
of the HCN 3--2 and HCO$^+$ 3--2 lines, 
as well as the maps of the velocity-integrated line ratio. 

\section{Line ratio in the ppV space}
\label{appendix:ppV}

In Sect.~\ref{subsect:resolved}, we discussed 
the spatially and kinematically symmetric structures 
characterized by $L'_\mathrm{HCN}/L'_\mathrm{HCO^+}$ 
exceeding the 90th percentile in each galaxy. 
This appendix provides a visualization of $L'_\mathrm{HCN}/L'_\mathrm{HCO^+}$ 
in the ppV space for all galaxies in which both HCN and HCO$^+$ are significantly detected, namely, nearly the whole sample 
except for UGC\,11763, UGC\,2982, and NGC\,5734. 

\begin{figure*}
\includegraphics[width=0.247\hsize]{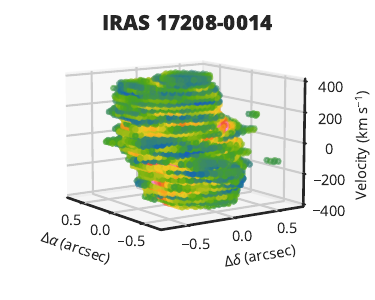}
\includegraphics[width=0.247\hsize]{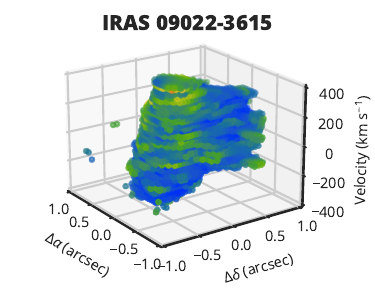}
\includegraphics[width=0.247\hsize]{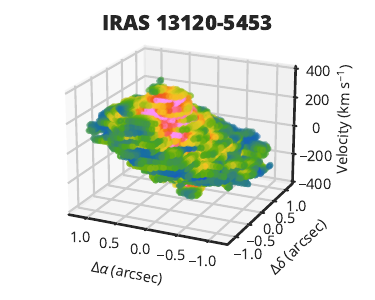}
\includegraphics[width=0.247\hsize]{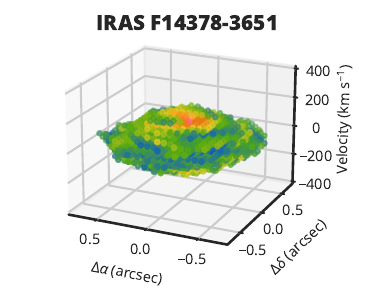}

\includegraphics[width=0.247\hsize]{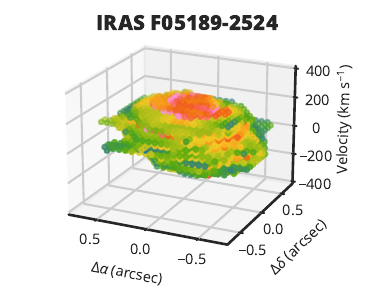}
\includegraphics[width=0.247\hsize]{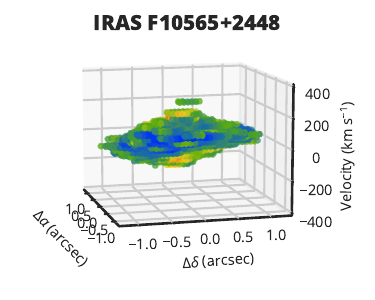}
\includegraphics[width=0.247\hsize]{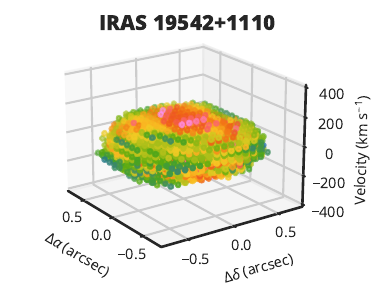}
\includegraphics[width=0.247\hsize]{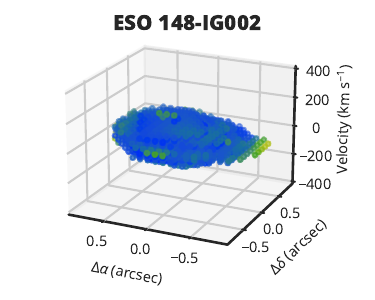}

\includegraphics[width=0.247\hsize]{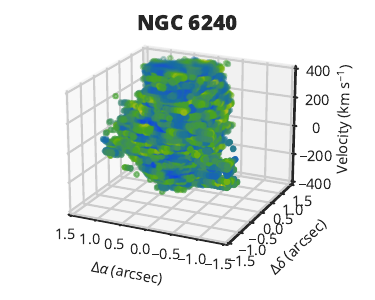}
\includegraphics[width=0.247\hsize]{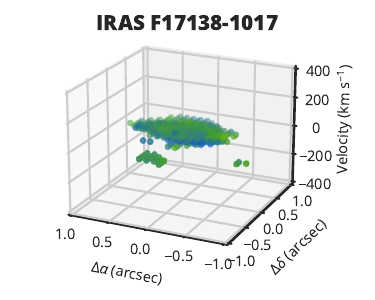}
\includegraphics[width=0.247\hsize]{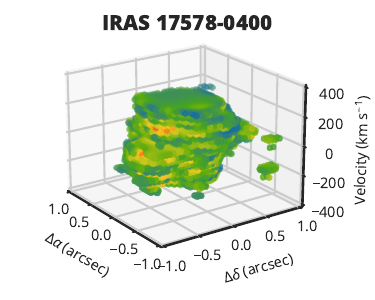}
\includegraphics[width=0.247\hsize]{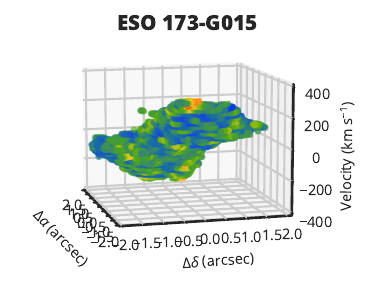}

\includegraphics[width=0.247\hsize]{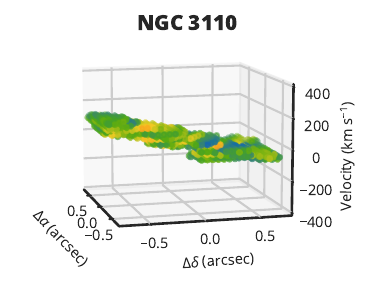}
\includegraphics[width=0.247\hsize]{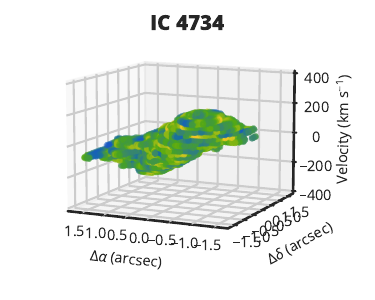}
\includegraphics[width=0.247\hsize]{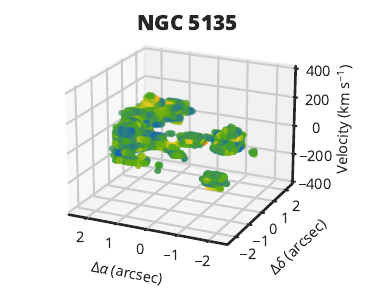}
\includegraphics[width=0.247\hsize]{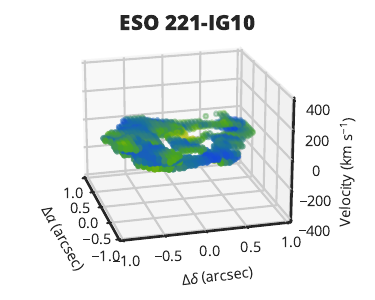}

\includegraphics[width=0.247\hsize]{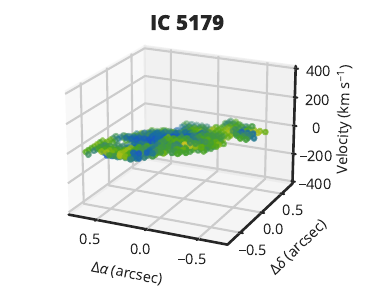}
\includegraphics[width=0.247\hsize]{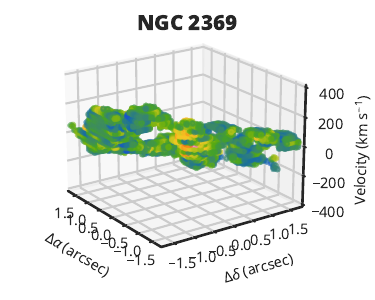}
\includegraphics[width=0.247\hsize]{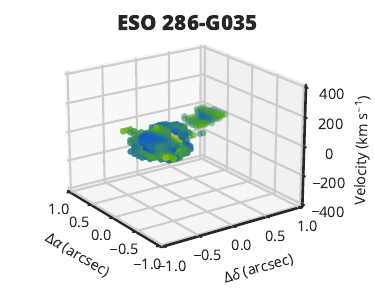}
\includegraphics[width=0.247\hsize]{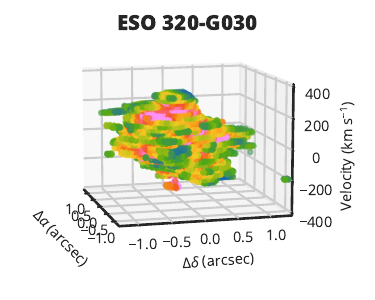}

\includegraphics[width=0.990\hsize]{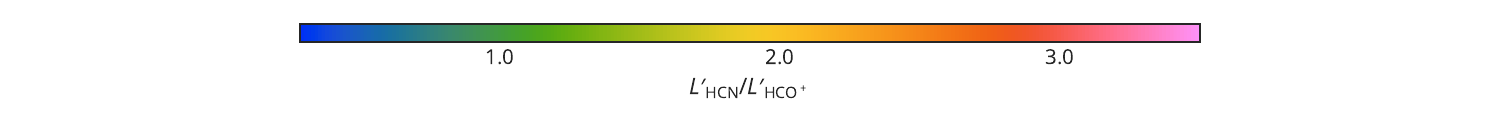}
\caption{Visualization of $L'_\mathrm{HCN}/L'_\mathrm{HCO^+}$ 
in the position-position-velocity space. 
The whole sample except for UGC\,11763, UGC\,2982, and NGC\,5734 
are presented. Each spaxel has a size of 
$0.05\arcsec\times0.05\arcsec\times20$ km s$^{-1}$. 
A 3$\sigma$ threshold clipping for both HCN and HCO$^+$ was applied. 
We only consider spaxels with the global aperture 
(see Sect.~\ref{subsect:spectra}) and with the velocity range 
listed in Table \ref{subsect:luminosities}. 
The colors represent $L'_\mathrm{HCN}/L'_\mathrm{HCO^+}$ 
in the range from 0.285 (blueish) to 3.5 (red-pinkish).}
\label{figure:ppV_whole}
\end{figure*}

Figure~\ref{figure:ppV_whole} presents the ppV plots for all spaxels 
where both HCN and HCO$^+$ are detected with greater than $3\sigma$ significance. 
Based on this figure, the sample appears to roughly split into two groups: 
one comprising cases where the velocity dispersion equals or exceeds 
any velocity gradient and another where the gas forms 
a low dispersion sheet exhibiting a velocity gradient. 
The former is predominantly found in merger cases, 
whereas the latter is characteristic of non-mergers 
(See Table~\ref{table:sample} for merger stages). 

\begin{figure*}
\includegraphics[width=0.247\hsize]{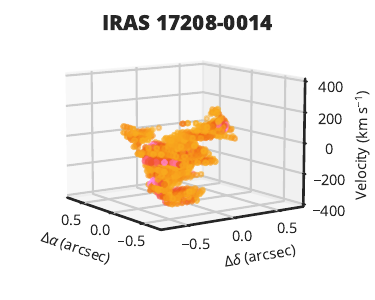}
\includegraphics[width=0.247\hsize]{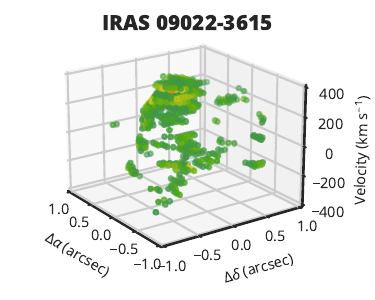}
\includegraphics[width=0.247\hsize]{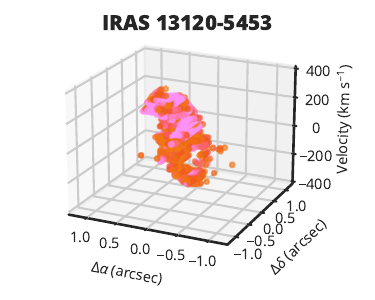}
\includegraphics[width=0.247\hsize]{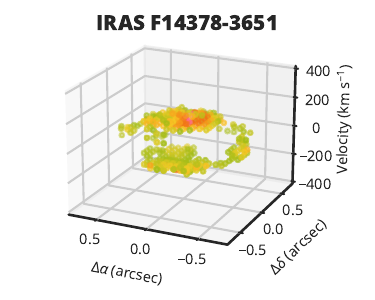}

\includegraphics[width=0.247\hsize]{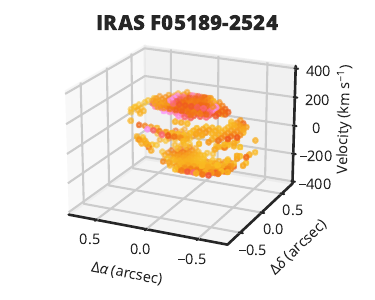}
\includegraphics[width=0.247\hsize]{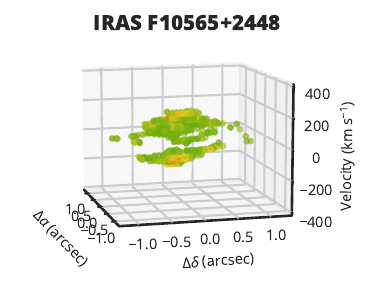}
\includegraphics[width=0.247\hsize]{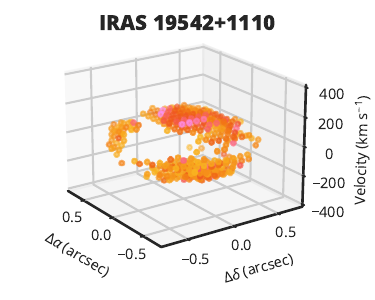}
\includegraphics[width=0.247\hsize]{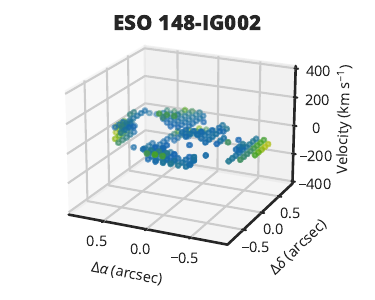}

\includegraphics[width=0.247\hsize]{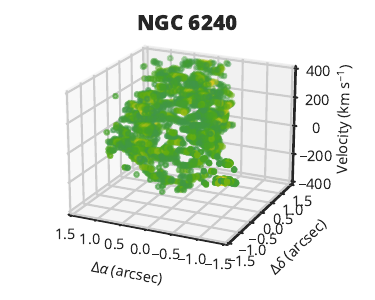}
\includegraphics[width=0.247\hsize]{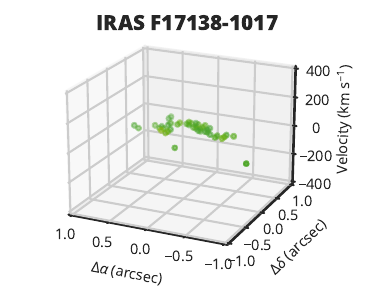}
\includegraphics[width=0.247\hsize]{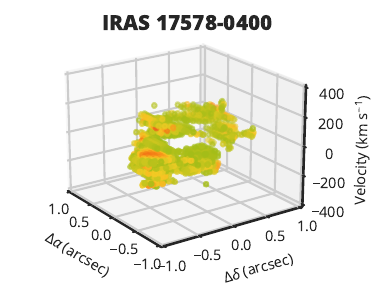}
\includegraphics[width=0.247\hsize]{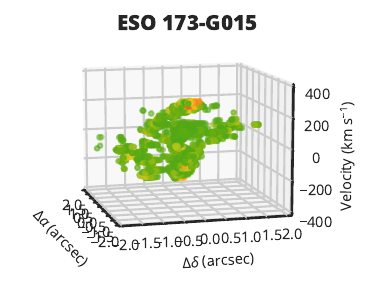}

\includegraphics[width=0.247\hsize]{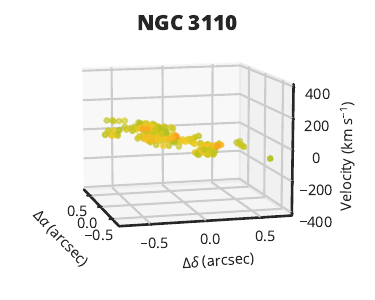}
\includegraphics[width=0.247\hsize]{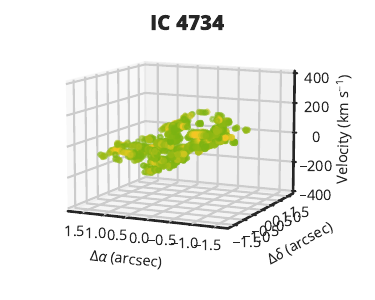}
\includegraphics[width=0.247\hsize]{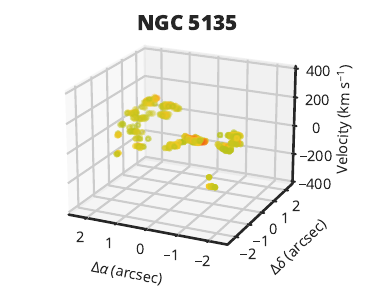}
\includegraphics[width=0.247\hsize]{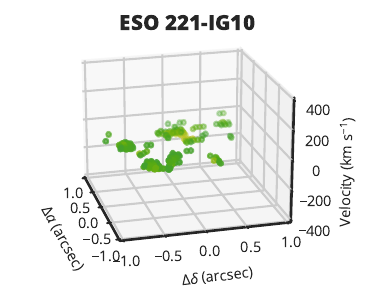}

\includegraphics[width=0.247\hsize]{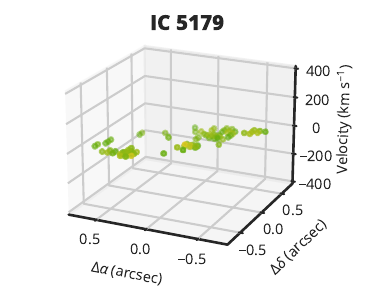}
\includegraphics[width=0.247\hsize]{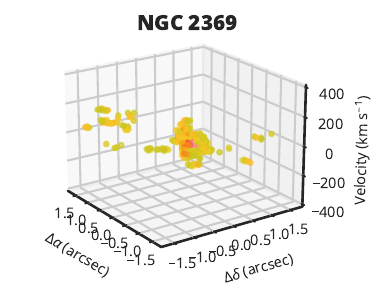}
\includegraphics[width=0.247\hsize]{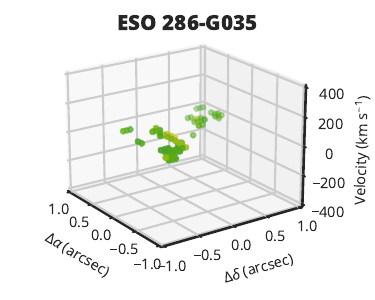}
\includegraphics[width=0.247\hsize]{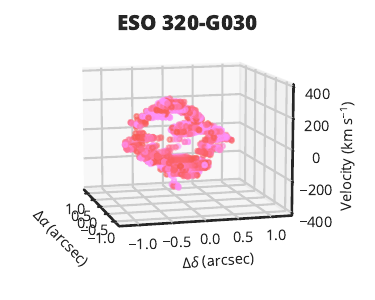}

\includegraphics[width=0.990\hsize]{colorscale.pdf}
\caption{Same as Fig.~\ref{figure:ppV_whole} but presenting spaxels 
with $L'_\mathrm{HCN}/L'_\mathrm{HCO^+}$ above the 90th percentile. }
\label{figure:ppV_90th}
\end{figure*}

Figure~\ref{figure:ppV_90th} presents the ppV plots from the same 
angle as Fig.~\ref{figure:ppV_whole} but includes only the spaxels 
with the line ratios higher than the 90th percentile. 
As discussed in the Sect.~\ref{subsect:resolved}, 
the spatially and kinematically symmetric structures 
appear as a filled bicone or a thin spherical shell 
in the galaxies with molecular out- and/or inflows 
previously found by CO and/or OH line observations. 
Some other galaxies also (marginally) exhibit symmetric structures, 
possibly suggesting the presence of out- or inflows.
(see Appendix~\ref{appendix:notes} for details). 

\section{Notes on individual galaxies}
\label{appendix:notes}

We briefly summarize previous reports of 
out- and inflows and discuss how they are related to 
the high $L'_\mathrm{HCN}/L'_\mathrm{HCO^+}$ regions 
(Fig.~\ref{figure:ppV_90th}) in individual galaxies: 

\begin{enumerate}
\item{\emph{IRAS\,17208$-$0014:}} 
\citet{Garcia-Burillo2015} identified molecular outflows 
by interferometric CO 2--1 line observations 
and subsequent kinematic modeling. 
According to them, the CO outflow is likely 
to be a collimated conical shape 
elongating in the SE(blueshifted)-NW(redshifted) direction 
(see Fig.~18 of \citealt{Garcia-Burillo2015}). In our analysis, 
the high $L'_\mathrm{HCN}/L'_\mathrm{HCO^+}$ structure 
appears to be a filled biconical shape whose direction of elongation 
is consistent with the CO results \citep{Garcia-Burillo2015}. 
Inflows, as indicated by the redshifted OH absorption feature 
\citep[$+$51\,km\,s$^{-1}$][]{Veilleux2013}, could also contribute to 
elevating $L'_\mathrm{HCN}/L'_\mathrm{HCO^+}$ near that velocity. 

\item{\emph{IRAS\,09022$-$3615:}} 
The presence of outflows in this galaxy is suggested 
by the OH absorption line. The OH 119\,$\mu$m absorption is shallow 
($\sim$10\% relative to the continuum), possibly indicating that 
the outflow is significantly collimated \citep{Gonzalez-Alfonso2017}. 
The high $L'_\mathrm{HCN}/L'_\mathrm{HCO^+}$ structure ambiguously 
appears to be a spherical shell, but its morphology seems asymmetric. 
It is uncertain if HCN enhancement is associated with the OH outflow. 

\item{\emph{IRAS\,13120$-$5453:}} 
In this galaxy, outflows have been found using the CO emission 
\citep{Fluetsch2019,Lutz2020} and the OH absorption 
\citep{Veilleux2013,Gonzalez-Alfonso2017}. 
The outflow parameters derived from these observations 
moderately disagree, probably due to limited signal-to-noise ratios, 
adopted morphology of the outflow, 
and excitation conditions for each molecular transition. 
In addition, \citet{Privon2017} reports line wings 
in the HCN 4--3 line as evidence for a high-velocity outflow. 
In our $L'_\mathrm{HCN}/L'_\mathrm{HCO^+}$ plots, 
enhancement of $L'_\mathrm{HCN}/L'_\mathrm{HCO^+}$ is seen 
in the central region at a wide range of velocities 
(roughly spanning [$-300$, $+300$]\,km\,s$^{-1}$), 
possibly in the form of a collimated and nearly pole-on outflow. 
The high line ratio structure shows velocity reversals, 
which may suggest the outflow is precessing. 
We did not see the high-velocity components ($>$300\,km\,s$^{-1}$) 
detected in the OH absorption, possibly due to the faintness 
of the HCN and HCO$^+$ emission. 

\item{\emph{IRAS\,F14378$-$3651:}} 
Molecular outflows have been detected through the blueshifted absorption feature 
in the far-infrared OH lines \citep{Veilleux2013,Gonzalez-Alfonso2017}. 
According to the model by \citet{Gonzalez-Alfonso2017}, the OH outflows 
are at velocities of 250\,km\,s$^{-1}$ and 600\,km\,s$^{-1}$, 
located 200--300\,pc away from the center. As for the CO observations, 
\citet{Lutz2020} found a slight flux excess at $\sim$250\,km\,s$^{-1}$ in 
the 1--0 transition, although it is not significant enough to be robust 
evidence for outflows. Our $L'_\mathrm{HCN}/L'_\mathrm{HCO^+}$ images 
show an enhancement in a wide-angle thin shell with a $\sim$250\,pc radius, 
with the maximum velocity of $\sim$200\,km\,s$^{-1}$ (uncorrected for 
inclination). The location of the HCN enhancement seems to be associated 
with the lower-velocity OH outflow. On the other hand, we did not 
see emission from the higher-velocity (600\,km\,s$^{-1}$) outflow 
in our data. To detect the higher-velocity component, higher sensitivity may be required. 

\item{\emph{IRAS\,F05189$-$2524:}} 
Similar to the case of IRAS\,F14378$-$3651, IRAS\,F05189$-$2524 shows 
an enhancement of $L'_\mathrm{HCN}/L'_\mathrm{HCO^+}$, 
in the form of a thin spherical shell in the ppV space. 
In terms of velocity, the HCN-enhanced shell agrees 
with the lower-velocity OH outflow at 200\,km\,s$^{-1}$ 
reported by \citet{Gonzalez-Alfonso2017}. 
The presence of the higher-velocity ($\gtrsim$500\,km\,s$^{-1}$) outflows is 
also known by the OH absorption \citet{Gonzalez-Alfonso2017} and 
the CO 3--2 emission \citep{Fluetsch2019}. In our observations, 
the line profiles of HCN and HCO$^+$ emission show small bumps 
near the velocity of the fast outflow (Figs.~\ref{figure:F05189-2524}). 
However, the morphology of bumps are asymmetric and are likely to be 
affected by emission from other molecular species. 
Thus, the high-velocity component is not confirmed in our data. 

\item{\emph{IRAS\,F10565$+$2448:}} 
The high $L'_\mathrm{HCN}/L'_\mathrm{HCO^+}$ regions appear to be 
a thin spherical shell with an ambiguous filled bicone. 
The highest $L'_\mathrm{HCN}/L'_\mathrm{HCO^+}$ is seen at velocities 
of $\sim$250\,km\,s$^{-1}$, which is consistent with the outflow velocity 
measured by OH absorption \citep{Veilleux2013,Gonzalez-Alfonso2017}. 
While CO line wings in the velocity range of 300--600\,km\,s$^{-1}$ 
indicate the presence of faster components \citep{Cicone2014}, 
we did not detect emission of HCN or HCO$^+$ in that velocity range. 
Higher sensitivity observations would be necessary 
to trace the fast components.
 
\item{\emph{IRAS\,19542$+$1110:}} 
Molecular outflows in this galaxy have been found 
through a blueshifted OH absorption feature at the velocity of 
$-$93\,km\,s$^{-1}$ \citep{Veilleux2013}. In our analysis, 
$L'_\mathrm{HCN}/L'_\mathrm{HCO^+}$ appears to be high in a thin 
spherical shell with a radius of $\sim$500\,pc and 
in a velocity range of about [$-$200, 200]\,km\,s$^{-1}$. 
Little is known about the morphology of the outflow so far, 
but the structure of $L'_\mathrm{HCN}/L'_\mathrm{HCO^+}$ suggests 
that the outflow shape could be a thin shell with a wide opening angle. 

\item{\emph{ESO\,148$-$IG002:}} 
Our data show that $L'_\mathrm{HCN}/L'_\mathrm{HCO^+}$ is 
relatively low throughout the galaxy 
($\sim$0.4 on average and $\sim$1 at the maximum; 
see Table \ref{table:ratio}). Although no molecular out- or inflows 
are known in this galaxy so far, the highest 10th percentile 
in $L'_\mathrm{HCN}/L'_\mathrm{HCO^+}$ is 
in the form of a thin spherical shell with a radius of $\sim$500\,pc 
and in a velocity range of about $\pm$20\,km\,s$^{-1}$, 
possibly suggesting the presence of a symmetrically shaped 
non-circular motion. 

\item{\emph{NGC\,6240:}} 
Due to the interaction between the two nuclei, 
the kinematic structure of NGC\,6240 appears to be quite complicated. 
The presence of outflows launched from each of the two nuclei 
was revealed by a detailed analysis of CO and [\ion{C}{i}] 
lines \citep{Cicone2014,Cicone2018,Saito2018} and 
an OH absorption feature \citep{Veilleux2013}. 
The high $L'_\mathrm{HCN}/L'_\mathrm{HCO^+}$ regions are, 
however, rather randomly distributed over the entire system. 
We think this could be because of the disturbance by the merger. 

The CO and H$_2$ line analysis \citep{Meijerink2013} 
revealed that a large fraction of the gas traced by 
the CO $J$=4--3 through 13--12 lines is affected by slow shocks, 
while only $\lesssim$1\% traced by the H$_2$ $v$$=$1--0 $S$(1) 
and $v$$=$2--1 $S$(1) lines is exposed to the high-velocity shocks 
($\sim$17--47\,km\,s$^{-1}$). Hence the shock velocity 
is slow ($\lesssim$10\,km\,s$^{-1}$) in the greater part of the system, 
no matter whether the shock is caused by a merger and/or the outflows. 
The shock velocity of $\lesssim$10\,km\,s$^{-1}$ is insufficient 
for HCN enhancement (see Sect.~\ref{subsect:chemistry} 
for dependencies of HCN enhancement on shock velocity). 

\item{\emph{IRAS\,F17138$-$1017:}} 
No dedicated study of the molecular gas distribution 
and kinematics was found for this galaxy. 
The spatial distribution of HCO$^+$ and HCN appears to be 
highly asymmetric and different between these two species. 
While the HCO$^+$ emission has a second peak 
$\sim$700\,pc to the south of the kinematic center, 
the HCN emission does not. Our results suggest that 
there are multiple components that have different chemical 
compositions in the system, although we cannot specify 
the physical origin of the chemical diversity. 

\item{\emph{IRAS\,17578$-$0400:}} 
For this galaxy, the analysis of the OH line profile 
suggests the presence of inflowing gas 
at $\sim$30\,km\,s$^{-1}$ \citepalias{Falstad2021}. 
The elongation of the HCN and HCO$^+$ emission at low velocity 
may hint at an outflow along its kinematic minor axis 
(\citetalias{Falstad2021}; \citealt{Yang2023}), 
as mentioned in Sect.~\ref{subsect:spectra}. 
Because of the difficulty in continuum subtraction 
due to the line forest, it is hard to quantitatively discuss 
$L'_\mathrm{HCN}/L'_\mathrm{HCO^+}$ on a spaxel-by-spaxel basis. 
We tentatively analyzed $L'_\mathrm{HCN}/L'_\mathrm{HCO^+}$ 
without continuum subtraction, but it is difficult to robustly 
conclude that the high $L'_\mathrm{HCN}/L'_\mathrm{HCO^+}$ regions 
are related to the out- and/or inflows. 
To further investigate the kinematics of IRAS\,17578$-$0400, 
we have conducted higher spatial resolution observations with ALMA. 
The detailed analysis of this observation 
will be presented in another paper (Yang et al.,~in prep.). 

\item{\emph{ESO\,173$-$G015:}} 
Although we did not find any publication on 
the presence of out- or inflows in this galaxy, 
the high $L'_\mathrm{HCN}/L'_\mathrm{HCO^+}$ regions 
were found in a symmetric shape. 
If we pick up the spaxels with ratios 
higher than the 97th percentile ($>$1.402), 
they appear more distinctly in a thin shell. 
Conducting a more detailed kinematic analysis would be worthwhile 
in order to see if there are any outflows.

\item{\emph{UGC\,11763:}} 
No publication on outflow signatures was found for this galaxy. 
The HCN and HCO$^+$ emission is too faint and compact 
to inspect the line ratio on a spaxel-by-spaxel basis. 
The linewidths of the HCN and HCO$^+$ lines seem to be 
much narrower than that of the CO 2--1 line 
\citep[e.g.,][the galaxy is referred to as PG2130+099]%
{MontoyaArroyave2023}, suggesting that higher sensitivity 
and higher spatial resolution would be needed to 
grasp the molecular composition in the extended diffuse gas. 

\item{\emph{NGC\,3110:}} 
The presence of a large-scale ($\gtrsim$ a few kiloparsecs) inflow is 
suggested by CO line observations \citep{Kawana2022}, 
but it is on much larger scales and not comparable 
to the HCN and HCO$^+$ line-emitting region of our data. 
We note that $L'_\mathrm{HCN}/L'_\mathrm{HCO^+}$ shows some variation 
in the ppV space, but its structure is not very symmetrical. 

\item{\emph{IC\,4734:}} 
We did not find any published out- or inflow signatures. 
$L'_\mathrm{HCN}/L'_\mathrm{HCO^+}$ is distributed rather uniformly 
throughout the galaxy. 

\item{\emph{NGC\,5135:}} 
There are no known molecular out- or inflows in this galaxy. 
High $L'_\mathrm{HCN}/L'_\mathrm{HCO^+}$ can be seen 
in some limited small regions, but it is hard to characterize their geometry. 

\item{\emph{ESO\,221$-$IG10:}} 
No out- or inflow signature is known for this galaxy. 
A slight enhancement of $L'_\mathrm{HCN}/L'_\mathrm{HCO^+}$ can be seen 
in the central region, but it is only present on the blueshifted side. 
Higher sensitivity observations and detailed analysis 
of kinematics would be advisable. 

\item{\emph{IC\,5179:}} 
For this galaxy, we found no published signature 
of out- or inflows. There are some spots where 
$L'_\mathrm{HCN}/L'_\mathrm{HCO^+}$ is 
elevated, but they appear rather random. 

\item{\emph{UGC\,2982:}} 
No publication on an out- or inflow signature was found. 
The HCN and HCO$^+$ emission is too faint 
to make any meaningful discussion. 

\item{\emph{NGC\,2369:}} 
Although there is no published out- or inflow signature, 
the high $L'_\mathrm{HCN}/L'_\mathrm{HCO^+}$ regions are 
concentrated in a few 100\,pc regions at the center of the galaxy. 
It would be worth investigating the origin of this concentration. 

\item{\emph{ESO\,286$-$G035:}} 
We did not find any publication on outflows in this galaxy. 
Moreover, $L'_\mathrm{HCN}/L'_\mathrm{HCO^+}$ does not vary much
and is almost uniformly distributed throughout the galaxy. 

\item{\emph{ESO\,320$-$G030:}} 
In this galaxy, a high-velocity (450\,km\,s$^{-1}$) 
molecular outflow was identified 
by high spatial resolution observation of 
the CO 2--1 line \citep{Pereira-Santaella2016}. 
The kinematic analysis of the cited study showed that the outflow has a size 
of 2.5\,kpc and that its blue- and redshifted regions lie 
in the northeast and southwest of the nucleus, respectively. 
Furthermore, \citet{Gonzalez-Alfonso2021} found 
a molecular inflow associated with the nuclear bar, 
which is traced by the CO 2--1 line at the radius of 230--460\,pc. 
They also pointed out that OH absorption features 
indicate the presence of inflowing gas at radii of 100--150\,pc. 
We note that $L'_\mathrm{HCN}/L'_\mathrm{HCO^+}$ largely varies 
over the galaxy in the form of a characteristic structure. 
Most notably, the high $L'_\mathrm{HCN}/L'_\mathrm{HCO^+}$ 
regions shape a thin spherical shell with a radius of $\sim$100\,pc. 
The velocity of the structure is roughly consistent 
with the inflows traced by the OH lines, indicating 
that HCN enhancement could be associated with these inflows. 
On the other hand, the high-velocity outflow 
found by \citet{Pereira-Santaella2016} is not obvious 
in $L'_\mathrm{HCN}/L'_\mathrm{HCO^+}$ 
in a spectrally resolved manner, 
but this is simply because we cannot calculate 
the ratio due to the faintness of the HCO$^+$ line 
at high velocities. In the velocity-integrated ratio map, 
we may have got a hint of the enhanced 
$L'_\mathrm{HCN}/L'_\mathrm{HCO^+}$ in the outflow 
(Fig.~\ref{figure:320-G030}). 
The northeastern part of the outflow exhibits 
a considerably high $L'_\mathrm{HCN}/L'_\mathrm{HCO^+}$ 
($\gtrsim$3), and the southwestern part also shows 
a moderately enhanced line ratio. 
Our subsequent ALMA observation with higher sensitivity 
and higher spatial resolution has clearly revealed 
the signature of outflow in the HCN line, 
and it will be presented in another paper (Gorski in prep.). 

\item{\emph{NGC\,5734:}} 
No published out- or inflow signature was found. 
We cannot discuss $L'_\mathrm{HCN}/L'_\mathrm{HCO^+}$ 
due to the faintness of the HCO$^+$ emission over the galaxy. 

\end{enumerate}

\end{appendix}

\end{document}